\documentclass[pdflatex,sn-vancouver-num]{sn-jnl}% Vancouver Numbered Reference Style
\usepackage{graphicx}%
\usepackage{multirow}%
\usepackage{amsmath,amssymb,amsfonts}%
\usepackage{amsthm}%
\usepackage{mathrsfs}%
\usepackage[title]{appendix}%
\usepackage{xcolor}%
\usepackage{textcomp}%
\usepackage{manyfoot}%
\usepackage{booktabs}%
\usepackage{algorithm}%
\usepackage{lmodern}%
\usepackage{algorithmicx}%
\usepackage{algpseudocode}%
\usepackage{listings}%
\usepackage{lmodern}
\usepackage{bbm} % bold math
\usepackage{bm}
\usepackage{mathtools}
\usepackage{dsfont} % identity operator
\usepackage{braket}
\usepackage{cancel}
\usepackage{slashed}
\usepackage{geometry}

\usepackage{ragged2e}
\usepackage{array}
\usepackage{tabularx}
\usepackage{longtable} % for tables that span multiple pages
\usepackage{booktabs}
\usepackage[most]{tcolorbox}

\usepackage{tikz}
\usetikzlibrary{decorations}
\usetikzlibrary{quantikz2}
\usepackage{pict2e,picture}
\usepackage{xcolor}

\newcommand{\ii}{\text{i}}

\newcommand{\eul}{\operatorname{e}}
\newcommand{\im}{\operatorname{i}}

\definecolor{mygreen}{RGB}{33, 128, 16}

\begin{document}

\title[Advanced Quantum Communication and Quantum Networks]{Advanced Quantum Communication and Quantum Networks}
\subtitle{From basic research to future applications}

\author[1]{\fnm{Bj\"orn} \sur{Kubala}}

\author[1]{\fnm{Alexander} \sur{Sauer}}

\author[1]{\fnm{Alessandro} \sur{Tarantola}}

\author[3]{\fnm{David} \sur{Fabian}}

\author[3]{\fnm{Anke} \sur{Ginter}}

\author[3]{\fnm{Olga} \sur{Kulikovska}}

\author[2]{\fnm{Fabio} \sur{Di Pumpo}}

\author[2]{\fnm{Johannes} \sur{Seiler}}

\author[2]{\fnm{Wolfgang P.} \sur{Schleich}}

\author*[1]{\fnm{Matthias} \sur{Zimmermann}}\email{matthias.zimmermann@dlr.de}

\affil[1]{\orgname{German Aerospace Center (DLR)}, \orgdiv{Institute of Quantum Technologies},  \orgaddress{\street{Wilhelm-Runge-Straße 10}, \postcode{89081} \city{Ulm}, \country{Germany}}}

\affil[2]{\orgdiv{Institut f{\"u}r Quantenphysik and Center for Integrated Quantum Science and Technology (IQ$^{\textrm{ST}}$)}, \orgname{Universit{\"a}t Ulm}, \orgaddress{\street{Albert-Einstein-Allee 11}, \postcode{89081} \city{Ulm},  \country{Germany}}}

\affil[3]{\orgname{Bundesdruckerei GmbH}, \orgaddress{\street{Kommandantenstr. 18},  \postcode{10969} \city{Berlin}, \country{Germany}}}

\abstract{Classical communication is the basis for many of our current and future technologies, such as mobile phones, video conferences, autonomous vehicles and particularly the internet. In contrast, quantum communication is governed by the laws of quantum mechanics. Due to this fundamental difference, it might offer enormous benefits for security applications,  more precise measurements, faster computations, and many other fields of application by interconnecting different quantum devices, such as quantum sensors, quantum computers, or quantum memories. This review provides an overview of the specific properties of quantum information networks. This includes the interfaces between the classical and the quantum regime, the transmission of the quantum information by physical implementations, and potential future applications of quantum networks. We aim to provide a starting point based on fundamental concepts of quantum information processing for further research on a future quantum internet.}


\maketitle

\pagebreak

\tableofcontents

\pagebreak

\section{Introduction} \label{sec:Intro}

\subsection{Quantum information networks} \label{ssec:Motivation} 

\subsubsection{Motivation}

Between the 1900s and 1980s, the first quantum revolution took place. Building on an improved quantum-mechanical understanding of atomic levels and the structure of matter, it paved the way for modern electronics and optics (including semiconductors, lasers, optical fibers, and more). Today, extensive experience with superpositions, entanglement, and other peculiar quantum effects is fueling a second quantum revolution \cite{Dowling.2003}, which will affect data acquisition (\emph{quantum sensing}), processing (\emph{quantum computing}) \cite{Bongs_PhiuZ_2025} and communication (\emph{quantum communication}). 

\emph{Quantum sensing} exploits unique quantum effects to produce innovative measurement devices with unprecedented sensitivity and accuracy \cite{Aslam2023}. The field has already deeply impacted metrology and has raised interest in the scientific community. Nevertheless, in this report, we focus on quantum communication and its intersection with quantum computers, and thus we will not discuss quantum sensing further in the following.

\emph{Quantum computing} has already gained fame because of its predicted ability to solve some specific problems exponentially faster than it is classically possible. Among these problems is prime factorization, cf.~\cite{Shor.1994, Shor.1997}, whose computational hardness underlies the security of the widely used RSA encryption scheme. Large enough quantum computers thus represent a looming threat to internet security\footnote{A recent report on the topic by the German Federal Office of Information Security can be found \href{https://www.bsi.bund.de/SharedDocs/Downloads/DE/BSI/Publikationen/Studien/Quantencomputer/Entwicklungstand_QC_V_2_1.html}{here}}. In addition, quantum computers also hold promise for groundbreaking improvements in a wide range of applications, from simulations in quantum chemistry to optimization, finance, and machine learning \cite{Dalzell_2025}. Even if today's available quantum computing devices are still far from offering computational resources for implementation \cite{Preskill_2018}, their disruptive potential is already foreseeable.

\emph{Quantum communication} represents various ways of coherently connecting distant quantum devices and opens the door to novel security concepts rooted in the very laws of quantum mechanics. Among them is \emph{quantum-secured classical communication}, embodied by protocols such as \emph{quantum key distribution} (QKD, \cite{Bennett.2014}) which enable safe exchange of classical, but crucially not quantum, information. While interesting in its own right, QKD is mainly discussed in this document to outline the current security context, and because it relies on quantum effects and hardware similar to those of genuinely quantum protocols. The main focus of this report remains on schemes based on the exchange of quantum information and their impact on future quantum communication.

When more than two parties are involved, quantum communication spills over to the discipline of \emph{quantum networks}. In this realm, unique applications with no classical counterpart, like money that cannot be forged and elections that cannot be rigged, become conceivable.

Quantum networks are approaching technological maturity, and a lot is known about each of their individual components. Yet, many questions still surround the interplay between them and with the existing classical infrastructure. This document aims at bridging this knowledge gap with a holistic approach to the topic that spans from foundational enabling technologies to the highest-level future applications. More precisely, we will try to: Provide an understanding of individual quantum network components and their interplay; Put quantum networks in the context of the existing classical infrastructure and current security landscape; Explore possible future applications while highlighting their strengths, limitations, and the challenges to their implementation.

\subsubsection{Overview of the current state of technologies} \label{sssec:OverviewAndSOTA}

The theoretical study of quantum networks is already quite advanced. The research activity has progressively changed from the discussion of few-node networks to the vision of a global quantum network, the \emph{quantum internet}. A lot of attention is being devoted to the design of such large networks and their components (repeaters \cite{Azuma.2023, Munro.2015}, entanglement distribution protocols \cite{Cirac.1997}, routing \cite{Hussein.2022}, etc.). A particular focus was placed on the development and implementation of security aspects as in QKD, cf.~\cite{Bennett.2014, Zapatero.2023} for quantum-secured classical communication, quantum key cards and tokens \cite{Pastawski2012}, various quantum authentication schemes, secure information retrieval from a database (``oblivious transfer"), quantum elections, etc. Nowadays, networks and quantum computation are often considered in combination in the field of \emph{Distributed Quantum Computing} (DQC). A particular type of DQC is blind quantum computing, which aims at allowing users to deploy quantum calculations on quantum computing ``servers", without the server knowing either the outputs or the calculation methods. Further additions to the design of quantum networks can be found in the topics of remote and distributed quantum sensing \cite{Zhang.2021}. This list, which could be continued, is intended to give an impression of the diversity and complexity of the topic of quantum networks.

Experimentally, we can distinguish two main lines of research that run in parallel while at very different technical readiness levels. On the one hand, QKD has reached commercial maturity \cite{Pljonkin.2018}, and is in the process of being integrated into the current infrastructure to provide a form of \emph{quantum-secured classical communication}. Urban QKD network prototypes are already in place \cite{Chen.2021, Du.2024, Sasaki.2011, Stucki.2011}, and the conceptual preparation of the interconnection of multiple local networks through satellite communication is in progress \cite{Bongs_PhiuZ_2025}. However, the practical implementation of genuinely quantum networks is in a very early stage. Much of the current debate still revolves around the optimal physical realizations of single network components, such as memories, quantum channels, and repeaters. Even if most agree with the usage of photons for transfer or information, linking \emph{stationary} and \emph{flying} qubits \cite{DiVincenzo.1997} also poses some challenges. The heterogeneous stationary-qubit platforms couple differently in the electromagnetic spectrum. Often, the desired telecom wavelengths can only be realized with frequency conversion \cite{Kumar.1990}, which, however, is accompanied by an increase in inefficiency and loss. 

This wide range of research approaches and the design of components illustrates the high complexity of quantum communication in general. Due to technological limitations, quantum networks existed long in the speculations of theorists only: The degree of control required to coherently store, manipulate, and transmit quantum systems was unthinkable until recently. However, as technology progresses, quantum information and quantum networks are rapidly transitioning from a theoretical field to an experimental one, and on to some initial approaches in commercial areas. As new ideas and proofs of concepts for networks emerge frequently, often focused on specific functionalities, it is hard to imagine a more fruitful time to design and discuss future applications. 

\subsection{Scope and structure of this report} \label{ssec:Structure}

In this review, we embrace an interdisciplinary perspective on quantum communication and quantum networks with a specific focus on future applications. We do not provide a detailed and fundamental introduction to the topic, but rather a general and broader overview. The most important theoretical principles are briefly introduced to provide a basis for the consideration of future applications of quantum networks. Thus, based on the fundamental knowledge of quantum information processing and transfer, first concepts are developed, and steps towards an experimental realization are worked out. The aim of this report is to provide a starting point for future investigations of the applications of quantum networks for advanced future communication. 

We divide the topic of quantum information transfer into three main parts: The transitions from the classical to the quantum regime and back again, the transfer of the quantum states via quantum channels, and finally general concepts for future applications of quantum networks. The review is more precisely organized as follows. 

Sect.~\ref{sec:Interface} consists roughly of three parts. We begin with a review of common \emph{qubit-encoding} schemes in Sect.~\ref{ssec:Encoding}.
Those schemes tell us how to store classical information using quantum states.
The second part, Sect.~\ref{ssec:Decoding}, relates to retrieving the information encoded in a quantum state.
We present important no-go results, such as the \emph{Holevo bound}, which represent crucial boundaries to what can be achieved with present or future technology. The third part, namely Sect.~\ref{ssec:QSDC}, is devoted to the discussion of a first example of application: \emph{Quantum Secure Direct Communication}. A summary of the section and some final thoughts appear in Sect.~\ref{ssec:InterfaceConclusions}. 

In Sect.\,\ref{sec:PhysicalBasis}, we switch gears and turn our attention to the physical realization, storage, and transfer of quantum information. 
After a short introduction to the topic, Sect.~\ref{ssec:carrier} surveys which physical systems can be used at all to encode quantum information, either for storage and processing at network nodes or during the transfer of information between them. That transfer can occur across different media, each with its advantages and disadvantages, as explained in Sect.\,\ref{ssec:Channels}. Physical realizations suitable for the \emph{storage} of quantum information are often inefficient for its \emph{transfer}, whence the need for hybrid approaches (namely, involving several different physical platforms for quantum information) discussed in Sect.\,\ref{ssec:hybrid}. Present-day realizations of quantum network nodes and their connection to quantum channels are finally discussed in Sect.~\ref{ssec:Nodes}.  The rather general point of view adopted so far is then specialized in Sect.~\ref{ssec:realization}, providing an experimentally realized example of the full transfer of quantum information between two network nodes. Sect.~\ref{ssec:PhysicalBasisConclusions} wraps the discussion up with a summary and conclusions.

Current and future applications of quantum networks are presented in Sect.~\ref{sec:Applications}. We start, in Sect.~\ref{ssec:QInternetVision}, with the vision of what a global quantum network, the \emph{quantum internet}, could look like, what it would likely be used for, and what protocols it could enable. Sect.~\ref{ssec:SecurityAndDQC} moves away from the previous theoretical considerations to provide examples of concrete applications. More precisely, it discusses: \emph{Blind quantum computing}, a paradigm allowing users to perform secret computations on untrusted servers; \emph{Oblivious transfer}, a protocol for secure retrieval of information from a database; \emph{Quantum key cards}, physical devices for digital authentication purposes. The presentation of these three applications is preceded by an introduction to the tightly linked fields of (quantum) internet security and distributed quantum computing. Finally, Sect.~\ref{ssec:VotingTheoryAndImpl} moves further away from theory and towards experiment by focusing on possible technological implementations of the \emph{traveling ballot protocol}, a voting scheme whose security hinges on the unforgeability of quantum information. The conclusions are reported in Sect.~\ref{ssec:ApplicationConclusions}.

In the summary and outlook, Sect.~\ref{sec:Conclusion}, we review first implementations of advanced quantum communication protocols in Sect.~\ref{sec:FirstImplementations}, and provide a roadmap towards the quantum internet and other major technological milestones in the realm of quantum networks in Sect.~\ref{sec:Roadmap}. \pagebreak

\section{Interface between classical and quantum information} \label{sec:Interface}
\begin{figure}[hbt]
	\centering
	\includegraphics[width=\textwidth]{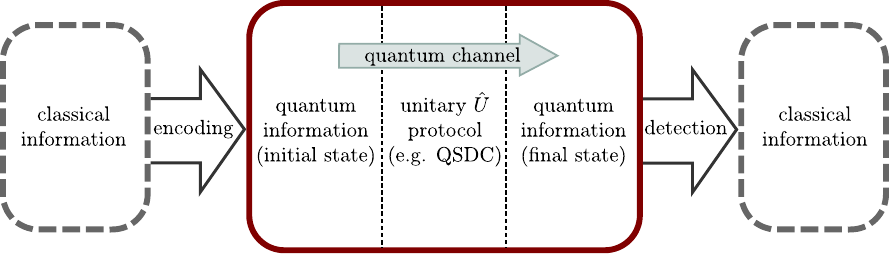}
	\caption{Scheme depicting the interface between classical and quantum information with purely quantum information processing and transfer via a quantum information channel. 
    Classical information has to be encoded first into quantum information. 
    To this end, various methods discussed in this section can be used and depend on the specific purpose.
    This encoded information then serves as initial state for some purely quantum unitary evolution $\hat{U}$ processing the encoded data.
    One example discussed in this section is quantum secure direct communication (QSDC).
    After processing, a final quantum state is achieved.
    Finally, this state is detected and classical information is recovered.}
	\label{fig:ClassQuantu}
    \end{figure}

Communication and information transfer are indispensable parts of our daily lives and therefore take place in the classical world. To make use of quantum systems for sending and processing information a transition from the classical regime to the quantum regime and vice versa is necessary (cf.~Fig.~\ref{fig:ClassQuantu}).
In this section, we analyze such interfaces between classical and quantum information. 
This includes the process of preparing a quantum state that encodes information as well as the final retrieval of the encoded information after the state has been transmitted and processed.
We examine the question of whether it is possible to use quantum states for more efficient information storage and transfer. Furthermore, we will take a closer look at the usage of quantum properties for advancements in security.

We first review how to encode classical information, which is typically given in the form of bits, into quantum states.
This corresponds to the left arrow in Fig.~\ref{fig:ClassQuantu}.
The most basic quantum state representing quantum information is given by the quantum counterpart to bits, i.e.\, qubits.
A qubit (quantum binary digit) is just a two-level quantum system.
There are various methods to perform encoding, which is why we will reduce them to two categories~\cite{Weigold2022,Khan2024,Rath2024}, basis encoding and superposition encoding.

We then look at the right arrow in Fig.~\ref{fig:ClassQuantu}, that is, the task of retrieving classical information from a given quantum system.
Here, we discuss the encoding schemes introduced in Sect.~\ref{ssec:Encoding} with respect to decoding, which triggers the question of whether some of them might allow reliably encoding several classical bits into a single qubit.
We will see that this effect is prevented by the Holevo bound for every encoding example, i.\,e. there is no gain in the number of encoded bits.
We also investigate a slightly different encoding task that aims to circumvent the restrictions imposed by Holevo's theorem.
First, we do not wish to recover all bits but only an arbitrary subset of fixed size.
Second, instead of recovering the original bits with certainty, we require the decoding procedure to be successful with a probability near unity.
It turns out that while the two modifications above allow us to encode more than one bit per qubit on average, the number of bits per qubit is only slightly larger than one, and thus the advantage is too small for practical applications.

Finally, in Sect.~\ref{ssec:QSDC}, we consider quantum secure direct communication (QSDC) \cite{Pan.2024} as a purely quantum way of data processing, eventually resulting in another quantum state, which has to be detected in order to recover classical information.
We will find that the discussion concerning the Holevo bound in Sect.~\ref{ssec:Decoding} can be directly transferred also to this protocol, ultimately preventing the transmission of more classical bits than qubits.
However, we will discuss the safety features of this method compared to other schemes. 
The outcome of our investigations is summarized in Sect.~\ref{ssec:InterfaceConclusions}.

\subsection{Encoding classical information into quantum states}
\label{ssec:Encoding}
We use the common bra-ket notation of quantum mechanics to denote states.
For the purpose of this text, state vectors are assumed to lie in $\mathbb{C}^d$ for some positive integer $d$.
Subscripts denote the system whose state is described by the vector.
For example the subscript $A$ in $\ket{0}_A$ emphasizes that the state belongs to Alice's quantum system but does not carry any further mathematical meaning.

A general quantum state can be considered as a superposition of some distinguished basis states $\ket{\Phi_i}$.
As a consequence, all encoded quantum states can be brought into the form
\begin{align}
\begin{split}
    \ket{\Phi}=\sum_{i=1}^d{c_i\ket{\Phi_i}},
\end{split}
\end{align}
where $c_i$ denote normalized coefficients ($|c_1|^2+\ldots+|c_d|^2=1$) and the states $\ket{\Phi_i}$ are connected to the encoded classical information.
The specific way in which classical information and the states $\ket{\Phi_i}$ are connected depends on the concrete method for encoding.
To this end, one can divide encoding methods into two categories, which are basis encoding and superposition encoding.
In the following section, we discuss both methods, where superposition encoding will be subdivided into several single methods.
For each method, we mention basic operations that need to be implemented to prepare the encoded states.

\subsubsection{Basis encoding}
For basis encoding, we choose $c_i$ to be unity for one specific $i$, while for all others it is zero~\cite{Cortese2018}.
As such, a natural choice for constructing $\ket{\Phi_i}$ are computational basis states.
In such a basis, the state 
\begin{align}
\begin{split}
    \ket{\Phi}=\ket{b_1}\otimes\cdots\otimes\ket{b_n},
\end{split}
\end{align}
is given by the tensor product of the basis states, where $n$ is the number of qubits.
Here, $\ket{b_k}$ is the basis vector for the $k$-th qubit space, with $b_k\in\{0,1\}$.
Hence, we assign to every classical bit one qubit, i.e. $k$ bits correspond to $k$ qubits.
In this sense, the number of bit sequences exactly equals the qubit space dimension.

As an example, consider the numbers 5 and 6.
Their binary equivalents are $101$ and $110$.
Thus, we can encode them into the quantum states $\ket{5}=\ket{1}\otimes\ket{0}\otimes\ket{1}$ and $\ket{6}=\ket{1}\otimes\ket{1}\otimes\ket{0}$.
Hence, we need to prepare two times three qubits in order to encode these numbers with basis encoding, which equals exactly the number of digits in the binary representations of 5 and 6.
From this example, one can immediately read off that there is no gain in information compared to the classical world since every bit needs exactly one qubit as a representative.

Normally, basis encoding is used for direct binary to qubit encoding, when arithmetic operations are mainly at the focus~\cite{Vedral1996,Leymann2020}.
Hence, the detection is straightforward by just measuring in the encoding basis and by that directly obtaining a determined result due to the orthonormality of the basis states.
Apart from arithmetic operations, another concrete example where this method can be applied is given in Sect.~\ref{ssec:QSDC} by quantum secure direct communication, where the justification for such a purely quantum information processing comes mainly from security reasons discussed there.

To produce states in a basis encoding we either have to be able to prepare $\ket{0}$ and $\ket{1}$ directly, or prepare just $\ket{0}$ and implement the $X$ gate, which is given by the action $X\ket{0} = \ket{1}$ and $X\ket{1} = \ket{0}$.

\subsubsection{Superposition encoding}
The category of superposition encoding can be split into several specific methods~\cite{Weigold2022,Khan2024,Rath2024}.
All of them have in common that they make use of a superposition of different $\ket{\Phi_i}$, in contrast to basis encoding.
In the following, we will present some basic types of superposition encoding and mention prominent examples for applications.

\paragraph{Basis superposition encoding}
This method uses the same basis states as for basis encoding above.
However, now we superimpose them
\begin{align}
\begin{split}
    \ket{\Phi}=\sum_i{c_i\ket{b_{1,i}}\otimes\cdots\otimes\ket{b_{n,i}}},
\end{split}
\end{align}
for the total encoded quantum state.
As one can observe, the number of bits still equals the number of qubits.
In this method, the (normalized) $c_i$ are usually equally distributed over the single superposition partners.

To encode the numbers 5 and 6 for the example from above, we find
\begin{align}
\begin{split}
    \ket{\Phi}=\frac{1}{\sqrt{2}}\left(\ket{101}+\ket{110}\right),
\end{split}
\end{align}
where $1/\sqrt{2}$ is the normalization.
Here we encoded both numbers in a single quantum state, whereas in the case of basis encoding we used two states.

The usage of this encoding is similar to basis encoding for computational purposes. 
However, it can have some speedups, e.g. in the context of the Grover or Shor algorithms.
The generalization of this method to higher dimensional data points is sometimes also referred to as quantum associative memory.

\paragraph{Amplitude encoding}
Amplitude encoding directly uses the amplitudes of each superposition partner to encode the desired real-valued prefactors~\cite{GonzalezConde2024,Khan2024,Rath2024}.
As such this method is also called wave function encoding~\cite{LaRose2020}, as it makes most obviously use of the superposition feature of a quantum state.
Hence, we use (normalized) $c_i$ to encode the classical information directly.
Thus, the $\ket{\Phi_i}$ can have a lower dimension than the associated bit space.

For the example from above, we obtain
\begin{align}
\begin{split}
    \ket{\Phi}=\frac{1}{\sqrt{11}}\left(\sqrt{5}\ket{0}+\sqrt{6}\ket{1}\right).
\end{split}
\end{align}
Here, $ 1/\sqrt{11} $ is the normalization.
We observe that we skipped the step to translate both numbers 5 and 6 into binaries and instead directly encoded them as prefactors of a single qubit.

As such, we have encoded six bits (three for each number) or two classical numbers into only one qubit.
Equivalently, one can state that we have encoded two classical features in $1=\log_2{2}$ qubits.
This law can be generalized to encoding $d$ classical features into $n=\log_2{d}$ qubits.
However, again, there is no deterministic detection in the computational basis and the number of measurements scales with the amplitude.

Amplitude encoding is frequently discussed in the context of quantum teleportation as well as quantum key distribution and quantum machine learning~\cite{Harrow2009,Duarte2019,LaRose2020}, such as for the BB84 protocol.
However, it is also vulnerable to e.g. dephasing noise~\cite{Nguyen2022}.

\paragraph{Angle encoding}
This method operates similar to amplitude encoding.
However, now the classical information is encoded into the relative phases~\cite{Khan2024,Rath2024} instead of real-valued prefactors.
These phase shifts are realized by two by two rotation matrices $\hat{R}$, acting on the qubits.
Here, $\hat{R}$ can be any Pauli rotation $\hat{R}^j\left(\theta\right)=\exp{\left(-i\theta\sigma_j/2\right)}$ depending on some rotation angle $\theta$, where $\sigma_j$ is the Pauli matrix in $j$-direction. The rotation matrices read
    \begin{align*}
    \hat{R}^x(\theta) &= \begin{bmatrix} \cos(\theta/2) & -i\cdot\sin(\theta/2) \\i\cdot\sin(\theta/2) & \cos(\theta/2) \end{bmatrix}, \\
	\hat{R}^y(\theta) &= \begin{bmatrix} \cos(\theta/2) & -\sin(\theta/2) \\ \sin(\theta/2) & \cos(\theta/2) \end{bmatrix}, \\
	\hat{R}^z(\theta) &= \begin{bmatrix} \exp(-i\theta/2) & 0 \\ 0 & \exp(i\theta/2) \end{bmatrix}.
	\end{align*}
In this sense, the information about the prefactor from basis superposition encoding is traded in for the angle and the specification of the rotation direction.

Hence, for a definite choice of $j$, the total state of an $n$-qubit system can be cast into the form
    \begin{align}
    \ket{\Phi}=\sum_i{c_i\bigotimes_{k=1}^n\hat{R}^j\left(\theta_k\right)\ket{0}_{k,i}}.
    \end{align}
Note that, similar as for basis superposition encoding, the dimensions of bits and qubits are equal.

For our example of encoding the numbers 5 and 6, we choose $c_i=1/\sqrt{2}$ and $\theta_k=\pi\cdot b_k$, where $b_k$ is the value of the $k$\textsuperscript{th} bit.
Thus, the encoded state reads
\begin{align}
\begin{split}
    \ket{\Phi}=\frac{1}{\sqrt{2}}\left(\hat{R}^x_1\left(0\right)\hat{R}^x_2\left(\pi\right)\hat{R}^x_3\left(0\right)\hat{R}^x_4\left(\pi\right)+\hat{R}^x_1\left(0\right)\hat{R}^x_2\left(\pi\right)\hat{R}^x_3\left(\pi\right)\hat{R}^x_4\left(0\right)\right)\ket{0000},
\end{split}
\end{align}
where we chose the rotations to act in $x$-direction.

Applications of this encoding technique can for instance be found in image processing~\cite{Yan2016}.
Here, the angle is used to store the information about a color by depicting the color information of one pixel in a suitable quantum imaging basis.
Other applications are quantum neural networks~\cite{Schuld2014}, data classification, or in order to reduce error rates in quantum circuits~\cite{Leymann2020,Beisel2022}.

\subsection{Retrieving classical information from quantum states}
\label{ssec:Decoding}

Once we have encoded our sequence of bits into quantum states we would like to perform useful tasks with them, and at a later stage decode them to get back the original information.
In this section we focus on the latter, that is, the ``detection'' arrow of Fig.~\ref{fig:ClassQuantu}.
Given a quantum system in some state we have to measure it to obtain classical information.
We restrict our attention to \emph{projective measurements}.
More general models of measurement exist but are not required for our purposes.
There are three important ingredients: The result of the measurement, the probability of that result occurring, and the state of the system after the measurement.
As to the results, we usually are not interested in their precise form but in the fact that they are distinguishable.
For this reason we will label the results by natural numbers or bit sequences.
A single measurement is described by a basis $\ket{0},\ldots,\ket{d-1}$ of the underlying $d$-dimensional state space, where $0,\ldots,d-1$ are the possible results of the measurement.
If the state of the system is described by the vector $\ket{\psi}$, then the probability of obtaining the outcome $j$ is given by $\left|\braket{\psi|j}\right|^2$, and the state of the system after the measurement is $\ket{j}$.
For a system with density matrix $\rho$, the probability is $\bra{j}\rho\ket{j}$ and the state after the measurement is again $\ket{j}$.

A short summary of this section is that the information storage capacities of quantum states are surprisingly limited.
To understand the subsequent no-go results one needs to be familiar with the information-theoretic notion of entropy.
The concept of thermodynamic entropy will not play a role in our discussions.
Among the best references on this subject is the original paper of Shannon \cite{Shannon.1948} that constitutes the beginnings of information theory.
A standard text on classical information theory is the book by Cover and Thomas \cite{Cover.2006}.
We refer to those works for a more sophisticated exposition of the concepts below.

\subsubsection{Information-theoretic description of entropy}

\paragraph{Entropy \textemdash~The expected amount of information}

The binary entropy of a random variable $X$ is defined as
	\begin{equation*}
	H(X) := \sum_{x}p_x \cdot \log_2\left(\frac{1}{p_x}\right),
	\end{equation*}
where the sum ranges over all possible values $x$ that $X$ can assume (with non-zero probability) and $p_x$ is the probability of the value $x$.
Note that $H(X)$ is always non-negative.
A common interpretation of the entropy is the shortest average length of a bit encoding of a random outcome described by $X$.
It is important to remark that this interpretation does not apply to a single outcome but a large number (formally: an infinite sequence) of independent and identically distributed instances of an experiment whose outcome is described by $X$.
This distinction is made clearer by the example of a series of coin tosses.
Imagine the coin is heavily biased so it comes up heads with probability $0.9$.
If the result of each coin toss must be encoded individually there is no other way than using one bit for each toss.
However, if one is allowed to encode multiple results at once, it is possible to do better.
For example one could encode the number of times head occurs between two appearances of tails.
Another useful, but not quite exact, interpretation of binary entropy is the expected number of yes-no questions required to determine $X$.
Think of the following game between Alice and Bob:
Alice randomly generates $X$.
Bob's task is to determine $X$ by asking yes-no questions about $X$ that Alice must answer truthfully.
If Bob plays optimally then on average the number of questions he has to ask lies between $H(X)$ and $H(X)+1$.
We refer to \cite{Cover.2006} for a proof of this fact.
Here, taking an average is essential.
In any single game, Bob could be lucky and guess the correct value with a single yes-no-question by asking for a specific value.
However, in most cases Bob will simply rule out that specific value and not learn anything else about $X$, so in the long run such a strategy is not useful unless with high probability $X$ has one of a few fixed values.

\paragraph{Conditional entropy \textemdash~Knowledge decreases uncertainty (on average)}
The conditional entropy of $X$ given $Y$ provides an answer to the question ``How many bits do I need to describe the value of X provided that I already know the value of Y?''.
	\begin{equation*}
	H(X|Y) := \sum_y p(y) \cdot \underbrace{\sum_x p(x|y) \cdot \log_2\left( \frac{1}{p(x|y)} \right)}_{\text{Entropy of }X\text{ given that }Y=y}
	\end{equation*}
Note that the conditional entropy describes the situation \emph{before} the value of $Y$ is revealed.
Therefore one takes the average over all values taken on by $Y$.
In the game of asking yes-no questions we can interpret $H(X|Y)$ as follows:
Again, Alice generates $X$ and Bob is required to find the value of $X$ via yes-no questions.
This time, however, before his first question, he is told a random value $Y$ that is in some way correlated with $X$.
This additional information may allow Bob to find $X$ with fewer attempts.
Now the average number of questions lies between $H(X|Y)$ and $H(X|Y)+1$, provided that Bob plays optimally.

\paragraph{Mutual information}
The mutual information of $X$ and $Y$, denoted by $I(X; Y)$, quantifies the information shared by $X$ and $Y$.
Imagine that describing the outcome $X$ requires four bits, and that, once we know $Y$, we only need three bits to determine $X$ due to existing correlations between $X$ and $Y$.
In that scenario $I(X;Y)$ would equal $1$ since knowing $Y$ saved us a single bit.
Formally the mutual information is defined as follows:
	\begin{equation*}
	I(X; Y) := \underbrace{H(X) - H(X|Y)}_{\text{Uncertainty removed from } X \text{ when } Y \text{ is revealed}}
	\end{equation*}
In the game of determining $X$ we can interpret $I(X;Y)$ as the average number of yes-no-questions saved when Bob is told the value of $Y$.

\paragraph{The entropy of quantum systems}
Also referred to as von Neumann entropy.
Given the state of a quantum mechanical system and a choice of measurement, the result of the measurement is described by a classical random variable.
The actual distribution depends on both the state and the measurement.
For example, measuring the state $\ket{+} = \frac{1}{\sqrt{2}}\ket{0}+\frac{1}{\sqrt{2}}\ket{1}$ in one of the bases $\ket{0}, \ket{1}$ and $\ket{+}, \ket{-}$ yields a uniform distribution in the former case and always results in the state $\ket{+}$ in the latter.
The entropy $S(\rho)$ of a state $\rho$ is the smallest possible classical entropy of the distribution over all measurements, which is attained when the system is measured in the eigenbasis of $\rho$.
	\begin{equation*}
	S(\rho) := \sum_{j=1}^N \lambda_j \log_2\left( \frac 1 {\lambda_j} \right) = -\mathrm{Tr}(\rho \log_2(\rho))
	\end{equation*}
Here $\lambda_1,\ldots,\lambda_N$ are the eigenvalues of the density matrix $\rho$.

\paragraph{The no-cloning theorem}
In quantum information theory there are several \emph{no-go theorems} that tell us that certain operations on quantum states cannot be realized.
The best-known of those results is the \emph{no-cloning theorem}, which states that it is impossible to copy an arbitrary unknown quantum state.
Here, unknown means that we are given a quantum system but do not have any information on its state, while copying means that we take another quantum system, whose state we do know, and prepare it in a way that makes it have the same state vector as the original system.
In symbols, we wish to implement the transformation $\ket{\psi}\otimes\ket{0} \mapsto \ket{\psi}\otimes\ket{\psi}$.
Such an operation does not preserve inner products and thus cannot be achieved by unitary transformations.
Of course, if we had a description of the system in terms of a state vector, we could prepare arbitrarily many copies of that state.
Similarly, if we knew that the system is in one of two orthogonal states, we could copy it as well by first measuring it in the basis given by the two possible states and then preparing copies of the state corresponding to the result.
The words ``arbitrary'' and ``unknown'' are crucial for the theorem.

\subsubsection{Holevo's bound}
\label{ssec:Holevo}
The bound we present in this section can be thought of as a quantitative no-go theorem, and occurs in the context of a fundamental question:
Can quantum states be used to improve the transmission of classical information?
The question is quite general, so let us specify what types of improvements we mean.
In the following discussion, we assume the absence of noise during transmission.
Given a sequence of, say, ten bits, Alice can send those bits to Bob over a classical channel.
Bob can then store all ten bits and read them at will.
If Alice is allowed to send qubits instead of bits, she can simply take ten qubits, fix a choice of basis, and prepare each of them in the basis state corresponding to the bit value of her classical bits.
This procedure clearly does not provide any advantage, but rather serves as a baseline to which more elaborate encodings are compared.
Is it possible for Alice and Bob to use the concepts of superposition and entanglement to transmit fewer than ten qubits and later retrieve all ten bits?
The short answer is ``not really''.
The Holevo bound provides a more sophisticated answer to that question.
Let us present the formal setting in which the bound is usually stated.
The information that Alice wants to send to Bob is described by a random variable $X$.
At first, this choice of description can be confusing because Alice's information may not be random at all.
Assuming that Alice and Bob have a general procedure that allows them to send arbitrary bits using qubits, that procedure must also work if Alice generates a sequence of bits at random.
It is instructive to remember that information-theoretic quantities such as entropies tell us what happens in the long run.
The situation of the bound below does occur many times with different people playing the roles of Alice and Bob.
In the long run, each possible bit sequence occurs with a certain relative frequency.
Those frequencies are described by $X$.

    \begin{table}[ht]
    \centering
    \begin{tabular}{m{0.9\textwidth}}
        \hline
        \textit{Abstract situation underlying Holevo's bound}  \\
        Alice wishes to send classical information to Bob via quantum states.
		\begin{enumerate}
		\item Alice generates a random element $X$ from a known finite set.
		\item Alice prepares her system the state $\rho_X$, where the assignment $X\mapsto \rho_X$ is known to both Alice and Bob in advance.
		\item Alice sends her system and nothing else to Bob.
		\item Bob measures the received system and gets a result $Y$.
		\item Bob estimates the value of $X$ using $Y$.
		\end{enumerate} \\
        \hline
    \end{tabular}
    \end{table}

It is important to point out that the only aspect of the procedure not known in advance is the actual information sent by Alice, that is, the value of $X$.
In particular Bob's choice of measurement cannot depend on $X$.
In the situation above one has
	\begin{equation}\label{eq:holevobound}
	\underbrace{I(X; Y)}_{\substack{\text{Mutual information of} \\ \text{message and measurement}}} \leq \underbrace{S\left(\sum_{x}p_x\rho_x\right)}_{\substack{\text{Entropy of the transmitted} \\ \text{system from Bob's p.o.v.}}} - \underbrace{\sum_{x}p_xS(\rho_x)}_{\substack{\text{Expected entropy} \\ \text{from Alice's p.o.v.}}},
	\end{equation}
where the sums are over all possible values $x$ that $X$ can assume.
Let us take a closer look at the individual terms appearing in the inequality.
Before the measurement, Bob does not know anything about $X$ apart from the probabilities of the possible values.
For this reason the state $\sum_xp_x\rho_x$, where $p_x$ denotes the probability that $X$ takes on the value $x$, describes Bob's knowledge of the system he received.
Alice knows the state she has prepared, so the entropy from her point of view is $S(\rho_x)$.
Since $\rho_x$ is prepared with probability $p_x$, the average entropy is $\sum_xp_xS(\rho_x)$.
If Alice was to send her information classically she would need to transmit $H(X)$ bits on average.
We can interpret $I(X;Y)$ as the number of required bits saved by first sending the quantum mechanical system to Bob.
In order to successfully transmit $X$ without any additional classical bits one needs $I(X;Y)=H(X)$.

We are not going to provide a proof of Eq.~\eqref{eq:holevobound} as we want to concentrate on its interpretation and consequences.
For a full proof, we refer to \cite{NielsenChuang}.

The right hand side of Eq.~\eqref{eq:holevobound} is always non-negative, whereas the entropy from Bob's point of view is always bounded by $\log_2(d)$, where $d$ is the dimension of the underlying state space.
If the transmitted system consists of $n$ qubits, then the Holevo bound implies $I(X;Y)\leq n$.
Therefore, if one wishes to encode classical information in quantum states and decode it later without errors, the number encoded bits must not exceed the number of available qubits.
A priori, it is conceivable that a clever use of superposition allows to store and transmit multiple bits in a single qubit.
The Holevo bound tells us that such an encoding always introduces uncertainty to the decoding procedure and quantifies that uncertainty using the concept of entropy.

\subsubsection{No partial reconstruction \textemdash~Quantum random access codes} \label{ssec:NoPartRec}
The Holevo bound forbids recovering more than $n$ bits of information from an $n$-qubit system.
It does \emph{not} instantly rule out the existence of a reliable encoding of more than $n$ bits in $n$ qubits which allows us to access only an arbitrary subset of a fixed size that does not exceed $n$.
Imagine, for example, we could encode five bits in four qubits such that it is possible to retrieve any four of the five bits at the expense of losing the remaining one.
In that sense all bits would be stored within the overall state of the two-level systems but the Holevo bound would not be immediately violated since the retrieved information never exceeds four bits.

A potential application of such an encoding would be private database retrieval, where a party wants to access an entry of a database of an untrustworthy owner subject to the following two conditions:
First, the accessing party must be able to obtain one desired entry of the database but not more, so the owner can be sure that only the previously specified amount of information was accessed.
Second, the owner must not know which entry was retrieved, so the query was indeed private.

Unfortunately, such encodings cannot exist unless the number of bits is only marginally larger than the number of qubits.
As with any impossibility result it is important to precisely state the underlying assumptions since the situation may be entirely different if a single of those assumptions is altered.
The purpose of this section is to formalize the idea of a partial reconstruction of bits as described above and provide quantitative bounds on how far beyond one bit per qubit we can get.
We emphasize that our goal is to present an impossibility result on the theoretical level.
We are aware that practically establishing a protocol comes with further obstructions such as dealing with noise.
The results below state that even if those additional obstructions were non-existent, one could not realize the desired encoding.
Before we state those results let us formalize the type of protocol we consider. 

    \begin{table}[h]
    \centering
    \begin{tabular}{m{0.9\textwidth}}
        \hline
        \textit{An abstract protocol for bit reconstruction}  \\
        This protocol depends on parameters $m,n,k\in\mathbb{N}$ where $m\geq n \geq k$.
		\begin{enumerate}
		\item Alice possesses a sequence of $m$ Bits $b_1,\ldots,b_m$ and $n$ two-level systems.
		\item Depending on her bit sequence Alice prepares her qubits in the overall state $\ket{\psi^{b_1\ldots b_m}}$.
			The assignment $b_1\ldots b_m \mapsto  \ket{\psi^{b_1\ldots b_m}}$ is known to both Alice and Bob in advance.
		\item Alice sends all $n$ qubits to Bob.
		\item Bob chooses $k$ out of $m$ positions, described by a $k$-element set $K\subset\{ 1,\ldots,m \}$.
		\item Bob measures the whole system in the basis $M_K$.
			The assignment $K\mapsto M_K$ is known to both Alice and Bob.
		\item Bob processes the result of his measurement and returns a sequence $\tilde b_1\ldots \tilde b_k$ of $k$ bits.
		\item The reconstruction is \emph{successful} if $\tilde b_1 \ldots \tilde b_k$ agrees with $b_K$.
		\end{enumerate} \\
        \hline
    \end{tabular}
    \end{table}

The quality of the protocol is measured in the minimum success probability
	\begin{equation*}
	\min_{b_1,\ldots,b_m\in\{ 0,1 \}, K\subset\{ 1,\ldots,m \}} P\left(\tilde b_1\ldots\tilde b_k = b_K\right),
	\end{equation*}
where $P$ describes the probabilities associated with Bob's measurement and potential post-processing in Steps 5 and 6.

The first impossibility result we present deals with the scenario in which we wish to reconstruct $k$ bits with certainty.

    \begin{table}[ht]
    \centering
    \begin{tabular}{m{0.9\textwidth}}
        \hline
        \textit{Impossibility of perfect reconstruction}  \\
        If $m>n$ the protocol above cannot be successful with unit probability, even in the case $k=1$. \\
        \hline
    \end{tabular}
    \end{table}

To make such a protocol work we would need to specify $2^m$ perfectly distinguishable states of our $n$-qubit system, which amounts to finding $2^m$ pairwise orthogonal vectors in a complex vector space of dimension $2^n$.
The size of a set of pairwise orthogonal vectors cannot exceed the dimension of the underlying space, and so $2^m \leq 2^n$.
Let us make this argument more precise.
For any two distinct bit sequences $b_1\ldots b_m$ and $b'_1\ldots b'_m$ there exists a position $j$ such that $b_j \neq b'_j$.
If such a $j$ is contained in the set $K$ chosen by Bob and the reconstruction is successful then a measurement in the basis $M_K$ can perfectly distinguish between  $\ket{\psi^{b_1\ldots b_m}}$ and $\ket{\psi^{b'_1\ldots b'_m}}$.
Thus the latter vectors must be orthogonal to each other.
Since $b_1\ldots b_m$ and $b'_1\ldots b'_m$ were arbitrary distinct sequences, the set $\{ \ket{\psi^{b_1\ldots b_m}} : b_1\ldots b_m\in \{ 0,1 \} \}$ consists of $2^m$ pairwise orthogonal vectors, which tells us that the dimension of the underlying state space is at least $2^m$.
This is possible only if $m \leq n$.

The above discussion shows that a reconstruction with unit probability can never beat a simple basis encoding.
However, it does not prevent the protocol from being successful with a probability of, say, $0.999$.
It turns out that relaxing the success probability to something slightly smaller than $1$ does not improve the situation.
This is best demonstrated in the special case $k=1$, which is known in the literature as a \emph{quantum random access code} \cite{ambainis_dense_2002}.
We point out that any upper bound on $p$ for the case $k=1$ immediately provides an upper bound for the case $k>1$ because being able to reconstruct more than one bit with high probability in particular allows one to access any single bit with at least the same probability of success.
The following result of Nayak \cite{nayak_optimal_1999} establishes that for $p$ close to $1$ in a random access encoding the number of bits can exceed the number of qubits merely by a small factor.

    \begin{table}[ht]
    \centering
    \begin{tabular}{m{0.9\textwidth}}
        \hline
        \textit{Limitations of random access codes (Nayak, 1999)}  \\
        Any $(m,n,p)$-random access code obeys the bound
		\begin{equation*}
		m \leq \frac{1}{1-H(p)}\cdot n,
		\end{equation*}
	where $H(p)$ denotes the base-two binary entropy function evaluated at $p$.
        \\
        \hline
    \end{tabular}
    \end{table}

\begin{figure}[hbt]
	\centering
	\includegraphics[width=0.7\linewidth]{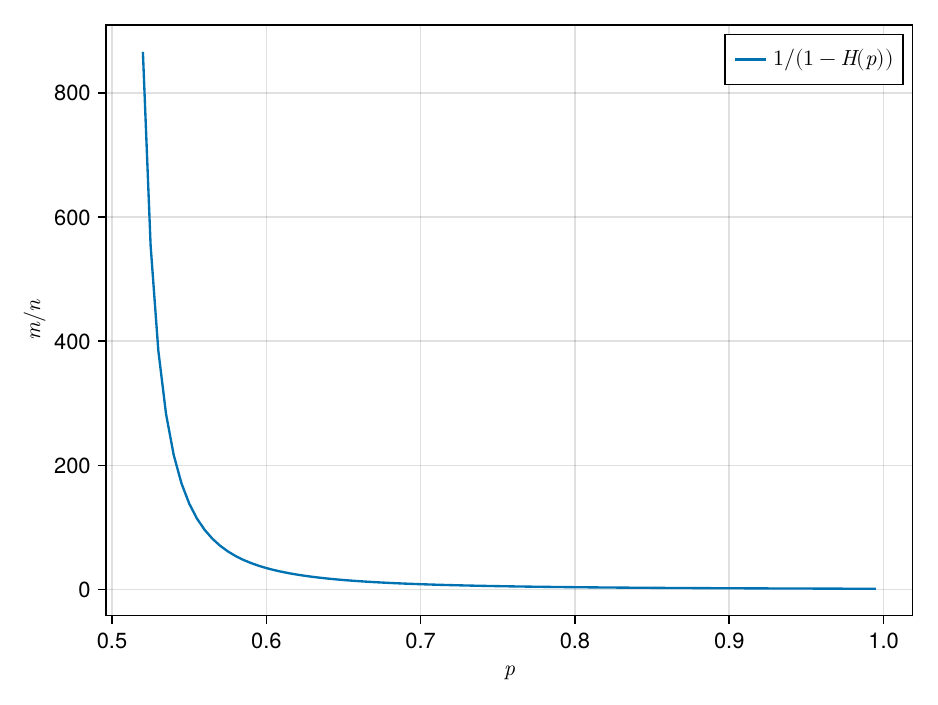}
	\caption{The quotient $m/n = 1/(1-H(p))$ from Nayak's theorem as a function of the success probability $p$ in the range $(1/2,1]$.
	The value at $p=1$ is $1$, which tells us that a perfect decoding requires $m=n$.}
	\label{fig:partial-reconstruction}
	\end{figure}

For example, if we want to retrieve a single bit with probability at least $0.9$ then $m$ must be smaller than $2n$, and if we increase the success probability to $0.99$ we already arrive at the bound $m < 1.088n$.
Figure \ref{fig:partial-reconstruction} displays the behaviour of the quotient $m/n = 1/(1-H(p))$ for varying values of $p$.
The first impossibility result stated above is a special case of the theorem for $p=1$.
Recall that randomly guessing individual bits has a success probability of $1/2$ per bit and thus provides a trivial $(m,n,1/2)$-random access encoding for any choice of $m$ and $n$.
This is reflected in the above theorem of Nayak by the fact that the term $1/(1-H(p))$ tends to infinity as $p\to 1/2$.

One final hope we might have is that the bounds above do not get much worse if instead of retrieving a single bit we want to reconstruct a tuple of $k$ bits, where $k \geq 2$.
A success probability of $0.5$ is useless for a single bit but could be useful when it applies to recovering one hundred bits at once.
The final impossibility result we present tells us that the situation does get worse for larger $k$.
It is essentially a variant of Nayak's theorem for qudits and, like the original theorem, is obtained by an inductive application of Fano's inequality \cite{fano1961transmission} and Holevo's bound.

    \begin{table}[ht]
    \centering
    \begin{tabular}{m{0.9\textwidth}}
        \hline
        \textit{Reconstruction of multiple bits}  \\
        For $k > 1$ any bit reconstruction protocol of the form specified above is successful with probability at least $p$ only if
		\begin{equation}\label{eq:multiple_bits}
		(p-H(p)/k)\cdot m \leq n+H(p)
		\end{equation}
        \\
        \hline
    \end{tabular}
    \end{table}

To deduce the bound \eqref{eq:multiple_bits} one does not need the full strength of the assumption that $k$ arbitrary bits can be reconstructed.
It suffices that, given an encoded bit string $b_1\ldots b_N$,  an arbitrary block $b_j\ldots b_{j+k-1}$ of $k$ contiguous bits can be recovered.
Therefore, it seems reasonable that even stronger bounds hold.
However, we believe that \eqref{eq:multiple_bits} is enough to rule out the bit reconstruction protocol from a practical point of view.

	\begin{figure}[hbt]
	\centering
	\includegraphics[width=0.7\linewidth]{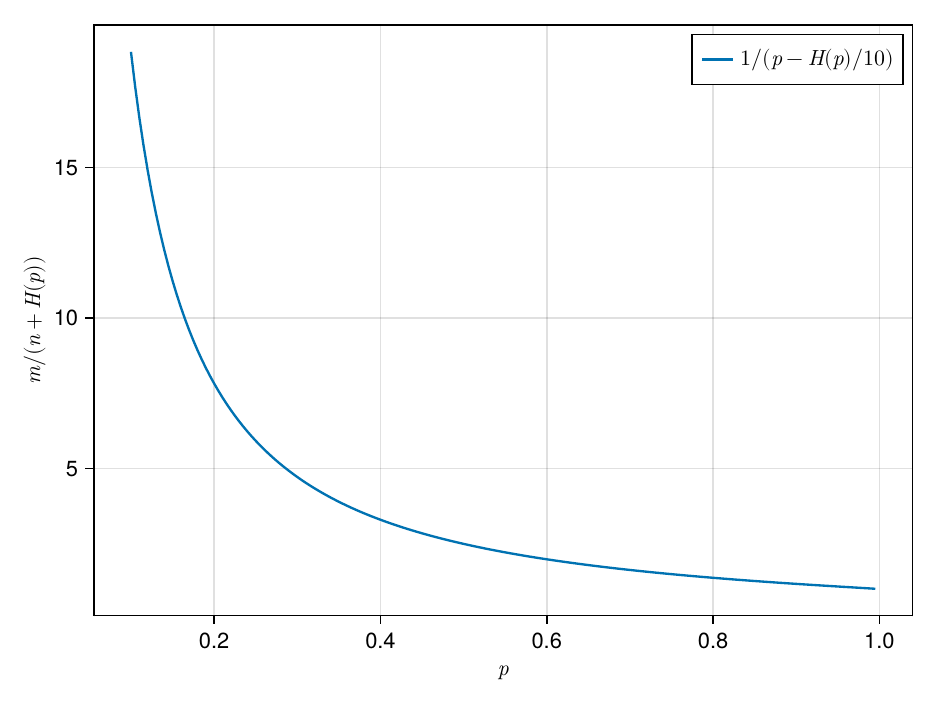}
	\caption{ The upper bound on the quotient $m/(n+H(p))$ given by \eqref{eq:multiple_bits} as a function of $p$ for $k=10$.}
	\label{fig:partial-reconstruction-2}
	\end{figure}

What are the consequences of the results above?
	\begin{enumerate}
	\item If some company tells us that it has developed a ``quantum'' database which allows the storage and retrieval of a huge number of bits with much fewer qubits we can respond that those claims are unfounded, even for a generous interpretation of ``retrieval''.
	\item More generally, for information storage and retrieval one cannot do much better than simply picking a base encoding.
	\item While the above bound initially seems helpful for secure applications as it can provide bounds on the information an eavesdropper is able to reconstruct from attacking a suitably defined protocol, one has to keep in mind that such bounds also apply to the intended receiver and thus reduce or even eliminate the usefulness of the protocol.
	\end{enumerate}
Can we further weaken our assumptions so that the restrictions presented above do not apply?
It clearly does not make any sense to store a bit if neither the bit itself nor any other bit depending on it is ever accessed.
Given bits $b_1,\ldots,b_m$, one could store them in $n$ qubits such that suitable measurements allow accessing $n$ derived bits $f_j = f_j(b_1,\ldots,b_m)$, $1\leq j \leq n$.
The discussion above does not forbid such a procedure.
However, one could simply classically compute the $f_j$ in advance and only send those bits.
Note that in both cases the $m$ original bits are inaccessible.
Still, there are two scenarios in which encoding $m$ bits and accessing $n$ processed bits can be useful:
First, data is not fully recorded classically but obtained from a stream of bits and directly stored in a quantum state.
Second, one does not necessarily have the computational resources required to determine the $f_j$.
We discuss the former scenario in this section.
The latter leads us to the area of \emph{Blind Quantum Computing}, which we present in more detail in Sect.\,\ref{sssec:BQC}.

\subsection{Quantum secure direct communication}
\label{ssec:QSDC}
Up to this point, we have explored how classical information can be encoded into quantum information, as well as how quantum information can be decoded back into classical form, including the limitations of these processes, in particular the different encoding schemes as well as the Holevo bound. In this section, we illustrate the combination of both techniques through an example of quantum secure direct communication.

With the advent of quantum computation, and in particular the introduction of Shor's algorithm \cite{Shor.1994} for prime factorization, the security of classical encrypting schemes have become at least theoretically threatened by quantum technology. In particular, classically asymmetric encryption schemes are often used to generate a key, which in turn is then used a by a symmetric encryption protocol. While the latter is often still secure, even with the advance of quantum technology, asymmetric encryption is based on one-way functions, which might be inverted using quantum computation. 

A possible solution to this problem is to use quantum mechanics itself to find new secure ways to either encrypt information directly or at least generate secure keys. The most common approach is to do the latter, often referred to as quantum key distribution (QKD, see \ref{sssec:Security}), where a classical key is created by virtue of quantum physics, such that eavesdropping is detected by both parties. Here, the encryption of the information is still performed by a classical symmetric encryption scheme, while only the key needed for this symmetric encryption is generated quantum mechanically. However, no encrypted information is transferred between Alice and Bob via a quantum channel.

In contrast to QKD, we are interested in transmitting the data securely via a quantum channel, and using the advantages of quantum mechanics directly. Such a scheme is often referred to as quantum secure direct communication (QSDC).

In the following sections, we review the general idea of QSDC for a bipartite state. In fact, we present the most basic case of an entangled qubit state, an ideal scheme and the conditions necessary to implement such a QSDC protocol. We note, that there are different protocols and implementations of QSDC. For simplicity, we only consider one basic example of QSDC, which brings out the general idea most clearly, while for different implementations we refer to Ref.~\cite{Pan.2024}. We then investigate the extension of QSDC to multipartite qubit systems, in particular tripartite systems. We introduce the different classes of entangled states and show how they perform under an extended QSDC protocol.

\subsubsection{Entanglement}
\label{ssec:entanglement}
A quantum state consisting of two or more subsystems is called entangled if it cannot be separated into a tensor product of the individual subsystems. For a bipartite state this means $\ket{\Psi}\neq\ket{\psi}\ket{\phi}$ for any states $\ket{\psi}$ and $\ket{\phi}$ on the individual subsystems. Prominent examples of entangled states are the Bell states $\ket{\Phi^{(\pm)}}=(\ket{00}\pm\ket{11})/\sqrt{2}$ and $\ket{\Psi^{(\pm)}}=(\ket{01}\pm\ket{10})/\sqrt{2}$. Entanglement is quantifiable by different entanglement measures, e.g., entanglement cost \cite{Hayden2001} or distillable entanglement \cite{Devetak2005}, which we will not discuss in detail here. Local operations on the individual subsystems of a quantum state cannot increase the degree of entanglement of the state. Therefore, entangling a state can only be achieved by some interaction between the particles. 

Consider a finite-dimensional Hilbert space $\mathbb{C}^n$, where a quantum state $\rho$ is defined as a density matrix, namely a Hermitian, positive semidefinite matrix with unit trace:
\begin{equation}
\rho = \rho^\dagger \,, \qquad \rho \geq 0 \,, \qquad \mathrm{Tr}(\rho) = 1.
\end{equation}
With density matrices $\rho(A)$ and $\rho(B)$ of $\mathbb{C}^{n_A}$ and $\mathbb{C}^{n_B}$ respectively, where $n_A$ and $n_B$ are the dimensions of the subsystems A and B, $\rho(A) \otimes \rho(B)$ is called a tensor product state. 
If the density matrix of such a bipartite system $\rho_{AB}$ can be written as a convex combination of tensor product states, it is called separable:
\begin{equation}
\rho_{AB} =  \sum_i p_i \rho_i(A) \otimes \rho_i(B), \quad p_i > 0, \quad \sum_i p_i = 1. 
\label{eq:separable}
\end{equation}
From these definitions it follows that both the set of all density matrices and the set of separable states are convex. 
A density matrix which cannot be written in the form \eqref{eq:separable} is called an entangled state. 
Over time, several ways of characterizing and measuring entanglement have emerged and their properties and relations have been studied, see e.g. \cite{Hayden2001,Devetak2005,Horodecki1998,Wootters1998,Peres1996,Horodecki1996,Horodecki1999,Cerf1999,Nielsen2001,Hiroshima2003,Vollbrecht2002,Bruss2019,Chen2003,Rudolph2003}.
The basic idea of these characterizations is to quantify the amount of entanglement contained in a quantum state in some way.
In this context it is useful to introduce the \textit{ebit}, the amount of entanglement contained in a maximally entangled two-qubit state, i.e. a Bell state like $\ket{\Phi^+} = (\ket{00} + \ket{11})/\sqrt{2}$.

\subsubsection{QSDC with bipartite states} \label{sssec:QSDC2}
The idea behind quantum secure direct communication is to transfer information using an entangled state between the sender Alice and the receiver Bob. The most common way to perform such a task is by employing a bipartite system. In this case, the QSDC provides the advantage that if the entangled state is established the information cannot be intercepted by an eavesdropper, since the information is only accessible through the entire state. However, only one subsystem is sent containing no useful information on its own. Intercepting and measuring this subsystem alone will therefore only provide random noise, but no information. 
\begin{figure}[hbt]
	\centering
	\includegraphics{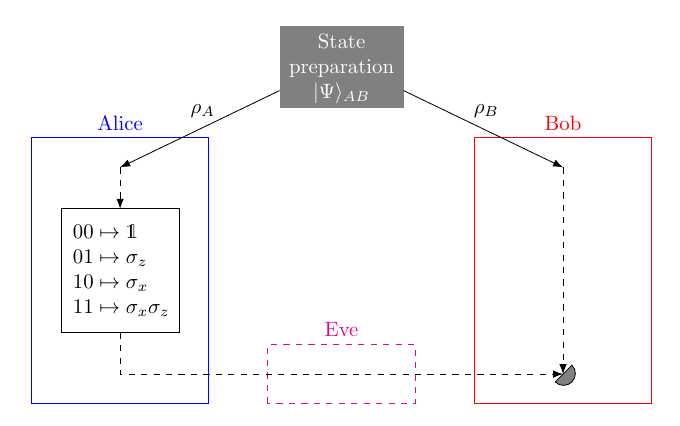}
	\caption{Schematic representation of the quantum secure direct communication protocol (QSDC) for a bipartite qubit system. The two qubits of the bipartite state are sent to Alice and Bob respectively. Alice then encodes her information by acting with either no operation or one of the three Pauli operators on her qubit. The complete state of the system is then described by one of the four Bell states. She then sends her qubit to Bob, who performs a Bell measurement on the complete two qubit state to decode Alice's information. As soon as the entangled two qubit state is established among Alice and Bob, an eavesdropper (Eve) can only access the qubit sent by Alice. Since this qubit is in a maximally mixed state for either encoded state, Eve cannot gain any information about the operation performed by Alice.}
	\label{fig:qsdc}
\end{figure}
In general there exist different possibilities for a protocol that fulfills this task. The idea of the most common QSDC protocol for a bipartite qubit state is depicted in Fig.\,\ref{fig:qsdc} and discussed below:

    \begin{table}[ht]
    \centering
    \begin{tabular}{m{0.9\textwidth}}
    \hline
    \textit{A QSDC protocol for bipartite qubit states}  \\
		\begin{enumerate}
        \item The sender (Alice) and the receiver (Bob) share a maximally entangled bipartite state $ \ket{\Phi^{(+)}} = (\ket{00}+\ket{11})/\sqrt{2} $.
        \item Alice encodes two bits of information into either of four different operations on her qubit, i.e. $ 00 \to \mathds{1} $, $ 01 \to \sigma_{z} $, $ 10 \to \sigma_{x} $ and $ 11 \to \sigma_{x}\sigma_{z} $.
        \item The resulting state is either of the four Bell states $\ket{\Phi^{(+)}}$, $\ket{\Phi^{(-)}}$, $\ket{\Psi^{(+)}}$ or $\ket{\Psi^{(-)}}$. Thus the four operations lead to perfectly distinguishable states.
        \item Alice sends her qubit to Bob.
        \item Bob measures which Bell state the entire system is in, thus reconstructing the information Alice has encoded.
		\end{enumerate} \\
    \hline
    \end{tabular}
    \end{table}
    
With one Bell pair two bits of classical information can therefore be transmitted from Alice to Bob. We note that this amount of information corresponds to the Holevo bound. In fact, two qubits can never transmit more than two bits of classical information. 

\paragraph{Security of this QSDC protocol}
The use of entanglement also manifests itself as the key advantage in terms of the security of the presented protocol. In order to transmit her information to Bob, Alice only sends her qubit along a possibly open quantum channel. The information about her two encoded bits, however, is shared across the entire bipartite state. Independent of her information encoded on the Bell state, the state of her qubit alone
\begin{equation}
	\rho_A = \frac{\mathds{1}_{A}}{2},
\end{equation} 
is always in a completely mixed single qubit state, which follows directly from tracing out Bob's qubit for any maximally entangled Bell state.

Hence, no information about the entire Bell state is accessible through a measurement on this qubit alone. If Eve performs a measurement on this qubit, she only disturbs the Bell state projecting it onto a pure separable state. Such a state is thus detectable from Bob's measurement statistics and Alice's knowledge of the encoded states. 
Thus, as soon as the maximally entangled state shared between Alice and Bob is verified, the ideal protocol is at least theoretically secure.

Moreover, if Eve performed a measurement on this state or replaced the qubit by a qubit of her own and sent it to Bob, the resulting two qubit state would be a separable two qubit state. As a result Eve could be detected by performing a correlation measurement on decoy qubits, similar to the procedure in most QKD protocols.

\paragraph{Experimental requirements}
Even though there is a theoretical security, experimentally this scheme poses many challenges. The requirements for a secure transmissions are: 
\begin{itemize}
	\item Maximally entangled state needs to be prepared and successfully distributed among the parties.
	\item The entangled state must be verified.
	\item The state must not decohere during the runtime of the protocol. In particular, Bob needs to be able to store his qubit during Alice's operations.
	\item A quantum channel must transmit Alice's qubit to Bob.
	\item Bell measurement has to be performed on both qubits.
\end{itemize}

These requirements are quite challenging, and often difficult to meet by current technologies. The first three conditions all depend on the possibility to create, distribute and verify a maximally entangled state. For a non-maximally entangled state the protocol does not lead to a set of mutually orthogonal set of states. In this case, Bob can only decrypt the message probabilistically. Moreover, a non-maximally entangled state will compromise the security of the protocol, as Alice's qubit then contains additional information about the encoded bits. The verification of the entanglement is thus not only crucial for Bob obtaining the correct result, but also for the security. 
The requirements on a quantum channel and its realization in practice is discussed in Sect.~\ref{sec:PhysicalBasis}". Finally, the Bell measurement requires single qubit Hadamard gates, as well as a two-qubit CNOT gate, and thus basic quantum computation elements.

\subsubsection{QSDC with tripartite states} \label{sssec:QSDC3}
So far, we have only considered the direct communication between two parties. While in general one can consider a network consisting only of peer-to-peer (P2P) connections, quantum mechanics also offers the possibility to entangle states between more than two parties at once.
Thus, we now want to extend the QSDC protocol discussed in the previous section to a multipartite system, and investigate the usefulness of such an extension. For simplicity, we first restrict ourselves to three parties, Alice, Bob and Charlie, where the first two now play the role of a sender, while Charlie is the receiver of the combined message. 

The aim of this section is to analyze whether this extension is useful for any practical purpose, and especially if an extension to a multipartite network will provide any advantage over point to point connections between two individual parties. Moreover, we want to determine any restrictions of such a shared multipartite network.

The idea of the adapted protocol is displayed schematically in Fig.~\ref{fig:qsdc-tripartite} and its procedure is as follows:
\begin{figure}[hbt]
	\centering
	\includegraphics[width=0.9\linewidth]{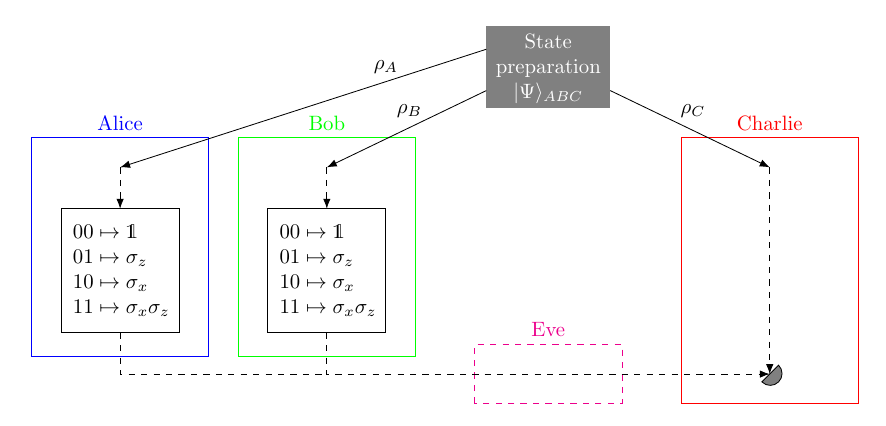}
	\caption{Schematic representation of the quantum secure direct communication protocol (QSDC) for a tripartite qubit system with two senders and one receiver. The three qubits of the tripartite state are sent to Alice, Bob and Charlie, respectively. Alice and Bob then each encode their information by acting with either no operation or one of the three Pauli operators on their respective qubit. Alice and Bob then send their qubits to Charlie, who performs a measurement on the complete three qubit state to decode the information.}
	\label{fig:qsdc-tripartite}
\end{figure}

    \begin{table}[ht]
    \centering
    \begin{tabular}{m{0.9\textwidth}}
    \hline
    \textit{A QSDC protocol for tripartite qubit states}  \\
		\begin{enumerate}
        \item The senders (Alice and Bob) and the receiver (Charlie) share a maximally entangled tripartite state $ \ket{\Psi} $.
        \item Alice and Bob each encode two bits of information into either of four different operations on their respective qubit, i.e. $ 00 \to \mathds{1} $, $ 01 \to \sigma_{z} $, $ 10 \to \sigma_{x} $ and $ 11 \to \sigma_{x}\sigma_{z} $.
        \item Alice and Bob send their respective qubits to Charlie.
        \item Bob measures the state of the entire system, in order to reconstruct the information Alice and Bob have encoded.
		\end{enumerate} \\
    \hline
    \end{tabular}
    \end{table}

The main difference from the bipartite case is that now two senders encode their bits, while only one receiver obtains the information. Form the Holevo bound it is directly clear, that since we consider a tripartite qubit state, only three bits of information can be read out by Charlie. However, the two senders each encode two bits and therefore a total of four bits. Hence, one bit of information has to be lost by this transmission even in the ideal case. 

\paragraph{Two different classes of entangled states}
For tripartite systems, i.e.\,quantum states consisting of three components, two distinct and inequivalent classes of entanglement exist. By that we mean, that the two classes of states cannot be transformed into each other by means of any local operations and classical communications (LOCC). Thus, these two classes can exhibit different behaviors under certain protocols. This is in contrast to bipartite qubit states, where only one class of entangled states exists. The maximally entangled state of these two classes are the Greenberger-Horn-Zeilinger (GHZ) state
\begin{equation}\label{eq:GHZ-state}
	\ket{\mathrm{GHZ}} = \frac{\ket{000}+\ket{111}}{\sqrt{2}}
\end{equation}
as well as the $ W $ state
\begin{equation}\label{eq:W-state}
	\ket{W} = \frac{\ket{001}+\ket{010}+\ket{100}}{\sqrt{3}}.
\end{equation}

The apparent difference between these two states is already visible when we trace out one of the three systems. Since both states are symmetric in all three subsystems, this is independent of the subsystem traced out.

For the GHZ class we immediately see that tracing out one of the subsystems directly leads us to the mixed state
\begin{equation}
	\rho_{\mathrm{GHZ,red}} = \frac{\ket{00}\bra{00}+\ket{11}\bra{11}}{2},
\end{equation}
which is a mixture of two separable states, and thus also a separable state. This behavior is similar to tracing out one subsystem of a maximally entangled Bell state, which also results in a mixed state without any entanglement.

In contrast, when we trace out one subsystem of the $W$ state, we find
\begin{equation}
	\rho_{\mathrm{W,red}} = \frac{\ket{00}\bra{00}+\left(\ket{01}+\ket{10}\right)\left(\bra{01}+\bra{10}\right)}{3},
\end{equation}
which is also mixed but remains entangled.

In the following we discuss the idea of our extended protocol for these two classes of states, separately.

\paragraph{QSDC with GHZ state}
The first class of states we discuss are the GHZ states, as defined by Eq.~\eqref{eq:GHZ-state}. The encoding operations employed by Alice and Bob in the second step of the protocol then lead to the states as given by Tab.~\ref{tab:GHZ-state-encoding}, where the state $\ket{\Psi_{abc}}$ for bit values $a,b,c$ is defined by
    \begin{equation}
    \ket{\Psi_{abc}} = \frac{\ket{a,b,0}+(-1)^c\ket{1-a,1-b,1}}{\sqrt{2}}.
    \end{equation}

\begin{table}[hbt]
    \centering
    \begin{tabular}{c|c c c c}
     Alice / Bob    & 00 & 01 & 10 & 11  \\ \hline
       00  & $\ket{\Psi_{000}}$ & $\ket{\Psi_{001}}$ & $\ket{\Psi_{010}}$ & $\ket{\Psi_{011}}$ \\ 
       01  & $\ket{\Psi_{001}}$ & $\ket{\Psi_{000}}$ & $\ket{\Psi_{011}}$ & $\ket{\Psi_{010}}$ \\
       10  & $\ket{\Psi_{100}}$ & $\ket{\Psi_{101}}$ & $\ket{\Psi_{110}}$ & $\ket{\Psi_{111}}$\\
       11  & $\ket{\Psi_{101}}$ & $\ket{\Psi_{100}}$ & $\ket{\Psi_{111}}$ & $\ket{\Psi_{110}}$
    \end{tabular}
    \caption{GHZ state after Alice and Bob performed their encoding operations on their respective qubit. The complete three qubit state transforms into 8 different mutually orthonormal states. Thus, by performing a measurement Charlie can recover three bits with certainty. These three bits correspond to the first bit of each Alice and Bob, as well as the parity bit of the second bit of Alice and Bob. Due to the Holevo bound it is clear, that three bits is the maximal number of bits such a protocol with a tripartite qubit system can transmit. In this sense, this protocol for the GHZ state is ideal.}
    \label{tab:GHZ-state-encoding}
\end{table}
As a result, we find eight different orthonormal states. Charlie can distinguish these states perfectly and thus obtains three bits of information. As discussed above, this is the maximal information accessible due to the Holevo bound of a tripartite qubit system. Hence, this protocol is ideal regarding the amount of transmittable data, and no other encoding would lead to an improvement. 

When we take a closer look at the encoding Tab.~\ref{tab:GHZ-state-encoding}, we find that Charlie is able to reconstruct the first bit of both Alice and Bob, while he can only reconstruct the parity, but not the individual second bits of both senders. Hence, the protocol has the possible advantage that Alice and Bob can send parity check bit via the channel to Charlie, without him knowing the individual bits.
\begin{table}[hbt]
	\centering
    \setlength\tabcolsep{0pt}
	\begin{tabular*}{\linewidth}{@{\extracolsep{\fill}} c|c c c c}
		A/B & 00 & 01 & 10 & 11  \\ \hline
		00  & $\ket{00}\bra{00}+\ket{11}\bra{11}$ & $\ket{00}\bra{00}+\ket{11}\bra{11}$ & $\ket{01}\bra{01}+\ket{10}\bra{10}$ & $\ket{01}\bra{01}+\ket{10}\bra{10}$ \\ 
		01  & $\ket{00}\bra{00}+\ket{11}\bra{11}$ & $\ket{00}\bra{00}+\ket{11}\bra{11}$ & $\ket{01}\bra{01}+\ket{10}\bra{10}$ & $\ket{01}\bra{01}+\ket{10}\bra{10}$ \\
		10  & $\ket{01}\bra{01}+\ket{10}\bra{10}$ & $\ket{01}\bra{01}+\ket{10}\bra{10}$ & $\ket{00}\bra{00}+\ket{11}\bra{11}$ & $\ket{00}\bra{00}+\ket{11}\bra{11}$ \\
		11  & $\ket{01}\bra{01}+\ket{10}\bra{10}$ & $\ket{01}\bra{01}+\ket{10}\bra{10}$ & $\ket{00}\bra{00}+\ket{11}\bra{11}$ & $\ket{00}\bra{00}+\ket{11}\bra{11}$
	\end{tabular*}
	\caption{State accessible to Eve, when Alice and Bob send their qubits to Charlie over an open quantum channel. In contrast to bipartite state, where each state is the maximally mixed single qubit state, and therefore no information about the operations can be accessed, here we find either of two orthogonal bipartite states. Thus, Eve can gain one bit of information about the bits encoded by Alice and Bob.}
	\label{tab:GHZ-state-Eve}
\end{table}
In order to investigate the security of the protocol with the GHZ state, we refer to Tab.~\ref{tab:GHZ-state-Eve}. Here, the combined state of Alice and Bob's qubits sent through the quantum channel to Charlie is displayed for all possible encodings. In contrast to the bipartite case, where all states reduce to the maximally mixed state, and are therefore indistinguishable for Eve, in the GHZ case, the eavesdropper has one of two possible mixed states, depending on the encoded state. Since these states are orthogonal to each other, Eve can theoretically divide the states into two different classes without being detected. From Tab.~\ref{tab:GHZ-state-Eve}, we determine that Eve actually can learn the parity of the first bit of Alice and Bob.

\paragraph{QSDC with W state}
For the $ W $ state we now apply the same protocol as for the GHZ state. The aim is to investigate the differences of this choice of state. When Alice and Bob perform their encoding operations, the three qubit state is transformed according to Tab.~\ref{tab:W-state-encoding}. 
\begin{table}[hbt]
	\centering
    \setlength\tabcolsep{0pt}
	\begin{tabular*}{\linewidth}{@{\extracolsep{\fill}} c|c c c c}
		A/B    & 00 & 01 & 10 & 11  \\ \hline
		00  & $\ket{W}$ & $\ket{001}+\ket{010}-\ket{100}$ & $\ket{101}+\ket{110}+\ket{000}$ & $\ket{101}+\ket{110}-\ket{000}$ \\ 
		01  & $\ket{001}-\ket{010}+\ket{100}$ & $\ket{001}-\ket{010}-\ket{100}$ & $\ket{101}-\ket{110}+\ket{000}$ & $\ket{101}-\ket{110}-\ket{110}$ \\
		10  & $\ket{011}+\ket{000}+\ket{110}$ & $\ket{011}-\ket{110}+\ket{000}$ & $\ket{111}+\ket{100}+\ket{010}$ & $\ket{111}-\ket{010}+\ket{100}$\\
		11  & $\ket{011}-\ket{000}+\ket{110}$ & $\ket{011}-\ket{000}-\ket{110}$ & $\ket{111}-\ket{100}+\ket{010}$ & $\ket{100}+\ket{010}-\ket{111}$
	\end{tabular*}
	\caption{State after Alice and Bob performed their encoding operations. In contrast to the GHZ state, the W state does not transform into eight different mutually orthogonal states. Thus, it is not possible for Charlie to discriminate among the different states. Instead, the resulting states only belong to either of two separable subspaces, allowing for the detection of at least one bit. Numerically, we find that the Holevo bound for the maximal decodeable information is given by 2.91 bits, which is slightly lower than the ideal value of 3 bits for the GHZ state.}
	\label{tab:W-state-encoding}
\end{table}

As we can directly see from the results, the states produced are no longer orthogonal to each other as in the GHZ case. Instead, while all states are mutually different from one another, the states only fall into two separated subspaces of the complete Hilbert space. 

In contrast to the GHZ case, the accessible information is no longer easily determined from this table. Instead, we use the Holevo bound to determine an upper bound for this ensemble, where we assume that all encodings happen with the same probability. In this case, the Holevo bound is given numerically by
\begin{equation}\label{eq:W-Holevo}
	\chi = 2.91 \ \mathrm{bits}
\end{equation} 
as the maximally deduceable information. Hence, as expected less then 3 bits of information are accessible through this scheme with the $W $ state. 

The question of whether we can improve this protocol by choosing different encoding bases for Alice and Bob is quite difficult to investigate analytically. However, numerical simulations have not provided any encoding bases that improves the value given by Eq.~\eqref{eq:W-Holevo}. Thus, we conclude that the accessible information for a $ W $ state cannot be improved over this limit.

\begin{table}[ht]
	\centering
	\renewcommand{\arraystretch}{2}
	\begin{tabular}{c|c c c c}
		A/B    & 00 & 01 & 10 & 11 \\ \hline  
		00  & $\displaystyle\frac{1}{3}\rho_{00}+\frac{2}{3}\rho_{\Psi^{(+)}}$ & $\dfrac{1}{3}\rho_{00}+\dfrac{2}{3}\rho_{\Psi^{(-)}}$ & $\dfrac{1}{3}\rho_{10}+\dfrac{2}{3}\rho_{\Phi^{(+)}}$ & $\dfrac{1}{3}\rho_{10}+\dfrac{2}{3}\rho_{\Phi^{(-)}}$  \\
		01  & $\dfrac{1}{3}\rho_{00}+\dfrac{2}{3}\rho_{\Psi^{(-)}}$ & $\dfrac{1}{3}\rho_{00}+\dfrac{2}{3}\rho_{\Psi^{(+)}}$ & $\dfrac{1}{3}\rho_{10}+\dfrac{2}{3}\rho_{\Psi^{(-)}}$ & $\dfrac{1}{3}\rho_{10}+\dfrac{2}{3}\rho_{\Phi^{(+)}}$  \\
		10  & $\dfrac{1}{3}\rho_{01}+\dfrac{2}{3}\rho_{\Phi^{(+)}}$ & $\dfrac{1}{3}\rho_{01}+\dfrac{2}{3}\rho_{\Phi^{(-)}}$  & $\dfrac{1}{3}\rho_{11}+\dfrac{2}{3}\rho_{\Psi^{(+)}}$ & $\dfrac{1}{3}\rho_{11}+\dfrac{2}{3}\rho_{\Psi^{(-)}}$ \\
		11  & $\dfrac{1}{3}\rho_{01}+\dfrac{2}{3}\rho_{\Phi^{(-)}}$ & $\dfrac{1}{3}\rho_{01}+\dfrac{2}{3}\rho_{\Phi^{(+)}}$ & $\dfrac{1}{3}\rho_{11}+\dfrac{2}{3}\rho_{\Psi^{(-)}}$ & $\dfrac{1}{3}\rho_{11}+\dfrac{2}{3}\rho_{\Psi^{(+)}}$
	\end{tabular}
	\caption{States accessible to Eve, when Alice and Bob send their qubits to Charlie over an open quantum channel. Here, the resulting states are combinations of the density operators  $\rho_{ij}=\ket{ij}\bra{ij}$ denoting separable states, and $\rho_{\Psi^{(\pm)}}= \ket{\Psi^{(\pm)}}\bra{\Psi^{(\pm)}}$ as well as $\rho_{\Phi^{(\pm)}} = \ket{\Phi^{(\pm)}}\bra{\Phi^{(\pm)}}$, which are the maximally entangled Bell states. The Holevo bound for these states is given by about $ 1.09 $bits, thus containing slightly more information than the GHZ protocol.}
	\label{tab:W-state-Eve}
\end{table}
We finally take a look at the security of this protocol. Analogously to the GHZ state, we therefore calculate the reduced system, where we traced out Charlie's qubit, as it is not accessible to Eve, when Alice and Bob send their qubit over an open quantum channel. The resulting states, depending on the operation of Alice and Bob are given in Tab.~\ref{tab:W-state-Eve}. As a result, the Holevo bound on the information Eve can reconstruct from the reduced bipartite mixed state over all possible results for Alice and Bob's qubits, where we assume that all encodings are equally likely, is 
\begin{equation}
    \chi_{E} = 1.09 \ \mathrm{bits} \,,
\end{equation}
which is slightly larger than one bit, and therefore larger than for the GHZ case.

\subsection{Conclusion} \label{ssec:InterfaceConclusions}
We have encountered several ways to encode a sequence of bits in a collection of qubits and have discussed the significance and impact of some important no-go theorems like the no-cloning theorem and the Holevo bound.
In particular, the latter specifies the limit for the bit-wise transfer rate for information by restricting the number of bits one can retrieve from a quantum system to one bit per qubit both when a clear reconstruction of the encoded information is needed and when a certain probability of error is allowed.
Moreover, we have seen that we cannot avoid those restrictions by adopting a relaxed interpretation of the word reconstruction.
Finally we took a look at how entanglement can be used for secure communication and which experimental requirements need to be fulfilled for such a communication scheme.
We explored quantum secure direct communication both with two and with three parties, and realized that the case of three parties does not offer a substantial advantage over peer-to-peer communication with two-party QSDC.

\pagebreak
%%%%%%%%%%%%%%%%%%%%%%%%%%%%%%%%%%%%%%%%%%%%%%%%%%%%%%%%%%%%%%%%%%%%%%%%%%%%%%%%%%%%%%%%%%%%%%%%%%%
%%%%%%%%%%%%%%%%%%%%%%%%%%%%%%%%%%%%%%%%%%%%%%%%%%%%%%%%%%%%%%%%%%%%%%%%%%%%%%%%%%%%%%%%%%%%%%%%%%%

\section{Physical basis of quantum information transfer} \label{sec:PhysicalBasis}
In the previous sections, we have discussed the basics of quantum information and communication from a point of view which almost exclusively relied on an abstract mathematical language of quantum states, density matrices, and qubits. Before discussing how these concepts can be used in applications, it is time to consider how quantum information is realized in actual physical systems, which different platforms are used to manipulate and store it, and most crucially how it is transferred between different parties.

Obviously, this is a vast field that covers virtually all areas of physics and ranges from fundamental physics questions to technological and material science topics. While many relevant topics may only be touched upon or even be left out completely, what this section will provide is (i) a brief overview of realizations and platforms, which can in principle be used for the transfer of quantum information, and (ii) some insight in the current state of the art for the most prominent systems and technologies.
Thereby, this section aims to equip the reader with some intuition on how experimentalists proceed to realize the abstract concepts and schemes discussed in other sections and thus gain some rough idea how long, how difficult, and how complex the technological advance required for a specific application may be.

To achieve this, we will start the section by approaching the question, how to transfer quantum information, from an unbiased, naive point of view, i.e., we will get the ball rolling with such simple, obvious questions as
\begin{itemize}
    \item[-] Presumably there need to be some kind of quantum information carrier: What possible physical form could these carriers take? What are the patent (dis)advantages of each carrier?
    \item[-] How are carriers actually moved between two distant points A and B?
    \item[-] What are these local ``nodes" A and B and what is happening there? How is quantum information stored there and how are carriers and storage coupled?
\end{itemize}

By tackling these questions, the naively proposed variety of carriers and platforms will boil down to the most prominent ones: photons in the optical or microwave frequency range. Concentrating on the latter platform, we will then discuss in greater depth a number of fundamental experiments. These constitute crucial building blocks for a minimal quantum network and at the same time illustrate the generic steps towards implementation other platforms also have to grapple with.

\subsection{Carriers of quantum information} \label{ssec:carrier}
The most common carriers of information in our classical, everyday world are electromagnetic waves (the various radio bands or microwaves
 for mobile phone networks, but also light for our visual perception). They can carry both analogue/continuous and discrete/digital information. Less common in the modern world is communication using a massive carrier medium (i.e. a letter), but we can easily imagine to transfer massive black or white balls to transmit classical bits. 
 While the quantum mechanical wave-particle duality precludes strict distinctions for quantum information, it can, nonetheless, be useful to group within this continuum methods that are better imagined in the image of a wave or a particle. As we will see below from the example of visible light/photons, which picture actually makes sense may depend not only on the carrier of information but also on the encoding.

Before opening the discussion on individual carriers of quantum information, two further remarks are in order:
\begin{itemize}
\item Certain carriers may be useless for long-range transmission of quantum information and still play a crucial role in hybrid systems and, for instance, constitute a bus between different platforms for memory qubits, processing qubits, and flying qubits.
\item Generally speaking, a carrier is suitable for quantum information if it can maintain a quantum coherent state for a long time, respectively over a long distance,  against all environmental influences causing decoherence (or even more damaging may effectively remove the carrier by scattering or absorption). Generically, the following properties of a specific platform are relevant: the energy or frequency; the strength and nature of coupling to the environment, e.g., through magnetic moment or charge; and the coupling of the degree of freedom carrying the quantum information to potential internal degrees of freedom. For wave excitations in a medium or on an interface, microscopic characteristics of the wave, such as dispersion relations and the lifetimes of quasiparticles are furthermore important. 
\end{itemize}

In the following sections, we will list all carriers, we can possibly imagine to play any role in quantum information transfer, be it short- or long-term or in far-fetched futuristic visions. We will briefly elaborate on each option to discuss specific advantages or problems, to find specific application niches, or to wholly discard them.   

\subsubsection{Waves and massless particles as wave excitations} 
Electromagnetic waves are not only the most ubiquitous classical information carrier, but also the most advanced option for quantum information: Namely, we will comprehensively discuss photons in the visible/infrared and the microwave range, which are widely used, while the more extreme ends of the electromagnetic spectrum are less promising. As the only other types of waves with some possible relevance to the transfer of quantum information, we will briefly comment on spin and on acoustic waves.

\paragraph{Optical photons \label{subsubsec:optical_photons}}
Two properties of an electromagnetic wave mainly govern, if and how it is suitable to carry quantum information: its carrier frequency and the degree of freedom, in which information is encoded. We will discuss their impact on the example of optical photons before later comparing them with the microwave regime. 

\subparagraph{Frequency} 
The frequency of an electromagnetic wave or, correspondingly, the energy of a single photon, $E=\hbar \omega = h \nu$, is crucial for a number of reasons:
\begin{itemize}
    \item Frequency determines the \emph{transmission properties} through open-air, see Sect.~\ref{subsec:free_propagation}, or other media or through waveguides. Low absorption and scattering allows long-distance transfer of visible and infrared light through open air, but there are limitations under certain atmospheric conditions (clouds) and underwater. This is in contrast to microwaves and radio frequencies traversing clouds, and to very low-frequency EM waves penetrating better through water. Glass fibers of various types (to be discussed in Sect.~\ref{subsubsec:optical_fiber} below) offer high-quality, low-loss waveguides in the visible and even more so in the infrared (``telecom wavelengths"). 
    
    \item The \emph{thermal background} noise in any transport channel is also affected by frequency. Specifically, there is an unavoidable thermal contribution from the Bose-Einstein occupation specifying the number of photons in a mode of frequency $\omega$ in equilibrium. One big advantage of optical photons over microwaves is that such thermal effects can be neglected at room temperature, since $\hbar \omega \gg k_B T$ at optical frequencies.
    
    \item \emph{Compatible energy scales}: The interface of the quantum carrier of information with local nodes (be it for storage, creation, or manipulation of information) is generically most efficient, strong, and easy for matching energy scales. Optical radiation thus is not optimally suited to directly couple to superconducting qubits, which naturally operate on microwave energies and interact with microwave photons. However, note that the bridging of energy scales by transducers is feasible and an important active field of research, as will be extensively discussed in Section \ref{ssec:hybrid} on hybrid systems. Optical coupling to other qubits (often with the aim of realizing quantum repeaters \cite{neuwirth2021quantum,sangouard2011quantum}) has already been shown in many experiments, for instance \cite{canteri2024photon,knaut2024entanglement}.
    
    \item \emph{Devices for manipulation and creation}: In a similar fashion, the frequency or wavelength of the electromagnetic carrier dictates which devices can be used to direct the flow of information and how the carrier is originally generated. For the short wavelengths of visible light, linear (geometric) optics devices allow easy focusing, reflection, beam-splitting, $\hdots$, while, for instance, free-space microwave communication may have to consider directional antenna, and x-ray frequencies require extremely complex setups for lensing. The same applies for waveguide propagation and the corresponding fundamental differences in coupler- or router-design depending on the frequency. Concerning sources, frequencies will often be chosen based on the availability of good lasers for coherent light, or high-quality single-photon or photon-pair sources.      
\end{itemize}

\subparagraph{Degree of freedom used for information encoding\label{par:photonic_encodings}}
Optical electromagnetic waves offer various degree of freedoms which can be used to encode quantum information. One issue to consider in choosing the encoding is what kind of quantum information is to be transferred, e.\,g., a collection of qubits, qudits, or a harmonic oscillator state with its basis of an infinite number of Fock states. For the transfer of qubits typically single photons are considered and most often the spin degree of freedom, respectively the \emph{polarization} (horizontal/vertical or left/right circular) is chosen. Another option is the \emph{temporal} waveform with time-bin encoding, where a qubit is given by the quantum mechanical superposition of a photon at an early/late time, or a dual-color scheme with a superposition of a photon at two different frequencies \cite{Lu:23}. The \emph{spatial} degree of freedom can be used in the form of orbital angular momentum states \cite{yao2011orbital,mair2001entanglement,Yao:06,chen2018mapping,zhou2015classical,Pecoraro_PRA_2019,Mirhosseini_2015}, or in so-called dual-rail or which-path encodings. Less common for optical waves (but common in the microwave domain, see \ref{ssec:realization} below) is using the quantum mechanical state of a particular electromagnetic mode, which can be seen in quantum optics as a harmonic oscillator state occupied by a superposition of, e.g., one and three photons.

\paragraph{Microwaves}
Turning now to the suitability of microwaves for information transfer, we can consider the same questions tackled above for optical electromagnetic waves. Some of these arise already for classical information transfer (in which microwave frequencies are prominently used), while others are specific for the quantum case.

\subparagraph{Frequency} 
We can again consider the four aspects related to frequency itemized above.
\begin{itemize}
\item \emph{Transmission properties: } Compared to the optical, microwaves have some advantage in free-space transmission through difficult atmospheric conditions, such as clouds or fog. Waveguides (though more bulky than optical fibres) exist both in multiple designs for room temperature transport (where they are optimized in radar engineering) and for cryogenic temperatures, where these are supplemented by various form of superconducting on-chip structures. 

\item \emph{Thermal background} radiation is more detrimental for microwaves due to frequencies being about three order of magnitudes smaller than in the optical case. Its effect can mostly be escaped by operating at milli-Kelvin temperature ranges. Consequently, the creation, storage, or readout of quantum information in microwave signals is performed within a cryogenic environment. Transfer has been demonstrated in specifically designed similarly cooled waveguides over distances of tens of meters. Free-space and room temperature transfer may be challenging but not infeasible \cite{casariego2023propagating,gonzalez2022open}. 

\item The \emph{compatibility} of microwaves with the natural energy scales of excitations of superconducting qubits allowing for easy, direct coupling makes microwave photons particularly attractive for carrying quantum information. 

\item There exist a variety of tools for \emph{manipulating} the transfer of quantum information via microwaves in a cryogenic environment, some of them will be discussed in the context of specific experiments in Sect.~\ref{ssec:realization} below. Corresponding devices operating at room temperature and for free-space communication are not always designed for a low-power, quantum signal. Specifically concerning sources of quantum signals low-temperature superconducting electronics devices offer generically stronger nonlinearities than optical devices boosting functionality and efficiency.
\end{itemize}

\subparagraph{Degree of freedom used for information encoding} 
There is slightly less variety in the degree of freedom of microwaves used for carrying quantum information. All the experiments we discuss below use the occupation of a harmonic oscillator mode. Experiments on quantum teleportation and Bell tests used quadratures of two-mode squeezed light similarly to the way polarization is used in the optical regime. Spin (polarization) and orbital angular momentum are typically not usable for quantum information transfer at microwave frequencies. The use of time-bin \cite{kurpiers2019quantum}, dual frequency and dual-rail encoding also seems feasible.

\paragraph{Electromagnetic waves at other frequencies} 
The prominent regimes of visible, infrared and microwave frequencies stretch across a wide range in the middle of the electromagnetic frequency band (with the notable gap at terahertz frequencies).
Going beyond this range to the extremes of the spectrum will presumably amplify the comparative strength and weaknesses of optical and microwave photons as carriers. Radio-waves at even lower frequencies than microwaves are well suited for classical communication and may even outperform microwaves under certain transmission conditions. However, lower frequencies come with more background radiation issues and inescapable thermal noise even at the lowest temperature affecting the creation of single-photon or other quantum states. To our knowledge, there are no current proposals for quantum information transfer based on RF frequencies. Higher than visible frequencies, such as x-ray, may transmit through material and for very long distances, but health hazards and the complexities of x-ray optics lessen their appeal for information transfer applications. 

Electromagnetic waves are not the only type of physical waves that can carry information. While they are distinguished by the fact that they can exist without a carrier, so that they can truly propagate through free space, this is by no means crucial for a wave's fundamental suitability to carry classical or quantum information and, indeed, in many applications electromagnetic waves propagating in a carrier medium or along a carrier waveguide are used. While it turns out that they are less relevant for quantum information transfer, for completeness, we will briefly comment on some other types of waves. 

\paragraph{Spin waves and magnonics}
Chains of coupled spins allow for low-energy excitations, which can propagate coherently for distances much longer than their wavelength with little dissipation. Many theory proposals have investigated the role of such chains for quantum state transfer between localized distant spin states \cite{Christiandl_PRL_04,bose2007quantum}. Mostly applications are seen as short-distance link or quantum bus mediating between different quantum registers of a quantum computer. Typically, proposals envision engineered ordered chains acting as fixed wire-like connections, while there are also alternatives, such as an interesting proposal that considers linking distant NV-centers via a disordered chain \cite{Yao_PRL_11}.

Spin waves may also have practical relevance for quantum networks as part of hybrid systems. In the emerging field of magnonics \cite{Barman_JOPCM_21}, which deals with the propagation and manipulation of spin-wave excitations (magnons) through periodic magnetic media, an important research direction is the coupling of magnons to other excitations and their consequent use as transducer of interaction and information between different platforms and possibly different types of qubits \cite{Lachance-Quirion_APX_19}. For instance, coherent coupling between magnons and microwave photons was realized \cite{Zhang_PRL_14} and exploited using a ferrimagnetic crystal of yttrium iron garnet (YIG) for the detection of a single magnon \cite{lachance2020entanglement}. An other proposal envisions hybrid quantum networks with magnonic and mechanical nodes \cite{Li_PRXQ_21}.  

\paragraph{Sound waves}
Sound waves (or their quanta called phonons) are a further familiar carrier of classical information, most plainly in humans' oral face-to-face communication. For purposes of quantum information transfer, sound can play a role in the form of surface acoustic waves (SAWs).

\subparagraph{Surface acoustic waves\label{par:SAW}}
Quantized wave packets of a surface acoustic wave traveling across a piezoelectric substrate have recently gained substantial interest as carriers of quantum information.
In experiments, very similar to the microwave quantum state transfer experiments we will describe in detail below in Sect.~\ref{ssec:realization}, such phonons can be controllably excited by one superconducting qubit, travel coherently over a distance much larger than their wavelength, and be routed to another qubit which is then excited. Both node-to-node transfer of quantum states and entanglement generation have been demonstrated \cite{Bienfait_Science_19,Dumur_NPJQI_21}. While similar to microwave photons in energy, propagation speed and wavelength are shorter, offering some practical advantage for controllability. The quantum information will typically be encoded in the state occupation of a certain itinerant mode, i.e., in the superposition of the mode being occupied or not rather than in an internal degree of freedom. An extensive toolbox of devices for manipulation has been successfully developed \cite{aref2016quantum}, owing largely to the strong coupling in piezoelectric materials. 

SAW phonons can also be confined in resonators, realizing the acoustic equivalent of circuit quantum electrodynamic \cite{manenti2017circuit} and allowing transfer of many ideas for quantum information processing from this field, and such resonators have been proposed as local storage nodes of quantum information \cite{satzinger2018quantum}. An alternative approach is the use of SAW phonons as flying qubits, as demonstrated by the creation of a time-bin encoded phononic qubit in \cite{Zivari_SciAdv_22}. Ultimately, schemes from optical quantum information processing may be copied in the SAW domain.

As in some of the cases discussed here, the long-term potential lies not in long-distance communication, but in short-distance interconnects, either directly between superconducting qubit registers within a quantum computer, or between qubits of different type in hybrid setups and transducers, see Sect.~\ref{ssec:hybrid}. SAW phonons can couple to NV centers \cite{bainbridge2023} and to optical quantum dots \cite{weiss2018interfacing,choquer2022quantum}. In fact, their wavelength in the $\sim 600\,\textrm{nm}$ regimes matches the optical wavelengths. 

Besides being the carrier of quantum information, surface acoustic waves can also play the role of an active medium to transport quantum information encoded in an electron \cite{Hermelin_Nature_11,mcneil2011demand,Takada_NatCom_19,Bertrand_NatNano_16}.

Finally, and somewhat disconnected from the other scenarios discussed in this section, phonons or more generally vibrational modes can be seen as carriers of information in the role they play as mediating coupling between qubits in certain types of linear ion traps \cite{wright2019benchmarking}.

\subsubsection{Massive particles}
Massive particles can also be used to store quantum information in various degrees of freedom, and, as we will discuss below, there are prominent and advanced quantum computing platforms using them. As such, they could also constitute the nodes of a quantum network. While long distance information transfer (or entanglement distribution followed by teleportation \cite{Daiss_Science_21,olmschenk2009quantum}) between such networks may still be done via photons, there is also the `natural' possibility to move the quantum information simply by moving the massive particle. As the massive particles used tend to interact more strongly with the environment than a photon, and as they obviously move much more slowly than photons, such transfer is restricted to shorter distances. Nonetheless, this type of transport can play an important role within a single quantum information processing unit, it may connect several modules of a quantum computer, or it may serve as an intermediate step in a hybrid setup.

\paragraph{Ions or neutral atoms\label{subsubsec:ions}}
Both ions and neutral atoms have been employed in various platforms for quantum information processing (see \ref{subsubsec:massive}), where they typically excel by their long coherence time. Since first experiments showed that these coherence properties can be preserved, when the particles are moved around in properly careful fashion, such shuttling has been an important part of proposed and realized quantum processing units. 

For neutral atoms, optical tweezers were used to move qubits within a single processing array \cite{beugnon2007two}, providing the ability to dynamically change interconnectivity and thereby
opening a path towards scalable quantum processing \cite{bluvstein2022quantum}. Truly free-flying massive qubits have also very recently been realized, when optical tweezers were used to `throw and catch single atoms' \cite{Hwang_Optica_23}.

For ions, shuttling can be used to entangle different regions within a trap for later teleportation \cite{wan2019quantum}. In complex 2D or 3D architectures they can provide connectivity on a large scale \cite{wan2020ion,pino2021demonstration} in integrated devices, and even a link between different chip modules has been demonstrated \cite{akhtar2023high} and is seen as a crucial breakthrough step towards commercial quantum computing \cite{lekitsch2017blueprint}. From the big-picture viewpoint on quantum information transfer, this covers extremely small distances but may be at the leading edge concerning fidelities and highly promising transfer rates. For example, the company ``Universal Quantum" claimed\footnote{Press release on the topic available at:\\ \url{https://universalquantum.com/knowledge-hub/universal-quantum-develops-key-enabler-of-million-qubit-quantum-computer}} $2400$ qubits/s and a fidelity of $99.999993\%$. 

\paragraph{Electrons and electron-like excitations\label{subsubsec:electrons}}
The situation for electrons is very similar to the one for ions and neutral atoms discussed here.  Again, the electron with its internal spin-degree of freedom constitutes a natural candidate for encoding quantum information, and from the very beginning of quantum information technology there have been efforts to make use of electrons as stationary qubits, typically by confining electrons in various types of ``quantum dot" structures in semiconductors (see Sect.~\ref{subsubsec:QD}) where spin-states (or sometimes other degrees of freedom) are used to encode quantum information. Such quantum dots and more complex configurations thereof can then also form the node of a quantum network, and transferring the quantum information between nodes by physically moving the electron is an obvious way to interconnect them. Again, as for ions and neutral atoms, while long-range connections are more feasible using various photonic schemes, short-range interconnects between different processing units, oftentimes called ``bus" in computing architectures, can be realized without the overhead of electron-photon forth- and back-transducers, if the electron is used as a moving carrier of quantum information.

Shuttling of electrons from one quantum dot to a distant one is possible without severely affecting the information encoded in its spin degree of freedom, limited mostly by the requirement to shuttle sufficiently fast as compared to the decoherence times of the stationary qubits. One scheme, evocatively called the ``bucket brigade approach", is transferring an electron through an array of quantum dots, swapping between neighboring dots by moving electronic states energetically up and down. It reached a rate three orders of magnitude faster than the dephasing times in Si \cite{Mills_NatCom_19}. A second scheme, dubbed ``conveyor-mode single-electron shuttling", requires less control lines and signals to define a dynamically moving quantum dot \cite{seidler2022conveyor,kunne2024spinbus}.

Others recently also propose the possibility of considering an electron flying qubit not merely as an information-carrying channel, but to manipulate or process the information on the fly \cite{Edlbauer_EPJ_22,Yamamoto_NatNano_12}. This is similar to some optical quantum computing proposals, where interference, statistics, and measurement are combined (Manifestly, electrons differ from photons in their fermionic statistics and their direct interaction). 

A recent review \cite{Edlbauer_EPJ_22} discusses three types of electrons flying somewhat more freely than in the shuttling scenarios above: hot ballistic electrons \cite{fletcher2013clock} typically in two dimensional electron gases (2DEG) or quantum Hall edge channels, minimal-excitation states of the Dirac-sea called Levitons
\cite{dubois2013minimal} in 2DEGs, edge states of graphene, and most prominently (though also less distinct from shuttling) the use of surface acoustic waves to emit and transfer electrons \cite{Hermelin_Nature_11,mcneil2011demand,Takada_NatCom_19} while conserving their spin state \cite{Bertrand_NatNano_16}. Typically, channels defined in conventional two-dimensional electron gases or quantum Hall edge channels are used, but more exotic proposals such as manipulating electrons floating on superfluid liquid Helium with SAWs exists \cite{Byeon_NatCom_21}.

\subsection{Channels for the transfer of quantum information} \label{ssec:Channels}
For the two most prominent possible carriers of quantum information, photons in the visible/infrared or microwave regime, the requirements for the transmission of quantum signals
are not very different from those for a low-loss classical channel. They are merely more stringent because the classical solution to overcome loss by amplification or by redundancy is restricted in the quantum or single-photon regime, for instance, by the no-cloning theorem and the technological difficulties of installing relay stations, namely quantum repeaters. Nevertheless, huge progress has been made in the field, mostly focused on the technological development of quantum key distribution, where free-space atmospheric, satellite, and fiber-based solutions for optical frequencies are becoming available. Here, we will only briefly touch the extensive field of QKD, but also comment on lesser-known channels to give a concise but rather complete view.

For some recent reviews covering both experimental state-of-the-art and conceptual advances on quantum teleportation, see, e.g. \cite{hu2023progress,erhard2020advances,pirandola2015advances}.

\subsubsection{Free propagation\label{subsec:free_propagation}}
Electromagnetic waves, and hence photons carrying quantum information, can freely propagate without a carrier medium, but can also be transferred through gas and liquids (or even some dense solids) with a certain attenuation strongly dependent on their frequency. For both optical and microwave photons, we will thus first consider the most relevant case of transmission through the low atmosphere as relevant for line-of-sight links, then comment on satellite communication, which can combine vertical transmission through the atmosphere with the near empty-space transfer between satellites (the latter could in principle for some applications also be replaced by the physical movement of a single satellite, if long-term quantum memories were available). Finally, water is the only other medium widely considered for the case of underwater communication.  

\subsubsection{Propagation through the atmosphere - line-of-sight versus satellite}
\paragraph{Optical photons}
Line-of-sight quantum communication via optical photons has reached considerable distances of more than 100 km \cite{ursin2007entanglement}.
The degrading effects of various mechanisms, such as beam wander, diffraction and turbulence-induced beam spreading, and background noise have been widely studied, see  \cite{klen2023numerical,ghalaii2022quantum,pirandola2021limits} for some examples. They depend strongly on wavelength and atmospheric conditions, such as humidity, dust or fog, and altitude. Indeed, in \cite{ursin2007entanglement} a low attenuation for a wavelength of $710\,\textrm{nm}$ of $0.07\,\textrm{dB/km}$ at higher altitudes of $\sim 2000\,\textrm{m}$  was exploited. Although losses are not purely absorptive, one is ultimately limited by an exponential dependence on transmission distance.

As the effective depth of the atmosphere for vertical ground-to-satellite communication is only about $10\,\textrm{km}$ a global quantum network could be based on optical quantum communication via satellites \cite{pirandola2021satellite,kaltenbaek2021quantum}. Satellite-to-ground quantum key distribution is realized by the Micius satellite orbiting at around $500\,\textrm{km}$ altitude \cite{Liao.2017,lu2022micius}. Loss estimates given in \cite{Liao.2017} are at $1200\,\textrm{km}$ distance 22 dB due to diffraction, 3-8 dB due to atmospheric absorption and turbulence, and less than 3 dB due to pointing error. Crucially, the QKD experiment uses a downlink protocol that reduces beam spreading compared to an uplink. Good atmospheric conditions and night-time operation remain important issues, see, e.g., \cite{abasifard2024ideal} for a discussion of optimizing daylight operations.

\paragraph{Microwave regime}
Free-space propagation of microwaves is well established for current classical communication tasks such as mobile and wireless local area networks. It benefits from very low atmospheric absorption losses of $6.3 \times 10^{-6}\,\textrm{dB/m}$  
in clear weather conditions and suffers less from unfavorable conditions such as fog and clouds than optical waves \cite{fesquet2023perspectives}. 
However, its quantum performance is limited by the background noise of thermal photons (the Bose-Einstein occupation yields about $1000$ photons of $5\,\textrm{Ghz}$ at room temperature, while these are completely negligible for optical frequencies). As the natural carrier frequency to couple to superconducting quantum devices operating in the same energy range, microwave quantum transfer and entanglement distribution have received interest for free space and guided transmission \cite{casariego2023propagating}. The in- and out-coupling from the cryogenic environment, in which the superconducting devices are operating, to free space and the careful antenna design are an understudied issue and experiments in the quantum regime are lacking. A first theory study \cite{gonzalez2022open} estimates a maximum distance of approximately 500 m for direct entanglement transmission in open air with parameters feasible for state-of-the-art experiments using an optimized antenna design \cite{gonzalez2022coplanar} and loss modeling, see also \cite{gonzalez2024wireless}. For inter-satellite communication with orbits, where background radiation is reduced to the 3 Kelvin of the cosmic microwave background, geometric losses become crucial, and large emitter and receiver antenna will be needed for long distances. Experimentally, an all-cryogenic, short distance experiment demonstrating microwave single-shot quantum key distribution \cite{fesquet2024demonstration}  
uses added noise to prove the feasibility of secure microwave quantum communication. The authors conclude that current technology allows open-air communication for local networks (up to $\sim 80\,\textrm{m}$) and estimate a reach of more than 1000 m for commercial superconducting cables under cryogenic conditions.

\subsubsection{Water \label{subsec:water}}
Underwater environments pose challenges for communication with electromagnetic waves, in general. Specifically, typical radio signals from satellites or planes can penetrate only a few meters of water. Seawater propagation in the optical regime is limited both by attenuation (which is about three orders of magnitude stronger than in the atmosphere) due to absorption and scattering by water molecules, suspended particles, and by air bubbles close to the surface; as well as by refractive index variations caused by random fluctuations of density, salinity, and temperature. 
Nevertheless, quantum communication in seawater is actively pursued and QKD experiments have been performed over tens of meters \cite{ji2017towards,feng2021experimental}; see \cite{Shelar_2023} for a review and \cite{zuo2021security} for modeling and performance analysis. The extreme environment of deep sea may, somewhat surprisingly, offer some advantages, as it is very quiet with respect to fluctuations in those parameters that cause reflective index changes, and it may actually be a favorable environment for preserving quantum information encoded in orbital angular momentum states of photons \cite{chen2020underwater}.
To demonstrate the feasibility of a quantum link between two different media, \cite{chakravartula2020implementation} performed quantum teleportation of photons across an air--water interface. 

\subsubsection{Guided transmission}
Instead of free propagation, guided transmission can be used for various electromagnetic waves. This has the generic advantage of nearly completely avoiding any spreading out of the transmitted energy from the target region. For free propagation this is caused in the low-wavelength domain of optical photons by diffraction or turbulence in the atmosphere, while for the longer wavelengths of microwaves, sender and receiver antenna dimensions have to be increased to improve the efficiency of transmission. Guided transmission along an optical fiber or any type of microwave waveguide pays for this advantage, however, by the fact that the electromagnetic field now necessarily couples to the matter of the guide material and can whereby incur losses, as well as by a potentially more complex dispersion and the issue of in- and out-coupling losses.

\paragraph{Optical fiber\label{subsubsec:optical_fiber}}
Optical fibers, where light is guided by total internal reflection at the boundaries between an inner core and a cladding resulting in a number of discrete modes, have become a basic technology for communication. This was reflected in awarding the Nobel Prize in Physics 2009 ``for groundbreaking achievements concerning the transmission of light in fibers for optical communication'' to Charles Kao for his 1960's work on ultrapure glass fibers. Modern developments include single-mode fibers with an attenuation as low as $0.1400\,\textrm{dB/km}$ \cite{khrapko2024quasi}; hollow-core and multi-core fibers; multiplexing in various degrees of freedom, such as polarization, wavelength, amplitude, and phase to increase data transfer capacity  \cite{essiambre2012capacity} and new materials such as variously doped silicates and sapphire. 

Based on a fundamentally different optical guiding principle than total internal reflection, namely on the creation of band gaps, are photonic crystal fibers 
\cite{knight1998photonic,russell2003photonic}, which allow for more flexibility in dispersion and band engineering and functionalization for sensing or amplification. In \cite{bozinovic2013terabit} such fibers were used for orbital angular momentum multiplexing. 

Concerning the transfer of quantum information considerable progress is continuously made: Both in increasing the length of transmission, for instance, from QKD over a standard telecom fiber of $122\,\textrm{km}$ in 2004 \cite{gobby2004quantum} to specific experimental demonstrations reaching nearly ten times that distance in 2023 \cite{liu2023experimental}.  
Moreover, QKD has moved from lab tests using spooled fibers to fiber deployed between two cities \cite{Chen.2021} and is even employing fibers carrying conventional telecommunications traffic \cite{thomas2024quantum}.

\paragraph{Microwave waveguides\label{par:mu_waveguides}}
The design and analysis of various forms of transmission lines and waveguides consisting of one or two conductors and in planar or three-dimensional variants is an essential part of any textbook on microwave engineering, where the propagation of classical electromagnetic waves of different nature (transverse electrical, transverse magnetical, $\hdots$) and dispersion characteristics are discussed \cite{pozar2012microwave}. 
For the quantum realm, the review \cite{casariego2023propagating} includes a brief summary of the most commonly used ways to guide microwaves over larger distances. 
These are commercially available superconducting Niobium-Titanium coaxial with $50\,\Omega$ characteristic impedance, which at cryogenic  temperatures below 10 K guide microwaves in the GHz range with absorption losses on the order of $10^{-3}\,\textrm{dB/m}$ \cite{kurpiers2017characterizing} or more bulky and niobium or aluminium rigid waveguides, where losses are lower, as the field is distributed over a larger volume and no inner dielectrics are used. 
Other (planar) waveguides are used for shorter on-chip connections. We will discuss a variety of short-range experiments on the transfer of quantum states in detail in Sect.~\ref{ssec:realization}, where more details and also some illustrations are given. At least two groups have reported links between cryogenic systems over $\sim 10\,\textrm{m}$ distances to establish local area superconducting quantum communication networks \cite{magnard2020microwave,yam2023cryogenic}, and there is a good theoretical understanding and modeling of basic issues and problems; see, e.g., \cite{vermersch2017quantum,xiang2017intracity,salari2024cryogenic}.
In addition to the series of experiments discussed in Sect.~\ref{ssec:realization} focusing on the physical transfer of a quantum state, two experiments closer to the QKD, entanglement distribution, and teleportation of continuous-variable states are noteworthy: a first short-range proof in \cite{fedorov2021experimental}, the subsequent demonstration of noise resilience \cite{fesquet2024demonstration}, as well as a time-bin based protocol 
\cite{kurpiers2019quantum}.

There exist a number of other physical platforms, some of which are already mentioned in Sect.~\ref{ssec:carrier} above in the context of carriers of information, which technically constitute channels for the transport of quantum information, however, over extremely short distances such that their possible use is restricted to connections between different functional units within a single quantum processing device.

\paragraph{Electron propagation channels in solid-state materials}
In Sect.\,\ref{subsubsec:electrons} on using electronic excitations as qubits several types of flying electron qubits were considered. These are performing a guided motion along different types of channels defined in a solid-state material \cite{Edlbauer_EPJ_22,fletcher2013clock,dubois2013minimal}. For instance, two-dimensional electron gases (2DEG) or quantum Hall edge channels are used in such a way. 
\paragraph{Active channels}
While electrons in the just-mentioned scenarios are freely propagating along predefined static channels, there are a number of scenarios, where electrons (but also other carriers) are actively transported along a dynamically manipulated channel. A prominent example is the use of surface acoustic waves (see also the next section and \ref{par:SAW}) moving an information-carrying electron along 2DEG or quantum Hall samples \cite{Hermelin_Nature_11,mcneil2011demand,Takada_NatCom_19,Bertrand_NatNano_16}, or around the surface of superfluid liquid Helium \cite{Byeon_NatCom_21}.
Electrons on liquid Helium can also be moved via clocked transfer along gate-defined paths using charge-coupled devices (CCDs) \cite{sabouret2008signal,bradbury2011efficient,jennings2024quantum}. Similar gate-defined active channels for electrons in semiconductors, called the ``bucket brigade approach" \cite{Mills_NatCom_19} and ``conveyor-mode single-electron shuttling" \cite{seidler2022conveyor,kunne2024spinbus} were also discussed in Sect.~\ref{subsubsec:electrons} above.
For trapped ions, the use of moving traps to establish links between modules \cite{akhtar2023high}, see also Sect.~\ref{subsubsec:ions}, can also be seen as an active channel.

\subsection{Hybrid approaches\label{ssec:hybrid}}
The combination of fundamentally different physical platforms to form a hybrid system is a popular concept in many quantum technological applications. It allows combining strengths and advantages of individual subcomponents while avoiding respective weaknesses and shortcomings. The concept is particularly promising in the quantum realm because in many cases, it is not possible for fundamental reasons to optimize systems in different aspects simultaneously, but rather optimization of one attribute (e.g., the coupling strength) is accompanied by the degradation of another (here the sensitivity to external perturbations and thus the decoherence time).

For the quantum technology considered here, the direct quantum information transfer between nodes of a quantum network, an often envisioned hybrid approach (see, e.g., \cite{Kumar_QST_19}) is the combination of: 
\begin{itemize}
\item Superconducting qubits with one- and two-qubit gates to perform quantum computing operations (with their advantages of controllable geometry, good scalability, high clock rate);
\item Optical photons for information transfer over long distances;
\item Possibly dedicated memory qubits, realized on another platform with longer coherence times.
\end{itemize}

The direct coupling of optical (i.e., high-energy photons) with superconducting qubits is typically not possible (roughly speaking, this is due to the energy mismatch and the fact that superconductivity can easily be destroyed by the energy input associated with optical light). These considerations lead to the approach of first coupling the superconducting qubits to itinerant microwave photons in order to then `convert' these into optical photons. This requires transducers from the optical to the microwave range (and back). Note, however, that many other hybrid approaches have been conceived for the tasks discussed here, e.g., the coupling of superconducting qubits to spin-based qubits in the solid state, quantum dot-magnon coupling combined with magnon-photon coupling, and many more with some already mentioned above.

We focus here on microwave-optical photon transducers (see \cite{Awschalom_PRXQ_21} for a more general overview) as (i) potentially technologically relevant in the short term (since both superconducting quantum computers and optical quantum communication are rather advanced) and (ii) as exemplary for the challenges and benefits of hybrid approaches, while also complementing our explanations of the physical realization of microwave quantum devices presented in more detail in Sect.~\ref{subsec:microwave_realization} below. 
Although there are good and recent reviews for this field \cite{Lambert_AQT_20,Lauk_QST_20}, on which much of the following is based, it should be noted that it is very dynamic, so that key figures and parameters can change quickly and new approaches are constantly emerging.

Before we delve into individual implementations, we briefly discuss general requirements.\\
Microwave optical photon transducers should:
\begin{itemize}
\item Operate in the single-photon regime, i.e., at the very lowest power levels (coming with associated requirements on low noise levels) and also work in a quantum coherent fashion;
\item Have a high quantum efficiency ($\eta \simeq 1$) (one output photon for each input photon);
\item Be bi-directional;
\item Operate at cryogenic temperatures of $\sim 10\,\textrm{mK}$ and require a lower energy input than the cooling powers available in that regime;
\item Operate sufficiently faster than decoherence times and  have a high bandwidth.
\end{itemize}

Let us first look at the \emph{general physical principles} on which a transducer is built before we see how these are implemented on different platforms.
The task of a \emph{quantum mechanical transducer} is the `faithful transfer' of quantum information content from a microwave mode, abstractly represented by a pair of bosonic operators $\hat{b}_\mu^{(\dagger)}$, into an optical mode, $\hat{a}_\nu^{(\dagger)}$ (and ultimately also in the opposite direction). 
In essence, this can be realized by a frequency mixing process caused by any kind of electromagnetic nonlinearity.
While there is no bare photon-photon interaction in the vacuum, in matter an effective nonlinearity is caused by light-matter interaction, but it generally is weak, particularly in optical frequency ranges.

Two generic strategies are followed to produce stronger effects. On the one hand, resonances, respectively optical and microwave resonators, of any kind can be used. A long dwell time of an excitation in a high-quality resonator allows the interaction of the photon to last for a long time. This can also be seen as multiple interactions occurring in each pass through the resonator.
Intuitively, one can also imagine that the entire energy quantum of a photon is being concentrated in a resonator, namely in a narrow frequency range and in a small spatial volume. Both lead to an increase in the interaction with another excitation which is similarly concentrated to the same volume and energy range. This is dubbed an ``increase in the density of states", a small ``mode volume" and a good ``overlap" of modes and in some contexts referred to as ``Purcell effect". Cavity enhancement of the coupling by a sharp resonance can often degrade competing requirements of large bandwidth and high operating speed.

Secondly, there are \emph{collective effects} that typically lead to a $\sqrt{N}$ increase. The number $N \gg 1$ can thereby denote the number of fully functional transducer units, e.g., in an ensemble of atoms, or it can simply be the number of photons in a certain (``auxiliary") mode, which is strongly classically driven. Typically, as in the opto-electro-mechanical case, a linearization in fluctuations around classical averages then leads to interaction terms where the coupling is increased by $\sqrt{N}$.

Another common feature of many transducer platforms is the competition between (two) different frequency mixing processes. Mechanisms to suppress unwanted processes and amplify desired ones are therefore part of many transducer designs.
Depending on field and context, these are referred to as sum/difference generation (typically in the classical frequency conversion regime) or as (anti-)Stokes processes and are described in the effective Hamiltonian by beam-splitter or squeezing terms.

The nonlinearities necessary for frequency conversion can either be an inherent bulk material property, possibly enhanced by the use of resonators as explained above, which is the case for electro-optic materials such as lithium niobate (LiNB0$_3$), they can stem from few-level systems, typically as clouds of atomic vapour, or they can be mediated by a third mode interacting  with both electromagnetic modes at microwave and at optical frequencies. This is the case for optomechanics, piezoelectric platforms, and magneto-optics.

We want to conclude by briefly explaining some of the features of the most prominent microwave-to-optical transducers:
\begin{itemize}
\item Electro-optical: Crystalline materials lacking inversion symmetry have a $\chi^{(2)}$-term in the electric susceptibility allowing for three-wave mixing. A photon of a pump field can thereby up-convert a microwave photon into an optical photon at the sum-frequency in a process enhanced by the large number of photons in a strong pump tone. Commercial LiNB0$_3$ electro-optic modulators (Pockels cells) reach conversion efficiencies of $\eta\sim 3\times 10^{-7}$ and  with GaAs $\eta \sim 10^{-5}$ was realized \cite{khan2007optical}. By high-quality cavity enhancement, optimization of the modal overlap, and suppression of competing processes efficiencies up to the percent range were reached \cite{fan2018superconducting} and  $\eta\sim 0.25$ is assumed to be achievable \cite{soltani2017efficient};

\item Magneto-optical:  An optical field can excite spin waves (magnons) in yttrium iron garnet (YIG), which can be tuned into resonance to a microwave field. Conversion efficiencies are currently very low with $\eta\sim10^{-10}$ \cite{hisatomi2016bidirectional};

\item Individual atoms and ensembles: atomic systems have both microwave and optical transitions, which can be used for transduction using a classical optical field to connect the differing energy ranges. Note that the long-lifetimes of atomic states make the same systems also a prime candidate for quantum memories. Microwave transitions coupling low-lying hyperfine transitions close to the electronic ground state have magnetic dipole character and are very weak, so a large ensemble with a corresponding $\sqrt{N}$ enhancement has to be used. Highly excited Rydberg transitions have an order of $10^6$ stronger coupling to microwaves due to a large electrical dipole moment, so that even single atoms can be used. With Rydberg atoms an efficiency of $\eta\sim 0.05$ has been reached and an improvement of up to $\eta\sim 0.7$  is deemed possible \cite{vogt2019efficient};

Atom-like impurities within crystals can similarly be used with rare-earth ion doping, with erbium a prime candidate both for transducers and for quantum memory cells. Using classical input fields $\eta\sim 10^{-5}$ has been reached and near-unity efficiency is claimed reachable by improving cavity enhancement and mode matching \cite{fernandez2019cavity};

\item Optomechanics and piezo-electrical: A mechanical resonator can couple to both microwave and optical electromagnetic fields and thus mediate a conversion process. Various geometries and platforms exist, for instance, a micromechanical membrane, whose motion influences both the resonances of an optical cavity and the capacitance of an LC-resonator in the microwave regime, reaching conversion efficiencies of $\eta\sim 0.1$ \cite{andrews2014bidirectional}, and optomechanical crystals optimizing the mode overlap \cite{mirhosseini2020superconducting}. For an efficient coupling between microwave excitations in microelectronic superconducting devices to mechanical degrees of freedom, interdigital transducers are used to excite surface  acoustic waves, see \ref{par:SAW}.
\end{itemize}

\subsection{Nodes for storage and processing of quantum information} \label{ssec:Nodes}
Nodes should be able to store quantum information and, depending on the complexity and level of the envisioned quantum network, allow some degree of processing of the stored information. In these aspects, the requirements and, in consequence, the proposed physical platforms for network nodes directly parallel physical realizations of stationary qubits for generic quantum computing purposes. However, nodes require a good interconnect to the transfer channels and the itinerant information carriers, so that quantum information can be written and retrieved from the node, or in alternative protocols entanglement can be built up between distant nodes to be eventually used for the quantum teleportation of information. That latter aspect alone is the essence of quantum memories and quantum repeaters.

As these two facets of nodes are each related to very wide, active fields of research on their own, we shall keep our discussion here quite brief. There is an enormous amount of original and review literature on various levels of expertise and technical depth readily available. What we will do is to give a very concise overview of different platforms, focusing on the respective properties and peculiarities specifically relevant for their use in quantum networks.

\subsubsection{Qubit platforms for network nodes}
We will very briefly discuss various qubits platforms starting with the ones, which have already been used for rudimentary network tasks or seem most promising for such use. We will specifically highlight realized minimal quantum network platforms establishing a connection between two distant nodes for state transfer or entanglement distribution. 

\paragraph{Trapped Ions\label{subsubsec:massive}} 
These have been one of the first platforms to successfully perform quantum information processing tasks and remain among the most advanced and promising platforms \cite{haffner2008quantum}.  
Ions are confined by electrical DC- and RF-fields defining so-called traps. The originally bulky three-dimensional Paul-traps developed in the 1950's have evolved into microfabricated traps just above a chip surface \cite{wright2019benchmarking}, where typically of the order of ten ions are confined in a linear array. This array can then function as a building block of more complex configurations: of 2D-arrays
\cite{holz20202d}, 
 or as part of a dynamically controlled modular structure with interaction and storage regions
\cite{pino2021demonstration,akhtar2023high,lekitsch2017blueprint}. 
Scaling towards a larger scale can be achieved (as discussed in Sect.~\ref{subsubsec:ions} above) via intermodule links that move ions from chip to chip, or by optically creating entanglement between distant chips \cite{olmschenk2009quantum,moehring2007entanglement,stephenson2020high}. 
Advanced integrated optics will also be crucial for larger devices
\cite{mehta2020integrated, niffenegger2020integrated}. 

Being fairly well isolated from many environmental noise sources, trapped-ion qubits excel with long coherence times (from seconds up to one hour \cite{wang2021single}). Lasers are used for cooling and most gate operations, while microwave gates also exist \cite{ospelkaus2011microwave,lekitsch2017blueprint}. 
Although gate times are slow on an absolute scale (compared to other qubits and classical computers), it is not only possible to perform a very high number (up to $10^6$) of operations within the coherence time, but also to achieve high fidelity of all operation steps \cite{harty2014high}.
Compared to competing technologies, trapped ions offer inherently good reproducibility (``each ion is the same"). While requiring a high-quality vacuum, micro-Kelvin cryogenics are not needed, and even room-temperature operation is possible.

\paragraph{Neutral atoms}
Neutral atoms can be trapped in optical lattices or arrays of tweezers created with the use of  spatial light modulators. A nice introduction on the creation of a perfectly loaded array can be found in a recent review on the field \cite{henriet2020quantum}. Besides the ability to create large arrays of various geometries \cite{barredo2016atom}, neutral atom arrays have the advantages of identical constituent qubits, individual addressability and long coherence times of the order of seconds, if quantum information is encoded in the hyperfine ground states of Rubidium atoms \cite{henriet2020quantum,saffman2019quantum}. Two-qubit gates are then executed by using highly excited ``Rydberg"-states of the atoms \cite{jaksch2000fast,saffman2010quantum}. With principal quantum numbers $\sim 100$ these offer high coupling and thereby fast gate times. The capabilities of neutral atom quantum computers were, for instance, shown in \cite{graham2022multi} and even logical qubits for error correction have been demonstrated \cite{reichardt2024logical}. 

On the road towards quantum networks, the interconnect from a flying photonic qubit to a neutral atoms has been used for storage and on-demand retrieval \cite{kalb2015heralded}. Also the transfer of an atom's quantum state and the creation of entanglement between distant atoms has been realized in systems using atoms trapped within optical cavities \cite{ritter2012elementary,hofmann2012heralded}. How to integrate these types of photonic links with processing arrays has been recently discussed in \cite{covey2023quantum}. Other approaches propose optical links between Rydberg atoms without cavities \cite{Grankin_PRA_18} or consider atomic ensembles, which are particularly promising for use as memories and quantum repeaters, see, e.g., \cite{sangouard2011quantum}.

\paragraph{Superconducting devices\label{par:SC_qubits}}
Qubits can be engineered in superconducting electronic devices that exploit the nonlinearity of a superconducting tunnel junction (Josephson junction). Nonlinearity, the prerequisite for any kind of logical circuit, is used to create an ``artificial atom" with addressable discrete energy levels with non-degenerate excitation energies \cite{martinis1985energy}.
The energy differences of the two levels used as a qubit are thereby in the GHz range, so that mK-temperatures \--- far below the superconducting transition temperature \--- are required for coherent operations. The greatest advantage of superconducting qubits is that coupling qubits in various (however fixed) geometries is relatively straightforwardly achieved by electrical connections of various types (galvanic, inductive, capacitive) and with all-electrical control and read-out the prospects for scalability are promising, and devices with $~20-50$ qubits are now common.
Starting from several basic designs \cite{bouchiat1998quantum,nakamura1999coherent,orlando1999superconducting,devoret2004superconducting}, the field rapidly developed to advanced schemes that aim to satisfy competing requirements. The operation speed should be increased, but it is limited by decoherence due to susceptibility to various noise sources. The goal is to improve the overall fidelity. For reviews, details of setups and theory, and current performance numbers, see \cite{clarke2008superconducting,vool2017introduction,kockum2019quantum,kjaergaard2020superconducting}. 
 
One modern development is the increased integration of complex superconducting microwave structures for read-out, control, and tunable coupling \cite{blais2021circuit,gu2017microwave}.  
This type of device was also used to investigate the controlled release of a quantum state from one qubit (or resonator) and its near-perfect recapture by another, which we discuss at length in Sect.~\ref{ssec:realization} below, where the reader will also find more details about the different device schemes and working principles.  

Two major avenues are investigated to break through the limitations of current systems beyond small-step optimization \cite{jurcevic2021demonstration}. Error-correcting codes on the one hand \cite{andersen2020repeated},  
 and qubits with inherent insensitivity to certain error types and autonomous error corrections, such as recently popular cat qubits \cite{reglade2024}.
The latter have also been proposed for an elaborate quantum network scheme using microwave-optical transducers \cite{Kumar_QST_19}.

\paragraph{Quantum dot based qubits \label{subsubsec:QD}}
To form quantum dots, electrons are confined to such a small region that their energy levels become discrete forming an `artificial atom', which eventually can be occupied by only a single electron (or hole) \cite{kastner1993artificial,kouwenhoven1998quantum,kouwenhoven1995coupled}. Lateral quantum dots defined by electrostatic confinement by metallic gates in a two-dimensional electron gas of high mobility formed in semiconductor heterostructures have been pursued as qubits for a long time \cite{loss1998quantum}. 
Quantum information is thereby typically stored in a spin degree of freedom (as charge or charge dipoles couple too strongly to the environment to keep their coherence for long) but other degrees of freedom (orbital, valley, $\hdots$) are also possible. 

Conventional architectures \cite{loss1998quantum} realize single-qubit gates with microwave pulses, two-qubit gates via the intrinsic Heisenberg exchange interaction between two dots, and allow for electrical readout via a spin-charge conversion process \cite{elzerman2004single}. Typical are gate times of nanoseconds \cite{zajac2018resonantly}, coherence times of milliseconds and fidelities of 99$\%$ \cite{veldhorst2014addressable,huang2019fidelity}, see tables in \cite{stano2022review,chatterjee2021semiconductor,gao2024advances}
and \cite{burkard2023semiconductor,bluhm2019semiconductor} for some modern reviews.

While early research has focused on GaAs structures \cite{hayashi2003coherent,petta2005coherent,chan2004few,van2002electron},  silicon has also been explored for a while \cite{lim2009electrostatically} and offers the huge technological advantage of compatibility with existing semiconductor technologies \cite{lim2009electrostatically} with exciting prospects for scaling and integration \cite{maurand2016cmos,ciriano2021spin,ruffino2022cryo}.
 Remaining challenges are the inherent fabrication-related disorder, which must be overcome by the fine-tuning of a huge number of gate parameters \cite{botzem2018tuning,teske2019machine}, and decoherence caused mainly by nuclear (bulk) spins and surface effects, as well as heat management and operating temperatures \cite{yang2020operation}.

Another recent development is the integration with superconducting microwave cavities \cite{petersson2012circuit,mi2017strong,borjans2020resonant,yu2023strong,burkard2020superconductor}. These can act as a bus for longer-range coupling or eventually for the out-coupling of the quantum information to microwave photons for longer-range information transfer in the microwave regime or optically after transduction, see Sect.~\ref{ssec:hybrid}. An alternative is the creation of a distant entanglement (and subsequent quantum teleportation of quantum information) between quantum dots via optical photons \cite{delteil2016generation,stockill2017phase,delteil2017realization}. Although the types of quantum dots that allow easy interconnections to optical photons differ from those used for quantum information processing discussed so far, it seems feasible to extend the approach to couple the two types locally \cite{kim2016optically}. 

\paragraph{Impurity-based qubits in solid state platforms}
Defects in a solid-state lattice can form bound states for a single electron (or hole) with a complex, atom-like level structure. They have thus sometimes been dubbed ``ions trapped in a solid state". Quantum information can be written into the electron spin of the orbital ground state of the defect to realize a qubit with long relaxation and coherence times. The information can also be transferred to nearby nuclear spins for long-term storage.

Prominent and well established platforms are different color centers in diamond, for instance, Nitrogen (NV) and Silicon Vacancies (SiV), where a nitrogen (silicon) atom replaces a carbon atom adjacent to a vacant site, and phosphor donors in silicon.

Since the first detection of magnetic resonance on a single NV center at room temperature in the fluorescence emission spectrum gained by confocal microscopy \cite{gruber1997scanning}, the platform has been widely investigated \cite{doherty2013nitrogen}. 
The platform offers quantum sensing \cite{barry2020sensitivity} of different physical quantities, and  complex pulse protocols were developed to decouple from certain noise sources and to realize gates by controlling the effective coupling between different spins in a system, where their physical interaction is always present. Effective coherence times up to a second have been reached at liquid nitrogen temperatures of 77 K \cite{bar2013solid}, and even at room temperature quantum information processing is possible \cite{wolfowicz2021quantum}. Scaling and the coupling of several NV centers is still a problem, and typically today's experiments use a single NV center's electron spin and surrounding nuclear spins. Nonetheless, the achieved complexity of $\sim 10$ qubits \cite{bradley2019ten} may be sufficient for use as a simple node of a quantum network. Optically interfacing the NV \cite{Awschalom.2018}, simple network tasks using distant NVs have been realized, such as Bell tests \cite{hensen2015loophole} and entanglement distribution \cite{kalb2017entanglement,humphreys2018deterministic}, and even first steps beyond a two-mode network \cite{pompili2021realization}. Other vacancies are also studied \cite{wolfowicz2021quantum,li2024heterogeneous}, for instance, realizing a telecom network using SiV \cite{knaut2024entanglement}, with each requiring different operating conditions and bringing distinct advantages and challenges.

Similarly to NVs in diamond, in a silicon host material  $^{31}\textrm{P}$ donor atoms come equipped with a fast qubit and a long-lived qubit, namely the electron and nuclear spin \cite{freer2017single}. Read-out and control was performed electronically by CMOS-compatible nanostructures \cite{pla2012single} and impressive coherence times \cite{saeedi2013room} and fidelities were reached. An advantage of the platform is the atomistically exact device fabrication \cite{schofield2025roadmap} through scanning probe lithography, by which a precise design of the coupling between several qubits was realized \cite{he2019two}, which is an exceedingly challenging task for platforms such as NVs. 
Phosphor donors in silicon do not interact strongly with light, so that other so-called radiation damage centers have received more attention for an immediate role in quantum networks \cite{bergeron2020silicon}. One of these, the T center in silicon, where one Si is substituted by a pair of carbon atoms, has emission in the telecommunication bands and has been proposed for such tasks as microwave-to-optical transduction by an ensemble \cite{PRXQuantum.4.020308}. Distribution of entanglement has been demonstrated as a `proof-of-concept' for T centres as a distributed quantum computing and networking platform \cite{afzal2024distributed}.

\paragraph{Cavities}
Cavities, oscillators or resonators can also play a role as nodes of a quantum network. Again, optical and microwave electromagnetic resonators and mechanical oscillations will be considered.

\subparagraph{Optical resonators} 
They are mostly used not to store quantum information, but to mediate the coupling between the traveling electromagnetic wave and qubits of various kinds \cite{reiserer2015cavity,reiserer2022colloquium}. As discussed in Sect.~\ref{ssec:hybrid}, the Purcell effect leads to an efficient coupling by focusing the density of states and enhancing the interaction time. Fabry–Perot cavities, where resonances develop between two mirrors are used in experiments with neutral atoms \cite{ritter2012elementary}. Solid state qubits can be integrated into Bragg mirrors or photonic crystals cavities, where the mode volume is reduced far below $\lambda^3$ \cite{choi2017self} to achieve good overlap and strong coupling.

\subparagraph{Microwave resonators} 
Similarly to optical cavities, they can be used as a mediator, but they can also function as a high-quality storage mode for quantum information. Both cases will appear in the experiments discussed in Sect.~\ref{ssec:realization}. It should be noted that for the microwave regime nonlinear resonators form a natural bridge between the two extremes of a qubit and a perfectly harmonic oscillator.

\subparagraph{Mechanical resonators} 
Two types of mechanical resonators should be mentioned. First, resonators for surface acoustic waves formed by interdigited Bragg-type  mirror, which have been proposed as a local storage node of quantum information \cite{satzinger2018quantum} and would mirror many of the functional principles of superconducting microwave realizations. In addition, there are various nanomechanical oscillators in different frequency ranges, which can offer long phonon lifetimes up to seconds and coherence times above 0.1 ms \cite{Riedinger_Nature_18,maccabe2020nano}.

\paragraph{Molecular spins}
Contrary to the preceding, one may design a platform for quantum information processing using a bottom-up approach. Single-molecule magnets are synthesized to have a large spin with a large spin anisotropy and long coherence times. Inherent advantages of this approach \cite{chiesa2024molecular,carretta2021perspective,leuenberger2001quantum} are the cheap fabrication of an enormous number of identical units by standard chemical synthesis, the variability of chemical engineering of molecules combining different functionalities in different parts of the molecule, the possibility of self-organized dimers \cite{ardavan2015engineering}, trimers, or arrays of molecular qubits, and the multilevel characteristics leading to qudits \cite{moreno2018molecular} with a larger Hilbert space.
Early demonstrations achieved individual addressability via scanning-tunneling methods \cite{moreno2021measuring} and simple qubit functionalities \cite{ardavan2015engineering,ferrando2016modular} and properties \cite{bader2014room,zadrozny2015millisecond} have been shown. The larger dimensionality of the qudit can be exploited to perform Grover's algorithm \cite{godfrin2017operating,leuenberger2001quantum} in a single molecule and to embed quantum error correction in single molecule magnets \cite{chiesa2020molecular}.
Molecular magnets have been coupled to on-chip superconducting resonators and microwave photons \cite{carretta2021perspective,chiesa2023blueprint} and optical addressability is being investigated for various molecular spin systems \cite{moreno2021measuring,bayliss2020optically,kumar2021optical}. 

\paragraph{Topological qubits}
Topological qubits are one of the latest additions to the zoo of proposed qubits. They are distinguished by fundamental built-in protection against certain control errors and perturbations based on their topological nature \cite{nayak2008non,lahtinen2017short}. This stems from the quantum statistical properties of anyons under exchange operations called braiding. Although many different platforms are pursued \cite{alicea2012new,karzig2017scalable,flensberg2021engineered}, the most prominent being Majorana fermions realized at the ends of ferromagnetic chains \cite{nadj2014observation} and Josephson junction arrays. Currently, the field is in its infancy. Many experimental results are debated, and most of the issues beyond the mere realization of a qubit and simple operations, such as the ones required for a quantum network, are largely unexplored.

\subsubsection{Quantum interconnects}
The transfer of quantum information from the carrier, which is traveling across the quantum channel, to the stationary node, where the information is stored and processed, is obviously a crucial part of a quantum network. However, the variety and width of techniques, protocols, and platforms is vast, and most attempts at a categorization will go through the various nodes, carriers, and channels introduced above. In fact, in the different sections above, we have already discussed interconnects to the extent that was necessary to gauge the suitability of certain platforms as part of a quantum network. Some general working principles are also found in Sect.\,\ref{ssec:hybrid} on hybrid systems. In addition, it should be mentioned that there are very active research areas where the basic functions of quantum interconnects are used for quantum memories and quantum repeaters \cite{heshami2016quantum,wei2022towards,sangouard2011quantum,muralidharan2016optimal,neuwirth2021quantum,pettit2023perspective,Azuma.2023}. 

In the following section, we will explain the experimental realization of the exemplary case of microwave quantum information transfer between superconducting nodes. We will see how the interconnect is engineered to allow devices to controllably emit and absorb quantum information.

\subsection[Example: microwave quantum state transfer]{Example of an experimental realization:\\ quantum state transfer in a superconducting microwave platform
\label{ssec:realization}}
In this section, we want to give an example of the realization of one rudimentary task of a quantum network in one of the platforms introduced above: the direct transfer of an arbitrary quantum state between two network nodes realized in the microwave domain. 
The experiments realize a minimal version of a quantum network as proposed in \cite{Cirac.1997}. Their network consists of spatially separated nodes where stationary qubits store and process quantum information, connected to cavities, and quantum communication channels through which quantum information is transmitted as itinerant flying qubits. The qubits can obviously also be extended to arbitrary (single-mode) quantum states (i.e. the state of a harmonic oscillator, or cavity).

The specific task, which we are discussing here, is then the following:
\begin{itemize}
    \item  An arbitrary quantum state is to be transferred from a quantum node (stationary qubit or memory cavity) to an itinerant quantum state, which is then received and stored in another node. The protocol should be independent of the state, which is assumed to be unknown.
    \item We consider the direct transfer of the state, i.e. no teleportation protocols in which, for example, entanglement is used as a resource and combined with local measurements and classical information transfer.
    \item We deal with realizations in the microwave domain, i.e., stationary qubits or cavities are superconducting electronic devices with microwave excitations, and itinerant states are microwave pulses (in waveguides or also in free space).
    \item We will also briefly discuss preparation and measurement of the states, which are not part of the protocol but are used for characterization of the experiments and the device performance.
    \item We will not discuss the crucial topic of security against interception attempts and only minimally touch on error correction and encoding issues.
\end{itemize}
The intended transfer process can be divided into a number of sub-problems: 
\begin{enumerate}
    \item how can a quantum state be completely emitted from memory (“release” or “pitch”) in a controlled manner at a certain point in time and, conversely, 
    \item how can an itinerant quantum state, which is encoded in a pulse and radiated onto a cavity, be transferred without loss into an excited state of this stationary node (“catch”).
    \item How can transmission and reception be combined and what steps are required to test and characterize the performance?
\end{enumerate}
In the following, we will explain how each of these was solved experimentally and finally combined in a number of seminal works \cite{yin2013catch,wenner2014catching,pfaff2017controlled,axline2018demand,kurpiers2018deterministic}. 

We start by introducing, in Sect.~\ref{subsec:devices}, the basic microwave components that are required for the realization of stationary and itinerant qubits and oscillators. Then, in Sect.~\ref{subsec:microwave_realization} we will address controlled emission and absorption (``pitch and catch") of a quantum state and the role of time-reversal protocols.
We can then discuss the full protocol in Sect.~\ref{subsec:full_transfer}, both for the transfer of generic single-mode states and for qubits.  
We will wrap up with a discussion of related and alternative tasks and the further developments building on the groundbreaking experiments in \ref{subsubsec:alternatives}.

\subsubsection{Fundamental microwave devices\label{subsec:devices}}
The microwave components used in the experiments discussed below are well-established circuit quantum electrodynamics (QED) devices, notably used in modern superconducting quantum computing designs. In this brief explanation of that hardware, we will draw on various reviews \cite{wendin2017quantum,vool2017introduction,krantz2019quantum,blais2020quantum,rasmussen2021superconducting,blais2021circuit}, in particular on a recently published article \cite{blais2021circuit}, from which some of the figures in the following are reproduced.

\begin{figure}[b]
	\centering
	\includegraphics[width=0.95\columnwidth]{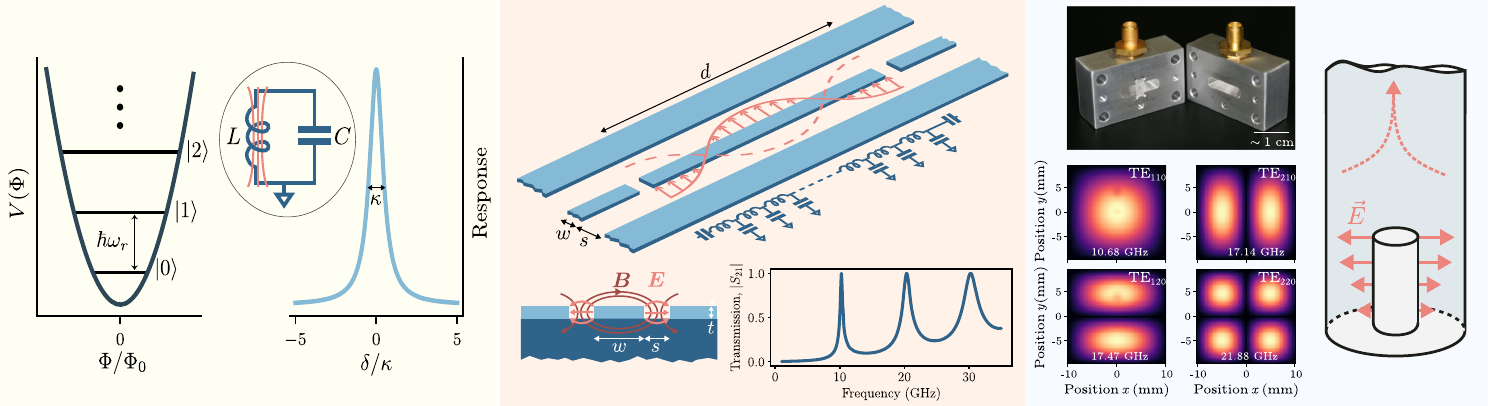} 
	\caption{Different variants of  superconducting microwave resonators used in quantum information transfer experiments. Energy scheme, lumped-element realization, and resonance curve of an electrodynamic resonator (left); coplanar waveguide as resonator with field distribution (center); 3D microwave cavity (right). Reproduced with permission from \cite{blais2021circuit}.
}
\label{fig:cavity_realizations}
\end{figure}

We are interested in three components: The qubit, resonator, and waveguide and their mutual coupling. Qubits and/or resonators will constitute the nodes of a quantum network, where quantum information is stored and manipulated, while the network links are provided by microwave excitations traveling along waveguides.
The qubits used are often transmon-type. These can be imagined as LC resonators that develop weak nonlinearity by substituting a linear inductance with a superconducting Josephson junction. Being largely insensitive to charge fluctuations, these qubits have long coherence times \cite{blais2021circuit}.
Fig.\,\ref{fig:cavity_realizations} shows different variants of a superconducting microwave resonator. Conceptually, this corresponds to an LC circuit and, indeed, it can also actually comprise these conventional elements (``lumped elements"). Other typical realizations are linear or 3D resonators. The former are short sections (of the spatial dimension of the microwave wavelength) of a coplanar waveguide that often consists of a central conductor between two outer ground planes (see Fig.\,\ref{fig:cavity_realizations}). Conductors and ground planes are thin superconducting layers on a substrate. Newer variants often use three-dimensional microwave cavities, which achieve particularly high qualities (long life time/sharp resonances). 

A qubit and resonator can be coupled in various ways: inductively, capacitively, galvanically, or via Josephson junctions. 
While the type of coupling affects the strength and detailed description of the coupling, the basic physics can often be reduced to a standard coupling Hamiltonian, the Jaynes-Cummings Hamiltonian, which describes the exchange of excitations between the resonator and qubit. A Hamiltonian of the same form is also obtained for the coupling between resonator (or qubit) and waveguide, with the difference that the waveguide has a dense mode spectrum. Instead of explicitly describing the dynamics of all these waveguide modes, one considers them as a bath and describes dissipative processes using the Lindblad equations or input-output theory (see \cite{blais2021circuit} for details and further references).

\subsubsection{Emission and reception of quantum microwave states 
\label{subsec:microwave_realization}}
The aim of this section is to explain the basic problems and solutions that arise in realizing the emission, transmission, and reception of quantum states. Considering the emission process first, the task is thus to emit from some memory at a certain point in time an arbitrary, unknown quantum state in a controlled manner; without any information remaining in the memory or otherwise being lost. 
Conversely, an itinerant quantum state, i.e., a certain occupied state of a spatio-temporal mode (e.g., a certain pulse shape occupied in superposition with one or two photons) shall be transferred to a stationary state (without, for example, a photon ever being reflected from the receiver cavity). 
To understand the problem more clearly, we first consider the natural decay of an excitation from a cavity or the conventional excitation of a cavity by a pulse. 

\paragraph{Natural decay} 
We assume that the cavities are coupled to a waveguide so strongly that the decay or excitation through this waveguide dominates all other potential decay processes. 
A single photon in the cavity then has a certain constant, small probability of escaping from the cavity in a certain small time interval. This is defined by the decay rate $\gamma$, which can be regarded as frequency-independent in the relevant range. For the probability of finding the photon still in the cavity, this yields an exponential decay with rate $\gamma$ (note that this is valid in the so-called Born-Markov limit under certain assumptions on the mode structure of the waveguide \cite{blais2021circuit}). 
For a classical excitation (coherent state), one finds the same exponential decay dynamics.
The pulse emitted in such an exponential decay process has its maximum intensity at its wavefront that is the first to reach a receiver cavity. To actually observe such an exponential decay, the initial state has to be created diabatically (instantaneously)  in the cavity at the desired emission time. Alternatively, the coupling between cavity and waveguide must be suddenly switched on, or steeply increased. Experimental realizations of both approaches will be discussed in the following.

\paragraph{Conventional excitation} 
Let us now consider the excitation of a cavity by an incident pulse. A continuous wave is partly reflected at the interface between the waveguide and cavity and partly transmitted into the cavity, whereby it becomes excited. The created excitation in turn continuously decays into the waveguide. Thereby, interference occurs between the `direct' reflection and the `re-emission'. 
If now a pulse, which originates from a natural decay process, as just described, impinges onto a receiver cavity, the wavefront strikes first with maximum intensity, so that the immediate direct reflection is also strong. However, re-emission will only slowly build up concomitant with increasing excitation of the cavity and becomes strongest at a later point in time when the direct reflection is already weak (see Fig.\,2 in \cite{wenner2014catching}). 

\begin{figure}[bt]
	\centering
	\includegraphics[width=0.9\columnwidth]
{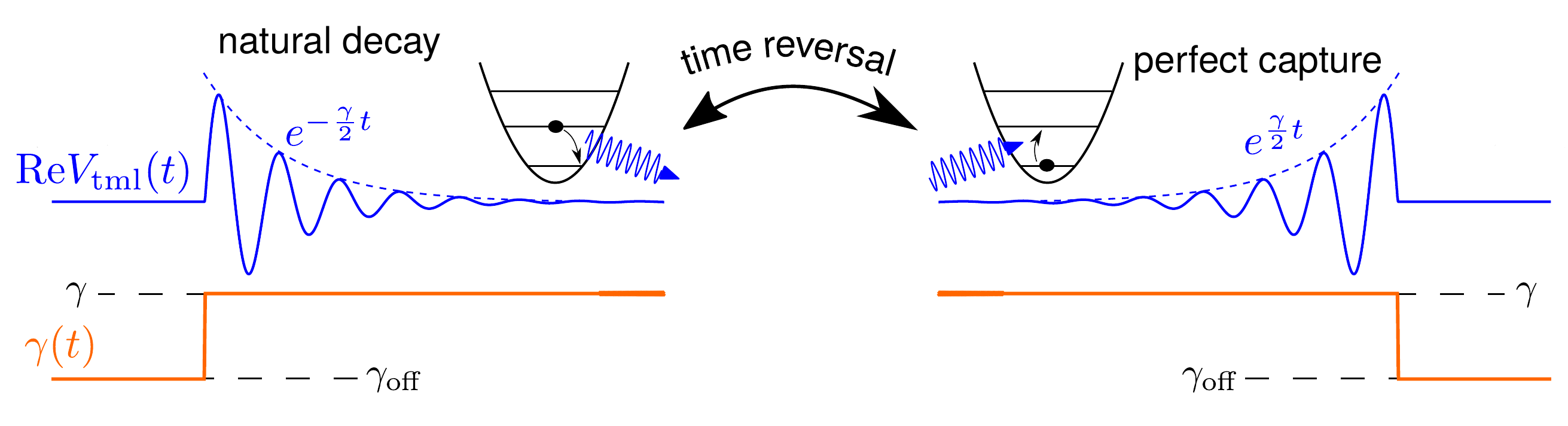}
	\caption{Time dependence of controllable coupling and drive pulses used in \cite{wenner2014catching} to demonstrate perfect capture of a pulse, which has the time-reversed shape of a pulse stemming from a natural decay process.
    } \label{fig:time-reversed_pulse}
\end{figure}

\paragraph{Time reversal} 
Imagine now that we had the natural waveform reversed with maximum intensity right at the end of the pulse: Direct reflection and re-emission would both be weak at the beginning and strong at the end of the pulse. 
In fact, in such a pulse, destructive interference of the two terms can completely suppress the reflection at all times. This effect can also be understood as a time reversal of the natural decay process. A pulse that is a time-reversed decay pulse is perfectly absorbed (see Fig.\,\ref{fig:time-reversed_pulse}). 
This pulse can be generated by an arbitrary waveform generator (AWG) to demonstrate the feasibility of perfect absorption (see below and \cite{wenner2014catching}). Going one step further, one can work with a time-reversal invariant pulse to realize complete emission of a quantum state from one cavity and perfect absorption of the same state by another cavity. 
As explained above, a time-dependent coupling can change the emitted pulse shape, e.g., to a time-reversal invariant shape, and from the same time-reversal argument as above it immediately follows that the time-reversed coupling will perfectly absorb this time-reversal invariant pulse.

\paragraph{Controllable coupling\label{par:controllable_coupling}} 
The above considerations have established that time-dependent coupling is an essential component of a transfer protocol. 
We now want to briefly outline two methods to realize such control which fundamentally differ in their basic idea. Having such different methods at hand could be of benefit when considering different applications and the transfer to other platforms and protocols. 

\begin{figure}[tb]
	\centering
	\includegraphics[width=0.9\columnwidth]{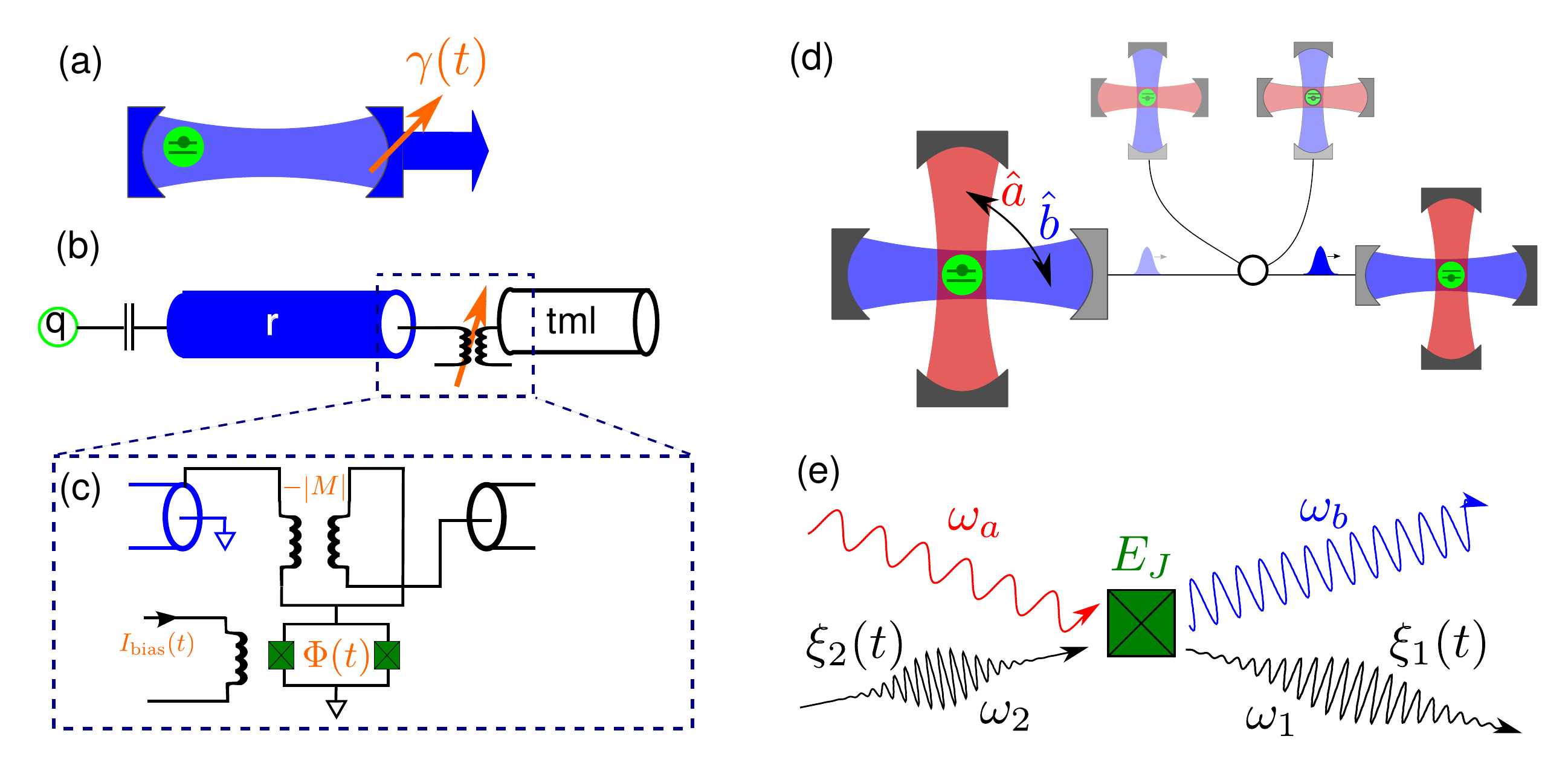}
	\caption{Controllable coupling of a cavity to a transmission line. Left side shows a direct control of coupling between storage cavity and waveguide in the schematic pictorial language of cavity QED (a) and circuit QED (b) and the experimental realization (c) of the controllable coupler employed in Ref.\,\cite{yin2013catch}.\\ Control of coupling was also realized \cite{pfaff2017controlled,axline2018demand} by a parametric driving scheme of a setup (d), where a Josephson qubit and storage $\hat{a}$ and communication cavity $\hat{b}$ are nonlinearly coupled, cf. Fig.\,\ref{fig:network_cavities}. The physical process described by the Hamiltonian $H_\textrm{conv}$ in the main text is sketched in (e).
    } \label{fig:controllable_coupler}
\end{figure}

\emph{Direct control:} In the first realized variant \cite{yin2013catch}, the coupling between stationary memory and waveguide is directly varied. 
Precise control and, in particular, good suppression of the coupling before the desired emission of the quantum state is achieved by combining a static negative mutual inductance and the (positive) inductance of a SQUID that can be controlled by a magnetic flux (see Fig.\,\ref{fig:controllable_coupler}). 
In \cite{yin2013catch} a controllable coupler of this type was used for the first time in a purely on/off mode to controllably emit a single photon from a cavity at a given time. Thereby, a qubit was used both to generate an initial excitation in a cavity and to confirm the successful controlled emission from the cavity by measuring the excitation remaining in the cavity after performing the protocol. The emitted quantum state can also be detected directly by quadrature measurements on the waveguide. The experiment additionally demonstrated phase-coherent excitation, the excitation and emission of superposition states, as well as the ability to realize a generic time-dependent control to emit arbitrary pulse shapes.
In a second experiment \cite{wenner2014catching} the same time-controlled coupler was used in on/off mode to capture a pulse shaped by an AWG. The arguments based on time-reversal invariance were confirmed, and the reflectionless reception of a time-reversed exponential pulse was shown.

\emph{Coupling through parametric driving: }  
The second variant uses established designs that combine two cavities and a transmon qubit. Effectively, a superconducting tunnel junction (Josephson junction) couples three modes in a nonlinear manner (see Figs.\,\ref{fig:controllable_coupler} and \ref{fig:network_cavities} and \cite{pfaff2017controlled}), as described by a Hamiltonian
$\hat{H} = E_J \cos{[\phi_a(\hat{a}+\hat{a}^\dagger) + \phi_b(\hat{b}+\hat{b}^\dagger) + \phi_c(\hat{c}+\hat{c}^\dagger)]}$, where parameters $\phi_{a/b/c}$ indicate zero-point fluctuations of phase variables of the three modes.
A transmon mode (c) is used to prepare a complex quantum state in a long-lived storage cavity (a), as well as for readout (see below), while a communication cavity (b) is strongly coupled to waveguides. 
If the system is periodically driven via additional control lines, the Hamiltonian mediates a four-wave mixing process: two photons of the drive signals with frequencies $\omega_1$  and $\omega_2$ interact with two photons of the cavities with frequencies $\omega_a$ and $\omega_b$ (see Fig.\,\ref{fig:controllable_coupler}). 
The symmetry of the Hamiltonian requires an even number of waves or photons to be involved, and an energy conservation argument (formally a rotating wave approximation) allows selecting a desired process by choosing the irradiated frequencies; for instance, the conversion of an $\omega_a$ to an $\omega_b$ photon by $\omega_1 - \omega_2 = \omega_a -\omega_b$. 
The effective conversion Hamiltonian, $\hat{H}_\textrm{conv}=g(t)\hat{a}\hat{b}^\dagger + g^*(t)\hat{a}^\dagger\hat{b}$ with $g(t)= E_J \phi_a^2\phi_b^2\xi_1^*(t)\xi_2(t)$ then describes the transfer of a photon from the storage cavity to the communication cavity, where the strength of the so-induced coupling of these two cavities is given by the product of the amplitudes of the two drive signals $\xi_{1/2}$. 
If these are now varied slowly (compared to the cavity frequencies), a fully time-controlled coupling between the memory cavity and the communication cavity is realized. The communication cavity, in turn, is coupled so strongly to the waveguide that in effect a controllable coupling between the memory mode and the waveguide is realized. That coupling then can be used to implement the full transmit and receive protocol discussed above.

\subsubsection{Transfer of quantum microwave states in a minimal quantum network\label{subsec:full_transfer}}
\begin{figure}[t]
	\centering
	\includegraphics[width=0.9\columnwidth]{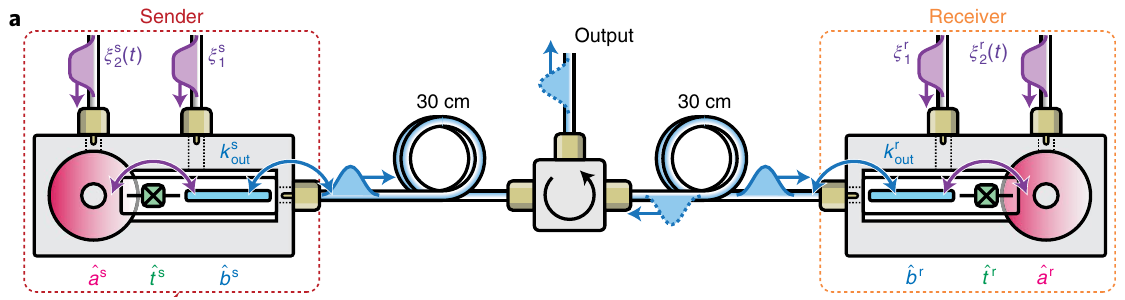} 
	\caption{Minimal version of a quantum network with two nodes, each consisting of qubit, storage-, and communication cavity, which was used in \cite{axline2018demand} for the direct transfer of quantum states. Reproduced with permission from Springer Nature.}
\label{fig:network_cavities}
\end{figure}

The minimal version of a quantum network with two nodes, each consisting of a qubit, storage, and communication cavity as described above (see Fig.\,\ref{fig:network_cavities}) was experimentally realized in \cite{axline2018demand}.  That experiment then fully demonstrated the proposal of \cite{Cirac.1997} for the deterministic direct transfer of a quantum state.  

The platform uses the controllable parametric coupling mechanism just discussed to emit and absorb a time-reversal symmetric pulse. It builds on a precursor experiment \cite{pfaff2017controlled} that focused on the emission part of the protocol. 
The full transfer experiment \cite{axline2018demand} then combined this emission with the absorption by a second node, see Fig.\ \ref{fig:network_cavities}.
The quality of transfer is already quite high for a first experiment with $93 \%$ of the energy of an incident pulse being received. 
The overall transfer efficiency is $\eta\sim0.7$. It is limited by errors of about $8\%$ introduced by undesired excitations of the transmon used for creation and detection, deviations in the pulse shape amounting to $2\%$ each for emission and absorption, and photon loss in the waveguide of $15\%$ (with a transmission distance of about $1\,\textrm{m}$!). Correcting for excitation errors a fidelity of $F\sim0.8$ is achieved for the transmission of arbitrary qubit states (superpositions of the Fock states $|0\rangle$ and $|1 \rangle$). This exceeds an important limit of $2/3$ for the reconstruction of a quantum state employing classical transmission protocols. 
First steps towards error correction were also explored in the same work. Furthermore, it was shown that  entanglement between the nodes can be generated by the partial emission of a quantum state. The achieved effective rate for generating entanglement of $\sim (100\,\mu\textrm{s})^{-1}$ is limited mainly by the time required for reset. Crucially, entanglement can be distributed across the network faster than it is destroyed by decay in each node, allowing for scaling of the network.

\paragraph{Qubit- versus cavity-realization\label{subsec:qubit_vs_resonator}}
Conceptually very similar experiments can also be done in platforms, where the transferable state is restricted to the two-level state space of a single qubit (instead of the, in principle, infinite number of Fock states of a resonator as in the heretofore discussed experiments). In \cite{kurpiers2018deterministic} the Wallraff group at ETH Z\"urich demonstrated such deterministic direct transfer of arbitrary qubit states as the basic building block of a quantum network. A $3$-level system (``qutrit" realized by a transmon) acts as a stationary qubit. 
In a cavity-assisted Raman process, pulses are used to transfer the occupation of the excited state of the qubit, first, into the second excited state and then in a time-controlled manner into a communication cavity from which photons are eventually emitted into the waveguide. 
In that manner, an effective time-dependent coupling, and thereby pulse shaping, are realized in the transmitter and receiver units. Similarly to the set of experiments described above using storage cavities, \cite{kurpiers2018deterministic} demonstrates and tests perfect absorption, the transfer of a complete set of qubit states and their fidelity, and the generation of entanglement by partial transfer.

The clearest distinction setting the qubit-realization apart from the cavity-realizations discussed above lies in the (obvious) limitation to qubit states and associated differences in state preparation and characterization. The key performance figures of the system are comparable, in many cases slightly superior to those discussed above. For example, the experiment achieved the following results:
\begin{itemize}
\item $67.5\%$ transfer efficiency for a protocol duration of $180\,\textrm{ns}$;
\item $23\%$ photon loss in the $\sim 1\,\textrm{m}$ waveguide and $98\%$ pure absorption efficiency;
\item an entanglement rate of $50\,\textrm{kHz}$ for a fidelity of $F=79\%$.
\end{itemize}
At the current stage of development, both experimental realizations seem equally suitable for the development of a quantum network.

\subsubsection{Recent developments and cross-connections 
\label{subsubsec:alternatives}}

In previous parts of this section, we introduced a line of seminal experiments, which step-by-step opened the door to quantum state transfer and entanglement distribution in superconducting microwave devices. In the years since these experiments have been performed, the field has progressed substantially. We will now briefly review some of these advances along three distinct directions: improving the node-to-node links, enhancing the network complexity, and increasing the node-to-node distance.

\begin{itemize}
\item{One line of advancement focused on improving the connection by removing circulators and creating a more direct coupling of nodes via resonant states in the cables or waveguides.  
In \cite{leung2019deterministic}, a hybridized dark mode, which is largely immune to decay in the coaxial cable, was used  for the transfer of single microwave photons and for Bell pair generation.
Similar direct resonator connections were combined with error correction via cat-state encoding in \cite{burkhart2021error} for error-detected state transfer and entanglement generation.
The increased fidelity by entanglement purification (and dynamical decoupling methods on the nodes) was also shown in \cite{yan2022entanglement}.
}
\item{Progress was also made in establishing more complex networks: Extending each node to contain three qubits, \cite{zhong2021deterministic} used direct bi-directional connects to transfer a three-qubit Greenberger–Horne–Zeilinger (GHZ) state from one node to another and created a six-qubit entangled state. Going beyond to the multi-node, multi-qubit case, \cite{niu2023low} connected five nodes each containing four qubits in a star topology by low-loss interconnects and demonstrated state transfer and the eventual entanglement of twelve qubits. }
\item{Finally, the separation between nodes was greatly increased: Some experiments focused on increasing the nominal length of the link, but kept nodes in a standard cryostat for now.
Two chips were connected by a spooled-up 64-meter-long cable bus with an  ultralow loss of $0.32\,\textrm{dB/km}$  to perform pitch-and-catch of modes with time-reversal symmetry and distribute entanglement in an experiment using newly-developed `gmon' couplers \cite{qiu2023deterministicquantumteleportationdistant}. Others established cryo-links between separate cryostats housing the nodes. Transfer between qbits over 5 m of aluminum waveguide kept at mK and entanglement distribution was shown in \cite{magnard2020microwave}, while continuous wave entanglement was distributed in a 6.6 m sized ``quantum local area network" in \cite{yam2023cryogenic}. On a more fundamental level, distances were pushed to 30 m, preparation, protocol, and read-out times were reduced, and fidelities improved to finally achieve ``Loophole-free Bell inequality violation with superconducting circuits" in the tour-de-force experiment of \cite{storz2023loophole}. }
\end{itemize}

We finish this section by briefly outlining connections to other protocols that can serve as an alternative to the direct transfer of a quantum state presented here.
Quantum teleportation (for references, see the review article \cite{hu2023progress}) conveys an unknown quantum state by first sharing a common entangled state as a resource between the sender and receiver via a quantum channel. The sender then performs Bell measurements between their part of the entangled state and the unknown qubit and classically transmits the results to the receiver, which obtains the teleported quantum state through the corresponding measurements. While there are major fundamental differences between transfer via teleportation or through direct physical transferral,
we have seen above that elements of a direct state transfer protocol can also be used to generate entangled states. Quantum teleportation protocols have been performed in the optical domain over long distances, as mentioned on many occasions above. Related protocols, such as remote state preparation \cite{bennett2001remote}, have also been realized in hybrid atom-photon systems \cite{rosenfeld2007remote}, which are seen as another promising potential building block for quantum networks. 

\subsection{Conclusion} \label{ssec:PhysicalBasisConclusions} 

This section aimed to bridge the gap between theory and experiment by analyzing promising physical realizations of quantum information transfer, processing, and storage, while highlighting strengths and weaknesses. Therefore, it provided a brief overview of how to proceed to realize the abstract concepts and schemes for future applications discussed in the other sections. Based on the variety of technological approaches presented and discussed, one can get an impression of the complexity of the realization and an insight into the current state of the technologies.  

We started with an overall view of very different quantum information carriers. Here we pointed out that the most advanced realizations for transferring quantum information are given by electromagnetic waves in particular in the optical regime but, for specific use cases, also in the microwave regime. The more extreme ends of the electromagnetic spectrum are less promising because of crucial limitations in range and practical feasibility. However, other approaches may give further possibilities to realize interfaces between different platforms, or processing and storage units of the information.
Especially massive particles are used in quantum processing units as quantum computing devices. Such processing units together with the even more important storage units present the nodes for a future quantum network. Physical implementations of these components, such as various qubit platforms, were briefly introduced, and the major advantages and shortcomings of each proposal were underlined. In particular, hybrid approaches are promising for future applications, as they combine heterogeneous platforms in an attempt to reap the benefits of each while mitigating drawbacks. 
For a quantum information transfer, the realization of the channels is essential. Due to developments in the establishment of quantum key distribution testbeds, huge progress has been made with regard to first realizations of free-space atmospheric, satellite, and fiber-based solutions for optical frequencies.

Finally, to give an example of the realization of the whole process of quantum information transfer, we illustrated the direct transfer of an arbitrary quantum state between two network nodes realized in the microwave domain. This included the emitting process from a memory, the transfer through a cavity, and the reception of quantum microwave states. 

Basic principles of a quantum network have been realized in various platforms and quick progress is continuously made for all important components. We stand on the brink of the quantum network era, and networks with a small number of nodes, of limited extent, or with limited functionality are becoming available for technological applications.

\pagebreak

\section{Quantum network applications} \label{sec:Applications}

In the context of designing future quantum networks, Sects.~\ref{sec:Interface} and \ref{sec:PhysicalBasis} focused on fundamental physical processes, interfaces to the classical regime, and the realization of quantum information transfer. In this section, we discuss concepts for applications that are enabled by such quantum networks for advanced quantum communication. Its subsections are ordered from the most general to the most specific. Thus, Sect.~\ref{ssec:QInternetVision} begins with the introduction to a \emph{vision} of a future global quantum network: the \emph{quantum internet}. Its structure, capabilities, and requirements are examined, embracing first a holistic viewpoint and specializing later to the minutiae of its technical components and primitive protocols. With the description of the enabling technology out of the way, Sect.~\ref{ssec:SecurityAndDQC} starts with concrete applications, more specifically in the realm of \emph{security and authentication}. The proposed examples of \emph{blind quantum computing}, \emph{database query} and \emph{quantum key cards} are preceded by a short introduction to the fields of internet security and distributed quantum computing, the latter being a crucial building block for two of the three aforementioned examples. Finally, Sect.~\ref{ssec:VotingTheoryAndImpl} describes in detail the theoretical basis and implementation possibilities for a \emph{quantum election protocol}.
The list of applications mentioned above is by no means exhaustive, yet it should provide a good glimpse into the near- and far-future opportunities created by the upcoming advent of quantum networks.

\subsection{Vision of a quantum internet} \label{ssec:QInternetVision}

Today, the internet is an essential part of our daily life, enabling anything from the most complicated industrial process to simple instant messaging between people. Its story traces back to 1969, when ARPANET, a rudimentary network connecting few American universities, was created in an attempt to share the computational resources scattered across different institutions. At the time, very few people (if any) could have predicted the shocking cultural, industrial, economic, and social revolution that ARPANET's legacy would bring about. Today, experts~\cite{Castelvecchi.2018} believe that we are witnessing the early beginnings of another revolution: The rise of the \emph{quantum internet}.

\begin{figure}[hbt]
	\centering
	\includegraphics[width=0.7\columnwidth]{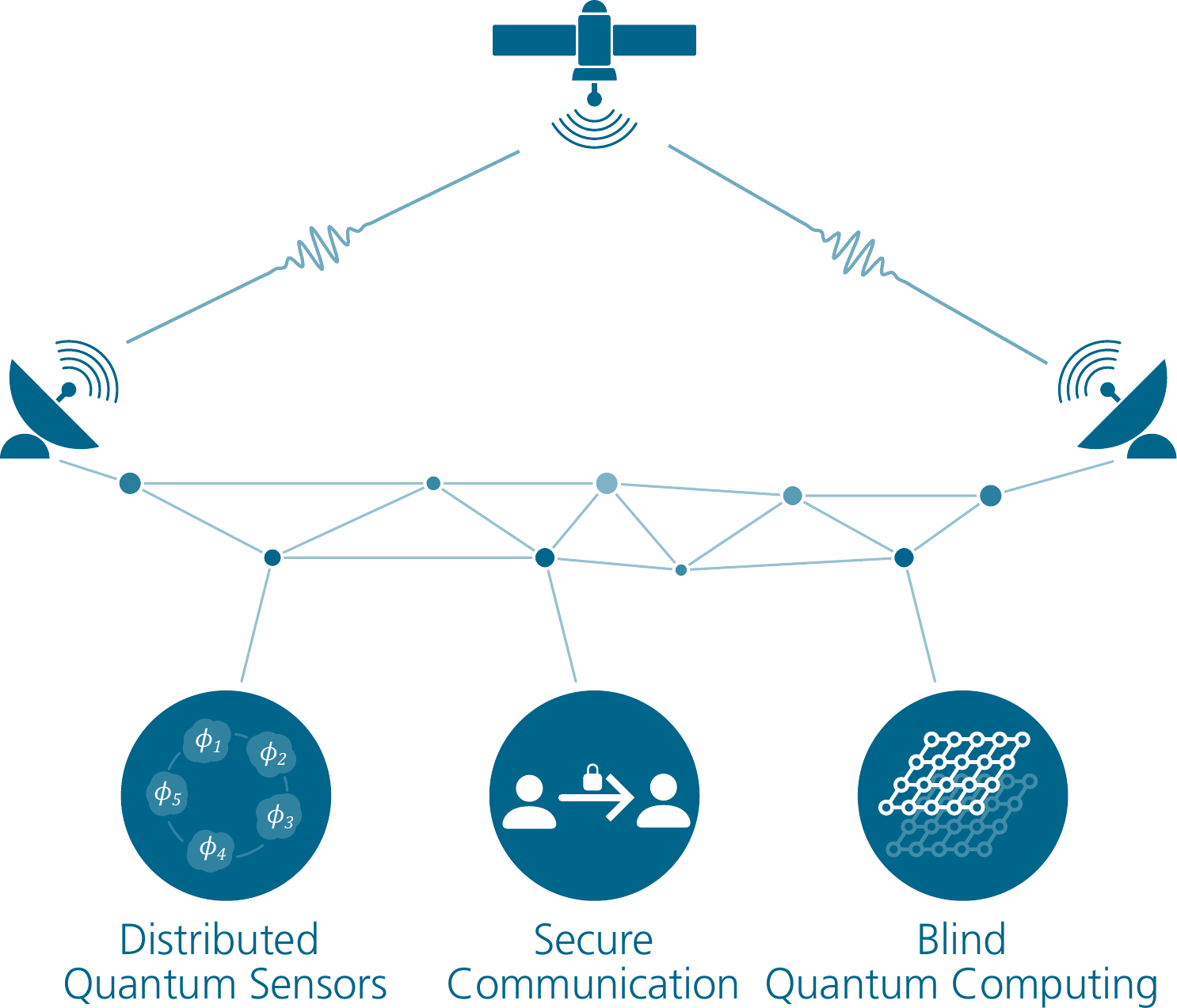} 
	\caption{Conceptual depiction of the structure of a quantum internet. Points represent nodes of the network and lines quantum channels. Examples of applications are reported at the bottom. Figure adjusted from \cite{Bongs_PhiuZ_2025}. \label{fig:QNetwork}}
\end{figure}

The quantum internet \cite{Rohde2021, Dur2017,Gyongyosi2022} is, in essence, a network consisting of systems that can generate, store, process, send, and receive quantum information. It is made up of three fundamentally distinct components \cite{Wehner2018} (see also the graphic representation in Fig.\,\ref{fig:QNetwork}):
\begin{itemize}
    \item \emph{End nodes}, which are essentially the parties that want to exchange quantum information. They can practically be anything from \emph{quantum sensors} \cite{Zhang.2021} to \emph{quantum memories} (more on these in Sect.\,\ref{sssec:ComponentsQI}) or fully fledged \emph{quantum computers};

    \item \emph{Quantum channels}, namely the physical supports over which quantum information is carried across nodes. Their physical realization has been discussed in detail in Sect.\,\ref{sec:PhysicalBasis}. We mention in passing that classical channels will still play an essential supporting role for many protocols.  
    In other words, the classical and quantum internet are expected to coexist and collaborate;

    \item \emph{Quantum repeaters}, whose role is to enable long-distance communication despite the fundamental limitation that quantum information cannot be simply copied or amplified without changing the quantum system. They are intrinsically different from classical repeaters and are thus at the center of intense theoretical research (cf.\,\cite{Munro.2015} and Sect.\,\ref{sssec:ComponentsQI} of this document).
\end{itemize}
Our interest in the quantum internet is justified by the following (incomplete) list of possible applications, which testify its disruptive potential. In the realm of cryptography, the quantum internet shall make it possible to provide long-term safe symmetric keys using \emph{quantum key distribution} protocols \cite{Bennett.2014} and secure authentication (for example, through quantum key cards, cf.\,Sect.\,\ref{sssec:QKeyCard}, or the protocol presented in \ref{sssec:Security}). Further potential applications deal with elections that cannot be rigged (see traveling ballot protocol, Sect.\,\ref{sssec:Elections}), or money that is prevented from being counterfeited \cite{Aaronson.2009, Lutomirski2009, Farhi.2012, Aaronson.2012, Gavinsky2012}. Moreover, it will bring distributed tasks that are classically hard or impossible, such as Byzantine agreement \cite{BenOr.2005} and leader election \cite{Ganz.2017}, within technological reach. Making distant quantum processors work in unison, according to the paradigms of \emph{distributed quantum computing} (see Sect.\,\ref{sssec:DQC}), will furthermore boost our computational power, hopefully expediting research in areas that benefit from a quantum treatment, such as quantitative finance, simulation of complex systems (e.g., for weather forecasting), and optimization \cite{Abbas.2024}. Finally, the end user will have safe access to quantum machinery via \emph{blind quantum computing} protocols, cf.\,Sect.\,\ref{sssec:BQC}.

The rest of this section elaborates on the cues given above and is organized more precisely as follows. In Sect.\,\ref{sssec:ComponentsQI}, we briefly examine the components of the quantum internet, their role, and the technological challenges that concern their physical implementation. In Sect.\,\ref{sssec:QTransfer}, we focus on \emph{quantum channels} and \emph{repeaters}, laying down examples of practical protocols for quantum communication and highlighting the pivotal role of entanglement swapping. In addition, we discuss the importance of network topology and design. 

We remark that the current Sect.\,\ref{ssec:QInternetVision} shall be seen as a preliminary to the more application-oriented Sects.\,\ref{ssec:SecurityAndDQC} and \ref{ssec:VotingTheoryAndImpl}, concerned respectively with \emph{security} and \emph{distributed-computing} applications, and with the example of \emph{quantum election scheme} in more detail. The quantum internet is indeed the crucial \emph{enabling technology} for all the concrete applications presented in those sections.

\subsubsection{Components of the quantum internet} \label{sssec:ComponentsQI}

This section delves a bit deeper into the single components that make up the quantum internet: \emph{end nodes, quantum channels}, and \emph{repeaters}. All three components benefit greatly from \emph{quantum memory}. Thus, we start with a definition of a quantum memory and an overview of its features. Then, we take a look at the three aforementioned components in a systematic manner.

The term \emph{quantum memory} \cite{Heshami.2016} simply denotes a device capable of coherently storing quantum information for a long enough period of time, namely much longer than the time required to perform the desired protocol, such as a quantum computation, one or more rounds of communication, or a subsystem measurement. The simple truth that neither computation nor communication can occur instantly is enough to justify the need for some form of \emph{information storage}, i.e.\,, of quantum memories. More pragmatically, we will see shortly how memories can enable specific protocols or make them impossible, depending on their availability at nodes of a network or repeaters in between. It should be remarked that the long coherence time required from quantum memories will almost certainly be achieved only through some form of error correction \cite{Terhal.2015}. Probably, fault-tolerant quantum memories will emerge earlier than fault-tolerant quantum computers, which also demand fault-tolerant gates. 

The overview of the three main components of the quantum internet and their realistic hardware implementation follows. In discussing each component, we often refer to the ``stages" of network development envisioned in \cite{Wehner2018}. A quantum network is said to have reached a certain stage when all its nodes, quantum channels, or repeaters possess a certain functionality. Our reference \cite{Wehner2018} identifies five main stages: \emph{Prepare and measure}, where all nodes are capable of preparing and measuring arbitrary quantum states; \emph{Entanglement distribution}, where maximally entangled pairs can be created between any two nodes; \emph{Quantum memory}, where all nodes have long storage times and can perform elementary processing on few qubits; \emph{Few-qubit fault-tolerant}, where each node is a fault-tolerant quantum computer, but small enough to be simulated classically; \emph{Quantum computing}, where each node can store and fault-tolerantly process more logical qubits than any classical machine can simulate. Although it is impossible to foresee what trajectory the development of quantum networks will follow, any envisioned path, e.g.~the one of \cite{Wehner2018}, represents a useful tool in organizing our overview of quantum-internet components.
\begin{enumerate}
    \item \emph{End nodes.} There are very few restrictions on what an end node could be and the functionalities it may have. The basic requirement is the ability to send and receive quantum information: Otherwise, it simply would not be part of a quantum network. However, this feature alone does not make the end node really functional. Most publications agree \cite{Wehner2018} in considering prepare-and-measure nodes, capable of preparing and measuring arbitrary quantum states, as the first technologically relevant stage of quantum-network development. Prepare-and-measure networks (namely, where all the nodes share this feature) are already capable of implementing protocols like quantum key distribution (QKD, see box in Sect.\,\ref{sssec:Security}). The addition of quantum memories takes networks to the homonymous stage. \emph{Blind quantum computing} becomes possible, as long as at least one node has the processing power to act as a server. Simple \emph{leader election protocols} and \emph{clock synchronization} \cite{Komar.2014} are also feasible. The few-qubit fault-tolerant stage enables the first, simple \emph{distributed quantum computations}. Finally, in the quantum computing stage, distributed algorithms of arbitrary width and depth can be executed, as well as classically hard network tasks like \emph{byzantine agreement} \cite{Lamport.1982}.

    \item \emph{Quantum channels.} The physical transfer of quantum information from one node to another has to take place in some medium, dubbed quantum channel in this context. The most promising information carriers for quantum communication are photonic qubits, and quantum channels for their propagation thus range from dark and multiplexed fibers, to free space. A lot of information on alternative channel types, their performance, advantages and disadvantages can be found in Sect.~\ref{sec:PhysicalBasis}, and we refrain from repeating it here. Let us only stress that no channel is perfectly lossless, and hence no advance in channel technology can remove the need for the next networks component: quantum repeaters. 

    \item \emph{Quantum repeaters.} The task of quantum repeaters is to enable quantum communication between nodes that are too far apart for direct, coherent information transfer through an ordinary quantum channel. Since quantum information cannot be amplified, these devices work, by necessity, completely differently from their classical counterparts. They may either extend coherence times of the transferred information, so that it reaches its destination unadulterated, or provide shared entanglement between the nodes, $A$ and $B$, that need to be connected. This entanglement, together with inexpensive classical communication, can later enable quantum communication via protocols like \emph{quantum teleportation}, to be discussed at length in Sect.\,\ref{sssec:QTransfer}. In realistic scenarios, the second alternative appears more feasible than the first one. As always, consider the example of photonic communication. A repeater could extend the range of viable communication by error-correcting a message encoded in a ``logical photonic qubit". Such a logical qubit would most likely consist in a multi-photon entangled state, that needs to be generated deterministically for practical usability. Photonic states of this kind are produced, currently, with extremely low success rates \cite{Schwartz.2016}, and no breakthroughs in this technology are foreseeable in the near future. By contrast, the paradigm of \emph{entanglement swapping} is theoretically well-established, and has been experimentally demonstrated (see e.g.\,\cite{Jia.2004}). Here is how entanglement swapping works, in a nutshell. More details will be provided in Sect.\,\ref{sssec:QTransfer}, while discussing means of entanglement distribution. Suppose you have two nodes, $A$ and $B$, and a repeater $R_1$ in between. Node $A$ ($B$) locally prepares a Bell state
    \begin{equation}
        \ket{\Phi^+}_{a_1, a_2} \coloneqq \frac{(\ket{00} + \ket{11})_{a_1, a_2}}{\sqrt{2}}
    \end{equation}
    on qubits $a_1, a_2$ ($ \ket{\Phi^+}_{b_1, b_2} $ on qubits $b_1, b_2$) and sends $a_2$ ($b_1$) to a repeater $R_1$. The repeater then performs a joint Bell-basis measurement on $ a_2, b_1 $. After the measurement, the residual qubits $ a_1, b_2 $ are maximally entangled, and more precisely in state $ \ket{\Phi^+}_{a_1, b_2} $. The process can then be iterated to allow creation of entangled pairs across arbitrarily large distances with high fidelity.

    There is a broad agreement, within the scientific community, that repeaters should be assumed \emph{trusted}, at least in the early stages of the quantum internet. That is to say, the first networks shall be designed under the assumption that repeaters only do what they were built for (entanglement distribution), and cannot be tampered with by malicious agents. However, it should be noted that this is a strong assumption and needs to be further developed, particularly with regard to security requirements. Satellites, which are much harder to manipulate than regular hardware on Earth, could be considered \emph{trusted} to a good approximation \cite{Bongs_PhiuZ_2025}, while always accounting for the vulnerabilities of free-space propagation. Of course, in the long run, network architectures should be designed to be resilient against untrusted nodes. 
\end{enumerate}
Further details on the structure of the quantum internet go beyond the scope of this brief introduction. However, a comment on the experimental state of the art is in order. Functional quantum networks, namely in the \emph{prepare and measure} stage, are in their early stages. However, networks of \emph{trusted repeaters} communicating to each other through QKD have been realized over metropolitan distances \cite{Peev.2009, Sasaki.2011, Stucki.2011}, and QKD itself is partially at the beginning of the stage of commercial maturity \cite{Pljonkin.2018}. The road towards a fully fledged quantum internet is still long and full of obstacles, but recent technological advancements \cite{Lvovsky.2009, Northup.2014, Awschalom.2018, reiserer2015cavity} suggest that the first (genuinely) quantum networks may see the light before the end of the decade.

\subsubsection{Quantum information transfer} \label{sssec:QTransfer}

All of the genuinely quantum applications involving more than one party require the coherent (and, possibly, secure) exchange of qubits among nodes. The discipline studying how this exchange is performed, both on theoretical and practical grounds, takes the name of \emph{Quantum Information Transfer} (QIT in the following) and is the subject of this section. Despite the striking similarities with \emph{Classical Information Transfer}, our body of knowledge on the latter cannot be easily translated to QIT, due to a series of inherent differences between the classical and quantum world, which we summarize in Tab.\,\ref{tab:QuantVsClass}.

\begin{table}[hbt]
    \centering
    \begin{tabular}{c|c}
        CLASSICAL & QUANTUM \\
        \hline
        Read at will & Information-disturbance tradeoff \\
        \hline
        Copy at will & No-cloning theorem \\
        \hline
        Information stored for centuries & Very short coherence times
    \end{tabular}
    \caption{Main differences between classical and quantum systems, from an information-theoretic perspective.}
    \label{tab:QuantVsClass}
\end{table}

The peculiar traits of quantum information when it comes to \emph{readout, cloning} and \emph{storage} look like severe limitations from a classical standpoint: Qubits are fragile, cannot be simply copied, and are destroyed by observation. In addition, they cannot generally carry more information than ordinary bits due to the Holevo bound, cf.\,Eq.\,\eqref{eq:holevobound} in Sect.\,\ref{ssec:Holevo}. Yet, from a different point of view, these limitations are precious resources: For example, the inability to copy or read a message without altering it is at the very heart of information-theoretic secure communication protocols. It is therefore worthwhile to embrace this alternative viewpoint in order, on the one hand, to explore the potential and limitations of quantum communication and, on the other, to enable the many groundbreaking applications proposed in this section. Among others: \emph{Distributed quantum computing} (cf.\,Sect.\,\ref{sssec:DQC}) and its variants like \emph{blind quantum computing} (cf.\,Sect.\,\ref{sssec:BQC}); authentication via \emph{quantum key cards} (cf.\,Sect.\,\ref{sssec:QKeyCard}); \emph{Quantum Elections} (cf.\,Sect.\,\ref{sssec:Elections}) and more. One should, finally, not forget that quantum information transfer is an integral part of the \emph{quantum internet} envisioned in Sect.\,\ref{ssec:QInternetVision}.

The section is organized, more specifically, as follows. In the first paragraphs, we examine the most basic case of quantum information transfer: Two parties and a single quantum link (\emph{QLink} for short). We propose three elementary protocols for single-qubit transfer: \emph{Direct Information Transfer} (DIT), \emph{Direct Quantum Teleportation} (DQTp) and \emph{Quantum Teleportation} (QTp). The first two bear the adjective ``direct" because realistic implementations thereof are likely to require physical exchange of qubits between the parties, as shall be seen later. For each protocol, we analyse the required resources and comment on their security against \textit{eavesdropping}, a theme of great relevance for higher-level applications (cf.\,Sect.\,\ref{sssec:Security}). We then stress how teleportation protocols require pre-existing shared entanglement, i.e., a phase of \emph{entanglement distribution} preceding the actual transfer. This phase is discussed next. In closing, we comment on the multi-party case and on how the topology of a network influences its communication requirements. 

As a quick remark before proceeding, we stress how this section differs from Sect.\,\ref{ssec:QSDC} and Sect.\,\ref{sec:PhysicalBasis}. The former does concern \emph{quantum information transfer}, but focuses on its security aspects only, neglecting resource efficiency. The latter investigates physical carriers of quantum information, whereas this section puts the emphasis on theoretical protocols. We also mention, in passing, that DQTp is not a standard nomenclature: To the best of our knowledge, this simple protocol has no widespread name in the literature.
\vspace{1\baselineskip}

\subsubsection{Single-qubit transfer methods} \label{sssec:1QbitTransfer}

As anticipated, we propose three protocols \cite{FelicitasBinder.03092024} to exchange a single qubit between two parties, Alice ($A$) and Bob ($B$). They are introduced by specifying their quantum circuit, and later compared from the standpoints of resource consumption and security. Throughout the discussion, we assume that there is a single quantum link between $A$ and $B$, namely remote two-qubit operations can only involve one (fixed) qubit per party. This setup mimics various current computing architectures, where only fixed subset of the available stationary qubits can be couple to flying ones for communication purposes.

\paragraph{Direct Information Transfer} 

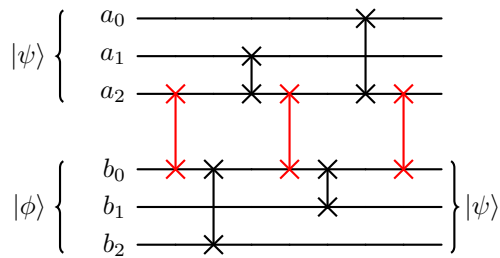
\begin{figure}[hbt]
    \centering
    \begin{quantikz}
        \lstick[3]{$\ket{\psi}$} \hspace{6pt} & \lstick{$a_0$} \wireoverride{n} & & & & & & \swap{2} & & & \wireoverride{n} \\
        & \lstick{$a_1$} \wireoverride{n} & & & \swap{1} & & & & & & \wireoverride{n} \\
        & \lstick{$a_2$} \wireoverride{n} & \swap[style=red]{2} & & \targX{} & \swap[style=red]{2} & & \targX{} & \swap[style=red]{2} & & \wireoverride{n} \\
        \wireoverride{n} \hspace{15pt} & \wireoverride{n} & \wireoverride{n} & \wireoverride{n} & \wireoverride{n} & \wireoverride{n} & \wireoverride{n} & \wireoverride{n} & \wireoverride{n} & \wireoverride{n} & \wireoverride{n} \\
        \lstick[3]{$\ket{\phi}$} \hspace{6pt} & \lstick{$b_0$} \wireoverride{n} & \targX[style=red]{} & \swap{2} & & \targX[style=red]{} & \swap{1} & & \targX[style=red]{} & \rstick[3]{$\ket{\psi}$} \\
        & \lstick{$b_1$} \wireoverride{n} & & & & & \targX{} & & & & \wireoverride{n} \\
        & \lstick{$b_2$} \wireoverride{n} & & \targX{} & & & & & & & \wireoverride{n} \\
    \end{quantikz}
    \caption{Direct Information Transfer (DIT), three qubits per party. The black (red) two-qubit gates are local (non-local). The notation $ j_k $, $(j=a,b; \, k=0,1,2)$, stands for $j$'s $k$-th qubit. Figure adapted from \cite{FelicitasBinder.03092024}.  \label{fig:DIT}}
\end{figure}

This simple protocol implements the classical idea of ``cut-pasting" information. It consists of literally swapping the qubits in $A$'s and $B$'s possession via a sequence of local and non-local SWAP gates. Its circuit representation is found in Fig.\,\ref{fig:DIT} for the three-qubit case. The generalization to higher qubit numbers is conceptually identical.
Notice that this method requires precisely $n$ non-local two-qubit gates for the transfer of $n$ qubits. Moreover, the latter must be physically sent from $A$ to $B$. This is, currently, of extreme technical difficulty even for short distances. One should also stress that very few guarantees exist on the security of the protocol: The traveling qubits may be intercepted and acted upon by an eavesdropper $E$ (Eve), and neither $A$ nor $B$ would be able to tell without measuring (and thus destroying) the received information. On the positive side, the protocol requires no ancillary qubits.

\paragraph{Direct Quantum Teleportation} 

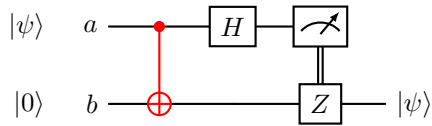
\begin{figure}[hbt]
    \centering
    \begin{quantikz}
        \lstick{$\ket{\psi}$} \hspace{6pt} & \lstick{$a$} \wireoverride{n} & \ctrl[style=red]{1} & \gate{H} & \meter{} \wire[d][1]{c} & \wireoverride{n} & \wireoverride{n} \\
        \lstick{$\ket{0}$} \hspace{6pt} & \lstick{$b$} \wireoverride{n} & \targ[style=red]{} & & \gate{Z} & \rstick{$\ket{\psi}$} 
    \end{quantikz}
    \caption{Direct Quantum Teleportation (DQTp). As in Fig.\,\ref{fig:DIT}, non-local gates are reported in red. Figure adapted from \cite{FelicitasBinder.03092024}. \label{fig:DQTp}}
\end{figure}

This method relies directly on quantum effects, as it employs entanglement as a communication resource. The circuit implementing the protocol, for two qubits, is depicted in Fig.\,\ref{fig:DQTp}.
We can divide it into two conceptually distinct phases. First, entanglement between the state $\ket{\psi}$, that we wish to transfer, and the ``recipient" qubit (prepared in $ \ket{0} $) is established via a CNOT. The second phase only involves local operations and classical communication: Alice measures her qubit in the $x$-basis (equivalently, applies $H$ and measures in the computational basis) and sends the measurement outcome to Bob over a classical channel. Bob then reconstructs the original state $\ket{\psi}$ by applying a $Z$ gate on his qubit, conditioned on the outcome of Alice's measurement. In greater detail, the protocol proceeds as follows:
\begin{equation}
    \begin{aligned}
    \ket{\psi}_a \otimes \ket{0}_b \equiv ( \alpha \ket{0} + \beta \ket{1})_a \otimes \ket{0}_b &\overset{CNOT}{\longrightarrow} (\alpha \ket{00} + \beta \ket{11})_{ab} \\
    &\overset{H}{\longrightarrow} \frac{1}{\sqrt{2}} \big( \alpha (\ket{00} + \ket{10}) + \beta ( \ket{01} - \ket{11} ) \big)_{ab} \\
    &\overset{\text{measurement}}{\longrightarrow}
    \begin{cases}
        (\alpha \ket{0} + \beta \ket{1})_b & \text{if} \ \ket{0}_a \ \text{measured} \\
        (\alpha \ket{0} - \beta \ket{1})_b & \text{if} \ \ket{1}_a \ \text{measured}
    \end{cases} \\
    &\overset{CZ}{\longrightarrow} (\alpha \ket{0} + \beta \ket{1})_b \,.
    \end{aligned}
\end{equation}
Several remarks are in order. First, the protocol still requires one non-local gate per transferred qubit, just like direct information transfer. However, the gate is a CNOT rather than a SWAP. It is customary to choose CNOT, rather than SWAP, as part of the elementary gate set for a universal quantum computer \cite{Boykin.2000}. Given this choice, a SWAP gate decomposes as three CNOTs \cite{Barenco.1995} and is thus three times more expensive, favoring direct quantum teleportation over direct information transfer. Second, this transfer method is \emph{unidirectional} by design. Bidirectional extensions require doubling the number of qubits of each party. Third, direct quantum teleportation still requires to execute a costly, possibly slow and (as of now) relatively error-prone non-local gate ``on the fly", making it unsuitable for applications that are time-sensitive or whose success can be hindered by a communication error.

The example of direct quantum teleportation also illustrates a unique quantum effect: The separation of \emph{information transfer} and \emph{communication}. The two are always paired in the classical realm: $A$ can only communicate a bit of information to $B$ by transferring it to them. The same is not always true in the quantum world, and particularly in this protocol. Indeed, consider the following realistic implementation of the CNOT in Fig.\,\ref{fig:DQTp}: Bob sends his qubit to Alice, who locally performs the CNOT and sends the qubit back. Even though a qubit was physically \emph{transferred} from $A$ to $B$ and back, no communication has taken place yet. The latter only occurs after Bob receives the classical bit encoding the result of Alice's measurement. This peculiar split between transfer and communication could also be phrased as follows. Place the start of the direct quantum teleportation protocol \emph{after} the application of the CNOT. Then, we could say that
\begin{equation}
    1 \ \text{qubit} = 1 \ \text{ebit} + 1 \ \text{bit} \,,
\end{equation}
that is, one qubit can be \textit{communicated} by spending one shared \emph{entanglement bit} (\emph{ebit}, see definition in Sect.\,\ref{ssec:entanglement}) and \emph{transferring} one (classical) bit. The possibility of splitting quantum communication into a phase of \emph{entanglement distribution} followed (in due time) by purely classical information transfer lies at the foundation of other communication schemes, like the \emph{quantum teleportation} that we shall discuss next.

We close the analysis of direct quantum teleportation by reporting the circuit for the two-qubit case. The extension to $n$ qubits follows by induction.
\begin{figure}[hbt]
    \centering
    \begin{quantikz}
        \lstick[2]{$\ket{\psi}$} \hspace{12pt} & \lstick{$a_0$} \wireoverride{n} & & & \swap{1} & & & & \meter{} \wire[d][3]{c} & \wireoverride{n} & \wireoverride{n} \\
        \wireoverride{n} \hspace{12pt} & \lstick{$a_1$} \wireoverride{n} & \ctrl[style=red]{1} & \gate{H} & \targX{} & \ctrl[style=red]{1} & \gate{H} & \meter{} \wire[d][1]{c} & \wireoverride{n} & \wireoverride{n} & \wireoverride{n} \\
        \lstick{$\ket{0}$} \hspace{12pt} & \lstick{$b_0$} \wireoverride{n} & \targ[style=red]{} & \swap{1} & & \targ[style=red]{} & & \gate{Z} & & \rstick[2]{$\ket{\psi}$} \\
        \lstick{$\ket{0}$} \hspace{12pt} & \lstick{$b_1$} \wireoverride{n} & & \targX{} & & & & & \gate{Z} & & \wireoverride{n} \\
    \end{quantikz}
    \caption{Direct Quantum Teleportation (DQTp), two qubits per party. Figure adapted from \cite{FelicitasBinder.03092024}. \label{fig:DQTp2QBits}}
\end{figure}
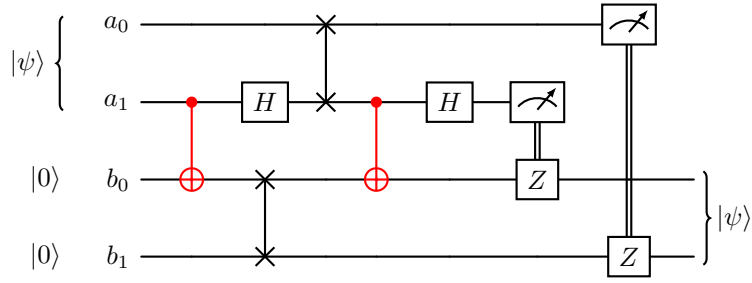

\paragraph{Quantum Teleportation} 

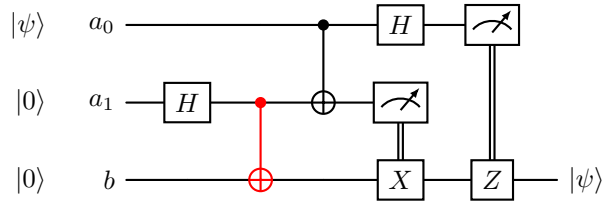
\begin{figure}[hbt]
    \centering
    \begin{quantikz}
        \lstick{$\ket{\psi}$} \hspace{12pt} & \lstick{$a_0$} \wireoverride{n} & & & \ctrl{1} & \gate{H} & \meter{} \wire[d][2]{c} & \wireoverride{n} & \wireoverride{n} \\
        \lstick{$\ket{0}$} \hspace{12pt} & \lstick{$a_1$} \wireoverride{n} & \gate{H} & \ctrl[style=red]{1} & \targ{} & \meter{} \wire[d][1]{c} & \wireoverride{n} & \wireoverride{n} & \wireoverride{n} \\
        \lstick{$\ket{0}$} \hspace{12pt} & \lstick{$b$} \wireoverride{n} & & \targ[style=red]{} & & \gate{X} & \gate{Z} & \rstick{$\ket{\psi}$} \\
    \end{quantikz}
    \caption{Quantum Teleportation (QTp). The first two gates turn qubits $a_1, b$ into a shared Bell pair, and could be omitted if shared entanglement is assumed as a prerequisite of the protocol. Figure adapted from \cite{FelicitasBinder.03092024}. \label{fig:QTp}}
\end{figure}

This is the protocol introduced in \cite{Bennett.1993}. Its circuit representation, for the single qubit case, is depicted in Fig.\,\ref{fig:QTp}.
Just like for direct quantum teleportation, it is useful to divide the protocol into two parts. In the first part, which comprises the first two time slices of the circuit in Fig.\,\ref{fig:QTp}, entanglement is established between Bob's qubit $b$ and Alice's ancillary qubit $a_1$ by preparing the joint Bell state \cite{NielsenChuang}
\begin{equation}
    \ket{\Phi^+}_{a_1,b} = \frac{(\ket{00} + \ket{11})_{a_1,b}}{\sqrt{2}} \,. \label{eq:BellPair}
\end{equation}
The second part uses this pre-established entanglement, local operations and classical communication to teleport Alice's qubit $a_0$, in state $\ket{\psi}$, onto Bob's qubit $b$. The full details are omitted here, but can be found in \cite{Bennett.1993, FelicitasBinder.03092024}.

Quantum Teleportation is much more flexible than direct quantum teleportation, despite sharing the paradigm of entanglement distribution followed by classical communication only. The latter protocol requires having $ \ket{\psi} $, the state to be transferred, available at the time of entanglement distribution. This may not be feasible, for example if $\ket{\psi}$ represents the result of a computation to be performed at a later time. Quantum teleportation poses no similar constraints: Alice can transfer any state at any time by exploiting the pre-established Bell pair, cf.\,Eq.\,\eqref{eq:BellPair}.

This protocol also offers advantages over direct information transfer (DIT), in spite demanding the ancillary qubit $a_1$. It is indeed much safer than DIT, because no transfer of qubits is required in the second part of the protocol, and the maximally entangled state $ \ket{\Phi^+}_{a_1,b} $ can be established securely in advance \cite{Pirker.2017}. In a nutshell, this is because all the information carried by $ \ket{\Phi^+}_{a_1,b} $ lies in its correlations, and cannot be accessed by intercepting and measuring the single transmitted qubit (be it $a_1$ or $b$): 
\begin{equation}
    \operatorname{Tr}_b \big( \ket{\Phi^+} \bra{\Phi^+}_{a_1,b} \big) = \frac{\mathds{1}_{a_1}}{2} \quad \text{and} \quad \operatorname{Tr}_{a_1} \big( \ket{\Phi^+} \bra{\Phi^+}_{a_1,b} \big) = \frac{\mathds{1}_{b}}{2} \,.
\end{equation}
On top of that, intercepting and reading both bits exchanged in the second part still says nothing on the transferred quantum state.

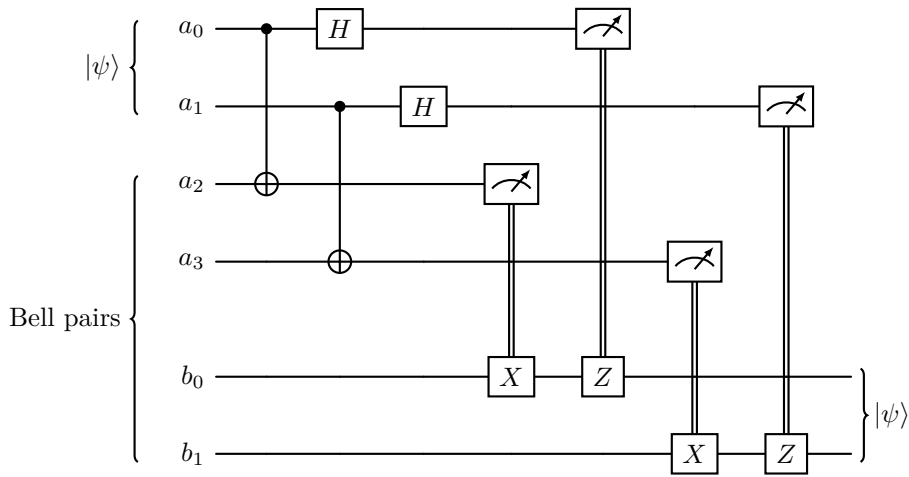
\begin{figure}[hbt]
    \centering
    \begin{quantikz}
        \lstick[2]{$\ket{\psi}$} \hspace{12pt} & \lstick{$a_0$} \wireoverride{n} & \ctrl{2} & \gate{H} & & & \meter{} \wire[d][5]{c} \\
        \hspace{12pt} & \lstick{$a_1$} \wireoverride{n} & & \ctrl{2} & \gate{H} & & & & \meter{} \wire[d][5]{c} \\
        \lstick[5]{Bell pairs} \hspace{12pt} & \lstick{$a_2$} \wireoverride{n} & \targ{} & & & \meter{} \wire[d][3]{c} \\
        \hspace{12pt} & \lstick{$a_3$} \wireoverride{n} & & \targ{} & & & & \meter{} \wire[d][3]{c} \\
        \hspace{12pt} \wireoverride{n} & \wireoverride{n} \\
        \hspace{12pt} & \lstick{$b_0$} \wireoverride{n} & & & & \gate{X} & \gate{Z} & & & \rstick[2]{$\ket{\psi}$} \\
        \hspace{12pt} & \lstick{$b_1$} \wireoverride{n} & & & & & & \gate{X} & \gate{Z} & \\
    \end{quantikz}
    \caption{Quantum Teleportation (QTp), two-qubit case. Bell pairs $\ket{\Phi^+}_{a_2,b_0}$ and $\ket{\Phi^+}_{a_3,b_1}$ are assumed to have been established prior to the beginning of the protocol. Figure adapted from \cite{FelicitasBinder.03092024}. \label{fig:QTp2QBits}}
\end{figure}

We close the overview of quantum teleportation by reporting the circuit representation of a two-qubit transfer, cf.\,Fig\,\ref{fig:QTp2QBits}. In the interest of readability, the phase of Bell-pair formation is omitted. The extension to $n$ qubits follows by induction.
 \vspace{1\baselineskip}

\paragraph{Comparison of the three methods}

We briefly compare the three protocols from the points of view of required \emph{computational resources} (in the form of ancillary qubits, gates and measurements) and \emph{security}.

As far as \emph{resources} are concerned, we draw Tab.\,\ref{tab:ResourcesComparison} from \cite{FelicitasBinder.03092024}. Inspecting this table reveals that direct- and quantum teleportation (DQTp and QTp) demand the same number of costly non-local gates \cite{Vidal.2002}, greatly outperforming direct information transfer (DIT). Moreover, DQTp requires no ancillary qubits, making it the most desirable method from the strict point of view of resource efficiency. However, security considerations and the superior flexibility of QTp may tilt the balance in its favor.

From a \emph{security} standpoint, QTp is indeed the better protocol. DIT is prone to all sorts of interceptions and undesired adulteration of the transferred qubits. DQTp is more resilient in this respect: The eavesdropper $E$ cannot intercept and read (through measurement) one half of the shared entangled pair, without Alice and Bob noticing. However, $E$'s interference destroys the message, effectively jamming communication. QTp overcomes both issues by, first, creating entangled pair securely, and later employing the shared entanglement for eavesdropper-proof communication.

It is widely believed that the enhanced security of QTp more than overcomes its higher qubit demand, making it the ideal candidate protocol for quantum information transfer. Based on this belief, theoretical and experimental research has been focusing mostly on this protocol. Just recently, teleportation between non-neighboring nodes of a network was experimentally demonstrated \cite{Hermans2022}, further backing its feasibility in real-world scenarios.
\begin{table}[hbt]
    \centering
    \begin{tabular}{|c|c|c|c|}
        \hline
       Resource  & DIT & DQTp & QTp \\
       \hline
       Ancillary qubits & 0 & 0 & n \\
       \hline
       Local universal gates & 6 (n-1) & 8 n - 6 & 9 n - 6 \\
       \hline
       Non-local universal gates & 3n & n & n \\
       \hline
    \end{tabular}
    \caption{Required resources for transferring $n$ qubits using the three aforementioned methods. The number of \emph{universal} gates is counted w.r.t. the elementary gate set $\{ H,T, \mathrm{CNOT} \}$ \cite{Boykin.2000}.}
    \label{tab:ResourcesComparison}
\end{table}

QTp uses entangled states as a resource for transferring quantum information for increased security and safety of the transfer.
Thereby, the concept of splitting the transfer of physical quantum states from the transfer of quantum information is utilized.
This makes it possible to repeat the physical transfer of quantum states until the required Bell pairs have been created without touching the quantum data to be transferred.
After the successful creation of Bell pairs, the quantum data can be teleported by sending classical bits, which can be sufficiently secured against losses and do not need to be secret.

\subsubsection{Entanglement distribution} \label{sssec:EntDist} 

As seen before, parties can utilize entangled states, \emph{ebits}, in combination with classical information to transfer quantum information.
These ebits can be exchanged beforehand and independently of the actual information transfer.
This independence removes the danger of losing information over a noisy channel, as the creation of ebits can be repeated until successful, without touching the information qubit.
This is important because we cannot simply copy quantum information to send it over the channel multiple times until success, which is possible for classical information over classical channels.
As it is impractical to connect all interested parties with all-to-all connections, some form of entanglement-distributing network has to be implemented.
For these, two main modes of operation can be identified, depending on whether or not quantum memories are available.

Without memories, after mapping out a path from A to B, all intermediate nodes of the network and the communicating parties have to perfectly coordinate and synchronize sending out and measuring their qubits.
These intermediate nodes could be fiber based for local networks or satellites and satellite terminals for intercontinental exchange of entanglement.
By performing appropriate measurements, the intermediate parties achieve \emph{entanglement swapping}, destroying their ends of two pairs of entangled qubits and creating entanglement between the two remaining qubits.
Chaining together enough steps of entanglement swapping results in the creation of an entangled state between A and B, which can then be used to transfer the quantum state.

\begin{figure}[hbt]
	\centering
	\includegraphics[width=0.9\columnwidth]{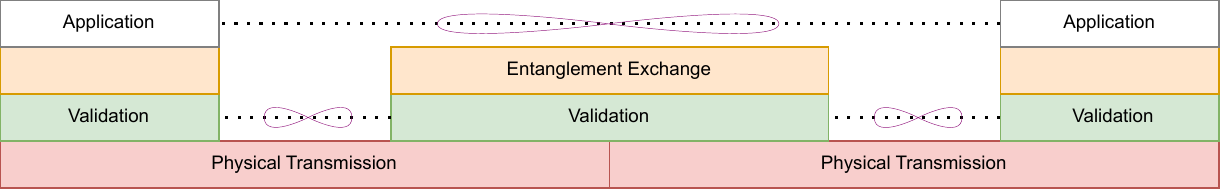}
	\caption{Creating entanglement between distant parties. In a first step, nodes which are connected by physical quantum channels exchange quantum information via the channel to create entangled pairs of qubits. By performing a joint measurement of the qubits at the central node and applying corrective operations, subject to the measurement results, at each of the two remaining nodes, an entangled state is created between the distant parties. This procedure can be layered to distribute entanglement between parties which are separated by many nodes in a quantum network. At each step, tests and entanglement purification \cite{Dur1999} can be applied to ensure the quality of the entanglement.} \label{fig:entanglement_swapping}
\end{figure}

With quantum memories, the communication network can be more flexible and powerful.
While the basic idea of creating entanglement between directly connected parties and then using entanglement swapping to connect any two parties of the network stays the same, there are several advantages.
First, the strict timing requirements are relaxed.
Every node of the network can work at its own speed while only delaying the communication instead of prohibiting it in case of imperfect synchronization.
Also, at each direct connection entanglement can be constantly accumulated for use in times of high demand, averaging out fluctuating demand.
Whenever two parties want to communicate, the intermediate nodes use their preshared entanglement to connect the parties via entanglement swapping \cite{Cacciapuoti2020, Rohde2021, Illiano2022, Li2024}.
This process needs some form of authorization or security measure, as otherwise the connecting nodes could collaborate to route the entanglement to a third party.
We give some ideas for securing connections in Sects.\,\ref{sssec:Security} and \ref{sssec:QKeyCard}.

\subsection{Applications for security and distributed computing} \label{ssec:SecurityAndDQC}

This section discusses some of the applications that will become feasible once their enabling technology, namely the quantum internet presented in Sect.\,\ref{ssec:QInternetVision}, reaches maturity. More specifically, we hereby focus on applications that concern (network) \emph{security}, \emph{distributed quantum computing} and a combination of the two.

The discussion of these themes is organized as follows. Sect.\,\ref{sssec:Security} consists in an introduction to the history and scope of \emph{internet security}. Primitives for more advanced protocols, such as QKD ``à la BB84", and basic applications, such as a simple QKD-based authentication scheme, are also presented. Sect.\,\ref{sssec:DQC} likewise introduces the other macrothematic area of \emph{Distributed Quantum Computing} (DQC), reporting in particular its motivation, state of the art and expected future developments. Sects.\,\ref{sssec:BQC}, \ref{sssec:Query} and \ref{sssec:QKeyCard} then deal with more specific applications. The first one (Sect.\,\ref{sssec:BQC}) is \emph{Blind Quantum Computing} (BQC), and sits right at the intersection between the two thematic areas. The other two (Sects.\,\ref{sssec:Query} and \ref{sssec:QKeyCard}), namely \emph{database query} and \emph{quantum key card}, lean more heavily towards the ``security" camp.

\subsubsection{Security and authentication} \label{sssec:Security}

\paragraph{A brief history of internet security} 
As already mentioned in Sect.\,\ref{ssec:QInternetVision}, the ancestor of modern internet, ARPANET, was founded in 1969 and consisted for decades in a very small network of trustworthy parties: American military facilities, public institutions and a few, selected universities. It is thus no wonder that the first communication and authentication protocols, among which the \emph{universal internet protocols} TCP/IP, where not designed with security concerns in mind: The primary focus at the time was on ``reliability, availability and performance" \cite{DeNardis.2007}.

As understandable as those priorities were, the lack of \emph{native} security features would eventually make itself felt. The first internet \emph{worm}, a self-replicating program that causes no direct harm but highly degrades performance of infected machines, was created in 1988 by the hands of Robert Morris, a graduate student at Cornell university. Morris was trialed for this act, and claimed having spread the worm in order to prompt action against the well-known security fallacies in the TCP/IP protocols. The episode certainly sparked interest in the medias and authorities, leading to what Morris had probably wished for: The creation of a task-force, the Computer Emergency Response Team (CERT), for addressing and preventing security breaches.

Morris may really have been moved by noble ideals, and his worm had the merit of dramatically raising awareness about internet security. At the same time, there is little doubt that this feat inspired less benign individuals, who saw the opportunity to capitalize on security fallacies to pursue personal gain or cause economic damage to others. The 90's and 00's were indeed decades of intense activity towards and against internet security: While the CERT, together with further American institutions and software companies, developed the first firewalls (implementing rudimentary user authentication), VPNs and public encryption schemes like RSA (based on factoring large numbers and still used to this day), anonymous ``hackers" were flooding the internet with viruses, spywares and phishing emails, disrupting e-commerce platforms with DDoS (Distributed Denial of Service) attacks, spreading concern in the society with the threat of \emph{cyberterrorism}.

Most internet attacks were eventually fended off, and there is no denying that the internet is now far safer than a few decades ago. At the same time, our history of fixing security issues via \emph{software} patches, and only after such liabilities had been exploited by malicious individuals, begs some questions: Would all these issues have surfaced, had we taken security seriously in the early days, when shaping the internet? Can we envision a new paradigm of security, that does not rely on \emph{a posteriori} software solutions, but is instead unbreachable to begin with?

\paragraph{Quantum internet and information-theoretic security} 
The first question is made extremely relevant by the early signs of the upcoming quantum internet revolution, which was discussed in Sects.\,\ref{ssec:QInternetVision} and \ref{sssec:ComponentsQI}. To avoid mistakes from the past, it is of paramount importance to start shaping the security landscape of this future network well in advance. This shall be done by proposing universal communication and security standards that grant safe interoperativity between potentially heterogeneous nodes. Such standards must moreover be envisioned in the light of the second question. Present-day cryptographic schemes rely on the inability of a malicious agent to perform the resource-intensive computations needed to decipher a given encrypted message. This notion of \emph{computational security} is shaken at the core by the advent of Quantum Computing (QC), because tasks that are classically hard may not be in the quantum realm: This is the case for prime factorization, where Shor's algorithm achieves an exponential speedup with respect to the best classical counterpart \cite{Shor.1994, Shor.1997}, raising concerns about the security of the widely used RSA asymmetric encryption scheme. New security schemes, both for the internet and its quantum extension, will have to be robust against quantum computing attacks. In other words, our focus shall shift from \emph{computationally}- to \emph{information-theoretically} secure protocols that cannot be broken, regardless of the available machinery or the kind of attack used. Ironically, \emph{information-theoretic security} is rooted in the very laws of quantum physics (``security by physicality") that made our current schemes obsolete. Nevertheless, achieving information-theoretical security is an ambitious aim with many requirements, but along the way much can be discovered and developed.

\paragraph{Overview} 
The aim of this section is thus to give a brief overview of the landscape of security and authentication in the quantum internet era. We start by listing the three main topics of current research: Secure classical communication via quantum key exchange; Secure classical communication via post-quantum cryptography; Secure quantum communication. Focusing on the first one, we provide a summary of the BB84 \cite{Bennett.2014} key distribution scheme and elucidate how it can be used to implement a simple authentication protocol. We close the section with some conclusions.

\paragraph{Some important topics in current literature} 
When considering a scenario where malicious agents have access to arbitrarily powerful quantum computers, researchers have mainly focused on the following:
\begin{enumerate}
    \item Enabling \emph{computationally safe classical communication} by designing encryption schemes that cannot be decrypted in a reasonable time frame, even with the most powerful (quantum) computer. The discipline studying such schemes has taken the name \emph{post-quantum cryptography} (PQC, see \cite{Hanafi.2025, Singh.2025} for recent reviews on the topic). This is seen by many as the most practical and the most important approach in the short term, to the point that the US National Institute of Standards and Technology (NIST) has recently finalized the selection of the first PQC standards for commercial use \cite{NationalInstituteofStandardsandTechnologyUS.2024, NationalInstituteofStandardsandTechnologyUS.2024b, NationalInstituteofStandardsandTechnologyUS.2024c, Alagic.2025}. Even earlier, in 2022, the US government had issued a memorandum \cite{Joseph.2022} advocating a generalized transition to quantum-resistant cryptography. Nevertheless, these schemes rely on mathematical assumptions of complexity and they do not grant long-term security. Due to fast and hardly predictable developments in new technologies such as AI and quantum technologies, cryptographic agility and other security layers become essential.
    
    \item Enabling \emph{safe classical communication} by encrypting messages using keys that have been distributed to separated parties through \emph{information-theoretically} secure schemes that hinge on the limitations of quantum mechanics. The most important example of such a scheme is \emph{Quantum Key Distribution}, or QKD for short, which we shall briefly describe below. Not only have some QKD protocols been proven theoretically secure, but experimental research on the subject is advanced. The first small QKD test networks have been performed over metropolitan distances \cite{Sasaki.2011, Zhang.2025, KADUM2025101774} and QKD devices are commercially available \cite{Pljonkin.2018}. Nevertheless, scaling of multi-user networks \cite{Horoschenkoff_2025}, distances between single nodes in the networks, security proofs \cite{tupkary2025qkdsecurityproofsdecoystate}, realizations of realistic use cases and further developments are still in the focus of current research. 

    \item Enabling \emph{theoretically safe quantum communication} via genuinely quantum protocols. Examples thereof follow. \emph{Quantum Secure Direct Communication} (QSDC) was introduced in the seminal reference \cite{Long.2002}. A description of the original protocol can be found in Sect.\,\ref{ssec:QSDC}, whence we hereby simply report that developments of that scheme can be found in \cite{Deng.2003}, and experimental realizations in \cite{Hu.2016, Huang.2022}. For more information, we redirect the reader to this review \cite{Pan.2023}. Alongside QSDC, some \emph{quantum authentication} protocols have been developed. The seminal reference in the field is generally considered \cite{Crepeau.1995}, whose scheme is based on oblivious transfer (cf.\,Sect.\,\ref{sssec:Query}). Ref.\,\cite{Lee2006} builds upon it to combine QSDC and authentication in a single protocol.
\end{enumerate}
Most of the schemes presented in point 3.~have been- or will be treated elsewhere in this work. PQC, on the other hand, leverages no quantum effects, and hence lies outside of the scope of the document. We thus focus on point 1., and in particular on QKD and its applications to authentication. This choice is also justified by the generalized conviction that QKD will be the dominant scheme in near-future technological applications. 

\paragraph{Short summary of QKD} 
The birth of QKD is traced back to the 1984 paper by Bennet and Brassard, reprinted in \cite{Bennett.2014} and colloquially known as BB84. The paper contains the best-known protocol for key distribution, which we quickly outline in Tab.~\ref{tab:BB84}. Our presentation is more akin to the one in \cite{SengW.loke.2024} than the original reference. 
\begin{table}[hbt]
    \centering
    \begin{tabular}{m{0.9\textwidth}}
        \hline
        \textit{Protocol: Quantum key distribution à la BB84}  \\
        \begin{itemize}
        \item Alice chooses uniformly at random an $n$-bit string $x=x_1 x_2...x_n \in \{0,1 \}^n$, and a ``basis string" $\theta = \theta_1 \theta_2...\theta_n \in \{0,1 \}^n$. She then prepares $n$ qubits in the respective states
        \begin{equation}
            \{ H^{\theta_1} \ket{x_1}, \dots , H^{\theta_n} \ket{x_n} \} \,,
        \end{equation}
        where $H$ is the Hadamard gate, acting on the computational basis like $ H \ket{0} = \ket{+} $, $H \ket{1} = \ket{-}$. The explicit correspondence between values $ (x_j, \theta_j) $ of the $j$-th entry of the random strings and the state of the $j$-th qubit is reported in the table below.
        \begin{center}
            \begin{tabular}{c|c|c|c|c}
                $(x_j, \theta_j)$ & $(0,0)$ & $(0,1)$ & $(1,0)$ & $(1,1)$  \\
                \hline
                qubit state & $\ket{0}$ & $\ket{1}$ & $\ket{+}$ & $\ket{-}$
            \end{tabular}
        \end{center}
        Alice now sends the prepared qubit states to Bob.

        \item Bob chooses at random a basis string $\theta' = \theta_1' ... \theta_n'$ and measures the qubits in the corresponding bases $\{\ket{0},\ket{1}\}$ or $\{\ket{+},\ket{-}\}$, obtaining outcome $x' = x_1'...x_n'$.

        \item Alice publicly announces $\theta$. The rounds where $ \theta_j \neq \theta_j' $ are discarded, the others kept. Let $S \coloneqq \{ j \ | \ \theta_j = \theta_j' \}$ denote the index set of the rounds that were kept.

        \item Alice picks a subset $ T \subset S $, $ |T| \simeq |S|/4 $, to test for eavesdropping. Since the same bases where used by $A$ and $B$, one expects $ x_j = x_j' $ for all $j \in S$. To account for possible imperfections in their quantum channel, $A$ and $B$ agree on an error threshold $ 0 \leq \varepsilon < 1 $ in advance. If, in this detection phase, the two find a number of errors
        \begin{equation}
            e = | \{ k \in T \ | \ x_k \neq x_k' \} |
        \end{equation}
        such that $ e/ |T| > \varepsilon $, the protocol is aborted. Alice and Bob know that their message suffered a ``measure-and-resend" attack.

        \item If the protocol succeeds,
        \begin{equation}
            x^\star = x_{i_1}...x_{i_{|S \setminus T|}} \,, \qquad i_j \in S \setminus T \ \text{for all} \ j=1,2,...,|S \setminus T| 
        \end{equation}
        is the shared secret key.
        \end{itemize}\\
        \hline
    \end{tabular} 
    \caption{Overview of the BB884 quantum key distribution protocol. \label{tab:BB84}}
\end{table}

The original protocol contained a further step of \emph{privacy amplification}, which we omit for conciseness. Despite relying on completely public quantum and classical channels, this protocol was proven theoretically secure. One should also mention that, after BB84, many (often improved) variants of QKD have appeared (e.g.\,\cite{Ekert.1991}). Our discussion of QKD will however end here, because our goal is not to analyze this scheme in itself and its security features, but rather show how it can be used to implement a simple authentication protocol. 

\paragraph{Example: Authentication with QKD} 
A simple, three-step classical authentication protocol can be improved by sharing secret keys via QKD, thus granting security. Consider, for example, the following setup \cite{Dusek.1999}. Alice and Bob have a pre-shared ``stash" of triples $ T^{(j)} = (s^{(j)}_1, s^{(j)}_2, s^{(j)}_3 ) $ of $n$-bit authentication strings $ s^{(j)}_k $, $k=1,2,3$, where $n$ can be tuned to achieved the desired level of security (the larger $n$, the safer the protocol). Assuming that they have access to an unjammable classical communication channel (an assumption that can easily be relaxed, see again \cite{Dusek.1999}), Alice and Bob can mutually authenticate as elucidated in Tab.~\ref{tab:QKDAuthentication}.
\begin{table}[hbt]
    \centering
    \begin{tabular}{m{0.9\textwidth}}
        \hline
        \textit{Protocol: QKD-enhanced three-step authentication}  \\
        \begin{itemize}
    
	    \item $A$ and $B$ tell each other the index $k_A$ ($k_B$) of the first ``unused" triple  $ T^{(k_A)} $ \big($ T^{(k_B)} $\big) in their pre-shared stash. If $ k_A \neq k_B $, they pick $ k \coloneqq \max \{ k_A, k_B \} $;

        \item $A$ sends $ s^{(k)}_1 $ to $B$;

        \item $B$ checks if the first identification string, that he just got from $A$, matches his own. If it does, he sends $ s^{(k)}_2 $ back to Alice. Otherwise, $B$ aborts the protocol and discards the triple $T^{(k)}$;

        \item $A$ checks if the second identification string, that she just got from $B$, matches her own. If it does, she sends $ s^{(k)}_3 $ back to Bob. Otherwise, $A$ aborts the protocol and discards the triple $T^{(k)}$;

        \item $B$ checks if the third identification string, that he just got from $A$, matches his own. If it does, the authentication is successful! Otherwise, $B$ aborts the protocol and discards the triple $T^{(k)}$;

        \item Upon authentication, Alice and Bob replenish their stash of shared identification triples via QKD.
        \end{itemize}\\
        \hline
    \end{tabular} 
    \caption{Overview of a simple, QKD-enhanced, 3-step authentication protocol. \label{tab:QKDAuthentication}}
\end{table}

Are all the three stages above really necessary? Yes, to prevent the following type of attack. Suppose Eve, $E$, wants to impersonate Alice or Bob. First, she would establish a communication with Alice, receive the first string $ s^{(k)}_1 $ and send a random second string $ \tilde{s}^{(k)}_2 $. Since, in all likelihood, $ \tilde{s}^{(k)}_2 \neq s^{(k)}_2 $, Alice would see through Eve's deception, abort the protocol and discard $T^{(k)}$. However, Eve has now obtained the first key. Eve could then try to impersonate Alice with Bob. She would send the correct first key (which she now has!), and receive $ s^{(k)}_2 $ from Bob. For the first two rounds of verification, Bob would thus be unaware of Eve's deception. However, the attack would be exposed when Eve provides the third, wrong, string $ \tilde{s}^{(k)}_3 $. At that point, Bob would discard $T^{(k)}$ too. Since both $A$ and $B$ have discarded the triple $T^{(k)}$ that Eve has information on, the information gained by Eve is now useless. 

\paragraph{Conclusions} 
Of course, with this simple example we are but scratching the surface of the authentication theme. Much more advanced protocols have been proposed, see e.g.\,the review \cite{Dutta2022} and the selected references \cite{Curty2001, Curty2002, Hong2017, Barnum2002, Nikolopoulos2020, Goorden2014}. Similarly, the research on secure direct communication schemes has evolved tremendously, see \cite{Pan.2023b} for a general overview, and so has that on QKD, cf.\,\cite{Zapatero.2023}. On top of these three general applications, there exists a number of rather specific tasks where quantum mechanics is known to provide security advantages. To name a few: 
\begin{itemize}
    \item The \emph{coin-flipping problem} \cite{Blum.1983}. Two parties, $A$ and $B$, that do not trust each other want to agree on a fair coin-flipping procedure over the phone. They consider a protocol ``$\varepsilon$-fair" (or $\varepsilon$-secure), $0 \leq \varepsilon \leq 1/2 $, if neither party can unbalance the probability of the toss towards heads or tails by more than $\varepsilon$. In other words, it is impossible for a cheating party to achieve
    \begin{equation}
        P[\text{heads}] > 1/2 + \varepsilon
    \end{equation}
    or 
    \begin{equation}
        P[\text{tails}] > 1/2 + \varepsilon \,.
    \end{equation}
    Classically, there exists no protocol with $ \varepsilon < 1/2 $ when assuming unbounded computational power of the cheating party \cite{Doescher.2002}. Quantum-mechanically, however, it is possible to achieve $ \varepsilon^\star = (\sqrt{2} - 1)/2 \leq \varepsilon \leq 1/2 $, and the lower bound $ \varepsilon^\star $ is tight \cite{Kitaev.2002};

    \item \emph{Secret sharing} \cite{Hillery.1999}. $A$ wants to share a secret with $B$ and $C$, in such a way that neither of the two can retrieve it by themselves, but they both can if they combine their knowledge. Classical solutions exist, but they are prone to eavesdropping (an unwanted party may discover the secret). Quantum solutions, e.g.\,the one in \cite{Cleve.1999}, amend this problem;

    \item \emph{Anonymous broadcasting}. One party among three wants to share a message (typically, one bit) with the others, without them knowing who transmitted the message. This three-party version is known as the ``dining cryptographers problem". It was introduced in \cite{Chaum.1988}, where a classical solution is also provided. Here, the advantage of a quantum protocol \cite{Huang.2022} lies in its immediate generalization to $N$ parties (for generic $N$), and again in its resilience to eavesdropping.
\end{itemize}
Despite the lively theoretical discussion, we are still very far from satisfactory (Q)Internet security. A coordinated effort towards the design, standardization and technological implementation of secure communication and authentication protocols is needed to ensure that the quantum internet does not suffer the same shortcomings of its classical predecessor. If we are far-sighted, the same quantum revolution that is disrupting the security protocols of the present will provide far safer ones in the future.

\subsubsection{Distributed Quantum Computing} \label{sssec:DQC}

\paragraph{What is DQC?} 
By \emph{Distributed Quantum Computing} (DQC for short) we mean the paradigm of sharing a common computational task across various, possibly distant, Quantum Processing Units (QPUs) \cite{Barral.2024, Caleffi.2024, Cuomo.2020, Denchev.2008}. It draws inspiration from the analogue field of \emph{distributed classical computing} (DCC), often sharing techniques and goals with it. Among the goals, the most obvious one is to increase the available computational power by making existing processors, dislocated over possibly large distances, work together to solve problems too complex to be tackled by any single one of them.

\paragraph{Motivations} 
Although goals are often shared, there exist many differences between DCC and DQC, even concerning their motivations. If, in the classical realm, there is little obstruction to the realization of a \emph{monolithic} supercomputer of arbitrary processing power, not the same holds in the quantum case. The scalability of monolithic quantum processors is hindered by the increasing requirement for qubit interconnectivity, as the qubit number increases. When qubits are locally hosted on a single physical platform, increasing connectivity indeed favours unwanted interactions, promoting decoherence \cite{vanMeter.2008}. Increasingly refined \emph{error correction protocols} (see \cite{Terhal.2015} and references therein) have been developed to mitigate these effects. However, they all leverage some form of redundancy of information, and thus require mapping multiple physical qubits to a single \emph{logical} one. The stronger the decoherence, the more sophisticated the error correction scheme, the larger the number of physical qubits required to make up a logical one. This entails that, in the monolithic paradigm, a linear increase in ``computational power" (here crudely measured by the number of logical qubits) demands a superlinear increase in the number of physical qubits, hindering scalability. This is where DQC exhibits its potential: Interconnecting small (hence, less noisy) QPUs allows the construction of a (delocalised) \emph{virtual quantum computer}, whose computational power scales linearly with the number of physical qubits, opening the door to tackling resource intensive tasks. 

On top of that, DQC offers unique advantages over its classical counterpart. First, it solves classically impossible distributed problems, like the GHZ and CHSH ``games" (cf.\,\cite{SengW.loke.2024}, Chapter 3 for details). Notice that, in theoretical computer science, distributed tasks are often phrased as ``games" that parties play against an adversarial arbiter. Successfully completing the task is tantamount to winning the game. Second, DQC is capable of performing some computational tasks \cite{Buhrman.2010} with a dramatic decrease in the number of bits exchanged among parties (\emph{communication overhead}). Such tasks include: The three-party problem; The distributed Deutsch-Josza problem \cite{Li.2023}; The intersection problem \cite{SengW.loke.2024}. Let us say a few words about each one. In the \emph{three-party problem}, Alice (A), Bob (B) and Charlie (C) each posses one n-bit string $x = x_1 x_2 ... x_n $, $x=a,b,c$ such that $ a \oplus b \oplus c = 1...1 $, and they are given the distributed task of evaluating the \emph{inner product} function
\begin{equation}
    f(a, b, c) = (a_1 \cdot b_1 \cdot c_1) \oplus (a_2 \cdot b_2 \cdot c_2) \oplus \dots \oplus (a_n \cdot b_n \cdot c_n) \,,
\end{equation}
where $\oplus$ denotes addition modulo $2$. Without discussing the distributed protocol, we report the conclusions of \cite{Cleve.1997}: this specific task requires, classically, the exchange of at least \emph{three} bits of information, whereas only \emph{two} are required by the optimal quantum protocol, which leverages a tripartite entangled state and local operations. The \emph{distributed Deutsch-Jozsa problem} is as follows. A and B are given the $n$-bit strings $a=a_1 a_2 ... a_n$ and $b = b_1 b_2 ... b_n$ respectively, with the promise that either the two strings are identical, or they differ in exactly $n/2$ positions. The task is to determine which alternative is realized. As argued in \cite{Buhrman.2010}, it turns out that $\log_2 n$ bits of communication are required when employing a quantum strategy that involves shared entanglement, whereas at least $0.007n$ bits are needed for a classical strategy. For large $n$, the quantum protocol thus exhibits an exponentially lower communication overhead. Lastly, in the \emph{intersection problem}, again treated in \cite{Buhrman.2010}, $A$ and $B$ have to compute
\begin{equation}
    g (a,b) =
    \begin{cases}
    1 \,, & \text{if} \ a_i = b_i \ \text{for at least one} \ i \\
    0 \,, & \text{otherwise} \,,
    \end{cases}
\end{equation}
where $a$ ($b$) denotes $A$'s ($B$'s) $n$-bit string. The optimal quantum protocol is based on Grover's search algorithm, and it can consequently be shown \cite{Buhrman.2010} that the required bits of communication scale like $\mathcal{O} (\sqrt{n})$ and $ \mathcal{O} (n) $ in the quantum and classical case, respectively.

Another benefit of the distributed paradigm is the possibility of merging different qubit platforms, each with its own strengths and weaknesses \cite{Ladd.2010}, into a single virtual machine. Algorithms may then be distributed so as to assign to each platform the computational task that best suits it \cite{Lee.2024, Xu.2022}. Furthermore, various distributed protocols have shown a reduction in circuit depth with respect to their monolithic counterparts \cite{Avron.2021}, suggesting \emph{parallelization} as a possible pathway towards noise reduction.

Besides granting all the advantages above, a fully developed \emph{distributed quantum computing environment} \cite{Cuomo.2020}, whose components we will analyze shortly, shall enable a plethora of concrete applications such as information-theoretic secure communication, quantum key distribution, quantum oblivious transfer (cf.\,Sect.\,\ref{sssec:Query}), quantum elections (cf.\,Sect.\,\ref{sssec:Elections}) and more. In the long run, quantum computing shall also be provided as a service, hence the need to design protocols that grant a client the faithfulness of the executed computations, while at the same time minimizing the information retained by the server. The discipline that studies such protocols, called \emph{Blind Quantum Computing} (BQC), is further discussed in Sect.\,\ref{sssec:BQC}. 

\paragraph{State of the art} 
We present the current status of Distributed Quantum Computing by tackling short- and long-term prospects separately. The field is indeed mature for real-life applications and implementation of distributed protocols on NISQ-era hardware. This is, first and foremost, thanks to the fact that distributed versions of most quantum algorithms with potential for quantum advantage have been developed. Among these, we mention: the Variational Quantum Eigensolver (VQE) \cite{DiAdamo.2021}; Deutsch-Jozsa \cite{Li.2023}; Simon's algorithm \cite{Tan.2022}; Grover \cite{Qiu.2024} and Shor \cite{Yimsiriwattana.2004}. The last two are of particular importance, because they grant significant speedups, with respect to the best classical algorithms, in the resolution of relevant problems. More precisely, Grover grants a quadratic speedup in the resolution of combinatorial/scheduling problems, while Shor an exponential speedup in the problem of prime factorization, of extreme importance for widely used encryption protocols like RSA \cite{Gidney.2021}.

\begin{figure}[hbt]
    \centering
    \begin{quantikz}
        \lstick[5]{n qubits} \hspace{6pt} & \lstick{$\ket{0}$} \wireoverride{n} & \gate{H} & \ctrl{5} & & & & \wireoverride{n} \dots & \wireoverride{n} & & \gate[5]{QFT^{-1}} & \meter{} \\
        & \lstick{$\ket{0}$} \wireoverride{n} & \gate{H} & & \ctrl{4} & & & \wireoverride{n} \dots & \wireoverride{n} & & & \meter{} \\
        & \lstick{$\ket{0}$} \wireoverride{n} & \gate{H} & & & \ctrl{3} & & \wireoverride{n} \dots & \wireoverride{n} & & & \meter{} \\
        & \lstick{\dots} \wireoverride{n} & \wireoverride{n} & \wireoverride{n} & \wireoverride{n} & \wireoverride{n} & \wireoverride{n} & \wireoverride{n} & \wireoverride{n} & \wireoverride{n} & \wireoverride{n} & \wireoverride{n} \\
        & \lstick{$\ket{0}$} \wireoverride{n} & \gate{H} & & & & & \wireoverride{n} \dots & \wireoverride{n} & \ctrl{1} & & \meter{} \\
        & \lstick{$\ket{\psi}$} \wireoverride{n} & \qwbundle{m} & \gate{ U^{2^0} } & \gate{ U^{2^1} } & \gate{ U^{2^2} } & & \wireoverride{n} \dots & \wireoverride{n} & \gate{U^{2^{n-1}}} & & \\
    \end{quantikz}
    \caption{Circuit representation of the QPE algorithm. The \emph{natural} partition into two registers containing $n$ and $m$ qubits each is manifest. A more interesting partition stems from distributing $ QFT^{-1} $. Figure adapted from \cite{FelicitasBinder.03092024}. \label{fig:QPE}}
\end{figure}
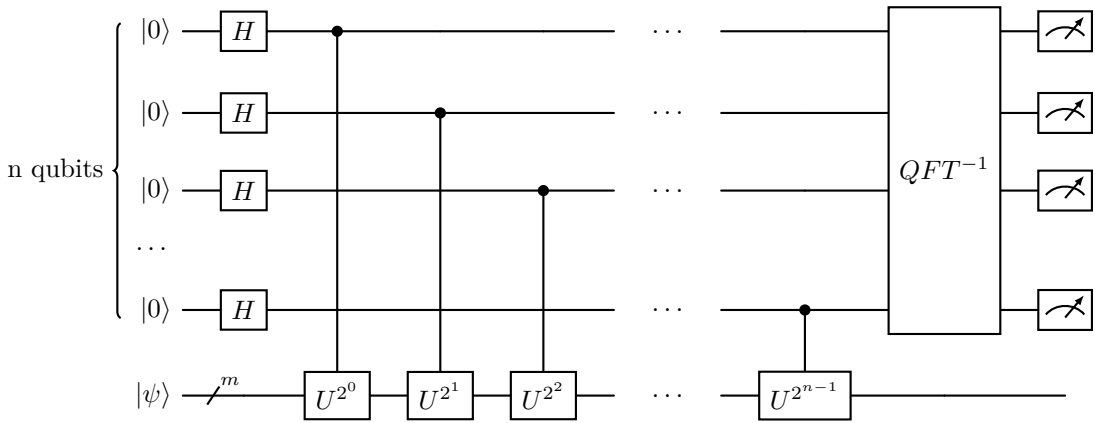

It is instructive to look at how a specific quantum algorithm gets distributed across different parties. We pick as an example \emph{Quantum Phase Estimation} (QPE), namely the crucial building block of Shor's algorithm for prime factorization. Let ust start by reviewing the setup and the monolithic implementation. An experimenter is given an $m$-qubit Hilbert space $ \mathcal{H} \cong \mathbb{C}^{2^m} $, a unitary $U$ on $ \mathcal{H} $ and an eigenvector $ \ket{\psi} \in \mathcal{H} $ of $U$, such that
\begin{equation}
    U = \eul^{2 \pi \im \theta} \ket{\psi} \,.
\end{equation}
His task is to determine the angle $\theta \in [0,1) $ with the best accuracy possible. An $n$-bit approximation of $\theta$, namely an $n$-bit string $j = j_0 j_1 ... j_{n-1}$ such that
\begin{equation}
    \theta = \frac{j}{2^n} + \mathcal{O} \left( 2^{-(n+1)} \right) \,, \label{eq:ThetaApprox}
\end{equation}
can be obtained with the well-known \emph{Quantum Phase Estimation} algorithm, which uses an ancillary $n$-qubit register, and whose circuit representation is reported in Fig.~\ref{fig:QPE}. Notice that, in Eq.\,\eqref{eq:ThetaApprox} and above, $j$ is used at once to denote the $n$-bit string $j = j_0 j_1 ... j_{n-1}$ and the number
\begin{equation}
    j = \sum_{k=0}^{n-1} j_k 2^k \,.
\end{equation}
A detailed analysis of how the algorithm works goes beyond the scope of this section. By contrast, we wish to discuss how the circuit in Fig.\,\ref{fig:QPE} can be distributed. We examine two possible splits \cite{FelicitasBinder.03092024}. The first one is the most immediate, and consists in hosting the $m$ qubits making up $\ket{\psi}$ and the $n$ ancillary qubits in separate nodes. Albeit interesting in principle, this split only involves distributing the controlled-$U$ gates, which is fairly simple because controlled gates act like the identity on the control qubit. A more interesting and instructive split is obtained by distributing the inverse Fourier transform, denoted by $QFT^{-1}$ in Fig.\,\ref{fig:QPE}. The key idea is to use the recursive representation of $QFT^{-1}$, cf.\,Fig.\,\ref{fig:QFTRecursive}, where
\begin{equation}
    R_k =
    \begin{pmatrix}
        1 & 0 \\
        0 & \eul^{2 \pi \im / 2^k}
    \end{pmatrix} \,.
\end{equation}
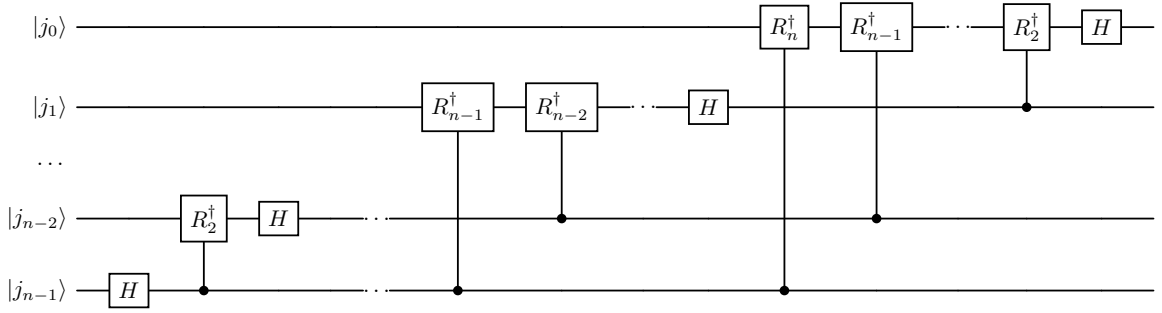
\begin{figure}[thb]
    \centering
    \resizebox{\textwidth}{!}{
    \begin{quantikz}
        \lstick{$\ket{j_0}$} & & & & & & & & & & \gate{R_n^\dagger} & \gate{R_{n-1}^\dagger} & \dots & \gate{R_2^\dagger} & \gate{H} & \\
        \lstick{$\ket{j_1}$} & & & & & & \gate{R_{n-1}^\dagger} & \gate{R_{n-2}^\dagger} & \dots & \gate{H} & & & & \ctrl{-1} & & \\
        \lstick{\dots} & \wireoverride{n} & \wireoverride{n} & \wireoverride{n} & \wireoverride{n} & \wireoverride{n} & \wireoverride{n} & \wireoverride{n} & \wireoverride{n} & \wireoverride{n} & \wireoverride{n} & \wireoverride{n} & \wireoverride{n} & \wireoverride{n} & \wireoverride{n} & \wireoverride{n} \\
        \lstick{$\ket{j_{n-2}}$} & & \gate{R_2^\dagger} & \gate{H} & & \dots & & \ctrl{-2} & & & & \ctrl{-3} & & & & \\
        \lstick{$\ket{j_{n-1}}$} & \gate{H} & \ctrl{-1} & & & \dots & \ctrl{-3} & & & & \ctrl{-4} & & & & & \\
    \end{quantikz}
    }
    \caption{Recursive representation of the inverse \emph{quantum Fourier transform}. It is \emph{recursive} in the following sense. Focus on the bottom $p$ wires of the circuit above. Those wires alone implement a $QFT^{-1}_p$ for the last $p$ qubits. The total $ QFT^{-1}_n $ transform can thus be seen as $QFT^{-1}_p$ followed by $QFT^{-1}_{n-p}$, where the (already transformed) last $p$ qubits serve as controls for $QFT^{-1}_{n-p}$. By iterating this reasoning, any integer partition of $n$ is seen to correspond to a partition of $QFT^{-1}_n$. Figure adapted from \cite{FelicitasBinder.03092024}.  \label{fig:QFTRecursive}}
\end{figure}

For any partition
\begin{equation}
    n = n_1 + n_2 + ... + n_k
\end{equation}
of $n$, there exists a partition of the qubits into $k$ separate nodes, where the first one hosts qubits $0,1,...,n_1 -1$, the second one qubits $n_1, n_1 + 1,...,n_1 + n_2 -1$, and so on. It should be clear from the circuit above that, for any such partition, the inverse Fourier transform can be distributed accordingly (see also explanation in the caption of Fig.\,\ref{fig:QFTRecursive}). When done naively, however, this entails a \emph{communication overhead}, here measured in the number of two-qubit gates across different nodes, that scales quadratically in the number $k$ of nodes. This is heuristically understood as follows. Inspect the circuit of Fig.\,\ref{fig:QFTRecursive}: There are $n-1$ two-qubit gates controlled by $(n-1)$-th qubit; $n-2$ two-qubit gates controlled by the $(n-2)$-th qubit, and so on. In total, we have
\begin{equation}
    \sum_{l=1}^{n-1} l = \frac{(n-1)n}{2} = \mathcal{O} (n^2) \label{eq:TwoQbitGates}
\end{equation}
two-qubit gates. If each qubit is taken as a node in a distributed network, then all $ n(n-1)/2 $ such gates connect separate nodes, namely the number of non-local gates scales quadratically in the number of nodes. A counting strategy similar to the one leading to Eq.\,\eqref{eq:TwoQbitGates} reveals that the scaling is quadratic in the number $k$ of nodes, even when a partition is chosen such that each node hosts more than a single qubit ($k<n$). Of course, such a scaling of the number of costly non-local gates does not seem to favour a distributed implementation of this protocol. However, less naive distributions of $QFT^{-1}$ exist, as we will remark at the very end of this section.

Aside from the purely theoretical envisioning of these distributed algorithms, proposals exist on how to implement them on current hardware, in spite of its limitations. For example, a prominent scheme utilizes circuit-cutting techniques and classical communication only to run an approximate version of any monolithic algorithm on noisy, distributed QPUs \cite{McClainGomez.2024}. Concrete experimental demonstrations are however still in their infancy. Most of the aforementioned algorithms require a number of qubits way beyond that of current machines to be relevant in applications. Take the example of Shor: Only recently \cite{Skosana.2021} has the number $21$ been factorized on an IBM (monolithic) quantum processor. When it comes to distributed computing, to the best of the authors' knowledge, state-of-the-art experiments showcase the implementation of remote gates and the presence of remote entanglement \cite{afzal2024distributed}, but cannot go much further.

Ideas for long-term applications of DQC are not lacking either. The increasing size and complexity of quantum networks will however demand a more systematic approach to circuit distribution and communication interfaces. On the first topic, \cite{Ying.2009} proposes an algebraic language to represent distributed circuits, which holds the promise of making circuit distribution and optimization dramatically more efficient via symbolic (rather than diagramatic) calculus. By contrast, \cite{Haner.2021} extends the classical ``Message Passing Interface" (MPI) to quantum architectures. More specifically, this framework treats classical and quantum communication separately. The first one is handled by the standard MPI, while the second one inherits all the classical MPI primitives (since using a qubit as a bit is always possible) while also allowing quantum-specific operations. In particular, quantum communication across nodes always requires shared maximally entangled pairs, in this model, and is realized via two elementary operations: 1) The \emph{fanout}, or \emph{entanglement copy}, which ``copies" the state of a qubit onto an ancilla $(\alpha \ket{0} + \beta \ket{1}) \otimes \ket{0} \mapsto \alpha \ket{00} + \beta \ket{11} $; 2) The ``move" operation, which transfers a qubit across nodes via quantum teleportation, cf.\,discussion in Sect.\,\ref{sssec:QTransfer}. Collective, network-wide operations are also available in QMPI, and the performance of a distributed algorithm can be evaluated (and to some extent optimized) via the SENDQ model, which accounts at once for the communication complexity and the delays produced by local operations. The introduction of such communication standards is instrumental to the creation of an industrial-ready \emph{distributed quantum computing ecosystem}. In the next paragraph, we comment on its envisioned structure and on the challenges that lie on the road towards its realization.

\paragraph{Far future of DQC} 
A future, global \emph{distributed quantum computing ecosystem} shall comprise four basic layers \cite{Cuomo.2020, Barral.2024, DiAdamo.2021, SengW.loke.2024}:
\begin{enumerate}
    \item The \emph{physical layer}, made up by the many QPUs delocalized across the planet, each with its own qubit carriers and gate architecture; 

    \item The \emph{network layer}, represented by a graph of classical and quantum links connecting QPUs, as well as shared entanglement for communication and computation purposes;

    \item The \emph{development layer}, where quantum programs are translated to machine language, distributed across available parties and optimized;

    \item The \emph{application layer}, where the source code implementing a given algorithm (e.g.\,the aforementioned Grover, Shor or Deutsch-Jozsa) resides.
\end{enumerate}
\begin{figure}[hbt]
    \centering
	\includegraphics[width=0.6\columnwidth]{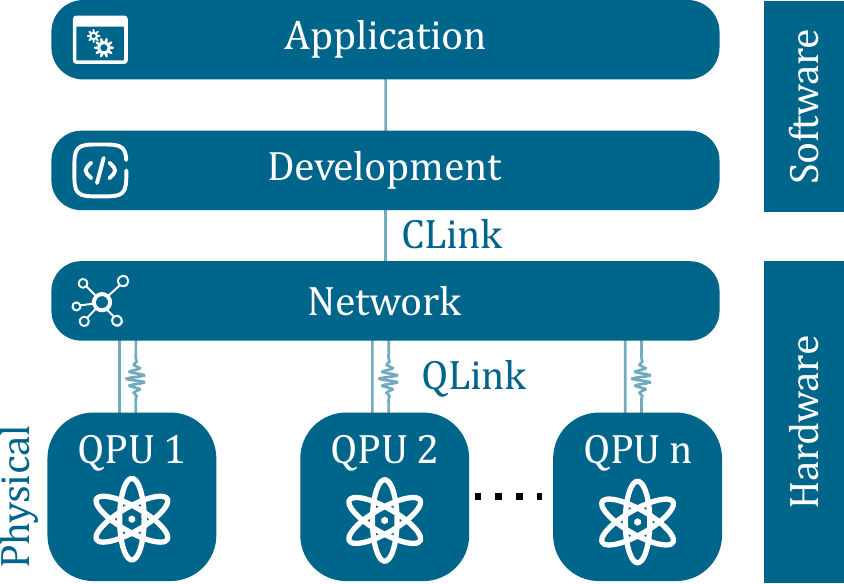} 
	\caption{The four layers of a distributed quantum computing ecosystem, see main text for a concrete definition of what belongs in which layer. Straight and wiggly light blue lines represent classical and quantum links, respectively. This figure is adapted from \cite{Barral.2024}, Fig.\,1. In producing this figure, open-licensed~.svg images by the following authors were modified and used: jcubic, Solar Icons, SVG Repo. \label{fig:DQCLayers}}
\end{figure}
The first two concern the actual \emph{hardware} that the network is based on. The other two are \emph{software} layers, concerned with the development of quantum algorithms and their mapping to the physical platforms for quantum computation. The practical implementation of each layer comes with a unique set of challenges, which we sketch below.
\begin{enumerate}
    \item \emph{Quantum processing units.} The main challenges in this context are noise, coherence time, fidelity and clock speed of quantum gates, scalability. So far, platforms that have low noise and gate error-rates have proven slow in clock speed, whereas fast platforms seem to be more error-prone. Since the task of achieving fast, scalable, fault-tolerant quantum computation concerns \emph{monolithic} rather than \emph{distributed} quantum computing, we will not provide further details here.

    \item \emph{Network layer.} There is almost unanimous consensus that, in most cases, quantum teleportation is the most promising means of communication. Alternative means of communication, e.g.\,through optical fibres, have been proposed \cite{Pellizzari.1997}, but they remain feasible only for some QC platforms, like trapped ions.

    If teleportation is to be used, a reliable way of producing and distributing maximally entangled qubit pairs will be required, as already discussed in Sect.\,\ref{sssec:QTransfer}. The distribution phase appears particularly challenging, and difficulties increase with the distance of the nodes to be connected. \emph{Quantum repeaters} and \emph{entanglement swapping} stand out as the most promising techniques for entanglement propagation, and they may even be implemented via low Earth orbit (LEO) satellite communication \cite{Djordjevic.2020}.

    \item \emph{Development layer.} This is the level where \emph{quantum compilers} bridge between the source code (in high-level programming language) and the physical machines. The task of compilers can be further broken down into two phases: \emph{Analysis}, where the source code is checked for syntax errors and translated into an intermediate language; and \emph{synthesis}, where the machine-level program undergoes the three further processes of \emph{optimization, verification} and \emph{qubit mapping} \cite{Barral.2024}.

    The \emph{synthesis} phase presents the arguably greatest technical challenges. The compiler must indeed be able to, first, distribute the quantum algorithm in agreement with all the constraints of the hardware, for example qubit connectivity. Moreover, the partitioning should be optimized so as to minimize both the number of qubits used for communication, and that of the costly non-local gates \cite{Vidal.2002}. Once a distributed algorithm is crafted, entanglement has to be propagated accordingly. Finally, the abstract logical qubits and gates have to be mapped to the physical qubits of the QPU.

    At the time of writing, no full-stack distributed compiler is available. We mention two prototypes: the one of \cite{Ferrari.2021}, based on \emph{circuit distribution} techniques; and \emph{Qurzon}, a \emph{circuit-cutting} based compiler developed in \cite{Chatterjee.2022}.

    Given the lack of well-developed options and the paramount prospective importance of quantum compilers in the era of the quantum internet, we expect this field of research to flourish in the upcoming years.

    \item \emph{Application layer.} This final layer is where research is perhaps most advanced, and we have already touched upon it in previous paragraphs. The main problems remain the identification of tasks where quantum algorithms outperform optimal classical ones, or where a distributed architecture is beneficial w.r.t.\,a monolithic one. Finally, there is great need for novel algorithm distribution schemes to base future compilers on.
\end{enumerate}
\paragraph{Closing remarks and words of caution} 
We close this short overview of distributed quantum computing with a couple caveats. The first one concerns the near-term viability of DQC: if the latter appears advantageous for large qubit numbers, as argued in the introduction to this section, the situation is not as clear-cut when few qubits are involved. Consider two different quantum-computing setups, one consisting in a monolithic QPU hosting 50 qubits, and the other one in two connected 25-qubits processors. Are the two capable of running the exact same algorithms? The expectation is that no, the distributed setup will not be able to reproduce calculations that involve all 50 qubits, because some of them (say, 5 per QPU) have to be reserved for storage of shared entanglement and communication purposes. At the same time, it may be argued that it is technologically simpler to realize two 30-qubit processors, with 5 communication qubits each (for a total of 50 ``computation" qubits), than it is to produce a single 50-qubits QPU. Which strategy is advantageous may heavily depend, in near-future applications, on the specifics of the hardware and experimental setup. The second caveat is that a distributed architecture (be it classical or quantum) is only beneficial as long as relatively little communication is required to perform the desired task. Indeed, if every node of an $N$-node network had to exchange a qubit with every other at each time-step of a depth $d$ protocol, the total number of exchanged qubits would be $d \cdot N!$, very quickly becoming unwieldy as $N$ grows. As a rule of thumb, we would consider an algorithm unsuitable for distribution if the number of exchanged qubits scales superlinearly in the number of network nodes. Luckily, algorithms with a linear scaling of the communication overhead exist. In fact, an example is QPE, albeit not in the ``naive" distributed version discussed above. The distribution scheme with linear scaling relies on the use of \emph{dynamic circuits}, where qubits are measured during the protocol, and the result of such measurements fed forward to the circuit for further computation \cite{Baumer.2024}.

\subsubsection{Blind Quantum Computing} \label{sssec:BQC}

In this section we give an overview of blind quantum computing.
In contrast to the focus on advantages of connected quantum computers regarding potential computational speedup in the previous section, here the main use case is secure computation. \\
Consider a client, Alice, who wants to perform some computation but lacks the resources to do so. However, the server Bob does have the resources to perform the computation. 
The problem now revolves around giving the Bob as little information as possible while still retrieving the correct result. 
In a classical setting, Bob knows the function or algorithm to be computed and the data is hidden by Alice with a suitable encryption scheme, e.g. homomorphic encryption \cite{rivest1978,gentry2009, acar2018}. 
It has been shown that in this scenario, classically it is in general impossible to completely hide the data for some algorithms \cite{Abadi1989}.\\
With a quantum system however not only the input and output data but also the algorithm itself can be completely hidden from the server \cite{Childs2005, Broadbent2009, Fitzsimons2017, Fitzsimons2017b}, with the maximal number of operations being the only information leaked.
To achieve this in practice, different approaches have been proposed, which we will discuss depending on the resources utilized by Alice.
These resources include the ability to perform single qubit gates, prepare quantum states or perform quantum measurements. 
In all these cases, Alice and Bob need a quantum channel to exchange single qubits.
A fully classical Alice may also instruct quantum servers to prepare quantum states, which remain unknown to the server. However, as Alice's part is purely classical, the blindness of the server may not exceed classical schemes, i.e. in this case the security is always based on some assumption on computational hardness \cite{Cojocaru2021}, or the amount of collaboration between compromised quantum servers. 
Thus, in the context of blind quantum computing, Alice needs some form of quantum abilities.

\begin{figure}[ht]
	\centering
	\includegraphics[width=0.8\columnwidth]{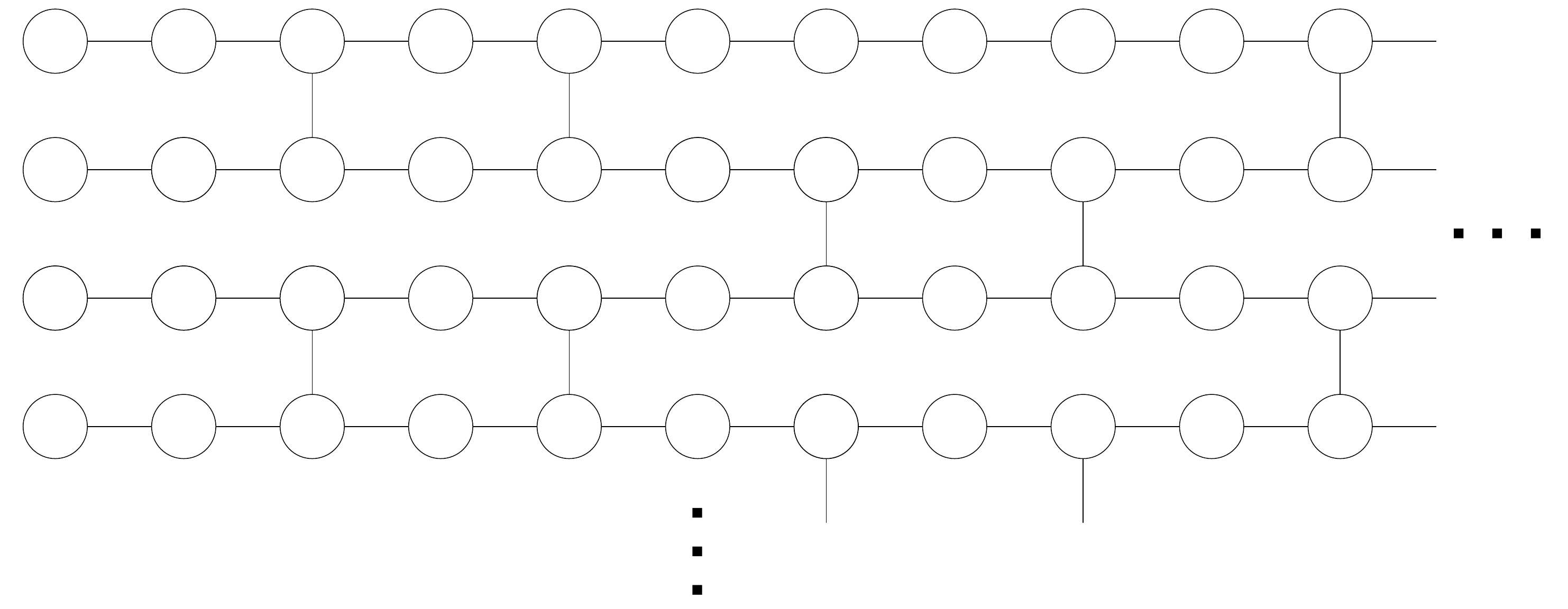} 
	\caption{The brickwork state for measurement based quantum computing as defined in \cite{Broadbent2009}. Each circle is one qubit initially in the $\ket{+}$ state, every line a Controlled-Z gate applied between the two connected qubits. The qubits are measured starting from the left. For each qubit the measurement angle depends on the algorithm and the results of already realized measurements.} 
    \label{fig:brickwork}
\end{figure}

\paragraph{Alice performs single qubit measurements}
One straightforward way to implement blind quantum computing utilizes measurement based quantum computing \cite{Raussendorf2001}.
In quantum theory there are two main types of operations on quantum states: unitary evolution, which is a deterministic and reversible transformation of a state, and measurements, which are probabilistic and force the state into one of the possible outcomes destroying any associated superposition.
While gate-based quantum computing relies on the former principle, evolving states continuously using deterministic and reversible gates until a final output state is created, measurement based quantum computing uses the latter.
To this end, a highly entangled resource state is created and its qubits are measured sequentially according to some algorithm.
The measurements performed on each qubit depend on the algorithm and need to include corrections based on the measurement results of previous qubits due to the probabilistic nature of the process.
By choosing a suitable resource state, e.g. the brickwork state as in \cite{Broadbent2009}, sketched in Fig. \ref{fig:brickwork}, every algorithm that does not exceed the size of the state can be implemented. Then, the only difference between the algorithms is the choice of measurement angle for each qubit.  

With this, blind quantum computing can be achieved by letting Bob create the required highly entangled cluster state. 
Instead of performing the measurements, Bob then sends the qubits one by one to Alice in a fixed order, who measures these qubits according to her algorithm \cite{Hayashi2015, Morimae2015}.
In this case, Alice only needs to be able to receive and measure single qubits.
As Bob prepares the resource state and Alice can perform her algorithm locally, no information about her data or computation is leaked.
While Bob does have access to all entangled qubits of the cluster state he has not sent to Alice, he would also need Alice's measurement results to obtain useful information from this state, similar to the situation in quantum teleportation. 
This setup requires Bob to send single qubits of a highly entangled state to Alice, which is experimentally demanding.

\paragraph{Alice prepares single qubits}
Similar to the previous method, Bob owns a measurement based quantum computer and creates a cluster state. 
Instead of initializing all qubits in the $\ket{+}$ state, he receives qubits from Alice, which are used as input states. 
For each of the input qubits of the cluster state, Alice chooses a random phase $\theta$, creating $\ket{\psi} =  1/\sqrt{2}(\ket{0} + e^{i \theta} \ket{1})$ \cite{Broadbent2009}.
These angles have to be suitably addressed during the measurement phases in the measurement based quantum computing protocol, which leads to changes in the measurement angles.

\begin{table}[ht]
    \centering
    \begin{tabular}{m{0.9\textwidth}}
        \hline
        \textit{Measurement based blind quantum computing}  \\
        \begin{itemize}
            \item Alice prepares states $\ket{\psi_i} =  \frac{1}{\sqrt{2}} \left( \ket{0} + e^{\ii \theta_i} \ket{1} \right)$ and sends them to Bob with instructions on how to set up a cluster state.
            \item Bob receives and stores states $\ket{\psi_i}$ and performs entangling Controlled-Z operations according to the cluster state architecture.
            \item Alice computes a measurement angle $\varphi_i\left(A, \theta_i, \{b_j\} \right)$ depending on the Algorithm $A$, her secret angle $\theta_i$, and previous measurement results $\{b_j\}$ and transmits it to Bob.
	        \item Bob performs measurement $\varphi_i$ on qubit $i$ and sends the result $b_i$ to Alice.
            \item  Alice and Bob repeat the previous two steps until all qubits have been measured.
        \end{itemize}\\
        \hline
    \end{tabular}
\end{table}

As Bob does not have access to $\theta$, even if he tried measuring the states sent by Alice, to him the required measurement angles communicated by Alice look random. 
Thus, he cannot infer any information about the operations he performed on the qubits. 
E.g., a CNOT operation or two single-qubit rotations may look exactly the same to him as the required measurement angles have been randomized by $\theta$.
Compared to sending qubits of the cluster state to Alice, Bob is more involved in this scenario, as he has to receive, entangle and measure the qubits according to Alice's specifications.
This is offset by the big advantage of dropping the requirement of transferring entangled qubits. 
Alice sends individual qubits, which are only entangled after arriving at Bob's device, where they stay until measured.
Thus, to realize such a scheme experimentally, quantum abilities are needed which are also relevant and sought after for other applications, e.g. quantum key distribution. 
A variation of this protocol has been implemented experimentally \cite{Barz2012}. 
However, in this implementation the entanglement was created by Alice before imprinting $\theta$ on the qubits, which makes the setup more demanding for Alice and requires multiple phase-stable optical connections to Bob.
Also, for this setup no clear path of scaling up the number of involved qubits is given as it relies on a multi-photon creation process instead of entangling individual qubits after their creation. 

\paragraph{Alice performs single qubit gates}
While blind quantum computing with measurement based quantum computers offers fully blind computations with reasonable quantum resources and an experimental realization seems possible within the next few years, most quantum computers which are currently in development utilize gate sets for computation.
In a first idea and proof-of-concept of blind quantum computing, Childs proposed a setup in which Alice has access to Pauli-X and -Z gates and an infrastructure to send, receive and store qubits \cite{Childs2005}. 
Bob runs a universal quantum computer and performs H, T and CNOT gates.
By randomly applying X and Z gates on her qubits before sending them to Bob for computation, Alice can completely hide her quantum states while still obtaining correct results by performing appropriate gate sequences to the qubits when retrieving them from Bob after computation \cite{Fisher2014, Li2018, Li2021}.\\
\begin{figure}[h]
	\centering
	\includegraphics[width=0.5\columnwidth]{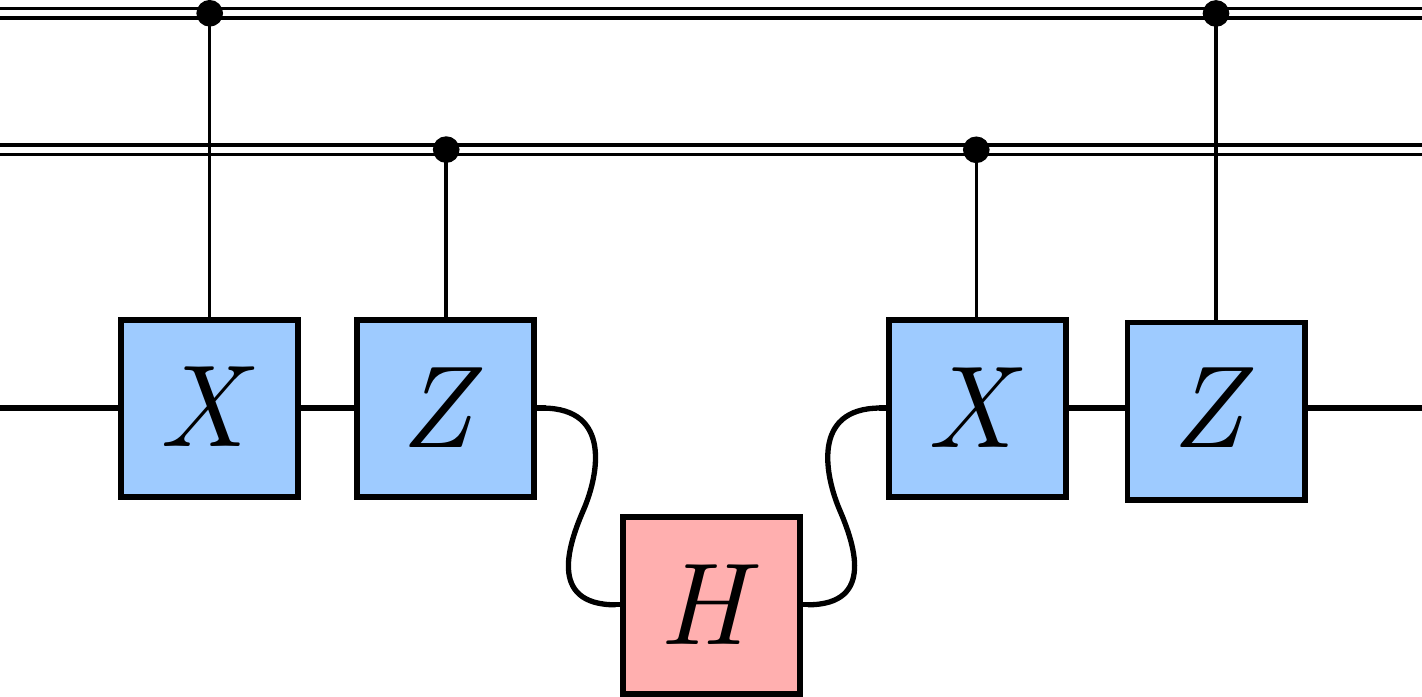} 
	\caption{By applying X- and Z-gates, Alice hides her quantum state from the quantum computer Bob, who performs the Hadamard (H-)gate. The two top lanes represent classical bits chosen randomly by Alice. Depending on her choice, she performs an X- and Z-gate before sending her qubit to Bob. As Bob does not know whether the gates have been applied, he gets no information about Alice's original state which from his point of view is equivalent to the maximally mixed state \cite{Childs2005}. After applying the H-gate, Bob sends the qubit back to Alice who again performs X- and Z-gates according to the same random bits as before. The whole process is equivalent to directly applying an H-gate to the original qubit for all choices of classical bits.} \label{fig:hiddenHgate}
\end{figure}
E.g., the sequence $Z^i X^j H Z^j X^i$ is identical to directly applying $H$ for all combinations of $i,j \in \{ 0, 1 \} $.
This can be seen from either explicit computation or the properties of the gates, with  $X^2 = I = Z^2 = H^2$ and $HXH = Z$.
Then 
\begin{equation}
    H = HZ^2 = H(HXH)Z = XHZ = ZHX.
\end{equation}
Without knowledge of $i$ and $j$ however, Bob cannot retrieve any information about Alice's original state as he only has access to the state after $Z^j X^i$ has been applied, see Fig.~\ref{fig:hiddenHgate}.
Suppose Bob does not apply the H-gate, but instead tries to obtain Alice's input state.
He needs to apply $X^i Z^j$ to undo the gates used by Alice.
He does not know $i$ and $j$ however, so with $\rho$ being the state he receives from Alice after applying her gates, he gets
\begin{equation}
    \rho_B = \frac{1}{4}\sum_{i=0}^1\sum_{j=0}^1 X^i Z^j \rho Z^j X^i = \frac{I}{4}.
\end{equation}
In a similar way, Alice can hide her states for a computation of T- and CNOT-gates, which is sufficient for universal quantum computing.  
By letting Bob run a fixed sequence of gates and only sending computational qubits at the right moment, using random qubits whenever the next gate is not needed at that step of the algorithm, Alice can also completely hide her algorithm while creating a linear overhead in gates needed for the computation \cite{Childs2005}.
Overall this method is very demanding, as the qubits have to be sent back and forth between Alice and Bob for every single gate in the quantum circuit. 
At the same time, Alice has to be able to store all qubits needed for the algorithm and apply the single qubit X- and Z-gates.
During all this, the entanglement between the qubits must be kept alive.

\paragraph{Partial blind quantum computing for gate based quantum computers}

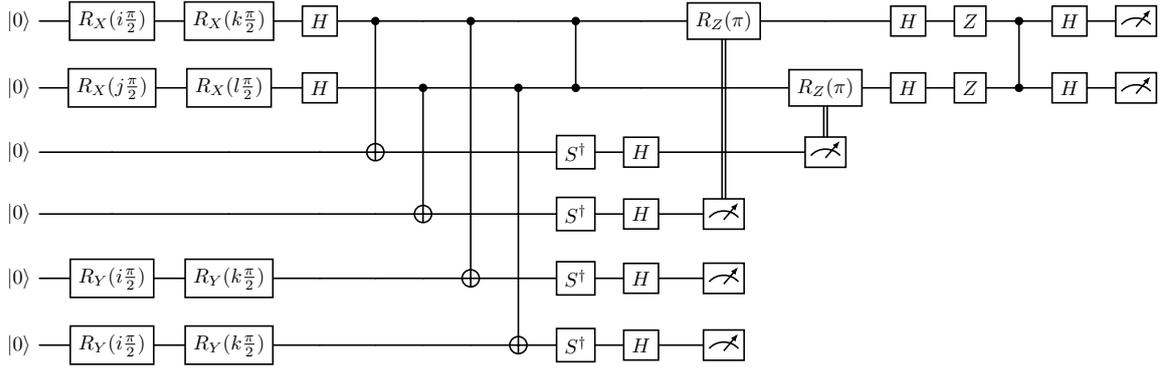
\begin{figure}[hbt]
    \centering
    \resizebox{\textwidth}{!}{
    \begin{quantikz}
        \lstick{$\ket{0}$} & \gate{R_X(i \frac{\pi}{2})}& \gate{R_X(k\frac{\pi}{2})} & \gate{H} & \ctrl{2} & & \ctrl{4} & & \ctrl{1} & & \gate{R_Z(\pi)}&  & \gate{H} & \gate{Z} & \ctrl{1} & \gate{H} & \meter{} \\
        \lstick{$\ket{0}$} & \gate{R_X(j\frac{\pi}{2})}& \gate{R_X(l\frac{\pi}{2})} & \gate{H} & & \ctrl {2} & & \ctrl{4} & \control{} & & &\gate{R_Z(\pi)} & \gate{H} & \gate{Z} & \control{} & \gate{H} & \meter{}  \\
        \lstick{$\ket{0}$} & & & & \targ{} & & & & \gate{S^\dagger} & \gate{H} & & \meter{} \vcw{-1} \\
        \lstick{$\ket{0}$} & & & & & \targ{} & & & \gate{S^\dagger} & \gate{H} & \meter{} \vcw{-3} \\
        \lstick{$\ket{0}$} & \gate{R_Y(i\frac{\pi}{2})}& \gate{R_Y(k\frac{\pi}{2})} & & & & \targ{} & & \gate{S^\dagger} & \gate{H} & \meter{} \\
        \lstick{$\ket{0}$} & \gate{R_Y(i\frac{\pi}{2})}& \gate{R_Y(k\frac{\pi}{2})} & & & & & \targ{} & \gate{S^\dagger} & \gate{H} & \meter{} \\
    \end{quantikz}
    }
    \caption{Example for a blind Grover search algorithm, taken from \cite{Consbruch2024}. The oracle, given by the $R_Z$ gates and the final outputs are hidden from the quantum computer. The first and second row of $R_X$ and $R_Y$ gates have to be performed by Alice and the oracle Oscar respectively, each randomly choosing $i,j,k,l \in \{1,3\}$ and another random bit determining whether performing the $R_X$ or $R_Y$ gates. \label{fig:blindGrover}}
\end{figure}

As the aforementioned method of performing Pauli-gates before and after every gate operation of the quantum computer requires a massive overhead in transferring qubits, we want to focus on schemes with preshared entanglement or limited exchange of quantum information during the computation, similar to the restrictions of distributed quantum computing.
In contrast to the measurement based approach, the gates do not emerge from additional information but have to be implemented by Bob, which makes hiding the algorithm challenging.\\
One option to hide the data and algorithm from Bob is by introducing additional qubits which are entangled to the original circuit qubits. 
While this increases the total number of qubits needed for an algorithm, the permanent sending of qubits between Alice and Bob known from other gate based schemes can be avoided.
When allowing Alice to provide the input states of the algorithm, it is even possible that Bob performs all gates and measurements on his quantum computer while no relevant information gets leaked to him.
In the context of a bachelor thesis we studied a relaxed scenario, where the Grover search algorithm shown in Fig. \ref{fig:blindGrover} is performed by Bob \cite{Consbruch2024}.
Instead of hiding all gates of the algorithm, resulting in the aforementioned massive overhead in entangled states, we focus on hiding the oracle of the Grover search algorithm and the output, thus giving Bob no information about the involved data.
However, Bob can infer that he performed a Grover algorithm which is not the case in fully blind quantum computing.

To hide the oracle and outputs of the algorithm, some input states are rotated by Alice and the oracle Oscar before sending the states to Bob, similar to the procedure used in measurement-based blind quantum computing.
The effect of these rotations is a change in the measurement outputs observed by Bob.
At the same time the oracle is hidden, as it can no longer be inferred by the application of the Z-gates, which also depend on the initial rotations. 
The algorithm is designed in a way that although Bob gets access to all six measurement results and the information whether the Z-gates of the oracle have been applied, for each oracle there is an equal amount of random initial rotations leading to these results, i.e. no choice of initial rotations or oracle is more likely than any other given Bob's information.

In conclusion, all three approaches can achieve blind quantum computing, each having different challenges and requirements for the involved parties.
While measurement-based blind quantum computing only needs one-way transfer of qubits and Alice's ability to measure or prepare single qubits, the gate-based approach requires Alice to perform some gates and back-and-forth communication of every qubit for each gate.
On the other hand, for gate-based quantum computing fewer qubits are sufficient as only one (or two) qubits are necessary to perform a gate while the resource state for measurement-based quantum computing requires many qubits for each gate.   
Which approach will be easiest to implement on a larger scale will depend on future developments in quantum computing and the transfer of qubits.
In the meantime, partially blind algorithms might be implemented like the one presented here.
By relaxing the requirements for gate-based blind quantum computing to only require transferring qubits once at the start of the protocol, the algorithm is no longer completely hidden, but in cases where all relevant information relies on some gates of the algorithm, this can still be enough for hiding its purpose.

\subsubsection{Database query} \label{sssec:Query}

One application of blind quantum computing are private database queries. 
Here, a client wants to retrieve an element of a database without having full access to it.
I.e. the client is not able to learn any information about the contents of the database besides the requested element.
Additionally, the owner of the database must not learn which element has been retrieved by the client, making the problem nontrivial.\\
One approach is implementing a Grover search with measurement based blind quantum computing and a third party supplying the oracle, called ``blind oracular quantum computation" \cite{Gustiani2021}.
Thus, the client does not get direct access to the oracle, as it is only communicated to the quantum computer in a hidden way with the previously discussed randomized measurement angles.
When constructing the oracles, it is important to always have exactly the same structure so the oracles only differ by the required measurement angles. 
Otherwise, the server could infer information about the oracle by analyzing size and structure of the cluster state it has to prepare \cite{Gustiani2021b}.\\
However, while some information remains hidden from the client, choosing the oracle depends on the item requested by the client.
So, while all information remains hidden from the quantum computer, the oracle does get information about the requested element and the database at the same time.

\paragraph{Oblivious transfer}
A protocol that is equivalent to private database queries as introduced above is 1-out-of-N-oblivious transfer.
One-out-of-N-oblivious transfer is a two-party protocol between a sender and a receiver, where N messages are sent encoded in a way such that the receiver can extract exactly one element \cite{Santos2022}.

\begin{table}[ht]
    \centering
    \begin{tabular}{m{0.9\textwidth}}
        \hline
        \textit{1-out-of-N oblivious transfer}  \\
        \begin{itemize}
	        \item The sender inputs N messages $m_i \in \{ 0,1 \}^l$.
            \item The receiver inputs a choice $j \in \{1,...,N\} $ and outputs $m_j$.
	        \item The sender gains no information about $j$.
	        \item The receiver gains no information about $m_i$ for $i\neq j$.
        \end{itemize}\\
        \hline
    \end{tabular}
\end{table}

This is a classical problem which can be solved using asymmetric public key schemes \cite{Naor1999}. 
For polynomial functions it has also been shown that it is possible to construct blind computing schemes from only this primitive \cite{Kilian1988}.\\
Utilizing quantum resources seems like a way to improve the security of this scheme.
Similar to the ideas presented in Sect.~\ref{ssec:Encoding}, the sender could encode its information in quantum states $\psi$ but now we do not impose any constraints on the number or dimension of the states.
Using measurements, the receiver then extracts some data from these quantum states while destroying them, thus being unable to get more information about the encoded messages.
Ideally, the receiver can extract exactly one message, while the sender does not know which message has been extracted.\\
However, it has been shown that -- in contrast to quantum key distribution -- quantum oblivious transfer cannot be made unconditionally secure \cite{Lo1997}.
Thus, we always need some additional assumptions about the abilities of an attacker or more parties have to be introduced.

\subsubsection{Quantum key card} \label{sssec:QKeyCard}

While quantum tokens are generally used for authentication via a network with potentially 
insecure transmission paths, there is also a special form of token where the verifying body is given at least temporary physical access. These can be, for example, keys for access to specially secured areas or bank cards for processing payments.
In these cases, the authorized person is in possession of a physical object on which information is stored that is used for authentication.
A distinction can be made between systems that store quantum information in the form of specific states and systems whose stored information is read out by a quantum physical process. It is also possible to divide them into local and distributed systems. In the case of local systems, such as a door lock that holds all the information required to determine the validity of a presented key, no communication with other parties is necessary during authentication. In distributed systems, e.g., as part of a payment process using a bank card, information is exchanged with third parties during verification, in this case the bank.

\textit{Quantum readout, qPUFs:}
In this method, the key is a classical object with properties that affect quantum states, a ``Physical Unclonable Function" (PUF).
This can be, for example, a fully or partially reflective material with a random irregular structure at the microscopic level \cite{Goorden2014, Nikolopoulos2017}. 
After production, quantum states of light (challenge) are prepared, the reflected or transmitted parts of the states (response) are measured and the resulting state pairs are stored.
Verification follows the same pattern, whereby only some of the challenge-response pairs are tested to see whether they match the expectation for the key.
The security of this variant is based on several assumptions \cite{Arapinis2021, Nikolopoulos2021, Skoric2010}. \\
Firstly, the challenges must contain overlapping quantum states so that a potential attacker cannot determine with certainty which challenge is to be answered and therefore cannot provide a suitable answer. 
For the same reason, the answers to similar challenges must differ sufficiently. 
In contrast to fully classical methods, the challenge-response pairs for a key can be publicly known due to their quantum properties.\\
Secondly, it must not be possible to produce a key with identical scattering behavior. 
Proof of this is only possible to a limited extent, as the key is a classic object and creating an exact copy can therefore only fail due to the technical capabilities of the attacker.
It must also be ensured that the function of the key cannot be carried out by quantum computers. 
However, as the key is defined via a finite number of transformations of quantum states, a universal quantum computer with sufficient memory and computing power can implement the same function.
This quantum readout method for PUFs is therefore relevant at best in a transitional period in which possibilities for generating and measuring quantum states are already comparatively easy to implement and widespread, but technological progress has not yet advanced far enough to copy physical objects sufficiently well or to simulate the function with quantum computers.   
Similar ideas have been proposed using quantum hardware as PUFs \cite{Phalak2021}, where the identification is tied to slightly different hardware specific errors during qubit manipulation. 
However, they suffer the same problem of potential simulation or imitation of the same errors by sufficiently advanced attackers.

\textit{Local testing of states:}
As unknown quantum states cannot be copied at will, concepts are conceivable in which a secret quantum state acts as a key. 
One initial idea is Wiesner's quantum money \cite{Wiesner1983}. Here, a bank first generates pairs of serial numbers and corresponding random bit strings. 
These are then used to prepare a quantum state which, together with the serial number, comprises a banknote. 
To check authenticity, the bank looks up the secret bit string associated with the serial number and carries out measurements on the quantum state, which only provide the expected results if the state has not been altered, e.g. by an attempt to copy it. 
One disadvantage of this procedure is that each location that can test the quantum state can also create any number of copies, as the random bit string is required for both actions. 
A check can therefore only be carried out by the bank. In addition, a secure database must be available for all issued banknotes. 
Many later proposed methods attempt to circumvent these disadvantages in various ways, e.g. by generating the bit string from the serial number with pseudo-random generators, whereby only a secret key needs to be stored, or with the help of asymmetric cryptography approaches \cite{Aaronson.2012, Gavinsky2012,Lutomirski2009,Schiansky2023,Bilyk2023,Goldwasser2012}. 
These enable verification without access to the information required to generate new valid quantum states. 
However, such a concept cannot be information-theoretically secure and is always based on assumptions about the capabilities of potential attackers instead. 
This is because the publicly known test procedure makes it possible in principle to test combinations of serial numbers and states until a valid combination is found. 
In contrast to testing by a central party, no external communication is necessary, so that no countermeasures can be taken, e.g. because a particularly large number of failed attempts are registered for a specific serial number or participant.  

\subsection{Quantum elections: Theory and implementation} \label{ssec:VotingTheoryAndImpl}

Below, we specialize our discussion to one application that is particularly relevant for the general public: How to enable secure ``correspondence voting" through quantum technologies. We start, in Sect.\,\ref{sssec:Elections}, with an overview of the inherent security \emph{requirements} of an election scheme. We then provide a short theoretical introduction to \emph{quantum election protocols} (Sect.\,\ref{sssec:QElProt}), and in particular to the seminal proposal in the field: the \emph{travelling ballot protocol}. Having laid down the theoretical foundation of the latter, we proceed to illustrate some practical implementations thereof, while highlighting their major technological roadblocks (Sect.\,\ref{sssec:ElectionImplementation}). Sect.\,\ref{sssec:ElectionSummary} closes our analysis of quantum elections with a summary and a feasibility assessment.

\subsubsection{Requirements on quantum voting schemes} \label{sssec:Elections}

\paragraph{Overview of voting schemes}
There exists a variety of  voting schemes and all require a different level of security. As some of them are vulnerable against quantum attacks, we explore in this subsection how the use of quantum technology can help to secure a particular category by using and implementing a so-called quantum voting scheme \cite{ Hillery2022, Jiang2012,Hillery2006,Bao2017,Horoshko2011,Shi2021,Vaccaro2007}. 

In principle, one has to distinguish  between \textit{open} and \textit{secret} ballots. In an open voting procedure, votes are cast publicly, for example by a show of hands. Here, the individual votes can be clearly identified. For example, these are made public by stating the name of the voter. In secret ballots, on the other hand, votes are cast anonymously, for example by collecting the votes of all eligible voters in a ballot box. Here it has to be guaranteed that the assignment of an individual vote to a specific eligible voter retrospectively is impossible.  

Accordingly, the different voting schemes go together with significant differences in the requirements for the election procedures. Open elections offer a high level of transparency and, in principle, any authorized voter (at least with a manageable number of eligible voters) and neutral bodies can directly check the individual votes and therefore the entire election result. This procedure is therefore very trustworthy and the result is easy to verify.
However, compared to secret ballots, it has the major disadvantage that voting is not anonymous. In fact, open voting procedures harbor the risk of effectively influencing eligible voters. For this reason, many relevant elections (for example, elections for members of the German Bundestag) are held by secret ballot. The cryptographic challenge here is to establish a correct and verifiable voting procedure that simultaneously ensures the anonymity of voters.

Particularly in elections with many eligible voters, anonymous electronic voting systems are sometimes used for this purpose. In most cases, these are based on the applicable assumptions regarding the computational complexity of certain methods and use, for example, discrete logarithms or integer factorization \cite{Vaccaro2007}. However, these procedures may be broken in the future if quantum computers are used. For this reason, this category of voting schemes could be secured by quantum election protocols in which quantum systems and protocols from the field of quantum communication are used. Importantly, they offer security against particular quantum-based attacks.

However, there are particular requirements for secure voting schemes that these quantum election protocols have to guarantee \cite{Jiang2012,Hillery2011}:
\begin{enumerate}
    \item Only votes of authorized voters can be taken into account.
\item Every interested voter must vote exactly once.
\item No one can control other voters' opinions about the issues to be voted upon.
\item No one can duplicate votes of voters (without being detected).
\item No one can change the votes of other voters (without being detected).
\item Every voter can confirm that her/his vote is taken into account.
\end{enumerate}

At the moment most quantum election protocols only satisfy a few of these requirements, in particular, in regard to points 4 and 5. For this reason, there are still several developments for quantum election protocols necessary until they could be used within practical settings. These contain advancements in regard to the requirements listed above as well as research towards their physical implementation. In the following, we focus on the latter one.

\subparagraph{Quantum election as a rudimentary quantum-network demonstration experiment\label{sssec:QElProt}}
Disregarding fundamental problems of fulfilling all requirements demanded from a secure voting scheme, we collect here the first considerations encountered when aiming to realize a specific simple quantum election protocol, namely Hillery's original traveling ballot proposal \cite{Hillery2006,Hillery2006b}.
Goal of this section is to exemplarily highlight the steps necessary for a demonstration experiment based on quantum information transfer, which goes beyond QKD (e.g., the chosen experiment should involve a true multi-party scenario), but which remains achievable within a limited time frame and with existing (e.g., within commercial or academic state-of-the-art) or soon anticipated technology.
For the reader of this review, the following is intended to serve beyond its immediate interest also as a template for the issues encountered in moving from abstractly formulated quantum information protocols to tangible physical devices and to the inevitable compromises when trying to attain desired functionality with available technological resources.

A first illustration of this role as an example is found in the considerations determining the choice of experiment to investigate. Quantum election protocols are an obvious choice for genuine multi-party applications, in particular those protocols, which have a clear dichotomy between a central election authority (which has considerable quantum resources) and potentially many voters with minimal quantum information processing abilities. The original traveling ballot protocol, which we will recapitulate below, is chosen based on its limited requirements on resources (requiring no multi-partite entanglement and a circular, one-way network for the transfer of comparatively simple quantum states) and despite known shortcomings (see below). Some of these may be partly overcome in second-generation refinements of a demonstrator experiment.

\subsubsection{The traveling ballot protocol}
\begin{figure}[hbt]
	\centering
	\includegraphics[width=0.5\columnwidth]{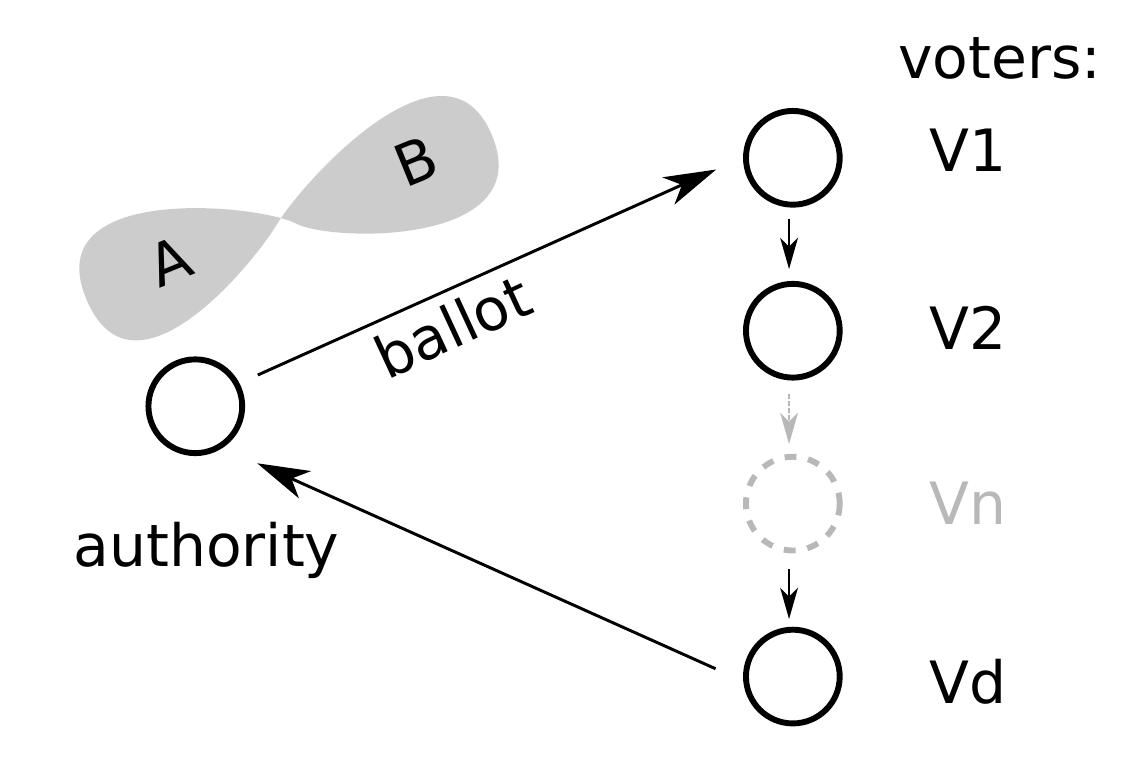} 
	\caption{Traveling ballot scheme according to \cite{Hillery2006}. 
    } \label{fig:traveling_ballot}
\end{figure}

In 2006 Hillery and coworkers proposed a simple, albeit imperfect, protocol exploiting quantum effects for elections \cite{Hillery2006,Hillery2006b}. Core of their ``traveling ballot" proposal is a ballot quantum state traveling from one voter to the next, see Fig.\,\ref{fig:traveling_ballot}, 
which for each voter looks identical before and after their vote and hence, taken in isolation, reveals no information about any vote. The ballot state is, however, entangled with a state kept with a central election ``authority" and 
it is the quantum correlations between these two states, which contain the information about the votes, which are changed by each participant's vote, and which are evaluated at the end of the protocol by the authority to obtain the vote count. The specific steps of the protocol are thus as follows:
\begin{enumerate}
    \item The authority prepares a bipartite entangled state of two $d$-level systems (called ``qudit", where $d \in \mathbb{N}$ is the number of voters participating in the election). The qudits are in an equal-weight superposition of each of their states, $0, 1,\hdots, d-1$, so that the state shared between authority qudit (A) and ballot qudit (B) is
    \begin{equation}
        | \Psi_0\rangle = \frac{1}{\sqrt{d}} \left( |0\rangle_A |0\rangle_B 
        + |1\rangle_A |1\rangle_B + \, \hdots\,+
        |d-1\rangle_A |d-1 \rangle_B  \right) \,;
    \end{equation}
    \item The authority keeps the authority qudit and sends the ballot qudit to the first voter (V1). The state the voter sees results from tracing over the authority part (which is unknown to V1) yielding a completely mixed qudit state, 
    \begin{equation}
        \rho_B =\textrm{Tr}_A\!\left\{|\Psi_0\rangle\langle\Psi_0 |  \right\}  = \frac{1}{d} \left( |0\rangle\langle 0|+ |1\rangle\langle 1| +\, \hdots\, + |d-1 \rangle\langle d-1| \right) \,;
    \end{equation}
    \item The first voter V1 casts his vote by applying one of two unitaries to the ballot qudit: the identity operation for a `NO' vote and a cyclic shift, \[|0\rangle_B \rightarrow U_\mathrm{YES} |0\rangle_B = |1\rangle_B,\; |1\rangle_B \rightarrow |2\rangle_B,\; \hdots,\; |d-1\rangle_B\rightarrow |0\rangle_B\]
    of the qudit base states for a `YES' vote and sends the state onwards to voter V2;
    \item V2 receives the ballot part of the state
    \begin{equation}
| \Psi_{V1}\rangle = \left\{
\begin{array}{lcl}
  \textrm{NO}&:& |\Psi_0 \rangle      \\  
  \textrm{YES}&:&   \frac{1}{\sqrt{d}} \left( |0\rangle_A |1\rangle_B 
        + |1\rangle_A |2\rangle_B + \, \hdots\,+
        |d-1\rangle_A |0 \rangle_B  \right)
\end{array}
\right. \;,
    \end{equation}
    which after tracing over the authority part yields again the completely mixed qudit state, irrespective of the vote of V1. \\
    Voter V2 executes his vote in the same manner as V1 and sends the resulting ballot qudit further on;
    \item After traveling through all voters, the last voter Vd sends the ballot back to the authority, which performs the vote count by a projective measurements on the states, $\sum_{n=0}^{d-1} |n\rangle_A |n+k\rangle_B$, to get the number $k$ of `YES' votes.
\end{enumerate}

As argued above, we investigate the traveling ballot protocol of Hillery here because of its conceptual simplicity and appeal, despite a number of known and well investigated weaknesses \cite{Arapinis2021}, some of which may be overcome by more advanced refinements. The most obvious of these weaknesses is, that any voter can obviously perform a YES voting operation multiple times, thus invalidating the correctness of the final vote tally. Moreover, two dishonest voters can collude to violate the privacy of the voters casting their vote in between the two of them by performing measurements on the voter part of the ballot \cite{Arapinis2021}. 

\subsubsection{Steps towards implementation} \label{sssec:ElectionImplementation}
At the core of the abstract traveling ballot protocol lies a bipartite entangled state of two qudits. The authority has to be able to prepare the entangled state and to execute a measurement in the combined Hilbert space of the two qudits, while the voters need to perform a relatively simple unitary operation on the ballot qudit. Starting point for any experimental realization thus is the physical implementation of that two-qudit state. That means choosing one particular encoding in some physical system with typically many more degrees of freedom. The steps and devices necessary for preparation, casting of a vote, and measurement will then depend on that chosen physical system and encoding scheme.

The obvious choice for a traveling ballot is some type of photonic state (be it at visible or at telecom wavelengths), as it offers facile and low-loss long-distance transfer of the ballot qudit between the voters, as well as the exploitation of comparatively simple quantum-optical devices for the manipulation by the voters. 
While, generally speaking, it is often favorable to realize stationary and itinerant (``flying") qubits or qudits in different fashion (``hybrid"), this would presumably offer little advantage for the current scheme: in particular, the stationary authority qudit needs to keep coherence only as long as the itinerant traveling ballot. In consequence, here, we consider only scenarios, where both qudits are implemented in identical manner.

\paragraph{Encoding the traveling ballot}
Without any claim on completeness, we will discuss four characteristically different qudit encodings: \emph{occupation number encoding, generalized cat states, multi-mode states, and multi-qubit encoding}. We will explain the reasoning leading to these encodings, discuss proposals for the physical implementation of preparing, manipulating, and measuring such states, and highlight challenges and limitations of each scenario.

\begin{figure}[t]
	\centering
	\includegraphics[width=0.9\columnwidth]{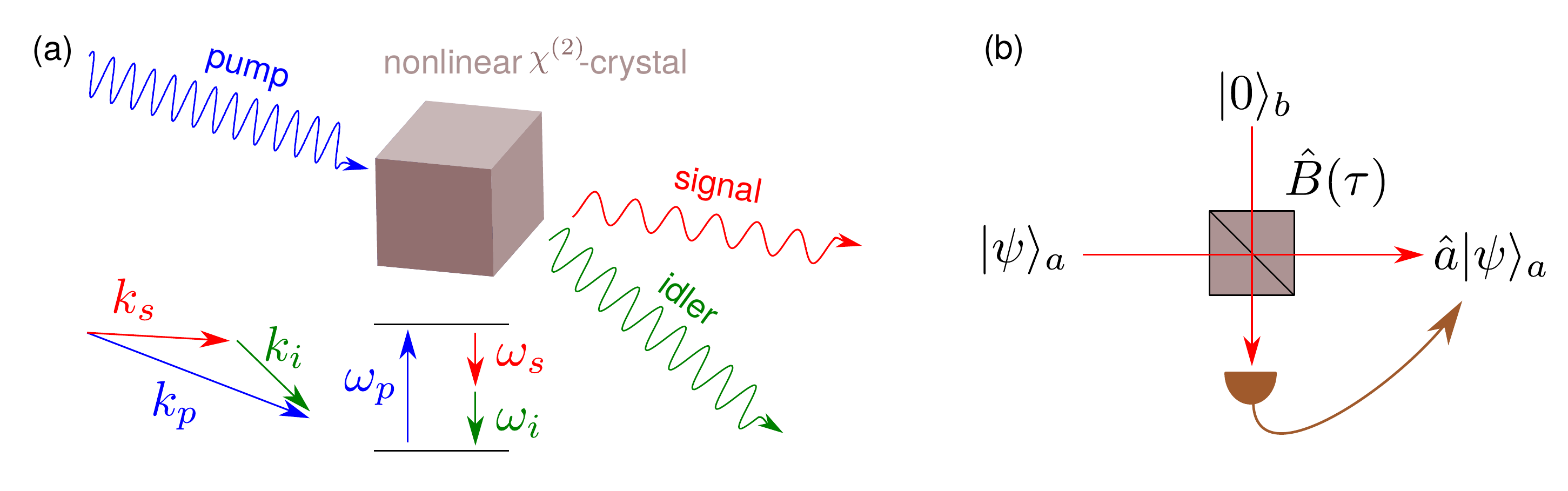}
    \caption{(a) Entanglement generation by optical downconversion. (b) Scheme for heralded single-photon subtraction by a beam splitter, see e.\,g.~\cite{Biagi2022}. A click from the on/off detector in the output mode heralds the application of the annihilation operator on the input state.
   } \label{fig:downconversion}
\end{figure}

\paragraph{Occupation number encoding}
The most immediate approach to the encoding of a photonic qudit, i.\,e., a $d$-level system, is using the first $d$ levels of easily available systems with infinitely many levels, namely the occupation number eigenstates of an optical mode.
Besides being a natural single-mode base for itinerant  photons (or stationary ones in the case of a resonant mode of a cavity), the occupation number encoding (also called Fock-state base) is a convenient choice for the most common way of creating bipartite entangled photonic states. This is routinely done by (non-degenerate) downconversion of a strong pump pulse around a frequency $\omega_p$ into two modes (called signal and idler) with $\omega_p=\omega_s + \omega_i$ in a crystal with $\chi^{(2)}$ nonlinearity, see Fig.\,\ref{fig:downconversion}.
The state created, 
\begin{align}
    \hat{S}(\xi)|0\rangle &=\exp{(\xi \hat{a}_s^\dagger \hat{a}_i^\dagger -\xi^* \hat{a}_s \hat{a}_i)} |n_s=0,\,n_i=0\rangle \\
    &= \sqrt{1-|\lambda|^2} \sum_n \lambda^n |n\rangle_s |n\rangle_i \quad\textrm{with}\quad \lambda = e^{i \textrm{arg}\xi} \tanh{|\xi|}
\end{align} 
called \emph{two-mode squeezed vacuum} would then be split between a authority and voters.

There are two major problems with the occupation number encoding approach: Firstly, while creating similar entanglement, the two-mode squeezed vacuum state clearly does not correspond to the desired equal-weight superposition. Indeed, typically squeezing is relatively small and the weights reduce strongly with higher occupation. One may envision nonlinear squeezing schemes which ameliorate this issue and, moreover, may lead to a sharper cut-off of weights at the desired level $d$. However, as a second problem, it is hard to imagine any simple, feasible scheme implementing the cyclic shift of the `YES' voting operation for the occupation number encoding.
In fact, a shift itself, is realizable by a standard optical device, a beam-splitter executing a (heralded) photon subtraction, see Fig. \,\ref{fig:downconversion}(b). 
If a quantum state $|\psi\rangle_a$ impinges on a highly transmittive beam-splitter with no input (`vacuum'$\;\equiv \;|0\rangle_b$) on its second input-port and a photon is detected on the reflective port, this photon for certain originates from the incoming quantum state, so that the transmitted state is given by the action of a photon annihilation operator on that state, $|\psi\rangle_a \rightarrow \hat{a}|\psi\rangle$, thus realizing a downward shift of each Fock state, $|n\rangle \rightarrow |n-1\rangle$.
Formally, one introduces the beam splitter operator
\begin{equation}
\hat{B}(\tau) = \exp{\left[\tau (\hat{a}\hat{b}^\dagger - \hat{a}^\dagger \hat{b}) \right]} \approx 1 + \tau (\hat{a}\hat{b}^\dagger - \hat{a}^\dagger \hat{b}) \;\textrm{for}\; r=\sin \tau \ll 1    \,,
\end{equation} 
so that the passage through the beam-splitter and subsequent detection of a photon in output-port b, 
\begin{align}
|\psi\rangle_a |0\rangle_b &\rightarrow 
 \left[ 1 + \tau (\hat{a}\hat{b}^\dagger - \hat{a}^\dagger \hat{b}) \right] |\psi\rangle_a |0\rangle_b = |\psi\rangle_a |0\rangle_b + \tau \hat{a} |\psi\rangle_a |1\rangle_b  \\
 \textrm{click in b, } ~_b\langle 1|\;: \quad \quad \quad \quad &\rightarrow \tau \hat{a} |\psi\rangle_a  \;\; _b\langle 1|1\rangle_b \sim \hat{a} |\psi \rangle_a 
\end{align}
results in photon subtraction.

However, the weights in the superposition of Fock states used in the occupation number encoding are changed and the vacuum contribution does not cyclically shift to state $|d-1\rangle$, but vanishes.  
Besides the incomplete implementation of the shift operation by the beam-splitter,  the operation is not deterministic, but is only successful with a small probability given by the reflection probability, which has to be small enough to avoid further unwanted correction terms to the simple photon subtraction result derived above.

Based on these results, we want to explore the option of using basis states for encoding, which have an inherent cyclical nature under a photon subtraction operation. Such states, called (generalized) Schr\"odinger cat states, seem to offer a simple way for implementing the voting operation.

\paragraph{Schr\"odinger Cat states}
In the famous Gedankenexperiment Schr\"odinger's cat is brought into a quantum mechanical superposition of two classical states (DEAD/ALIVE), which are complete opposites. Cat states accordingly are quantum superpositions of two quantum states, which are `nearly classical', namely coherent states of a harmonic oscillator, and which are classical opposites, meaning their amplitudes are reversed. 

Correspondingly, starting from the definition of  coherent states with a complex amplitude $\alpha$,
\begin{equation}
| \alpha \rangle \propto \sum_n \frac{\alpha^n}{\sqrt{n!} } | n \rangle  \quad \longrightarrow \quad 
|c_\pm \rangle = \frac{1}{{\cal{N}}_\pm}\left( |\alpha \rangle \pm |-\alpha\rangle\right) \,,
\end{equation}
we find the two conventional cat states $\ket{c_\pm}$. Its easy to see, that the cat state $\ket{c_\pm}$ contains only even/odd Fock states (and correspondingly has parity $\pm 1$). The two cats are thus orthogonal. Note, however, that the opposite-sign coherent states are only approximately orthogonal (for $|\alpha| \rightarrow \infty$).

As the coherent state is an eigenstate of the photon annihilation operator, $\hat{a} | \alpha \rangle = \alpha | \alpha \rangle$, cat states cyclically transform into each other under destruction of a photon
\begin{align*} 
\hat{a} |c_+\rangle   & \propto \hat{a} \left( |\alpha\rangle  + |-\alpha\rangle \right) & \hat{a} |c_+\rangle &\propto \hat{a} \left( |\alpha\rangle  - |-\alpha\rangle \right)\\
& = \alpha |\alpha\rangle -\alpha |-\alpha\rangle \quad \propto |c_- \rangle & & = \alpha |\alpha\rangle -(-\alpha) |-\alpha\rangle \quad \propto |c_+ \rangle \;.
\end{align*}
It is this intrinsic cyclic nature under photon destruction, which makes cat states appealing for encoding the traveling ballot state, $| \Psi_0\rangle = \frac{1}{\sqrt{d}} \left( |0\rangle_A |0\rangle_B 
        + |1\rangle_A |1\rangle_B + \, \hdots\,+
        |d-1\rangle_A |d-1 \rangle_B  \right)$.
        
\begin{figure}[hbt]
	\centering
	\includegraphics[width=0.7\columnwidth]{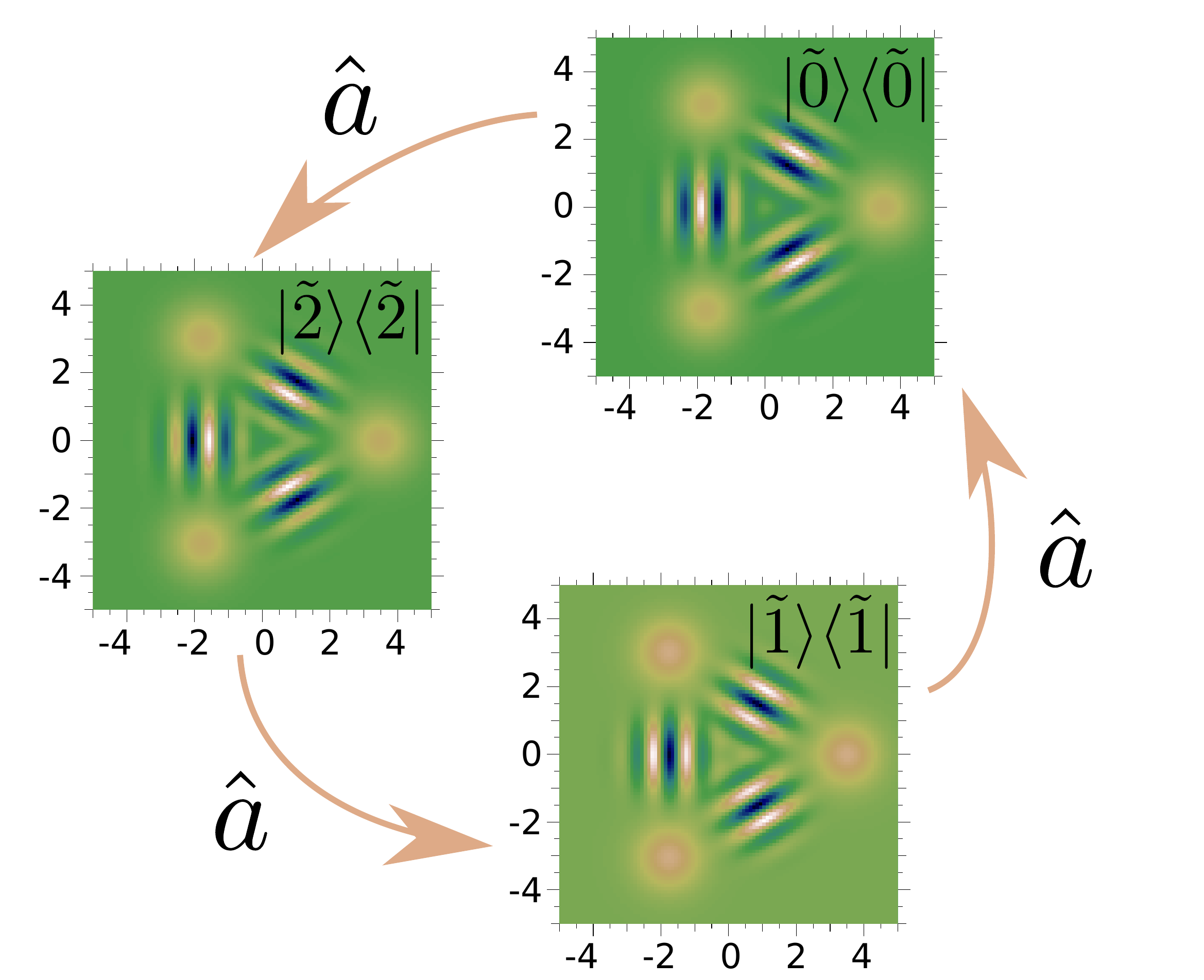} 
	\caption{Three-legged cat states composed from Fock states $|n\rangle$ with different remainders from division by $d=3$.} \label{fig:3leggedcats}
\end{figure}

Clearly, an encoding will require two electromagnetic modes (for authority and voter part of the ballot) and $d$-differing cat-like states. Such \emph{generalized d-legged Cat states} are achieved by splitting the coherent state into $d$ components, each containing Fock-states $\ket{n}$ with $n\; \textrm{mod}\; d = 0,\, 2\, \hdots,\,d-1$.
Examplary for $d=3$, these are superpositions of the three coherent states (forming a triangle or a three-legged star in phase space): 
\[
    \ket{e^{\frac{2\pi i}{3} 0} \alpha}\,,\;   \ket{e^{\frac{2\pi i}{3} 1} \alpha}\,,\; \ket{e^{\frac{2\pi i}{3} 2} \alpha}\,.
\]
with different phase-prefactors of the three yielding the different 3-legged cats.
\begin{align*}
    \ket{\tilde{0}} &=&   \ket{e^{\frac{2\pi i}{3} 0} \alpha}+&&\ket{e^{\frac{2\pi i}{3} 1} \alpha}+&& \ket{e^{\frac{2\pi i}{3} 2} \alpha} \\
    \ket{\tilde{1}} &=& \ket{e^{\frac{2\pi i}{3} 0} \alpha} +&&  e^{\frac{2\pi i}{3} 1\cdot 2}  \ket{e^{\frac{2\pi i}{3} 1} \alpha}+&& e^{\frac{2\pi i}{3} 2\cdot 2} \ket{e^{\frac{2\pi i}{3} 2} \alpha}  \\
    \ket{\tilde{2}} &=& \ket{e^{\frac{2\pi i}{3} 0} \alpha} +&&  e^{\frac{2\pi i}{3} 1\cdot 1}  \ket{e^{\frac{2\pi i}{3} 1} \alpha} +&&  e^{\frac{2\pi i}{3} 2\cdot 1}\ket{e^{\frac{2\pi i}{3} 2} \alpha} \;. \\
\end{align*}
The 3-legged cats of different parity and their cyclic connection by photon subtraction are shown in Fig.\,\ref{fig:3leggedcats}.

Concerning the generation of the starting entangled state,
\begin{align}
        | \Psi_0\rangle &=
         \frac{1}{\sqrt{d}} \left( |0\rangle_A |0\rangle_B 
        + |1\rangle_A |1\rangle_B + \, \hdots\,+
        |d-1\rangle_A |d-1 \rangle_B  \right) \\
        &\rightarrow  \frac{1}{\sqrt{d}} \left( |\tilde{0}\rangle_A |\tilde{0}\rangle_B 
        + |\tilde{1}\rangle_A |\tilde{1}\rangle_B + \, \hdots\,+
        |\tilde{d-1}\rangle_A |\tilde{d-1} \rangle_B  \right) \;,
    \end{align}
the situation is more complex. For a start, there are established optical schemes for the (heralded) generation of the standard (two-legged, single-mode) cat states. These use the very same photon subtraction methods discussed above for the cyclic manipulation on a single-mode squeezed vacuum state \cite{Dakna1997} (or even start from Fock-states \cite{ourjoumtsev2007}).  There are also schemes for the generalized cat states with $d>2$ legs, e.g. for $d=4$ in \cite{Hastrup2020}.

For the microwave regime, much attention has been given to generating conventional and generalized cat-states \cite{vlastakis2013} in a stationary (not itinerant) mode for application as a high-performance qubit \cite{leghtas2015,mirrahimi2014,ofek2016,lescanne2020,reglade2024,Grimm2020}. Two major schemes exist: one which realizes an effective two-photon decay and a two-photon drive of a high-quality `storage' cavity that is coupled to a leaky cavity via a (nonlinear) Josephson-junction element \cite{mirrahimi2014}; the other using an adiabatic connection between Fock-like ground and excited state of a Kerr-resonator to even and odd cat state of the same Kerr-resonator with an additional squeezing term \cite{Puri2017,Grimm2020}. 
This concept is also extendable to two modes \cite{resch2025fast}, which could constitute the authority's and the voters' part of the traveling ballot, if the scheme can be adapted to deliver an itinerant state.

Preparing a generalized cat-state, however, is not enough; the proposed encoding of the traveling ballot requires a superposition of cat-states. At first glance, this seems to entail an additional, hard task. However, it may not be necessary to be able to create cat states to create superpositions thereof. Namely, the superposition of cats corresponding to the two-party traveling ballot state is approximately a superposition of coherent states:
\begin{align}
    \ket{0,\,0} +  \ket{1,\,1}  \rightarrow & \ket{c^\alpha_+,\, c^\beta_+} + \ket{c^\alpha_-,\, c^\beta_-} \\
    &\approx \ket{\alpha,\, \beta} + \ket{-\alpha,\, -\beta} 
\end{align}

A similar result can be found for three-legged cats and (presumably) also for the general d-legged case. A naive estimation of the error introduced by the approximation leading to the second line suggests a minimization by using larger coherent states, but a detailed analysis of the security risk of the protocol associated with this error is still lacking. 
In principle, a superposition of coherent states can be created by a displacement operation, which depends on the state of a qubit \cite{vlastakis2013}, i.e., by passing a coherent state through a cavity, which is strongly detuned by a qubit in a superposition state. More elaborate schemes using two modes have also been proposed, e.\,g., in Ref.\,\cite{Cardoso2021}. Most interesting are experiments, which deterministically create cat states by reflecting coherent states of a cavity, in which an atom in a superposition state is trapped \cite{hacker2019deterministic}. The superposition induces phases on the reflected light and after measurement on the atoms an itinerant cat state is produced. Extensions to the two-mode case and to generalized cats seem feasible.

\emph{In conclusion}, we estimate that the creation of the initial state required for the cat-encoded traveling ballot scheme is rather complex, but there are several promising avenues showing considerable recent progress, so that the approach seems hard but feasible at least for a small number of voters. 

The two encodings considered so far use a single optical mode for the realization of the voters' part of the traveling ballot and thereby can be considered as minimizing the requirements on the transfer of that state. However, as seen above, creation, vote casting, and read-out of the vote count can be challenging. This suggests to investigate if other encodings with larger overhead with respect to the quantum state transfer may nonetheless be favorable with respect to the latter tasks.

\begin{figure}[tb]
	\centering
	\includegraphics[width=0.7\columnwidth]{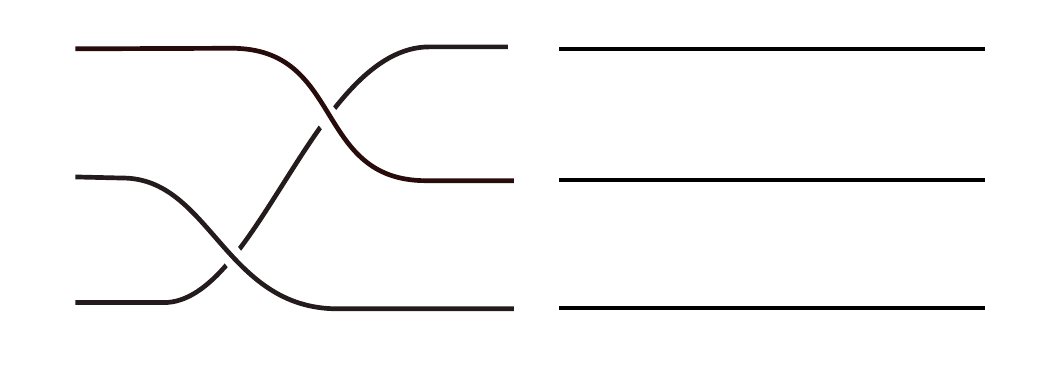} 
	\caption{The two voting operations. YES / cyclical shift, NO / identity operation } \label{fig:tritter}
\end{figure}

\paragraph{Multi-Mode encodings}
A particularly simple encoding uses a single photon in one of $d$ modes as pseudo qudit realization,
\begin{align*}
 |0\rangle_V \; &\widehat{=} \; |1,\,0,\,0 \rangle \\
 |1\rangle_V \; &\widehat{=} \; |0,\,1,\,0 \rangle \\
|2\rangle_V \; &\widehat{=} \; |0,\,0,\,1 \rangle \;.
\end{align*}
The nature of these modes is, in principle, arbitrary, but for the fact that (sufficiently stable) superpositions between the states can be created, as well as entanglement between the authority and voter qudit. Spatial, time-bin, frequency, orbital degree of freedoms, and many more are feasible, see \ref{par:photonic_encodings}, but for the explanation of the scheme and a possible first demonstrator experiment we envision $d$ different parallel-running fibers. This makes for a conceptually particularly simple voting procedure, see Fig.\,\ref{fig:tritter}, where the voter has to be able to switch between to fibre configurations (with or without cyclic permutation between input and output), which could be implemented in robust integrated optics devices, realizable, for instance, by laser writing in silica \cite{Wang2020}. 
Note, that the theoretical anonymity of the protocol relies on the fact, that the reduced voting state looks identical independent of the vote, i.e. whether the photon has undergone a cyclic permutation or not, meaning transmissions along the different photon paths have to leave the photons indistinguishable which, in practice, will place hard requirements on lines, switch, and the voting device.

\begin{figure}[t]
	\centering
	\includegraphics[width=0.7\columnwidth]{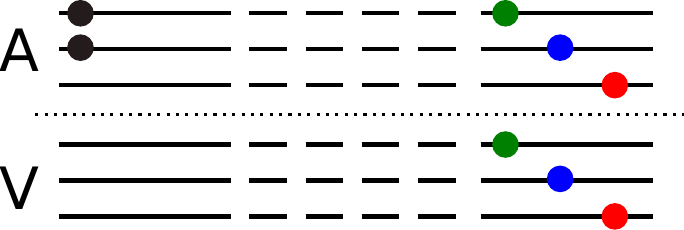}
	\caption{Creating the initial state in a multi-mode encoding. A network of beam-splitters and phase shifters (realized, for instance, in integrated optical devices \cite{Wang2020}), can transfer an input of two indistinguishable single-photons in two individual modes into the required initial state 
    $[ |0\rangle_V|0\rangle_A + |1\rangle_V|1\rangle_A +|2\rangle_V|2\rangle_A ]/\sqrt{3} \; \widehat{=} \; [ \textcolor{mygreen}{|1,\,0,\,0 \rangle_A |1,\,0,\,0 \rangle_V} + \textcolor{blue}{|0,\,1,\,0 \rangle_A |0,\,1,\,0 \rangle_V }+ \textcolor{red}{|0,\,0,\,1 \rangle_A |0,\,0,\,1 \rangle_V} ]/\sqrt{3}$ of $d=3$ superposed states with entanglement between the authority and the voter modes (upper and lower part of the device). The device complexity for $d=3$ voters is comparable to a $2d=6$ mode boson sampler with two photon input. \label{fig:sampler} }
\end{figure}

Creation of the entangled starting state is possible via a multi-mode linear optical device, similar to the ones used for boson-sampling or in some approaches to optical quantum computing. As indicated in Fig.\,\ref{fig:sampler}, one starts with two indistinguishable single-photons in two individual modes (which may stem from deterministic or heralded single-photon sources) which passing through some complex Mach-Zehnder like structure with precisely designed or tuned beam-splitters and phase shifters realizes the desired state at the output. We speculate that the read-out can, in principle, be realized by a device of similar complexity transforming the final state into one, where (correlation) measurements by photon counters on different outputs can realize the desired measurement (details remain to be investigated). 
In conclusion, while challenging creation, voting, and measurement seem feasible in the multi-mode encoding for a limited number of voters and require for the authority a device similar, but somewhat less demanding than a few-mode boson-sampler or the state preparation component of an optical quantum computer and for each voter a considerably simpler switchable linear-optics device. 

\paragraph{Multi-qubit encodings}
As a final encoding, we consider using multiple qubits to realize a qudit; based on the notion, of making best use of technology developed for generic quantum computing tasks. For simplicity, we will consider four voters, so that the voting part of the traveling ballot can be encoded in two qubits, 
\begin{align*}
 |0\rangle_V \; &\widehat{=} \; |00 \rangle \\
|1\rangle_V \; &\widehat{=} \; |01 \rangle \\
|2\rangle_V \; &\widehat{=} \; |10 \rangle \\
|3\rangle_V \; &\widehat{=} \; |11 \rangle \;.
\end{align*}
Creating the entangled starting state $|\Psi_0\rangle$ then requires four (bipartitely) entangled qubits, where the two qubits corresponding to the voter qudit have to be transferred. It is the latter part, which  goes beyond or stretches the limits of the capabilities of most current platforms in which minimal size quantum computers have already been realized. The same reasoning holds for the final measurement of the vote count. 

How out-coupling and transfer can be implemented will be highly platform dependent, as will be the details of the physical nature of the itinerant two qubit state. Exploiting multiple degrees of freedom at the same time (e.g., time-bin and polarization) single photon transfer may be sufficient. Also, both, teleportation and state transfer, seem possible options.

The voting manipulation, while a simple operation on the two qubit state (namely, $+1 \textrm{ mod } d$), nonetheless requires two-qubit gates. Again, any minimal quantum computer could perform the task, but for the challenges of in- and out-coupling of the states. A purely optical operation on the itinerant qubits (as investigated in all the other encodings above) may be favorable, and could be based on optical CNOT gates \cite{OBrien2003,Okamoto2011,Stolz2022}.

\subsubsection{Prospects of implementation} \label{sssec:ElectionSummary}
Above, we presented an explorative, incomplete investigation of the issues arising when considering the physical implementation of an abstract quantum election scheme, namely the traveling ballot protocol. Some interesting and relevant questions could not be fully resolved within the limitations of the current work. Nonetheless, we will try here to draw some preliminary conclusions and pinpoint the pieces missing for a more definite and complete picture.

Mostly left out of from the analysis above, are considerations of the vote counting part of the protocol. To some extent, the implementation of that last part of the protocol is separable from the preceding parts, i.e., we could test the validity of the final state, which results after all votes have been cast, without having the capability to implement the necessary projective measurements. In contrast, any experimental verifiability of the voting operation is unthinkable without the ability to generate the initial state (a partial test with an unentangled state corresponding only to the voter part of $|\psi_0\rangle$ is possible though). Going through the various considered encodings, the implementation of the vote count appears particularly challenging for the occupation number encoding and similarly far from straightforward for the cat state encoding. 
More favorable is the multi-mode encoding, where a read-out device of comparable complexity and design as the integrated optics device for creation sketched in Fig.\,\ref{fig:sampler} is envisioned, where the single-photon sources providing the two photon input state have to be swapped for (possibly single-photon) detectors on the output side.
Multi-qubit encoding requires few-qubit quantum computing capabilities for generating the input (and the for the vote casting) and equivalent capacities (but now also including the measurement operations of a minimal quantum computer) are expected to be sufficient for an implementation of the vote count. 

Important other issues omitted from the above analysis are (i) a detailed security analysis of the impact of an imperfect implementation of the different substeps of the protocol (generation, transfer, vote casting, and vote count), and (ii) a consideration of the scalability of the various encoding proposals to larger number of voters, which is obviously a crucial question for any real-world application.

We conclude this section with a preliminary assessment of the feasibility of implementing the traveling ballot protocol employing the encodings presented here based on the considerations put forward above.
\begin{itemize}
    \item For the \emph{occupation number encoding} there is no clear path to a short-term implementation with available technologies.
    \item  \emph{Cat state encoding} is appealing by its inherent cyclic nature, but further work is needed in how to harness the heralded  voting manipulation, on realizing an efficient creation mechanism, and on the implementation of the vote count. 
    \item For the \emph{multi-mode encoding}, implementation of a small-scale demonstration seems relatively straightforward, but it is unsettled if the multi-rail modes shown above are, indeed, most favorable and more research should probably be done on time-bin or frequency-based modes.
    \item \emph{Multi-qubit encodings} also seem feasible, but require the considerable resources of minimal-scale quantum computers not only for the authority, but also for each voter. Moreover, these must be able to in-/out-couple multi-qubit states for the transfer between the voters, which currently is still an extremely tough demand. Implementing a quantum voting scheme in this manner seems more attractive as a possible demonstrator experiment for the capabilities of minimal quantum computers with these abilities, than as a sole motivation for their development.
\end{itemize}
Finally, it is worth recalling that in the future there may well be alternative better encodings than the ones investigated here.

\subsection{Conclusion} \label{ssec:ApplicationConclusions}

This section aimed at showing the disruptive potential of quantum networks by discussing, in very concrete terms, some of their most impactful applications. We began, in Sect.~\ref{ssec:QInternetVision}, by exploring how local networks could be upgraded to the planetary-scale quantum internet, as well as the opportunities and challenges that come with such an upgrade. Many of those challenges concern security, which is why Sect.~\ref{ssec:SecurityAndDQC} focused on security applications, presented on a theoretical level. We descended to the physical level in Sect.~\ref{ssec:VotingTheoryAndImpl}, where a quantum election protocol was fleshed out in full detail, from theoretical conception to practical implementation.

Hopefully, the examples above have conveyed some of the many reasons to timely investigate quantum networks and their applications. Irrespective of whether the latter are developed to tackle the security threats posed by the advent of quantum computers, to guide future developments of the telecommunication infrastructure, or to provide new services to the public, the untapped potential of this research field should not be underestimated. \pagebreak

\section{Summary and outlook} \label{sec:Conclusion}

In this document, we have compiled what is available and needed for advanced quantum communication. We have mainly focused on the \emph{network} aspects of this upcoming quantum revolution, thus investigating how emerging quantum technologies shall connect to each other and could be interconnected to the classical world. More precisely, we tackled this question by first reviewing the theoretical foundations and unique features of quantum information, cf.~Sect.~\ref{sec:Interface}. Later, in Sect.~\ref{sec:PhysicalBasis}, we listed and discussed multiple platforms for the coherent transfer of quantum information, highlighting the strengths and weaknesses of each one. Finally, in Sect.~\ref{sec:Applications}, we explored some prospects and applications of future quantum networks. Now, we take another look at what science has achieved so far, what is within reach, what challenges still lie ahead, and what steps need to be taken to overcome them. This assessment will take the form of an outlook on the next steps towards a fully fledged quantum internet, the applications enabled by each new breakthrough, and the major roadblocks along the way.

\subsection{First implementations for advanced quantum communication} 
\label{sec:FirstImplementations}
The coherent transfer of quantum information has been experimentally demonstrated with several types of carrier (cf.\,Sect.\,\ref{sec:PhysicalBasis} for references and an in-depth review). We mention, in order of increasing distance achieved: Directly shuttled ions ($\sim$1 cm), microwave photons ($\sim$10 m), optical photons in fibers ($\sim$100 km), optical photons in free space ($\sim$1200 km, or enough to reach LEO satellites). Various cryptographic protocols based on quantum communication are at a late development stage. In particular, QKD has reached commercial maturity \cite{Pljonkin.2018}, and prototypes of trusted repeater networks over metropolitan distances have been realized \cite{Peev.2009, Sasaki.2011, Stucki.2011, Castelvecchi.2018}. Enlarging such networks to a global scale will most likely require the use of satellites as repeaters \cite{Sidhu.2021}. Encouraging progress has been made on this front: The Chinese satellite Micius was the first to achieve QKD using a trusted intermediate station \cite{Liao.2017}. Similar experiments, such as Eagle-1, are in preparation in Europe, and Germany is particularly active in the field with its QUBE and QUBE-II satellites \cite{Hutterer.2018}. The entanglement distribution (see the end of Sect.\,\ref{sssec:QTransfer}) remains challenging, mostly due to the lack of long-lived quantum memories. Nevertheless, there are ground-based demonstrations \cite{Du.2024}. In fact, entangled pairs have even been distributed from space, again by Micius \cite{Yin.2017}. 

The list above roughly exhausts what has been achieved so far. However, more is achievable within a short time-frame and with near-term technology. For instance, a protocol as follows should indeed become feasible within the next years (to know how each step is likely to be implemented, see Sect.\,\ref{ssec:realization}): Translating a (stationary) superconducting qubit hosted on a QPU to a microwave photon, acting as flying qubit; Transducing the photon to visible wavelengths; Propagating the visible photon, across free space, to a satellite repeater; Relaying the signal back to Earth and intercepting it with a receiving station; Transducing back to microwaves and coupling to a different, distant superconducting QPU. For purposes different from distributed quantum computation, the flying qubit could also be measured and transmitted to a designated recipient. Optical ground stations that receive photonic qubits for QKD and potentially advanced quantum communication are already being tested, for example, at the DLR Institute of Communications and Navigation, Oberpfaffenhofen\footnote{https://www.dlr.de/en/latest/news/2022/04/new-optical-ground-station-inaugurated-at-dlr-site-oberpfaffenhofen}.

\subsection{Path towards future quantum networks}
\label{sec:Roadmap}
\begin{figure}[hbt]
    \centering
	\includegraphics[width=0.9\columnwidth]{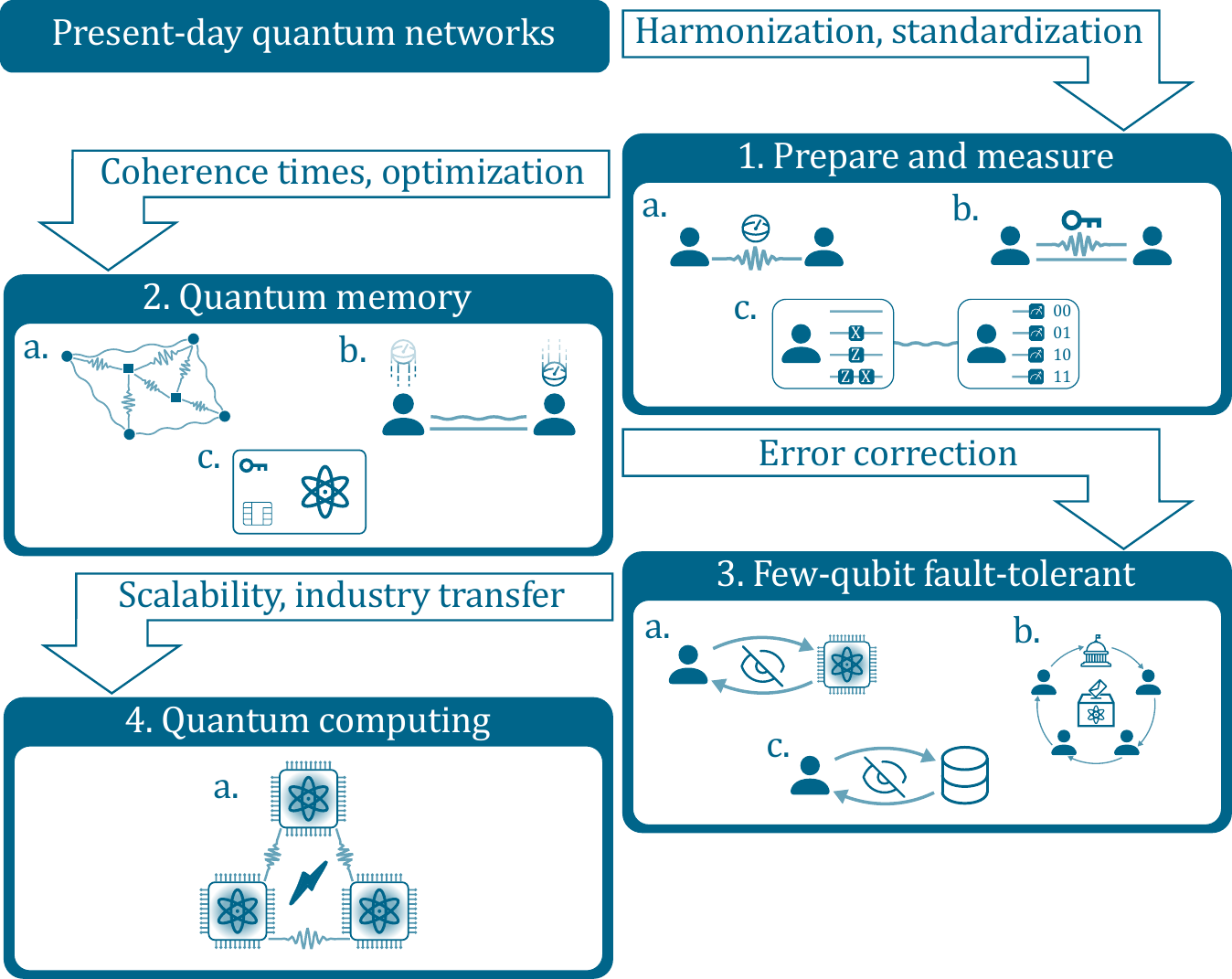} 
	\caption{Graphical depiction of the content of Tab.~\ref{tab:Outlook}. Boxes represent network stages, while the arrows linking them report the challenges from one stage to the next. Within each box, protocols pertaining that stage are represented. More specifically, those protocols are: 1.a.~Single-qubit transfer; 1.b.~Quantum key distribution; 1.c.~(Bipartite) quantum secure direct communication; 2.a.~Advanced entanglement distribution; 2.b.~(On-demand) quantum teleportation; 2.c.~Quantum key card; 3.a.~Blind quantum computing; 3.b.~Quantum elections; 3.c.~Database query; 4.a.~Distributed quantum computing. \\ 
    In producing this figure, open-licensed~.svg images by the following authors were modified and used: weareheroes, SVG Repo, Carbon Design, Icooon Mono, Sebastian Luhn, Noah Jacobus, Dazzle UI, Solar Icons.\label{fig:Outlook_v1}} 
\end{figure}

Having discussed, in previous paragraphs, the present and near future of quantum networks, we now shift the focus to their far future. First, we identify four upcoming network development stages, characterized by each network node possessing the functionality implicit in the stage name: 1.~prepare and measure; 2.~quantum memory; 3.~few-qubit fault tolerant; 4.~quantum computing. With respect to Ref.~\cite{Wehner2018}, from which our terminology is drawn, we group together the ``entanglement distribution" and ``quantum memory" stages under the latter name, because we believe that practical entanglement distribution schemes will require some form of quantum memory. Second, we place every application described in this review in the stage where it becomes implementable. Third, we highlight, for each stage, the (not only technological) challenges that must be overcome to even get there. While the results are shown in Tab.~\ref{tab:Outlook} and graphically in Fig.~\ref{fig:Outlook_v1}, we will take a closer look at why certain applications can be found there and what challenges led to this.

\begin{longtable}{p{0.25\textwidth} | p{0.45\textwidth} | p{0.2\textwidth} }
        \hline
        \begin{center} \textbf{Network stage} \end{center} & \begin{center} \textbf{Application or protocol} \end{center} & \begin{center} \textbf{Main challenges} \end{center} \\
        \hline
        Prepare and measure
        &\textit{Bipartite quantum secure direct communication} (Sect.~\ref{sssec:QSDC2})

        \textit{Single-qubit transfer protocols} (Sect.~\ref{sssec:1QbitTransfer})

        \textit{QKD and QKD-based authentication} (Sect.~\ref{sssec:Security})
        
        &\textit{Harmonization, standardization} 
        \\
        \hline
        Quantum memory & \textit{Advanced entanglement distribution schemes} (Sect.~\ref{sssec:EntDist}) 

        \textit{Multi-party quantum secure direct communication} (Sect.~\ref{sssec:QSDC3})
        
        \textit{On-demand quantum teleportation} (Sect.~\ref{sssec:1QbitTransfer})

        \textit{Quantum key card} (Sect.~\ref{sssec:QKeyCard})
        
        & \textit{Coherence times, optimization}
        \\
        \hline
        Few-qubit fault tolerant & \textit{Blind quantum computing} (Sect.~\ref{sssec:BQC})

        \textit{Database query} (Sect.~\ref{sssec:Query})

        \textit{Quantum elections} (Sect.~\ref{ssec:VotingTheoryAndImpl})
        
        & \textit{Error correction}
        \\
        \hline
        Quantum computing & \textit{Distributed quantum computing} (Sect.~\ref{sssec:DQC})
        
        & \textit{Scalability, industry transfer}
        \\
        \caption{For each stage of quantum-network development, the table reports the main applications that become attainable and the challenges that need to be overcome to even achieve that development stage. \label{tab:Outlook}}
\end{longtable}

\begin{enumerate}
    \item \emph{Prepare and measure}. Two of the three mentioned protocols, namely \emph{single-qubit transfer} and \emph{QKD}, require at most state preparation and measurement. Thus, there is no doubt about their belonging to this stage. Concerns could be raised, on the other hand, about \emph{bipartite QSDC} that requires both the establishment of shared Bell pairs between the sender and receiver nodes and basic quantum computing capabilities for the sender. In a ``star" topology, however, a single central sender node capable of preparing entangled qubit pairs, sending half of them and manipulating the other half, could one-way communicate with a plethora of receiver nodes. Most of the network (the receivers) would thus only be demanded to measure capabilities, making the protocol viable at this stage.

    As far as challenges go, we identify the following two. \emph{Harmonization}: Sect.~\ref{ssec:hybrid} suggested that heterogeneous qubit platforms are likely to coexist in future quantum networks. While this could represent an advantage, interoperability of different stationary- and flying-qubit platforms is required to grant functionality of the network. \emph{Standardization}: As just stated, heterogeneity of the physical supports of quantum information may be a strength of future network architectures. Yet, for technology to spread beyond the academic bubble, the variety of solutions adopted must be limited through international agreements and standardization efforts. This challenge, despite being mentioned here, applies to all phases of network development. In later stages, though, the standardization shall progressively shift from hardware to software (e.g.~for communication protocols, interfaces to existing classical software, etc.).

    \item \emph{Quantum memory}. We call ``advanced" those \emph{entanglement distribution schemes} that can produce $n$-party entangled states for arbitrary $n$. Many current proposals for their implementation leverage \emph{distillation} of multiple entangled pairs, easier to produce. Since the quality of the first established pair must be preserved until the last one is ready, memories are required. \emph{Multi-party QSDC} uses such advanced entanglement distribution schemes as a primitive, and thus automatically belongs to this stage. \textit{On-demand quantum teleportation} refers to the variant of the quantum teleportation (QTp) protocol, cf.~Sect.~\ref{sssec:1QbitTransfer}, where Bell pairs are established in advance and stored for later use. It is the most useful variant of QTp, but long-term storage of entangled pairs requires quantum memories. The same long-term storage of quantum states is at the basis of the concept of \emph{quantum key cards}. In the protocol of Sect.~\ref{sssec:QKeyCard}, a central authority has some quantum computing capabilities, but all authenticating users get away with simply storing their ``token state" for long enough.

    An obvious challenge in achieving this development stage is \emph{coherence time}. Quantum information is typically very short-lived and storing it for long periods of time is technically challenging. Quantum memories and thus the applications that rely on them cannot exist without long coherence times. It should be remarked that actual quantum memories will most certainly require some form of \emph{error correction}, which we address in the next point. A second challenge is represented by \emph{optimization}. At this stage, the need to optimize design aspects of quantum networks (e.g.~repeater placement, entanglement distribution policies, latency times, etc.) starts being felt. Optimization becomes even more relevant, the more advanced the network becomes.

    \item \emph{Few-qubit fault-tolerant.} In \emph{blind quantum computing} (BQC), clients with limited quantum computing capabilities outsource complex computations to a powerful server, but they are assured that the server knows neither the circuit it executes nor its outcome. The server is assumed to be a proper fault-tolerant quantum computer, but clients only need to perform very basic quantum computations, placing this application firmly in this network stage. The same reasoning applies to \emph{database query}, should be considered a special case of BQC, where the server only outputs an item from a list without performing calculations. Finally, in quantum election schemes like the \emph{traveling ballot protocols}, voters cast votes by applying a simple unitary to a quantum state provided by the electoral authorities. Simple quantum computing capabilities are thus sufficient for the voters, whereas the central-authority node may be more advanced.

    \emph{Error correction} is unsurprisingly the main challenge here. Transitioning from noisy intermediate-scale to fault-tolerant quantum machinery demands selecting, perfecting and implementing some of the many error-correction proposals. Different hardware platforms may be best-suited to different correction schemes, and the need for \emph{harmonization} and \emph{standardization} thus persists here.

    \item \emph{Quantum computing}. When quantum computers become widespread, their computational power could be combined via the paradigm of \emph{distributed quantum computing} (DQC). Genuine DQC can only take place if all or most end nodes have strong quantum computing capabilities. It is thus a very advanced distributed protocols that will emerge in the last stages of network development. 
    
    The first roadblock we mention is \emph{scalability}. Even after error-correction is achieved, what distinguishes early fault-tolerant from fully fault-tolerant quantum computation is the ability, in the latter era, to \emph{scale} the number of logical qubits without exponential overhead in the number of physical ones. This \emph{scalability} issue is a serious technological hindrance, and no clear path towards its solution is yet known, to the best of the authors' knowledge. Finally, a major challenge is represented by \emph{industry transfer}. The establishment of a global quantum network by public institutions only seems unlikely. To enable the involvement of the private sector, though, timely and effective technology and knowledge transfer from research to industry is required. While, on this front, the outlook seems rather positive, with more and more companies investing increasing amounts in quantum technologies, a deliberate and coordinated institutional effort remains desirable.
\end{enumerate}
This review attempted to provide a fruitful glimpse into the inner workings of quantum networks, their interplay with related disciplines, and their potential. The authors hope that the many applications reported in this document, as well as its holistic approach to the subject, will serve as inspiration for further scientific investigation, for the operational design of the future quantum-network infrastructure, and for the development of future quantum-network protocols.

\subsection*{Acknowledgments}
All authors acknowlegde fruitful discussions with various participants of the Qu-GoV project during the workshops in Ulm and Berlin. AS, AT, and MZ acknowledge insightful contributions by Alexander von Consbruch in regard to blind quantum computing and by Felicitas Binder in regard to single-qubit transfer methods in the context of their bachelor theses.

\pagebreak

\backmatter

\section*{List of abbreviations}

\begin{tabular}{l l}
    2DEG \qquad & Two-Dimensional Electron Gas \\
    ARPANET \qquad & Advanced Research Projects Agency NETwork \\
    AWG \qquad & Arbitrary Waveform Generator \\
    BB84 \qquad & It refers to the Bennet and Brassard QKD protocol, published in 1984 \\
    BQC \qquad & Blind Quantum Computing \\
    CCD(s) \qquad & Charge Coupled Device(s) \\
    CERT \qquad & Computer Emergency Response Team \\
    CHSH game \qquad & Clauser-Horne-Shimony-Holt, an imaginary game to certify quantum correlations \\
    CMOS \qquad & Complementary Metal-Oxide-Semiconductor \\
    CNOT \qquad & Controlled-NOT, a logical operation both in classical and quantum computation \\
    DC \qquad & Direct Current \\
    DCC \qquad & Distributed Classical Computing \\
    (D)DoS \qquad & (Distributed) Denial of Service \\
    DLR \qquad & Deutsches Zentrum f\"ur Luft- und Raumfahrt, a.k.a. the German Aerospace Agency \\
    DIT \qquad & Direct Information Transfer \\
    DQC \qquad & Distributed Quantum Computing \\
    DQTp \qquad & Direct Quantum Teleportation \\
    EM \qquad & Electro-magnetic \\
    GaAs \qquad & Gallium-Arsenicum, a metallic alloy \\
    GHZ \qquad & Greenberg-Horn-Zeilinger state, a common tri-partite entangled state \\
    LC-resonator \qquad & A capacitor-inductor resonating circuit \\
    LEO \qquad & Low Earth Orbit (typically, referring to satellites) \\
    LOCC \qquad & Local Operations and Classical Communication \\
    LiNB$\text{O}_3$ \qquad & Lithium niobate \\
    (Q)MPI \qquad & (Quantum) Message-Passing Interface \\
    NISQ \qquad & Noisy Intermediate-Scale Quantum (computing) \\
    NIST \qquad & National Institute of Standards and Technology, a US institution \\
    NV \qquad & Nitrogen-vacancy. It refers to a popular kind of defect centers in diamond \\
    P2P \qquad & Peer-to-Peer \\
    PQC \qquad & Post Quantum Cryptography \\
    PUF \qquad & Physical Unclonable Function \\
    QC \qquad & Quantum Computing or Quantum Computation, used interchangeably \\
    QED \qquad & Quantum Electrodynamics \\
    QFT \qquad & Quantum Fourier Transform. Not to be confused with Quantum Field Theory \\
    QIT \qquad & Quantum Information Transfer \\
    QKD \qquad & Quantum Key Distribution  \\
    QLink \qquad & Quantum Link \\
    QPE \qquad & Quantum Phase Estimation \\
    QPU \qquad & Quantum Processing Unit \\
    QSDC \qquad & Quantum Secure Direct Communication \\
    QTp \qquad & Quantum Teleportation \\
    RF \qquad & Radio-frequency \\
    RSA \qquad & ``Rivest-Shamir-Adleman", a family of public-key (asymmetric) cryptosystems \\
    SAW \qquad & Surface Acoustic Waves \\
    SiV \qquad & Silicon-vacancy. It refers to a popular kind of defect centers in diamond \\
    SQUID \qquad & Superconducting QUantum Interference Device \\
    SWAP \qquad & A classical (quantum) operation swapping the value of two bits (qubits) \\
    TCP/IP \qquad & Transmission Control Protocol/Internet Protocol \\
    T-center \qquad & A radiation damage center in silicon \\
    VPN \qquad & Virtual Private Network \\
    VQE \qquad & Variational Quantum Eigensolver \\
    YIG \qquad & Yttrium Iron Garnet \\
\end{tabular}

\section*{Declarations}

\subsection*{Availability of data and materials}
Not applicable.

\subsection*{Competing interests}
There are no competing interests.

\subsection*{Funding}
This article was initiated as part of the Qu-Gov project of the Bundesdruckerei GmbH commissioned by the German Federal Ministry of Finance on the subject of ``Physical principles of quantum information, communication, and processing".

\subsection*{Authors' contributions} 
All authors contributed equally to the definition of the content, structure and goals of the manuscript. While a fruitful exchange of ideas between co-authors inspired the entire document, during the writing phase the labor was divided according to individual preferences and expertise, with the following results. The general Sects.~\ref{sec:Intro}, \ref{sec:Conclusion} (Introduction, Summary and Outlook) and the structure were mostly covered by AT, AG, OK and MZ. DF, FDP, JS and WPS focused on the interface between classical and quantum information, Sect.~\ref{sec:Interface}. The hardware overview of Sect.~\ref{sec:PhysicalBasis} is almost exclusively BK's work. Finally, Sect.~\ref{sec:Applications} (on quantum networks applications) saw contributions by BK, AS, AT and MZ. The joint research on the topic of the article has been initiated by WPS and the Bundesdruckerei GmbH.

All authors read and approved the final manuscript.

\newpage

\bibliography{references} % common bib file

@inproceedings{Aaronson.2009,
 author = {Aaronson, Scott},
 title = {Quantum Copy-Protection and Quantum Money},
 pages = {229--242},
 publisher = {IEEE},
 isbn = {978-0-7695-3717-7},
 booktitle = {2009 IEEE 24th Annual Conference on Computational Complexity},
 year = {2009},
 address = {Piscataway, NJ},
url ={https://doi.org/10.1109/CCC.2009.42},
 doi = {10.1109/CCC.2009.42}
}

@inproceedings{Aaronson.2012,
 author = {Aaronson, Scott and Christiano, Paul},
 title = {Quantum money from hidden subspaces},
 pages = {41--60},
 publisher = {ACM},
 isbn = {9781450312455},
 series = {ACM Conferences},
 editor = {Karloff, Howard},
 booktitle = {Proceedings of the 44th symposium on Theory of Computing},
 year = {2012},
 address = {New York, NY},
url ={https://doi.org/10.1145/2213977.2213983},
 doi = {10.1145/2213977.2213983}
}

@article{Abadi1989,
 author = {Abadi, Mart{\'i}n and Feigenbaum, Joan and Kilian, Joe},
 year = {1989},
 title = {On hiding information from an oracle},
 pages = {21--50},
 volume = {39},
 number = {1},
 issn = {00220000},
 journal = {J Comput Syst Sci},
 doi = {10.1016/0022-0000(89)90018-4}
}

@article{abasifard2024ideal,
 author = {Abasifard, Mostafa and Cholsuk, Chanaprom and Pousa, Roberto G. and Kumar, Anand and Zand, Ashkan and Riel, Thomas and Oi, Daniel K. L. and Vogl, Tobias},
 year = {2024},
 title = {The ideal wavelength for daylight free-space quantum key distribution},
 volume = {1},
 number = {1},
 journal = {APL Quantum},
 doi = {10.1063/5.0186767}
}

@article{Abbas.2024,
 author = {Abbas, Amira and Ambainis, Andris and Augustino, Brandon and B{\"a}rtschi, Andreas and Buhrman, Harry and Coffrin, Carleton and Cortiana, Giorgio and Dunjko, Vedran and Egger, Daniel J. and Elmegreen, Bruce G. and Franco, Nicola and Fratini, Filippo and Fuller, Bryce and Gacon, Julien and Gonciulea, Constantin and Gribling, Sander and Gupta, Swati and Hadfield, Stuart and Heese, Raoul and Kircher, Gerhard and Kleinert, Thomas and Koch, Thorsten and Korpas, Georgios and Lenk, Steve and Marecek, Jakub and Markov, Vanio and Mazzola, Guglielmo and Mensa, Stefano and Mohseni, Naeimeh and Nannicini, Giacomo and O'Meara, Corey and Tapia, Elena Pe{\~n}a and Pokutta, Sebastian and Proissl, Manuel and Rebentrost, Patrick and Sahin, Emre and Symons, Benjamin C. B. and Tornow, Sabine and Valls, V{\'i}ctor and Woerner, Stefan and Wolf-Bauwens, Mira L. and Yard, Jon and Yarkoni, Sheir and Zechiel, Dirk and Zhuk, Sergiy and Zoufal, Christa},
 year = {2024},
 title = {Challenges and opportunities in quantum optimization},
 pages = {718--735},
 volume = {6},
 number = {12},
 journal = {Nat. Rev. Phys.},
 doi = {10.1038/s42254-024-00770-9}
}

@article{acar2018,
author = {Acar, Abbas and Aksu, Hidayet and Uluagac, A. Selcuk and Conti, Mauro},
title = {A Survey on Homomorphic Encryption Schemes: Theory and Implementation},
year = {2018},
issue_date = {July 2019},
publisher = {Association for Computing Machinery},
address = {New York, NY, USA},
volume = {51},
number = {4},
issn = {0360-0300},
url = {https://doi.org/10.1145/3214303},
doi = {10.1145/3214303},
journal = {ACM Comput Surv},
month = jul,
articleno = {79},
numpages = {35}
}

@article{afzal2024distributed,
 author = {Afzal, Francis and Akhlaghi, Mohsen and Beale, Stefanie J. and Bedroya, Olinka and Bell, Kristin and Bergeron, Laurent and Bonsma-Fisher, Kent and Bychkova, Polina and Chaisson, Zachary M. E. and Chartrand, Camille and others},
 year = {2024},
 title = {Distributed Quantum Computing in Silicon},
 journal = {arXiv},
 eprint = {arXiv:2406.01704},
}

@article{akhtar2023high,
 author = {Akhtar, M. and Bonus, F. and Lebrun-Gallagher, F. R. and Johnson, N. I. and Siegele-Brown, M. and Hong, S. and Hile, S. J. and Kulmiya, S. A. and Weidt, S. and Hensinger, W. K.},
 year = {2023},
 title = {A high-fidelity quantum matter-link between ion-trap microchip modules},
 pages = {531},
 volume = {14},
 number = {1},
 journal = {Nat. Commun.},
 doi = {10.1038/s41467-022-35285-3}
}

@article{Alagic.2025,
 author = {Alagic, Gorjan and Bros, Maxime and Ciadoux, Pierre and Cooper, David and Dang, Quynh and Dang, Thinh and Kelsey, John and Lichtinger, Jacob and Liu, Yi-Kai and Miller, Carl and Moody, Dustin and Peralta, Rene and Perlner, Ray and Robinson, Angela and Silberg, Hamilton and Smith-Tone, Daniel and Waller, Noah},
 year = {2025},
 title = {Status report on the fourth round of the NIST post-quantum cryptography standardization process},
journal ={{National Institute of Standards and Technology (U.S.) NIST Internal Report (IR) NIST IR 8545}},
 doi = {10.6028/NIST.IR.8545}
}

@article{alicea2012new,
 author = {Alicea, Jason},
 year = {2012},
 title = {New directions in the pursuit of Majorana fermions in solid state systems},
 pages = {076501},
 volume = {75},
 number = {7},
 journal = {Reports on Progress in Physics},
 doi = {10.1088/0034-4885/75/7/076501}
}

@article{ambainis_dense_2002,
 author = {Ambainis, Andris and Nayak, Ashwin and Ta-Shma, Amnon and Vazirani, Umesh},
 year = {2002},
 title = {{Dense quantum coding and quantum finite automata}},
 pages = {496--511},
 volume = {49},
 number = {4},
 journal = {J. ACM},
 doi = {10.1145/581771.581773}
}

@article{andersen2020repeated,
 author = {Andersen, Christian Kraglund and Remm, Ants and Lazar, Stefania and Krinner, Sebastian and Lacroix, Nathan and Norris, Graham J. and Gabureac, Mihai and Eichler, Christopher and Wallraff, Andreas},
 year = {2020},
 title = {Repeated quantum error detection in a surface code},
 pages = {875--880},
 volume = {16},
 number = {8},
 issn = {1745-2473},
 journal = {Nat. Phys.},
doi={10.1038/s41567-020-0920-y}
}

@article{andrews2014bidirectional,
 author = {Andrews, Reed W. and Peterson, Robert W. and Purdy, Tom P. and Cicak, Katarina and Simmonds, Raymond W. and Regal, Cindy A. and Lehnert, Konrad W.},
 year = {2014},
 title = {Bidirectional and efficient conversion between microwave and optical light},
 pages = {321--326},
 volume = {10},
 number = {4},
 issn = {1745-2473},
 journal = {Nat. Phys.},
 doi = {10.1038/nphys2911}
}

@article{Arapinis2021,
 author = {Arapinis, Myrto and Delavar, Mahshid and Doosti, Mina and Kashefi, Elham},
 year = {2021},
 title = {Quantum Physical Unclonable Functions: Possibilities and Impossibilities},
 pages = {475},
 volume = {5},
 issn = {2521-327X},
 journal = {Quantum},
 doi = {10.22331/q-2021-06-15-475}
}

@article{ardavan2015engineering,
 author = {Ardavan, Arzhang and Bowen, Alice M. and Fernandez, Antonio and Fielding, Alistair J. and Kaminski, Danielle and Moro, Fabrizio and Muryn, Christopher A. and Wise, Matthew D. and Ruggi, Albert and McInnes, Eric J. L. and others},
 year = {2015},
 title = {Engineering coherent interactions in molecular nanomagnet dimers},
 pages = {1--7},
 volume = {1},
 number = {1},
 journal = {npj Quantum Information},
 doi = {10.1038/npjqi.2015.12}
}

@inbook{aref2016quantum,
   title={Quantum Acoustics with Surface Acoustic Waves},
   ISBN={9783319240916},
   ISSN={2364-9062},
   DOI={10.1007/978-3-319-24091-6_9},
   booktitle={Superconducting Devices in Quantum Optics},
chapter={9},
   publisher={Springer International Publishing},
   author={Aref, Thomas and Delsing, Per and Ekström, Maria K. and Kockum, Anton Frisk and Gustafsson, Martin V. and Johansson, Göran and Leek, Peter J. and Magnusson, Einar and Manenti, Riccardo},
   year={2016},
   pages={217–244} 
}

@article{Aslam2023,
   title = {Quantum sensors for biomedical applications},
   volume = {5},
   issn = {2522-5820},
   url = {https://doi.org/10.1038/s42254-023-00558-3},
   doi = {10.1038/s42254-023-00558-3},
   journal = {Nat. Rev. Phys.},
   author = {Aslam, Nabeel and Zhou, Hengyun and Urbach, Elana K. and Turner, Matthew J. and Walsworth, Ronald L. and Lukin, Mikhail D. and Park, Hongkun},
   year = {2023},
   pages = {157-169} 
}

@article{Avron.2021,
 author = {Avron, J. and Casper, Ofer and Rozen, Ilan},
 year = {2021},
 title = {Quantum advantage and noise reduction in distributed quantum computing},
 volume = {104},
 number = {5},
 issn = {2469-9934},
 journal = {Phys Rev A},
 doi = {10.1103/PhysRevA.104.052404}
}

@article{Awschalom.2018,
 author = {Awschalom, David D. and Hanson, Ronald and Wrachtrup, J{\"o}rg and Zhou, Brian B.},
 year = {2018},
 title = {Quantum technologies with optically interfaced solid-state spins},
 pages = {516--527},
 volume = {12},
 number = {9},
 issn = {1749-4885},
 journal = {Nature Photonics},
 doi = {10.1038/s41566-018-0232-2}
}

@article{Awschalom_PRXQ_21,
 author = {Awschalom, David and Berggren, Karl K. and Bernien, Hannes and Bhave, Sunil and Carr, Lincoln D. and Davids, Paul and Economou, Sophia E. and Englund, Dirk and Faraon, Andrei and Fejer, Martin and Guha, Saikat and Gustafsson, Martin V. and Hu, Evelyn and Jiang, Liang and Kim, Jungsang and Korzh, Boris and Kumar, Prem and Kwiat, Paul G. and Lon{\v{c}}ar, Marko and Lukin, Mikhail D. and Miller, David A.B. and Monroe, Christopher and Nam, Sae Woo and Narang, Prineha and Orcutt, Jason S. and Raymer, Michael G. and Safavi-Naeini, Amir H. and Spiropulu, Maria and Srinivasan, Kartik and Sun, Shuo and Vu{\v{c}}kovi{\'c}, Jelena and Waks, Edo and Walsworth, Ronald and Weiner, Andrew M. and Zhang, Zheshen},
 year = {2021},
 title = {Development of Quantum Interconnects (QuICs) for Next-Generation Information Technologies},
 url = {https://link.aps.org/doi/10.1103/PRXQuantum.2.017002},
 pages = {017002},
 volume = {2},
 number = {1},
 journal = {PRX Quantum},
 doi = {10.1103/PRXQuantum.2.017002}
}

@article{axline2018demand,
 author = {Axline, Christopher J. and Burkhart, Luke D. and Pfaff, Wolfgang and Zhang, Mengzhen and Chou, Kevin and Campagne-Ibarcq, Philippe and Reinhold, Philip and Frunzio, Luigi and Girvin, S. M. and Jiang, Liang and others},
 year = {2018},
 title = {On-demand quantum state transfer and entanglement between remote microwave cavity memories},
 pages = {705--710},
 volume = {14},
 number = {7},
 issn = {1745-2473},
 journal = {Nat. Phys.},
 doi = {10.1038/s41567-018-0115-y}
}

@article{Azuma.2023,
 author = {Azuma, Koji and Economou, Sophia E. and Elkouss, David and Hilaire, Paul and Jiang, Liang and Lo, Hoi-Kwong and Tzitrin, Ilan},
 year = {2023},
 title = {Quantum repeaters: From quantum networks to the quantum internet},
 volume = {95},
 number = {4},
 issn = {0034-6861},
 journal = {Rev. Mod. Phys.},
 doi = {10.1103/RevModPhys.95.045006}
}

@article{bader2014room,
 author = {Bader, Katharina and Dengler, Dominik and Lenz, Samuel and Endeward, Burkhard and Jiang, Shang-Da and Neugebauer, Petr and {van Slageren}, Joris},
 year = {2014},
 title = {Room temperature quantum coherence in a potential molecular qubit},
 pages = {5304},
 volume = {5},
 number = {1},
 journal = {Nat. Commun.},
 doi = {10.1038/ncomms6304}
}

@phdthesis{Bainbridge2023,
author = {Bainbridge, James},
title = {Quantum control of acoustic waves},
year = {2023},
school = {{Macquarie University}},
month = {8},
type = {PhD thesis},
doi = {10.25949/23961813.v1}
}

@article{Bao2017,
 author = {Bao, Ning and {Yunger Halpern}, Nicole},
 year = {2017},
 title = {Quantum voting and violation of Arrow's impossibility theorem},
 volume = {95},
 number = {6},
 issn = {2469-9934},
 journal = {Phys. Rev. A},
 doi = {10.1103/PhysRevA.95.062306}
}

@article{bar2013solid,
 author = {Bar-Gill, Nir and Pham, Linh M. and Jarmola, Andrejs and Budker, Dmitry and Walsworth, Ronald L.},
 year = {2013},
 title = {Solid-state electronic spin coherence time approaching one second},
 pages = {1743},
 volume = {4},
 number = {1},
 journal = {Nat. Commun.},
 doi = {10.1038/ncomms2771}
}

@article{Barenco.1995,
 author = {Barenco, A. and Bennett, C. H. and Cleve, R. and DiVincenzo, D. P. and Margolus, N. and Shor, P. and Sleator, T. and Smolin, J. A. and Weinfurter, H.},
 year = {1995},
 title = {Elementary gates for quantum computation},
 pages = {3457--3467},
 volume = {52},
 number = {5},
 issn = {1050-2947},
 journal = {Phys. Rev. A},
 doi = {10.1103/physreva.52.3457}
}

@article{Barman_JOPCM_21,
author = {Barman, Anjan and Gubbiotti, Gianluca and Ladak, S and Adeyeye, A O and Krawczyk, M and Gräfe, J and Adelmann, C and Cotofana, S and Naeemi, A and Vasyuchka, V I and Hillebrands, B and Nikitov, S A and Yu, H and Grundler, D and Sadovnikov, A V and Grachev, A A and Sheshukova, S E and Duquesne, J-Y and Marangolo, M and Csaba, G and Porod, W and Demidov, V E and Urazhdin, S and Demokritov, S O and Albisetti, E and Petti, D and Bertacco, R and Schultheiss, H and Kruglyak, V V and Poimanov, V D and Sahoo, S and Sinha, J and Yang, H and Münzenberg, M and Moriyama, T and Mizukami, S and Landeros, P and Gallardo, R A and Carlotti, G and Kim, J-V and Stamps, R L and Camley, R E and Rana, B and Otani, Y and Yu, W and Yu, T and Bauer, G E W and Back, C and Uhrig, G S and Dobrovolskiy, O V and Budinska, B and Qin, H and van Dijken, S and Chumak, A V and Khitun, A and Nikonov, D E and Young, I A and Zingsem, B W and Winklhofer, M},
 year = {2021},
 title = {The 2021 Magnonics Roadmap},
 pages = {413001},
 volume = {33},
 number = {41},
 journal = {J. Phys.: Condens. Matter},
 doi = {10.1088/1361-648X/abec1a}
}

@inproceedings{Barnum2002,
  title={{Authentication of quantum messages}},
  author={Barnum, Howard and Cr{\'e}peau, Claude and Gottesman, Daniel and Smith, Adam and Tapp, Alain},
  booktitle={The 43rd Annual IEEE Symposium on Foundations of Computer Science, 2002. Proceedings.},
  pages={449--458},
  year={2002},
  organization={IEEE}
}

@article{Barral.2024,
title = {Review of Distributed Quantum Computing: From single QPU to High Performance Quantum Computing},
journal = {Comput Sci Rev},
volume = {57},
pages = {100747},
year = {2025},
issn = {1574-0137},
doi = {https://doi.org/10.1016/j.cosrev.2025.100747},
url = {https://www.sciencedirect.com/science/article/pii/S1574013725000231},
author = {David Barral and F. Javier Cardama and Guillermo Díaz-Camacho and Daniel Faílde and Iago F. Llovo and Mariamo Mussa-Juane and Jorge Vázquez-Pérez and Juan Villasuso and César Piñeiro and Natalia Costas and Juan C. Pichel and Tomás F. Pena and Andrés Gómez}
}

@article{barredo2016atom,
 author = {Barredo, Daniel and de L{\'e}s{\'e}leuc, Sylvain and Lienhard, Vincent and Lahaye, Thierry and Browaeys, Antoine},
 year = {2016},
 title = {An atom-by-atom assembler of defect-free arbitrary two-dimensional atomic arrays},
 pages = {1021--1023},
 volume = {354},
 number = {6315},
 journal = {Science},
doi={10.1126/science.aah3778}
}

@article{barry2020sensitivity,
 author = {Barry, John F. and Schloss, Jennifer M. and Bauch, Erik and Turner, Matthew J. and Hart, Connor A. and Pham, Linh M. and Walsworth, Ronald L.},
 year = {2020},
 title = {Sensitivity optimization for NV-diamond magnetometry},
 pages = {015004},
 volume = {92},
 number = {1},
 issn = {0034-6861},
 journal = {Rev. Mod. Phys.},
 doi = {10.1103/RevModPhys.92.015004}
}

@article{Barz2012,
 author = {Barz, Stefanie and Kashefi, Elham and Broadbent, Anne and Fitzsimons, Joseph F. and Zeilinger, Anton and Walther, Philip},
 year = {2012},
 title = {Demonstration of blind quantum computing},
 pages = {303--308},
 volume = {335},
 number = {6066},
 journal = {Science (New York, N.Y.)},
 doi = {10.1126/science.1214707}
}

@article{Baumer.2024,
  title = {Quantum Fourier Transform Using Dynamic Circuits},
  author = {B\"aumer, Elisa and Tripathi, Vinay and Seif, Alireza and Lidar, Daniel and Wang, Derek S.},
  journal = {Phys Rev Lett},
  volume = {133},
  issue = {15},
  pages = {150602},
  numpages = {7},
  year = {2024},
  month = {Oct},
  publisher = {American Physical Society},
  doi = {10.1103/PhysRevLett.133.150602},
  url = {https://link.aps.org/doi/10.1103/PhysRevLett.133.150602}
}

@article{bayliss2020optically,
 author = {Bayliss, S. L. and Laorenza, D. W. and Mintun, P. J. and Kovos, B. D. and Freedman, Danna E. and Awschalom, D. D.},
 year = {2020},
 title = {Optically addressable molecular spins for quantum information processing},
 pages = {1309--1312},
 volume = {370},
 number = {6522},
 journal = {Science},
 doi = {10.1126/science.abb9352}
}

@inproceedings{Beisel2022,
 author = {Beisel, Martin and Barzen, Johanna and Leymann, Frank and Truger, Felix and Weder, Benjamin and Yussupov, Vladimir},
 title = {Patterns for Quantum Error Handling},
 pages = {22--30},
 publisher = {{Xpert Publishing Services (XPS)}},
 isbn = {9781612089539},
 booktitle = {Proceedings of the 14th International  Conference on Pervasive Patterns and Applications (PATTERNS  2022)},
 year = {2022},
 address = {Wilmington, DE},
}

@article{Bennett.1993,
 author = {Bennett, C. H. and Brassard, G. and Cr{\'e}peau, C. and Jozsa, R. and Peres, A. and Wootters, W. K.},
 year = {1993},
 title = {Teleporting an unknown quantum state via dual classical and Einstein-Podolsky-Rosen channels},
 pages = {1895--1899},
 volume = {70},
 number = {13},
 issn = {0031-9007},
 journal = {Phys. Rev. Lett.},
 doi = {10.1103/PhysRevLett.70.1895}
}

@article{Bennett.2014,
 author = {Bennett, Charles H. and Brassard, Gilles},
 year = {2014},
 title = {Quantum cryptography: Public key distribution and coin tossing},
 pages = {7--11},
 volume = {560},
 issn = {03043975},
 journal = {Theor. Comput. Sci.},
 doi = {10.1016/j.tcs.2014.05.025}
}

@article{bennett2001remote,
 author = {Bennett, Charles H. and DiVincenzo, David P. and Shor, Peter W. and Smolin, John A. and Terhal, Barbara M. and Wootters, William K.},
 year = {2001},
 title = {Remote state preparation},
 pages = {077902},
 volume = {87},
 number = {7},
 issn = {0031-9007},
 journal = {Phys. Rev. Lett.},
 doi = {10.1103/PhysRevLett.87.077902},
 url = {https://link.aps.org/doi/10.1103/PhysRevLett.87.077902}
}

@inproceedings{BenOr.2005,
 author = {Ben-Or, Michael and Hassidim, Avinatan},
 title = {Fast quantum byzantine agreement},
 pages = {481--485},
 publisher = {{ACM Press}},
 isbn = {1581139608},
 editor = {Gabow, Hal and Fagin, Ronald},
 booktitle = {Proceedings of the 37th Annual ACM Symposium on Theory of Computing},
 year = {2005},
 address = {New York, NY},
url = {https://doi.org/10.1145/1060590.1060662},
 doi = {10.1145/1060590.1060662}
}

@article{bergeron2020silicon,
 author = {Bergeron, L. and Chartrand, C. and Kurkjian, A. T.K. and Morse, K. J. and Riemann, H. and Abrosimov, N. V. and Becker, P. and Pohl, H-J and Thewalt, M. L.W. and Simmons, S.},
 year = {2020},
 title = {Silicon-integrated telecommunications photon-spin interface},
 pages = {020301},
 volume = {1},
 number = {2},
 journal = {PRX Quantum},
 doi = {10.1103/PRXQuantum.1.020301}
}

@article{Bertrand_NatNano_16,
 author = {Bertrand, B. and Hermelin, S. and Takada, S. and Yamamoto, M. and Tarucha, S. and Ludwig, A. and Wieck, A. D. and B{\"a}uerle, C. and Meunier, T.},
 year = {2016},
 title = {Fast spin information transfer between distant quantum dots using individual electrons},
 pages = {672--676},
 volume = {11},
 number = {8},
 journal = {Nature Nanotechnology},
 doi = {10.1038/nnano.2016.82}
}

@article{beugnon2007two,
 author = {Beugnon, J{\'e}r{\^o}me and Tuchendler, Charles and Marion, Harold and Ga{\"e}tan, Alpha and Miroshnychenko, Yevhen and Sortais, Yvan R. P. and Lance, Andrew M. and Jones, Matthew P. A. and Messin, Ga{\'e}tan and Browaeys, Antoine and Grangier, Philippe},
 year = {2007},
 title = {Two-dimensional transport and transfer of a single atomic qubit in optical tweezers},
 pages = {696--699},
 volume = {3},
 number = {10},
 issn = {1745-2473},
 journal = {Nat. Phys.},
 doi = {10.1038/nphys698}
}

@article{Biagi2022,
 author = {Nicola Biagi and Saverio Francesconi and Alessandro Zavatta and Marco Bellini},
 year = {2022},
 title = {Photon-by-photon quantum light state engineering},
 url = {https://www.sciencedirect.com/science/article/pii/S0079672722000398},
 pages = {100414},
 volume = {84},
 issn = {0079-6727},
 journal = {Prog. Quantum Electron.},
 doi = {10.1016/j.pquantelec.2022.100414}
}

@article{Bienfait_Science_19,
 author = {Bienfait, A. and Satzinger, K. J. and Zhong, Y. P. and Chang, H-S and Chou, M-H and Conner, C. R. and Dumur, {\'E}. and Grebel, J. and Peairs, G. A. and Povey, R. G. and Cleland, A. N.},
 year = {2019},
 title = {Phonon-mediated quantum state transfer and remote qubit entanglement},
 pages = {368--371},
 volume = {364},
 number = {6438},
 journal = {Science},
 doi = {10.1126/science.aaw8415}
}

@article{Bilyk2023,
 author = {Bilyk, Andriyan and Doliskani, Javad and Gong, Zhiyong},
 year = {2023},
 title = {Cryptanalysis of three quantum money schemes},
 volume = {22},
 number = {4},
 issn = {1570-0755},
 journal = {Quantum Inf. Process.},
 doi = {10.1007/s11128-023-03919-0}
}

@article{blais2020quantum,
  title={Quantum information processing and quantum optics with circuit quantum electrodynamics},
  author={Blais, Alexandre and Girvin, Steven M and Oliver, William D},
  journal={Nat. Phys.},
  volume={16},
  number={3},
  pages={247--256},
  year={2020},
  publisher={Nature Publishing Group UK London},
  doi={10.1038/s41567-020-0806-z}
}

@article{blais2021circuit,
 author = {Blais, Alexandre and Grimsmo, Arne L. and Girvin, Steven M. and Wallraff, Andreas},
 year = {2021},
 title = {Circuit quantum electrodynamics},
 pages = {025005},
 volume = {93},
 number = {2},
 issn = {0034-6861},
 journal = {Rev. Mod. Phys.},
doi={10.1103/RevModPhys.93.025005}
}

@inproceedings{bluhm2019semiconductor,
 author = {Bluhm, Hendrik and Schreiber, Lars R.},
 title = {Semiconductor spin qubits---A scalable platform for quantum computing?},
 pages = {1--5},
 booktitle = {2019 IEEE International Symposium on Circuits and Systems (ISCAS)},
 year = {2019}
}

@article{Blum.1983,
 author = {Blum, Manuel},
 year = {1983},
 title = {Coin flipping by telephone a protocol for solving impossible problems},
 pages = {23--27},
 volume = {15},
 number = {1},
 issn = {0163-5700},
 journal = {ACM SIGACT News},
 doi = {10.1145/1008908.1008911}
}

@article{bluvstein2022quantum,
 author = {Bluvstein, Dolev and Levine, Harry and Semeghini, Giulia and Wang, Tout T. and Ebadi, Sepehr and Kalinowski, Marcin and Keesling, Alexander and Maskara, Nishad and Pichler, Hannes and Greiner, Markus and Vuleti{\'c}, Vladan and Lukin, Mikhail D.},
 year = {2022},
 title = {A quantum processor based on coherent transport of entangled atom arrays},
 pages = {451--456},
 volume = {604},
 number = {7906},
 journal = {Nature},
 doi = {10.1038/s41586-022-04592-6}
}

@article{Bongs_PhiuZ_2025,
author = {Bongs, Kai and Fuchs, Christian and Laiho, Kaisa and Moll, Florian and Wölk, Sabine and Zimmermann, Matthias},
title = {{Weltweite Vernetzung der Quantentechnologien: Quantenkommunikation via Satellit}},
journal = {Phys. unserer Zeit},
volume = {56},
number = {3},
pages = {116-123},
doi = {https://doi.org/10.1002/piuz.202401725},
year = {2025}
}

@article{borjans2020resonant,
 author = {Borjans, Felix and Croot, X. G. and Mi, Xiao and Gullans, M. J. and Petta},
 year = {2020},
 title = {Resonant microwave-mediated interactions between distant electron spins},
 pages = {195--198},
 volume = {577},
 number = {7789},
 journal = {Nature},
doi={10.1038/s41586-019-1867-y}
}

@article{bose2007quantum,
 author = {Bose, Sougato},
 year = {2007},
 title = {Quantum communication through spin chain dynamics: an introductory overview},
 pages = {13--30},
 volume = {48},
 number = {1},
 issn = {0010-7514},
 journal = {Contemporary Physics},
 doi = {10.1080/00107510701342313}
}

@article{botzem2018tuning,
 author = {Botzem, Tim and Shulman, Michael D. and Foletti, Sandra and Harvey, Shannon P. and Dial, Oliver E. and Bethke, Patrick and Cerfontaine, Pascal and McNeil, Robert P. G. and Mahalu, Diana and Umansky, Vladimir and others},
 year = {2018},
 title = {Tuning methods for semiconductor spin qubits},
 pages = {054026},
 volume = {10},
 number = {5},
 journal = {Phys. Rev. Applied},
doi={10.1103/PhysRevApplied.10.054026}
}

@article{bouchiat1998quantum,
 author = {Bouchiat, Vincent and Vion, Denis and Joyez, Philippe and Esteve, Daniel and Devoret, M. H.},
 year = {1998},
 title = {Quantum coherence with a single Cooper pair},
 pages = {165},
 volume = {1998},
 number = {T76},
 journal = {Phys. Scr.},
doi={10.1238/Physica.Topical.076a00165}
}

@article{Boykin.2000,
 author = {Boykin, P.Oscar and Mor, Tal and Pulver, Matthew and Roychowdhury, Vwani and Vatan, Farrokh},
 year = {2000},
 title = {A new universal and fault-tolerant quantum basis},
 pages = {101--107},
 volume = {75},
 number = {3},
 issn = {00200190},
 journal = {Inf. Process. Lett.},
 doi = {10.1016/S0020-0190(00)00084-3}
}

@article{bozinovic2013terabit,
 author = {Bozinovic, Nenad and Yue, Yang and Ren, Yongxiong and Tur, Moshe and Kristensen, Poul and Huang, Hao and Willner, Alan E. and Ramachandran, Siddharth},
 year = {2013},
 title = {Terabit-scale orbital angular momentum mode division multiplexing in fibers},
 pages = {1545--1548},
 volume = {340},
 number = {6140},
 journal = {Science},
 doi = {10.1126/science.1237861}
}

@article{bradbury2011efficient,
 author = {Bradbury, F. R. and Takita, Maika and Gurrieri, T. M. and Wilkel, K. J. and Eng, Kevin and Carroll, M. S. and Lyon, S. A.},
 year = {2011},
 title = {Efficient clocked electron transfer on superfluid helium},
 pages = {266803},
 volume = {107},
 number = {26},
 issn = {0031-9007},
 journal = {Phys. Rev. Lett.},
 doi = {10.1103/PhysRevLett.107.266803}
}

@article{bradley2019ten,
 author = {Bradley, Conor E. and Randall, Joe and Abobeih, Mohamed H. and Berrevoets, Remon C. and Degen, Maarten J. and Bakker, Michiel A. and Markham, Matthew and Twitchen, Daniel J. and Taminiau, Tim H.},
 year = {2019},
 title = {A ten-qubit solid-state spin register with quantum memory up to one minute},
 pages = {031045},
 volume = {9},
 number = {3},
 journal = {Phys. Rev. X},
 doi = {10.1103/PhysRevX.9.031045}
}

@inproceedings{Broadbent2009,
 author = {Broadbent, Anne and Fitzsimons, Joseph and Kashefi, Elham},
 title = {Universal Blind Quantum Computation},
 pages = {517--526},
 publisher = {{IEEE Computer Society}},
 isbn = {9780769538501},
 series = {FOCS '09},
 booktitle = {Proceedings of the 2009 50th Annual IEEE Symposium on Foundations of Computer Science},
 year = {2009},
 address = {USA},
 doi = {10.1109/FOCS.2009.36},
 url = {https://doi.org/10.1109/FOCS.2009.36},
}

@book{Bruss2019,
 author = {Bru{\ss}, Dagmar and Leuchs, Gerd},
 year = {2019},
 title = {Quantum information},
 address = {Weinheim},
 publisher = {Wiley-VCH},
 isbn = {9783527413539},
 doi = {10.1002/9783527805785}
}

@article{Buhrman.2010,
 author = {Buhrman, Harry and Cleve, Richard and Massar, Serge and de Wolf, Ronald},
 year = {2010},
 title = {Nonlocality and communication complexity},
 pages = {665--698},
 volume = {82},
 number = {1},
 issn = {0034-6861},
 journal = {Rev Mod Phys},
 doi = {10.1103/RevModPhys.82.665}
}

@article{burkard2020superconductor,
 author = {Burkard, Guido and Gullans, Michael J. and Mi, Xiao and Petta, Jason R.},
 year = {2020},
 title = {Superconductor--semiconductor hybrid-circuit quantum electrodynamics},
 pages = {129--140},
 volume = {2},
 number = {3},
 journal = {Nat. Rev. Phys.},
doi={10.1038/s42254-019-0135-2}
}

@article{burkard2023semiconductor,
  title = {Semiconductor spin qubits},
  author = {Burkard, Guido and Ladd, Thaddeus D. and Pan, Andrew and Nichol, John M. and Petta, Jason R.},
  journal = {Rev. Mod. Phys.},
  volume = {95},
  issue = {2},
  pages = {025003},
  numpages = {58},
  year = {2023},
  publisher = {American Physical Society},
  doi = {10.1103/RevModPhys.95.025003},
  url = {https://link.aps.org/doi/10.1103/RevModPhys.95.025003}
}

@article{burkhart2021error,
 author = {Burkhart, Luke D. and Teoh, James D. and Zhang, Yaxing and Axline, Christopher J. and Frunzio, Luigi and Devoret, Michel H. and Jiang, Liang and Girvin, Steven M. and Schoelkopf, Robert J.},
 year = {2021},
 title = {Error-detected state transfer and entanglement in a superconducting quantum network},
 pages = {030321},
 volume = {2},
 number = {3},
 journal = {PRX Quantum},
 doi = {10.1103/PRXQuantum.2.030321},
 url = {https://link.aps.org/doi/10.1103/PRXQuantum.2.030321}
}

@article{Byeon_NatCom_21,
 author = {Byeon, H. and Nasyedkin, K. and Lane, J. R. and Beysengulov, N. R. and Zhang, L. and Loloee, R. and Pollanen, J.},
 year = {2021},
 title = {Piezoacoustics for precision control of electrons floating on helium},
 pages = {4150},
 volume = {12},
 number = {1},
 journal = {Nat. Commun.},
 doi = {10.1038/s41467-021-24452-7}
}

@article{Cacciapuoti2020,
 author = {Cacciapuoti, Angela Sara and Caleffi, Marcello and Tafuri, Francesco and Cataliotti, Francesco Saverio and Gherardini, Stefano and Bianchi, Giuseppe},
 year = {2020},
 title = {Quantum Internet: Networking Challenges in Distributed Quantum Computing},
 pages = {137--143},
 volume = {34},
 number = {1},
 issn = {0890-8044},
 journal = {IEEE Network},
 doi = {10.1109/MNET.001.1900092}
}

@article{Caleffi.2024,
 author = {Caleffi, Marcello and Amoretti, Michele and Ferrari, Davide and Illiano, Jessica and Manzalini, Antonio and Cacciapuoti, Angela Sara},
 year = {2024},
 title = {Distributed quantum computing: A survey},
 pages = {110672},
 volume = {254},
 issn = {13891286},
 journal = {Comput Networks},
 doi = {10.1016/j.comnet.2024.110672}
}

@article{canteri2024photon,
 author = {Canteri, M. and Koong, Z. X. and Bate, J. and Winkler, A. and Krutyanskiy, V. and Lanyon, B. P.},
 journal = {arXiv},
 year = {2024},
 title = {A photon-interfaced ten qubit quantum network node},
 eprint = {arXiv:2406.09480},
}

@article{Cardoso2021,
 author = {Cardoso, Fernando R. and Rossatto, Daniel Z. and Fernandes, Gabriel PLM and Higgins, Gerard and Villas-Boas, Celso J.},
 year = {2021},
 title = {Superposition of two-mode squeezed states for quantum information processing and quantum sensing},
 pages = {062405},
 volume = {103},
 number = {6},
 issn = {2469-9934},
 journal = {Phys. Rev. A},
doi = {10.1103/PhysRevA.103.062405}
}

@article{carretta2021perspective,
 author = {Carretta, Stefano and Zueco, David and Chiesa, Alessandro and G{\'o}mez-Le{\'o}n, {\'A}lvaro and Luis, Fernando},
 year = {2021},
 title = {A perspective on scaling up quantum computation with molecular spins},
 volume = {118},
 number = {24},
 journal = {Appl. Phys. Lett.},
 doi = {10.1063/5.0053378}
}

@article{casariego2023propagating,
 author = {Casariego, Mateo and {Zambrini Cruzeiro}, Emmanuel and Gherardini, Stefano and Gonzalez-Raya, Tasio and Andr{\'e}, Rui and Fraz{\~a}o, Gon{\c{c}}alo and Catto, Giacomo and M{\"o}tt{\"o}nen, Mikko and Datta, Debopam and Viisanen, Klaara and Govenius, Joonas and Prunnila, Mika and Tuominen, Kimmo and Reichert, Maximilian and Renger, Michael and Fedorov, Kirill G. and Deppe, Frank and {van der Vliet}, Harriet and Matthews, A. J. and Fern{\'a}ndez, Yolanda and Assouly, R. and Dassonneville, R. and Huard, B. and Sanz, Mikel and Omar, Yasser},
 year = {2023},
 title = {Propagating quantum microwaves: towards applications in communication and sensing},
 pages = {023001},
 volume = {8},
 number = {2},
 journal = {Quantum Sci. and Technol.},
 doi = {10.1088/2058-9565/acc4af}
}

@article{Castelvecchi.2018,
 author = {Castelvecchi, Davide},
 year = {2018},
 title = {The quantum internet has arrived (and it hasn't)},
 pages = {289--292},
 volume = {554},
 number = {7692},
 journal = {Nature},
 doi = {10.1038/d41586-018-01835-3}
}

@article{Cerf1999,
 author = {Cerf, N. J. and Adami, C. and Gingrich, R. M.},
 year = {1999},
 title = {{Reduction criterion for separability}},
 pages = {898--909},
 volume = {60},
 number = {2},
 issn = {1050-2947},
 journal = {Phys. Rev. A},
 doi = {10.1103/PhysRevA.60.898}
}

@article{chakravartula2020implementation,
 author = {Chakravartula, Venkatesh and Samiappan, Dhanalakshmi and Kumar, R. and Manjari, A. P.},
 year = {2020},
 title = {Implementation of quantum teleportation of photons across an air -- water interface},
 volume = {52},
 number = {7},
 issn = {0306-8919},
 journal = {Opt. Quantum Electron.},
 doi = {10.1007/s11082-020-02449-8}
}

@article{chan2004few,
 author = {Chan, I. H. and Fallahi, P. and Vidan, A. and Westervelt, R. M. and Hanson, M. and Gossard, A. C.},
 year = {2004},
 title = {Few-electron double quantum dots},
 pages = {609},
 volume = {15},
 number = {5},
 journal = {Nanotechnology},
doi={10.1088/0957-4484/15/5/035}
}

@article{Chatterjee.2022,
 author = {Chatterjee, Turbasu and Das, Arnav and Mohtashim, Shah Ishmam and Saha, Amit and Chakrabarti, Amlan},
 year = {2022},
 title = {Qurzon: A Prototype for a Divide and Conquer-Based Quantum Compiler for Distributed Quantum Systems},
 volume = {3},
 number = {4},
 journal = {SN Comput Sci},
 doi = {10.1007/s42979-022-01207-9}
}

@article{chatterjee2021semiconductor,
 author = {Chatterjee, Anasua and Stevenson, Paul and de Franceschi, Silvano and Morello, Andrea and de Leon, Nathalie P. and Kuemmeth, Ferdinand},
 year = {2021},
 title = {Semiconductor qubits in practice},
 pages = {157--177},
 volume = {3},
 number = {3},
 journal = {Nat. Rev. Phys.},
doi={10.1038/s42254-021-00283-9}
}

@article{Chaum.1988,
 author = {Chaum, David},
 year = {1988},
 title = {The dining cryptographers problem: Unconditional sender and recipient untraceability},
 pages = {65--75},
 volume = {1},
 number = {1},
 issn = {0933-2790},
 journal = {J Cryptology},
 doi = {10.1007/BF00206326}
}

@article{Chen2003,
 author = {{Kai Chen} and {Ling-An Wu}},
 year = {2003},
 title = {A matrix realignment method for recognizing entanglement},
 pages = {193--202},
 volume = {3},
 number = {3},
 journal = {Quantum Inf. Comput.},
 doi = {10.26421/QIC3.3-1}
}

@article{chen2018mapping,
 author = {Chen, Yuan and Gao, Jun and Jiao, Zhi-Qiang and Sun, Ke and Shen, Wei-Guan and Qiao, Lu-Feng and Tang, Hao and Lin, Xiao-Feng and Jin, Xian-Min},
 year = {2018},
 title = {Mapping Twisted Light into and out of a Photonic Chip},
 pages = {233602},
 volume = {121},
 number = {23},
 issn = {0031-9007},
 journal = {Phys. Rev. Lett.},
 doi = {10.1103/PhysRevLett.121.233602}
}

@article{chen2020underwater,
 author = {Chen, Yuan and Shen, Wei-Guan and Li, Zhan-Ming and Hu, Cheng-Qiu and Yan, Zeng-Quan and Jiao, Zhi-Qiang and Gao, Jun and Cao, Ming-Ming and Sun, Ke and Jin, Xian-Min},
 year = {2020},
 title = {Underwater transmission of high-dimensional twisted photons over 55 meters},
 volume = {1},
 number = {1},
 journal = {PhotoniX},
 doi = {10.1186/s43074-020-0002-5}
}

@article{Chen.2021,
 author = {Chen, Jiu-Peng and Zhang, Chi and Liu, Yang and Jiang, Cong and Zhang, Wei-Jun and Han, Zhi-Yong and Ma, Shi-Zhao and Hu, Xiao-Long and Li, Yu-Huai and Liu, Hui and Zhou, Fei and Jiang, Hai-Feng and Chen, Teng-Yun and Li, Hao and You, Li-Xing and Wang, Zhen and Wang, Xiang-Bin and Zhang, Qiang and Pan, Jian-Wei},
 year = {2021},
 title = {Twin-field quantum key distribution over a 511 km optical fibre linking two distant metropolitan areas},
 pages = {570--575},
 volume = {15},
 number = {8},
 issn = {1749-4885},
 journal = {Nature Photonics},
 doi = {10.1038/s41566-021-00828-5}
}

@article{chiesa2020molecular,
 author = {Chiesa, Alessandro and Macaluso, Emilio and Petiziol, Francesco and Wimberger, Sandro and Santini, Paolo and Carretta, Stefano},
 year = {2020},
 title = {Molecular nanomagnets as qubits with embedded quantum-error correction},
 pages = {8610--8615},
 volume = {11},
 number = {20},
 journal = {J. Phys. Chem. Lett.},
 doi = {10.1021/acs.jpclett.0c02213}
}

@article{chiesa2023blueprint,
 author = {Chiesa, A. and Roca, S. and Chicco, S. and de Ory, M. C. and G{\'o}mez-Le{\'o}n, A. and Gomez, A. and Zueco, David and Luis, F. and Carretta, S.},
 year = {2023},
 title = {Blueprint for a molecular-spin quantum processor},
 pages = {064060},
 volume = {19},
 number = {6},
 journal = {Phys. Rev. Appl.},
 doi = {10.1103/PhysRevApplied.19.064060}
}

@article{chiesa2024molecular,
 author = {Chiesa, Alessandro and Santini, Paolo and Garlatti, Elena and Luis, Fernando and Carretta, Stefano},
 year = {2024},
 title = {Molecular nanomagnets: a viable path toward quantum information processing?},
 pages = {034501},
 volume = {87},
 number = {3},
 journal = {Reports on Progress in Physics},
 doi = {10.1088/1361-6633/ad1f81}
}

@article{Childs2005,
 author = {Childs, A. M.},
 year = {2005},
 title = {Secure assisted quantum computation},
 pages = {456--466},
 volume = {5},
 number = {6},
 issn = {15337146},
 journal = {Quantum Inf Comput},
 doi = {10.26421/qic5.6-4}
}

@article{choi2017self,
 author = {Choi, Hyeongrak and Heuck, Mikkel and Englund, Dirk},
 year = {2017},
 title = {Self-similar nanocavity design with ultrasmall mode volume for single-photon nonlinearities},
 pages = {223605},
 volume = {118},
 number = {22},
 issn = {0031-9007},
 journal = {Phys. Rev. Lett.},
 doi = {10.1103/PhysRevLett.118.223605}
}

@article{choquer2022quantum,
 author = {Choquer, Michael and Weis, Matthias and Nysten, Emeline D. S. and Lienhart, Michelle and Machnikowski, PaweL and Wigger, Daniel and Krenner, Hubert J. and Moody, Galan},
 year = {2022},
 title = {Quantum Control of Optically Active Artificial Atoms With Surface Acoustic Waves},
 pages = {1--17},
 volume = {3},
 journal = {IEEE Trans Quantum Eng},
 doi = {10.1109/TQE.2022.3204928}
}

@article{Christiandl_PRL_04,
 author = {Christandl, Matthias and Datta, Nilanjana and Ekert, Artur and Landahl, Andrew J.},
 year = {2004},
 title = {Perfect State Transfer in Quantum Spin Networks},
 url = {https://link.aps.org/doi/10.1103/PhysRevLett.92.187902      ,},
 pages = {187902},
 volume = {92},
 number = {18},
 journal = {Phys. Rev. Lett.},
 doi = {10.1103/PhysRevLett.92.187902}
}

@article{Cirac.1997,
 author = {Cirac, J. I. and Zoller, P. and Kimble, H. J. and Mabuchi, H.},
 year = {1997},
 title = {Quantum State Transfer and Entanglement Distribution among Distant Nodes in a Quantum Network},
 pages = {3221--3224},
 volume = {78},
 number = {16},
 issn = {0031-9007},
 journal = {Phys. Rev. Lett.},
 doi = {10.1103/PhysRevLett.78.3221}
}

@article{ciriano2021spin,
 author = {Ciriano-Tejel, Virginia N. and Fogarty, Michael A. and Schaal, Simon and Hutin, Louis and Bertrand, Benoit and Ibberson, Lisa and Gonzalez-Zalba, M. Fernando and Li, Jing and Niquet, Yann-Michel and Vinet, Maud and others},
 year = {2021},
 title = {Spin readout of a CMOS quantum dot by gate reflectometry and spin-dependent tunneling},
 pages = {010353},
 volume = {2},
 number = {1},
 journal = {PRX Quantum},
doi={10.1103/PRXQuantum.2.010353}
}

@article{clarke2008superconducting,
 author = {Clarke, John and Wilhelm, Frank K.},
 year = {2008},
 title = {Superconducting quantum bits},
 pages = {1031--1042},
 volume = {453},
 number = {7198},
 journal = {Nature},
doi={10.1038/nature07128}
}

@article{Cleve.1997,
 author = {Cleve, Richard and Buhrman, Harry},
 year = {1997},
 title = {Substituting quantum entanglement for communication},
 pages = {1201--1204},
 volume = {56},
 number = {2},
 issn = {2469-9934},
 journal = {Phys Rev A},
 doi = {10.1103/PhysRevA.56.1201}
}

@article{Cleve.1999,
 author = {Cleve, Richard and Gottesman, Daniel and Lo, Hoi-Kwong},
 year = {1999},
 title = {How to Share a Quantum Secret},
 pages = {648--651},
 volume = {83},
 number = {3},
 issn = {0031-9007},
 journal = {Phys Rev Lett},
 doi = {10.1103/PhysRevLett.83.648}
}

@article{Cojocaru2021,
 author = {Cojocaru, Alexandru and Colisson, L{\'e}o and Kashefi, Elham and Wallden, Petros},
 year = {2021},
 title = {On the Possibility of Classical Client Blind Quantum Computing},
 pages = {3},
 volume = {5},
 number = {1},
 journal = {Cryptography},
 doi = {10.3390/cryptography5010003}
}

@phdthesis{Consbruch2024,
 author = {von Consbruch, Alexander},
 year = {2024},
 title = {Blind quantum computing for gate-based architectures},
 school = {{Georg-August-Universit{\"a}t G{\"o}ttingen}},
 type = {Bachelor's Thesis}
}

@article{Cortese2018,
      title={{Loading Classical Data into a Quantum Computer}}, 
      author={John A. Cortese and Timothy M. Braje},
      year={2018},
      journal={arXiv},
      eprint={arXiv:1803.01958},
}

@book{Cover.2006,
 author = {Cover, Thomas M. and Thomas, Joy A.},
 year = {2006},
 title = {Elements of information theory},
 address = {Hoboken, NJ},
 edition = {Second edition},
 publisher = {Wiley-Interscience},
 isbn = {9780471241959},
 series = {A Wiley-Interscience publication},
 doi = {10.1002/047174882X}
}

@article{covey2023quantum,
 author = {Covey, Jacob P. and Weinfurter, Harald and Bernien, Hannes},
 year = {2023},
 title = {Quantum networks with neutral atom processing nodes},
 pages = {90},
 volume = {9},
 number = {1},
 journal = {npj Quantum Inf.},
doi={10.1038/s41534-023-00759-9}
}

@incollection{Crepeau.1995,
 author = {Cr{\'e}peau, Claude and Salvail, Louis},
 title = {Quantum Oblivious Mutual Identification},
 pages = {133--146},
 volume = {921},
 publisher = {Springer},
 isbn = {978-3-540-59409-3},
 series = {Lecture Notes in Computer Science},
 editor = {Guillou, Louis C.},
 booktitle = {Advances in cryptology - EUROCRYPT '95},
 year = {1995},
 address = {Berlin and Heidelberg},
url = {https://doi.org/10.1007/3-540-49264-X_11},
 doi = {10.1007/3-540-49264-X_11}
}

@article{Cuomo.2020,
 author = {Cuomo, Daniele and Caleffi, Marcello and Cacciapuoti, Angela Sara},
 year = {2020},
 title = {Towards a distributed quantum computing ecosystem},
 pages = {3--8},
 volume = {1},
 number = {1},
 issn = {2632-8925},
 journal = {IET Quantum Commun},
 doi = {10.1049/iet-qtc.2020.0002}
}

@article{Curty2001,
 author = {Curty, Marcos and Santos, David J.},
 year = {2001},
 title = {Quantum authentication of classical messages},
 volume = {64},
 number = {6},
 issn = {1050-2947},
 journal = {Phys Rev A: At Mol Opt Phys},
 doi = {10.1103/PhysRevA.64.062309}
}

@article{Curty2002,
 author = {Curty, Marcos and Santos, David J. and P{\'e}rez, Esther and Garc{\'i}a-Fern{\'a}ndez, Priscila},
 year = {2002},
 title = {Qubit authentication},
 volume = {66},
 number = {2},
 issn = {1050-2947},
 journal = {Phys Rev A: At Mol Opt Phys},
 doi = {10.1103/PhysRevA.66.022301}
}

@article{Daiss_Science_21,
author = {Severin Daiss  and Stefan Langenfeld  and Stephan Welte  and Emanuele Distante  and Philip Thomas  and Lukas Hartung  and Olivier Morin  and Gerhard Rempe },
 year = {2021},
 title = {A quantum-logic gate between distant quantum-network modules},
 pages = {614--617},
 volume = {371},
 number = {6529},
 journal = {Science},
 doi = {10.1126/science.abe3150}
}

@article{Dakna1997,
 author = {Dakna, Mohammed and Anhut, Tiemo and Opatrn\'y, T. and Kn{\"o}ll, Ludwig and Welsch, D-G},
 year = {1997},
 title = {Generating Schr{\"o}dinger-cat-like states by means of conditional measurements on a beam splitter},
 pages = {3184},
 volume = {55},
 number = {4},
 issn = {2469-9934},
 journal = {Phys. Rev. A},
 doi = {10.1103/PhysRevA.55.3184}
}

@article{delteil2016generation,
 author = {Delteil, Aymeric and Sun, Zhe and Gao, Wei-bo and Togan, Emre and Faelt, Stefan and Imamo{\u{g}}lu, Ata{\c{c}}},
 year = {2016},
 title = {Generation of heralded entanglement between distant hole spins},
 pages = {218--223},
 volume = {12},
 number = {3},
 issn = {1745-2473},
 journal = {Nat. Phys.},
doi={10.1038/nphys3605}
}

@article{delteil2017realization,
  title = {Realization of a Cascaded Quantum System: Heralded Absorption of a Single Photon Qubit by a Single-Electron Charged Quantum Dot},
  author = {Delteil, Aymeric and Sun, Zhe and F\"alt, Stefan and Imamo\ifmmode \breve{g}\else \u{g}\fi{}lu, Atac},
  journal = {Phys. Rev. Lett.},
  volume = {118},
  issue = {17},
  pages = {177401},
  numpages = {5},
  year = {2017},
  publisher = {American Physical Society},
  doi = {10.1103/PhysRevLett.118.177401}
}

@incollection{DeNardis.2007,
 author = {DeNardis, Laura},
 title = {A history of internet security},
 pages = {681--704},
 publisher = {Elsevier},
 isbn = {9780444516084},
 editor = {de Leeuw, Karl and Bergstra, Jan A.},
 booktitle = {The history of information security},
 year = {2007},
 address = {Amsterdam and Heidelberg},
url = {https://doi.org/10.1016/B978-044451608-4/50025-0},
 doi = {10.1016/B978-044451608-4/50025-0}
}

@article{Denchev.2008,
 author = {Denchev, Vasil S. and Pandurangan, Gopal},
 year = {2008},
 title = {Distributed quantum computing},
 pages = {77--95},
 volume = {39},
 number = {3},
 issn = {0163-5700},
 journal = {ACM SIGACT News},
 doi = {10.1145/1412700.1412718}
}

@article{Deng.2003,
 author = {Deng, Fu-Guo and Long, Gui Lu and Liu, Xiao-Shu},
 year = {2003},
 title = {Two-step quantum direct communication protocol using the Einstein-Podolsky-Rosen pair block},
 volume = {68},
 number = {4},
 issn = {1050-2947},
 journal = {Phys. Rev. A},
 doi = {10.1103/PhysRevA.68.042317}
}

@article{Devetak2005,
 author = {Devetak, Igor and Winter, Andreas},
 year = {2005},
 title = {{Distillation of secret key and entanglement from quantum states}},
 pages = {207--235},
 volume = {461},
 number = {2053},
 issn = {1364-5021},
 journal = {Proc. R. Soc. London, Ser. A},
 doi = {10.1098/rspa.2004.1372}
}

@incollection{devoret2004superconducting,
 author = {Devoret, Michel H. and Martinis, John M.},
 title = {Superconducting qubits},
editor = {Daniel Estève and Jean-Michel Raimond and Jean Dalibard},
series = {Les Houches},
publisher = {Elsevier},
volume = {79},
pages = {443-485},
year = {2004},
booktitle = {Quantum Entanglement and Information Processing},
issn = {0924-8099},
doi = {https://doi.org/10.1016/S0924-8099(03)80036-7}
}

@article{DiAdamo.2021,
 author = {DiAdamo, Stephen and Ghibaudi, Marco and Cruise, James},
 year = {2021},
 title = {Distributed Quantum Computing and Network Control for Accelerated VQE},
 pages = {1--21},
 volume = {2},
 journal = {IEEE Trans Quantum Eng},
 doi = {10.1109/TQE.2021.3057908}
}

@incollection{Divincenzo.1997,
 author = {Divincenzo, D. P.},
 title = {Topics in Quantum Computers},
 pages = {657--677},
 publisher = {Springer},
 isbn = {978-90-481-4906-3},
 series = {NATO ASI Series, Series E},
 editor = {Sohn, Lydia L. and Kouwenhoven, Leo P. and Sch{\"o}n, Gerd},
 booktitle = {Mesoscopic Electron Transport},
 year = {1997},
 address = {Dordrecht},
 doi = {10.1007/978-94-015-8839-3_18}
}

@article{Djordjevic.2020,
 author = {Djordjevic, Ivan B.},
 year = {2020},
 title = {On Global Quantum Communication Networking},
 volume = {22},
 number = {8},
 journal = {Entropy (Basel, Switzerland)},
 doi = {10.3390/e22080831}
}

@article{Doescher.2002,
author = {D\"{o}scher, C. and Keyl, M.},
title = {{An introduction to quantum coin tossing}},
journal = {Fluctuation Noise Lett},
volume = {02},
number = {04},
pages = {R125-R137},
year = {2002},
doi = {10.1142/S0219477502000944},
URL = {https://doi.org/10.1142/S0219477502000944}
}

@article{doherty2013nitrogen,
 author = {Doherty, Marcus W. and Manson, Neil B. and Delaney, Paul and Jelezko, Fedor and Wrachtrup, J{\"o}rg and Hollenberg, Lloyd C. L.},
 year = {2013},
 title = {The nitrogen-vacancy colour centre in diamond},
 pages = {1--45},
 volume = {528},
 number = {1},
 journal = {Phys. Rep.},
 doi = {10.1016/j.physrep.2013.02.001}
}

@article{Dowling.2003,
 author = {Dowling, Jonathan P. and Milburn, Gerard J.},
 year = {2003},
 title = {Quantum technology: the second quantum revolution},
 pages = {1655--1674},
 volume = {361},
 number = {1809},
 issn = {1364-503X},
 journal = {Philos. Trans. R. Soc. A},
 doi = {10.1098/rsta.2003.1227}
}

@article{Du.2024,
title = {{Entanglement distribution over 155 km metropolitan fiber using a CMOS-compatible silicon chip}},
journal = {Newton},
volume = {1},
number = {10},
pages = {100303},
year = {2025},
issn = {2950-6360},
doi = {https://doi.org/10.1016/j.newton.2025.100303},
url = {https://www.sciencedirect.com/science/article/pii/S2950636025002956},
author = {Jinyi Du and Xingjian Zhang and George F.R. Chen and Hongwei Gao and Dawn T.H. Tan and Alexander Ling},
}

@incollection{Duarte2019,
 author = {Duarte, Denio and St{\aa}hl, Niclas},
 title = {Machine Learning: A Concise Overview},
 pages = {27--58},
 publisher = {{Springer International Publishing}},
 isbn = {978-3-319-97556-6},
 editor = {Said, Alan and Torra, Vicen{\c{c}}},
 booktitle = {Data Science in Practice},
 year = {2019},
 address = {Cham},
 doi = {10.1007/978-3-319-97556-6_3}
}

@article{dubois2013minimal,
 author = {Dubois, J. and Jullien, T. and Portier, F. and Roche, P. and Cavanna, A. and Jin, Y. and Wegscheider, W. and Roulleau, P. and Glattli, D. C.},
 year = {2013},
 title = {Minimal-excitation states for electron quantum optics using levitons},
 pages = {659--663},
 volume = {502},
 number = {7473},
 journal = {Nature},
 doi = {10.1038/nature12713}
}

@article{Dumur_NPJQI_21,
 author = {Dumur, {\'E}. and Satzinger, K. J. and Peairs, G. A. and Chou, M.-H. and Bienfait, A. and Chang, H.-S. and Conner, C. R. and Grebel, J. and Povey, R. G. and Zhong, Y. P. and Cleland, A. N.},
 year = {2021},
 title = {Quantum communication with itinerant surface acoustic wave phonons},
 pages = {173},
 volume = {7},
 number = {1},
 journal = {npj Quantum Inf.},
 doi = {10.1038/s41534-021-00511-1}
}

@article{Dur1999,
 author = {D{\"u}r, W. and Briegel, H.-J. and Cirac, J. I. and Zoller, P.},
 year = {1999},
 title = {Quantum repeaters based on entanglement purification},
 pages = {169--181},
 volume = {59},
 number = {1},
 issn = {1050-2947},
 journal = {Phys. Rev. A},
 doi = {10.1103/PhysRevA.59.169}
}

@article{Dur2017,
 author = {D{\"u}r, Wolfgang and Lamprecht, Raphael and Heusler, Stefan},
 year = {2017},
 title = {Towards a quantum internet},
 pages = {043001},
 volume = {38},
 number = {4},
 issn = {0143-0807},
 journal = {Eur. J. Phys.},
 doi = {10.1088/1361-6404/aa6df7}
}

@article{Dusek.1999,
 author = {Du{\v{s}}ek, Miloslav and Haderka, Ond{\v{r}}ej and Hendrych, Martin and My{\v{s}}ka, Robert},
 year = {1999},
 title = {Quantum identification system},
 pages = {149--156},
 volume = {60},
 number = {1},
 issn = {1050-2947},
 journal = {Phys Rev A: At Mol Opt Phys},
 doi = {10.1103/PhysRevA.60.149}
}

@article{Dutta2022,
 author = {Dutta, Arindam and Pathak, Anirban},
 year = {2022},
 title = {A short review on quantum identity authentication protocols: how would Bob know that he is talking with Alice?},
 volume = {21},
 number = {11},
 issn = {1570-0755},
 journal = {Quantum Inf Process},
 doi = {10.1007/s11128-022-03717-0}
}

@article{Edlbauer_EPJ_22,
 author = {Edlbauer, Hermann and Wang, Junliang and Crozes, Thierry and Perrier, Pierre and Ouacel, Seddik and Geffroy, Cl{\'e}ment and Georgiou, Giorgos and Chatzikyriakou, Eleni and Lacerda-Santos, Antonio and Waintal, Xavier and Glattli, D. Christian and Roulleau, Preden and Nath, Jayshankar and Kataoka, Masaya and Splettstoesser, Janine and Acciai, Matteo and {Da Silva Figueira}, Maria Cecilia and {\"O}ztas, Kemal and Trellakis, Alex and Grange, Thomas and Yevtushenko, Oleg M. and Birner, Stefan and B{\"a}uerle, Christopher},
 year = {2022},
 title = {Semiconductor-based electron flying qubits: review on recent progress accelerated by numerical modelling},
 pages = {21},
 volume = {9},
 number = {1},
 journal = {EPJ Quantum Technol.},
 doi = {10.1140/epjqt/s40507-022-00139-w}
}

@article{Ekert.1991,
 author = {Ekert, A. K.},
 year = {1991},
 title = {Quantum cryptography based on Bell's theorem},
 pages = {661--663},
 volume = {67},
 number = {6},
 issn = {0031-9007},
 journal = {Phys Rev Lett},
 doi = {10.1103/PhysRevLett.67.661}
}

@article{elzerman2004single,
 author = {Elzerman, J. M. and Hanson, R. and {van Willems Beveren}, L. H. and Witkamp, B. and Vandersypen, L. M.K. and Kouwenhoven, Leo P.},
 year = {2004},
 title = {Single-shot read-out of an individual electron spin in a quantum dot},
 pages = {431--435},
 volume = {430},
 number = {6998},
 journal = {Nature},
doi={10.1103/PhysRevLett.94.196802}
}

@article{erhard2020advances,
 author = {Erhard, Manuel and Krenn, Mario and Zeilinger, Anton},
 year = {2020},
 title = {Advances in high-dimensional quantum entanglement},
 pages = {365--381},
 volume = {2},
 number = {7},
 journal = {Nat. Rev. Phys.},
 doi = {10.1038/s42254-020-0193-5}
}

@article{essiambre2012capacity,
 author = {Essiambre, Ren{\'e}-Jean and Tkach, Robert W.},
 year = {2012},
 title = {Capacity Trends and Limits of Optical Communication Networks},
 pages = {1035--1055},
 volume = {100},
 number = {5},
 issn = {0018-9219},
 journal = {Proc. IEEE},
 doi = {10.1109/JPROC.2012.2182970}
}

@article{fan2018superconducting,
author = {Linran Fan  and Chang-Ling Zou  and Risheng Cheng  and Xiang Guo  and Xu Han  and Zheng Gong  and Sihao Wang  and Hong X. Tang },
title = {{Superconducting cavity electro-optics: A platform for coherent photon conversion between superconducting and photonic circuits}},
journal = {Sci. Adv.},
volume = {4},
number = {8},
pages = {eaar4994},
year = {2018},
doi = {10.1126/sciadv.aar4994},
URL = {https://www.science.org/doi/abs/10.1126/sciadv.aar4994},
eprint = {https://www.science.org/doi/pdf/10.1126/sciadv.aar4994},
}

@book{fano1961transmission,
	author = {Fano, Robert M.},
	year = {1961},
	title = {{Transmission of Information: A Statistical Theory of Communication}},
	publisher = {MIT Press, New York and John Wiley},
    address = {London}
}

@inproceedings{Farhi.2012,
author = {Farhi, Edward and Gosset, David and Hassidim, Avinatan and Lutomirski, Andrew and Shor, Peter},
title = {Quantum money from knots},
year = {2012},
isbn = {9781450311151},
publisher = {Association for Computing Machinery},
address = {New York, NY, USA},
url = {https://doi.org/10.1145/2090236.2090260},
doi = {10.1145/2090236.2090260},
booktitle = {Proceedings of the 3rd Innovations in Theoretical Computer Science Conference},
pages = {276–289},
numpages = {14},
location = {Cambridge, Massachusetts},
series = {ITCS '12}
}

@article{fedorov2021experimental,
 author = {Fedorov, Kirill G. and Renger, Michael and Pogorzalek, Stefan and {Di Candia}, Roberto and Chen, Qiming and Nojiri, Yuki and Inomata, Kunihiro and Nakamura, Yasunobu and Partanen, Matti and Marx, Achim and Gross, Rudolf and Deppe, Frank},
 year = {2021},
 title = {Experimental quantum teleportation of propagating microwaves},
 pages = {eabk0891},
 volume = {7},
 number = {52},
 journal = {Sci. Adv.},
 doi = {10.1126/sciadv.abk0891}
}

@phdthesis{FelicitasBinder.03092024,
 author = {Binder, Felicitas},
 year = {03/09/2024},
 title = {Distributed quantum algorithms via single quantum links},
 address = {Ulm},
 school = {{Universit{\"a}t Ulm}},
 type = {Bachelor thesis}
}

@article{feng2021experimental,
 author = {Feng, Zhao and Li, Shangbin and Xu, Zhengyuan},
 year = {2021},
 title = {Experimental underwater quantum key distribution},
 pages = {8725--8736},
 volume = {29},
 number = {6},
 journal = {Opt. Express},
 doi = {10.1364/OE.418323}
}

@article{fernandez2019cavity,
  title = {{Cavity-enhanced Raman heterodyne spectroscopy in ${\mathrm{Er}}^{3+}:{\mathrm{Y}}_{2}{\mathrm{SiO}}_{5}$ for microwave to optical signal conversion}},
  author = {Fernandez-Gonzalvo, Xavier and Horvath, Sebastian P. and Chen, Yu-Hui and Longdell, Jevon J.},
  journal = {Phys. Rev. A},
  volume = {100},
  issue = {3},
  pages = {033807},
  numpages = {9},
  year = {2019},
  publisher = {American Physical Society},
  doi = {10.1103/PhysRevA.100.033807},
  url = {https://link.aps.org/doi/10.1103/PhysRevA.100.033807}
}

@article{ferrando2016modular,
 author = {Ferrando-Soria, Jes{\'u}s and {Moreno Pineda}, Eufemio and Chiesa, Alessandro and Fernandez, Antonio and Magee, Samantha A. and Carretta, Stefano and Santini, Paolo and Vitorica-Yrezabal, Inigo J. and Tuna, Floriana and Timco, Grigore A. and others},
 year = {2016},
 title = {A modular design of molecular qubits to implement universal quantum gates},
 pages = {11377},
 volume = {7},
 number = {1},
 journal = {Nat. Commun.},
 doi = {10.1038/ncomms11377}
}

@article{Ferrari.2021,
 author = {Ferrari, Davide and Cacciapuoti, Angela Sara and Amoretti, Michele and Caleffi, Marcello},
 year = {2021},
 title = {Compiler Design for Distributed Quantum Computing},
 pages = {1--20},
 volume = {2},
 journal = {IEEE Trans Quantum Eng},
 doi = {10.1109/TQE.2021.3053921}
}

@article{fesquet2023perspectives,
 author = {Fesquet, F. and Kronowetter, F. and Renger, M. and Chen, Q. and Honasoge, K. and Gargiulo, O. and Nojiri, Y. and Marx, A. and Deppe, F. and Gross, R. and Fedorov, K. G.},
 year = {2023},
 title = {Perspectives of microwave quantum key distribution in the open air},
 volume = {108},
 number = {3},
 issn = {1050-2947},
 journal = {Phys. Rev. A},
 doi = {10.1103/PhysRevA.108.032607}
}

@article{fesquet2024demonstration,
 author = {Fesquet, Florian and Kronowetter, Fabian and Renger, Michael and Yam, Wun Kwan and Gandorfer, Simon and Inomata, Kunihiro and Nakamura, Yasunobu and Marx, Achim and Gross, Rudolf and Fedorov, Kirill G.},
 year = {2024},
 title = {Demonstration of microwave single-shot quantum key distribution},
 pages = {7544},
 volume = {15},
 number = {1},
 journal = {Nat. Commun.},
 doi = {10.1038/s41467-024-51421-7}
}

@article{Fisher2014,
 author = {Fisher, K. A. G. and Broadbent, A. and Shalm, L. K. and Yan, Z. and Lavoie, J. and Prevedel, R. and Jennewein, T. and Resch, K. J.},
 year = {2014},
 title = {Quantum computing on encrypted data},
 pages = {3074},
 volume = {5},
 journal = {Nat. Commun.},
 doi = {10.1038/ncomms4074}
}

@article{Fitzsimons2017,
 author = {Fitzsimons, Joseph F.},
 year = {2017},
 title = {Private quantum computation: an introduction to blind quantum computing and related protocols},
 volume = {3},
 number = {1},
 journal = {npj Quantum Inf},
 doi = {10.1038/s41534-017-0025-3}
}

@article{Fitzsimons2017b,
 author = {Fitzsimons, Joseph F. and Kashefi, Elham},
 year = {2017},
 title = {Unconditionally verifiable blind quantum computation},
 volume = {96},
 number = {1},
 issn = {1050-2947},
 journal = {Phys Rev A: At Mol Opt Phys},
 doi = {10.1103/PhysRevA.96.012303}
}

@article{flensberg2021engineered,
 author = {Flensberg, Karsten and von Oppen, Felix and Stern, Ady},
 year = {2021},
 title = {Engineered platforms for topological superconductivity and Majorana zero modes},
 pages = {944--958},
 volume = {6},
 number = {10},
 journal = {Nature Reviews Materials},
 doi = {10.1038/s41578-021-00336-6}
}

@article{fletcher2013clock,
 author = {Fletcher, J. D. and See, P. and Howe, H. and Pepper, M. and Giblin, S. P. and Griffiths, J. P. and Jones, G. A. C. and Farrer, I. and Ritchie, D. A. and Janssen, T. J. B. M. and Kataoka, M.},
 year = {2013},
 title = {Clock-controlled emission of single-electron wave packets in a solid-state circuit},
 pages = {216807},
 volume = {111},
 number = {21},
 issn = {0031-9007},
 journal = {Phys. Rev. Lett.},
 doi = {10.1103/PhysRevLett.111.216807}
}

@article{freer2017single,
 author = {Freer, Solomon and Simmons, Stephanie and Laucht, Arne and Muhonen, Juha T. and Dehollain, Juan P. and Kalra, Rachpon and Mohiyaddin, Fahd A. and Hudson, Fay E. and Itoh, Kohei M. and McCallum, Jeffrey C. and others},
 year = {2017},
 title = {A single-atom quantum memory in silicon},
 pages = {015009},
 volume = {2},
 number = {1},
 journal = {Quantum Science and Technology},
 doi = {10.1088/2058-9565/aa63a4}
}

@article{Ganz.2017,
 author = {Ganz, Maor},
 year = {2017},
 title = {Quantum leader election},
 volume = {16},
 number = {3},
 pages ={73},
 issn = {1570-0755},
 journal = {Quantum Inf. Process.},
 doi = {10.1007/s11128-017-1528-8}
}

@article{gao2024advances,
doi = {10.1088/1742-6596/2809/1/012028},
year = {2024},
publisher = {IOP Publishing},
volume = {2809},
number = {1},
pages = {012028},
author = {Gao, Zhuoqing and Amaratunga, Gehan and Wang, Xiaozhi and Ma, Boyang},
title = {Advances in Electron-Based Qubits: A Review},
journal = {J. Phys.: Conf. Ser.}
}

@inproceedings{Gavinsky2012,
 author = {Gavinsky, Dmitry},
 title = {Quantum Money with Classical Verification},
 pages = {42--52},
 publisher = {IEEE},
 isbn = {978-0-7695-4708-4},
 booktitle = {2012 IEEE 27th Conference on Computational Complexity},
 year = {2012},
url ={https://doi.org/10.1109/CCC.2012.10},
 doi = {10.1109/CCC.2012.10}
}

@inproceedings{gentry2009,
author = {Gentry, Craig},
title = {Fully homomorphic encryption using ideal lattices},
year = {2009},
isbn = {9781605585062},
publisher = {Association for Computing Machinery},
address = {New York, NY, USA},
url = {https://doi.org/10.1145/1536414.1536440},
doi = {10.1145/1536414.1536440},
booktitle = {Proceedings of the Forty-First Annual ACM Symposium on Theory of Computing},
pages = {169–178},
numpages = {10},
location = {Bethesda, MD, USA},
series = {STOC '09}
}

@article{ghalaii2022quantum,
 author = {Ghalaii, Masoud and Pirandola, Stefano},
 year = {2022},
 title = {Quantum communications in a moderate-to-strong turbulent space},
 volume = {5},
 number = {1},
 journal = {Commun Phys},
 doi = {10.1038/s42005-022-00814-5}
}

@article{Gidney.2021,
 author = {Gidney, Craig and Eker{\aa}, Martin},
 year = {2021},
 title = {How to factor 2048 bit RSA integers in 8 hours using 20 million noisy qubits},
 pages = {433},
 volume = {5},
 issn = {2521-327X},
 journal = {Quantum},
 doi = {10.22331/q-2021-04-15-433}
}

@article{gobby2004quantum,
 author = {Gobby, C. and Yuan, Z. L. and Shields, A. J.},
 year = {2004},
 title = {Quantum key distribution over 122 km of standard telecom fiber},
 pages = {3762--3764},
 volume = {84},
 number = {19},
 issn = {0003-6951},
 journal = {Appl. Phys. Lett.},
 doi = {10.1063/1.1738173}
}

@article{godfrin2017operating,
 author = {Godfrin, Cl{\'e}ment and Ferhat, Abdelkarim and Ballou, Rafik and Klyatskaya, Svetlana and Ruben, Mario and Wernsdorfer, Wolfgang and Balestro, Franck},
 year = {2017},
 title = {Operating quantum states in single magnetic molecules: implementation of Grover's quantum algorithm},
 pages = {187702},
 volume = {119},
 number = {18},
 issn = {0031-9007},
 journal = {Phys. Rev. Lett.},
 doi = {10.1103/PhysRevLett.119.187702}
}

@book{Goldwasser2012,
 author = {Goldwasser, Shafi},
 year = {2012},
 title = {Proceedings of the 3rd Innovations in Theoretical Computer Science Conference},
 url = {http://dl.acm.org/citation.cfm?id=2090236},
 address = {New York, NY},
 publisher = {ACM},
 isbn = {9781450311151},
 series = {ACM Conferences},
 institution = {{Association for Computing Machinery and ACM Special Interest Group on Algorithms and Computation Theory}},
 doi = {10.1145/2090236}
}

@article{gonzalez2022coplanar,
 author = {Gonzalez-Raya, Tasio and Sanz, Mikel},
 year = {2022},
 title = {Coplanar Antenna Design for Microwave Entangled Signals Propagating in Open Air},
 pages = {783},
 volume = {6},
 journal = {Quantum},
 doi = {10.22331/q-2022-08-23-783}
}

@article{gonzalez2022open,
 author = {Gonzalez-Raya, Tasio and Casariego, Mateo and Fesquet, Florian and Renger, Michael and Salari, Vahid and M{\"o}tt{\"o}nen, Mikko and Omar, Yasser and Deppe, Frank and Fedorov, Kirill G. and Sanz, Mikel},
 year = {2022},
 title = {Open-Air Microwave Entanglement Distribution for Quantum Teleportation},
 volume = {18},
 number = {4},
 journal = {Phys. Rev. Appl.},
 pages = {044002},
 numpages = {30},
 publisher = {American Physical Society},
 doi = {10.1103/PhysRevApplied.18.044002}
}

@article{gonzalez2024wireless,
author = {Gonzalez-Raya, Tasio},
journal= {arXiv},
year= {2024},
 title = {Wireless Microwave Quantum Communication},
eprint={arXiv:2401.08708}
}

@article{GonzalezConde2024,
 author = {Gonzalez-Conde, Javier and Watts, Thomas W. and Rodriguez-Grasa, Pablo and Sanz, Mikel},
 year = {2024},
 title = {Efficient quantum amplitude encoding of polynomial functions},
 pages = {1297},
 volume = {8},
 issn = {2521-327X},
 journal = {Quantum},
 doi = {10.22331/q-2024-03-21-1297}
}

@article{Goorden2014,
 author = {Goorden, Sebastianus A. and Horstmann, Marcel and Mosk, Allard P. and {\v{S}}kori{\'c}, Boris and Pinkse, Pepijn W. H.},
 year = {2014},
 title = {Quantum-secure authentication of a physical unclonable key},
 pages = {421},
 volume = {1},
 number = {6},
 journal = {Optica},
 doi = {10.1364/OPTICA.1.000421}
}

@article{graham2022multi,
 author = {Graham, T. M. and Song, Y. and Scott, J. and Poole, C. and Phuttitarn, L. and Jooya, K. and Eichler, P. and Jiang, X. and Marra, A. and Grinkemeyer, B. and others},
 year = {2022},
 title = {Multi-qubit entanglement and algorithms on a neutral-atom quantum computer},
 pages = {457--462},
 volume = {604},
 number = {7906},
 journal = {Nature},
doi={10.1038/s41586-022-04603-6}
}

@article{Grankin_PRA_18,
 author = {Grankin, A. and Guimond, P. O. and Vasilyev, D. V. and Vermersch, B. and Zoller, P.},
 year = {2018},
 title = {Free-space photonic quantum link and chiral quantum optics},
 url = {https://link.aps.org/doi/10.1103/PhysRevA.98.043825               ,},
 pages = {043825},
 volume = {98},
 number = {4},
 journal = {Phys. Rev. A},
 doi = {10.1103/PhysRevA.98.043825}
}

@article{Grimm2020,
 author = {Grimm, Alexander and Frattini, Nicholas E. and Puri, Shruti and Mundhada, Shantanu O. and Touzard, Steven and Mirrahimi, Mazyar and Girvin, Steven M. and Shankar, Shyam and Devoret, Michel H.},
 year = {2020},
 title = {Stabilization and operation of a Kerr-cat qubit},
 pages = {205--209},
 volume = {584},
 number = {7820},
 journal = {Nature},
doi = {10.1038/s41586-020-2587-z}
}

@article{gruber1997scanning,
 author = {Gruber, A. and Drabenstedt, A. and Tietz, C. and Fleury, L. and Wrachtrup, Joerg and von Borczyskowski, C.},
 year = {1997},
 title = {Scanning Confocal Optical Microscopy and Magnetic Resonance on Single Defect Centers},
 pages = {2012--2014},
 volume = {276},
 number = {5321},
 journal = {Science},
doi={10.1126/science.276.5321.2012}
}

@article{gu2017microwave,
 author = {Gu, Xiu and Kockum, Anton Frisk and Miranowicz, Adam and Liu, Yu-xi and Nori, Franco},
 year = {2017},
 title = {Microwave photonics with superconducting quantum circuits},
 pages = {1--102},
 volume = {718},
 journal = {Phys. Rep.},
doi={10.1016/j.physrep.2017.10.002}
}

@article{Gustiani2021,
 author = {Gustiani, Cica and DiVincenzo, David P.},
 year = {2021},
 title = {Blind oracular quantum computation},
 pages = {045022},
 volume = {6},
 number = {4},
 journal = {Quantum Sci. Technol.},
 doi = {10.1088/2058-9565/ac13c8}
}

@article{Gustiani2021b,
 author = {Gustiani, Cica and DiVincenzo, David P.},
 year = {2021},
 title = {Blind three-qubit exact Grover search on a nitrogen-vacancy-center platform},
 volume = {104},
 number = {6},
 issn = {1050-2947},
 journal = {Phys. Rev. A},
 doi = {10.1103/PhysRevA.104.062422}
}

@article{Gyongyosi2022,
 author = {Gyongyosi, Laszlo and Imre, Sandor},
 year = {2022},
 title = {Advances in the quantum internet},
 pages = {52--63},
 volume = {65},
 number = {8},
 issn = {0001-0782},
 journal = {Commun. ACM},
 doi = {10.1145/3524455}
}

@article{hacker2019deterministic,
 author = {Hacker, Bastian and Welte, Stephan and Daiss, Severin and Shaukat, Armin and Ritter, Stephan and Li, Lin and Rempe, Gerhard},
 year = {2019},
 title = {Deterministic creation of entangled atom--light Schr{\"o}dinger-cat states},
 pages = {110--115},
 volume = {13},
 number = {2},
 issn = {1749-4885},
 journal = {Nat. Photonics},
 doi = {10.1038/s41566-018-0339-5}
}

@article{haffner2008quantum,
title = {{Quantum computing with trapped ions}},
journal = {Phys. Rep.},
volume = {469},
number = {4},
pages = {155-203},
year = {2008},
issn = {0370-1573},
doi = {10.1016/j.physrep.2008.09.003},
url = {https://www.sciencedirect.com/science/article/pii/S0370157308003463},
author = {H. Häffner and C.F. Roos and R. Blatt},
}

@article{Hanafi.2025,
 author = {Hanafi, Basil and Ali, Mohammad},
 year = {2025},
 title = {Analyzing the research impact in post quantum cryptography through scientometric evaluation},
 volume = {28},
 number = {1},
 pages ={32},
 journal = {Discov. Comput.},
 doi = {10.1007/s10791-025-09507-3}
}

@inproceedings{Haner.2021,
 author = {H{\"a}ner, Thomas and Steiger, Damian S. and Hoefler, Torsten and Troyer, Matthias},
 title = {Distributed quantum computing with QMPI},
 pages = {1--13},
 publisher = {IEEE},
 isbn = {9781450384421},
 editor = {de Supinski, Bronis R. and Hall, Mary and Gamblin, Todd},
 booktitle = {Proceedings of SC21: the International Conference for High Performance Computing, Networking, Storage and Analysis: St. Louis, Missouri, November 14-19, 2021},
 year = {2021},
 address = {Piscataway, NJ},
 doi = {10.1145/3458817.3476172}
}

@article{Harrow2009,
 author = {Harrow, Aram W. and Hassidim, Avinatan and Lloyd, Seth},
 year = {2009},
 title = {Quantum Algorithm for Linear Systems of Equations},
 url = {https://link.aps.org/doi/10.1103/PhysRevLett.103.150502}   ,
 pages = {150502},
 volume = {103},
 number = {15},
 journal = {Phys. Rev. Lett.},
 doi = {10.1103/PhysRevLett.103.150502}
}

@article{harty2014high,
 author = {Harty, T. P. and Allcock, D. T.C. and Ballance, C. J̃ and Guidoni, L. and Janacek, H. A. and Linke, N. M. and Stacey, D. N. and Lucas, D. M.},
 year = {2014},
 title = {High-fidelity preparation, gates, memory, and readout of a trapped-ion quantum bit},
 pages = {220501},
 volume = {113},
 number = {22},
 issn = {0031-9007},
 journal = {Phys. Rev. Lett.},
 doi = {10.1103/PhysRevLett.113.220501},
}

@article{Hastrup2020,
 author = {Hastrup, Jacob and Neergaard-Nielsen, Jonas Schou and Andersen, Ulrik Lund},
 year = {2020},
 title = {Deterministic generation of a four-component optical cat state},
 pages = {640--643},
 volume = {45},
 number = {3},
 journal = {Opt. Lett.},
 doi = {10.1364/OL.383194}
}

@article{hayashi2003coherent,
 author = {Hayashi, Toshiki and Fujisawa, Toshimasa and Cheong, Hai-Du and Jeong, Yoon Hee and Hirayama, Yoshiro},
 year = {2003},
 title = {Coherent manipulation of electronic states in a double quantum dot},
 pages = {226804},
 volume = {91},
 number = {22},
 issn = {0031-9007},
 journal = {Phys. Rev. Lett.},
doi={10.1103/PhysRevLett.91.226804}
}

@article{Hayashi2015,
 author = {Hayashi, Masahito and Morimae, Tomoyuki},
 year = {2015},
 title = {Verifiable Measurement-Only Blind Quantum Computing with Stabilizer Testing},
 doi = {10.1103/PhysRevLett.115.220502},
  journal = {Phys Rev Lett},
  volume = {115},
  issue = {22},
  pages = {220502},
  numpages = {5},
  publisher = {American Physical Society},
  url = {https://link.aps.org/doi/10.1103/PhysRevLett.115.220502}

}

@article{Hayden2001,
 author = {Hayden, Patrick M. and Horodecki, Michal and Terhal, Barbara M.},
 year = {2001},
 title = {{The asymptotic entanglement cost of preparing a quantum state}},
 pages = {6891--6898},
 volume = {34},
 number = {35},
 issn = {0305-4470},
 journal = {J. Phys. A: Math. Gen.},
 doi = {10.1088/0305-4470/34/35/314}
}

@article{he2019two,
 author = {He, Yu and Gorman, S. K. and Keith, Daniel and Kranz, Ludwik and Keizer, J. G. and Simmons, M. Y.},
 year = {2019},
 title = {A two-qubit gate between phosphorus donor electrons in silicon},
 pages = {371--375},
 volume = {571},
 number = {7765},
 journal = {Nature},
 doi = {10.1038/s41586-019-1381-2}
}

@article{henriet2020quantum,
 author = {Henriet, Lo\{\textquotedbl}$\backslash$ic and Beguin, Lucas and Signoles, Adrien and Lahaye, Thierry and Browaeys, Antoine and Reymond, Georges-Olivier and Jurczak, Christophe}

@article{hensen2015loophole,
 author = {Hensen, Bas and Bernien, Hannes and Dr{\'e}au, Ana\{\textquotedbl}$\backslash$is E. and Reiserer, Andreas and Kalb, Norbert and Blok, Machiel S. and Ruitenberg, Just and Vermeulen, Raymond F. L. and Schouten, Raymond N. and Abell{\'a}n, Carlos and others}

@article{Hermans2022,
 author = {Hermans, S. L. N. and Pompili, M. and Beukers, H. K. C. and Baier, S. and Borregaard, J. and Hanson, R.},
 year = {2022},
 title = {Qubit teleportation between non-neighbouring nodes in a quantum network},
 pages = {663--668},
 volume = {605},
 number = {7911},
 journal = {Nature},
 doi = {10.1038/s41586-022-04697-y}
}

@article{Hermelin_Nature_11,
 author = {Hermelin, Sylvain and Takada, Shintaro and Yamamoto, Michihisa and Tarucha, Seigo and Wieck, Andreas D. and Saminadayar, Laurent and B{\"a}uerle, Christopher and Meunier, Tristan},
 year = {2011},
 title = {Electrons surfing on a sound wave as a platform for quantum optics with flying electrons},
 pages = {435--438},
 volume = {477},
 number = {7365},
 journal = {Nature},
 doi = {10.1038/nature10416}
}

@article{Heshami.2016,
 author = {Heshami, Khabat and England, Duncan G. and Humphreys, Peter C. and Bustard, Philip J. and Acosta, Victor M. and Nunn, Joshua and Sussman, Benjamin J.},
 year = {2016},
 title = {Quantum memories: emerging applications and recent advances},
 pages = {2005--2028},
 volume = {63},
 number = {20},
 issn = {0950-0340},
 journal = {J. Mod. Opt.},
 doi = {10.1080/09500340.2016.1148212}
}

@article{heshami2016quantum,
 author = {Heshami, Khabat and England, Duncan G. and Humphreys, Peter C. and Bustard, Philip J. and Acosta, Victor M. and Nunn, Joshua and Sussman, Benjamin J.},
 year = {2016},
 title = {Quantum memories: emerging applications and recent advances},
 pages = {2005--2028},
 volume = {63},
 number = {20},
 issn = {0950-0340},
 journal = {Journal of modern optics},
 doi = {10.1080/09500340.2016.1148212}
}

@article{Hillery.1999,
 author = {Hillery, Mark and Bu{\v{z}}ek, Vladim{\'i}r and Berthiaume, Andr{\'e}},
 year = {1999},
 title = {Quantum secret sharing},
 pages = {1829--1834},
 volume = {59},
 number = {3},
 issn = {1050-2947},
 journal = {Phys Rev A: At Mol Opt Phys},
 doi = {10.1103/PhysRevA.59.1829}
}

@article{Hillery2006,
 author = {Hillery, Mark},
 year = {2006},
 title = {Quantum voting and privacy protection: first steps},
 journal = {SPIE Newsroom},
 doi = {10.1117/2.1200610.0419}
}

@article{Hillery2006b,
 author = {Mark Hillery and M{\'a}rio Ziman and Vladim$\backslash$'{\i}r Bu{\v{z}}ek and Martina Bielikov{\'a}},
 year = {2006},
 title = {Towards quantum-based privacy and voting},
 url = {https://www.sciencedirect.com/science/article/pii/S0375960105014738},
 pages = {75--81},
 volume = {349},
 number = {1},
 issn = {0375-9601},
 journal = {Phys. Lett. A},
 doi = {10.1016/j.physleta.2005.09.010}
}

@article{Hillery2011,
 author = {Bonanome, Marianna and Bu{\v{z}}ek, Vladim$\backslash$'{\i}r and Hillery, Mark and Ziman, M{\'a}rio},
 year = {2011},
 title = {Toward protocols for quantum-ensured privacy and secure voting},
 url = {https://link.aps.org/doi/10.1103/PhysRevA.84.022331}  ,
 pages = {022331},
 volume = {84},
 number = {2},
 journal = {Phys. Rev. A},
 doi = {10.1103/PhysRevA.84.022331}
}

@article{Hillery2022,
 author = {Hillery, Mark and Bergou, J{\'a}nos A. and Wei, Tzu-Chieh and Santra, Siddhartha and Malinovsky, Vladimir},
 year = {2022},
 title = {Broadcast of a restricted set of qubit and qutrit states},
 pages = {042611},
 volume = {105},
 number = {4},
 issn = {2469-9934},
 journal = {Phys. Rev. A},
 doi = {10.1103/physreva.105.042611}
}

@article{Hiroshima2003,
 author = {Hiroshima, Tohya},
 year = {2003},
 title = {{Majorization criterion for distillability of a bipartite quantum state}},
 pages = {057902},
 volume = {91},
 number = {5},
 issn = {0031-9007},
 journal = {Phys. Rev. Lett.},
 doi = {10.1103/PhysRevLett.91.057902}
}

@article{hisatomi2016bidirectional,
  title = {{Bidirectional conversion between microwave and light via ferromagnetic magnons}},
  author = {Hisatomi, R. and Osada, A. and Tabuchi, Y. and Ishikawa, T. and Noguchi, A. and Yamazaki, R. and Usami, K. and Nakamura, Y.},
  journal = {Phys. Rev. B},
  volume = {93},
  issue = {17},
  pages = {174427},
  numpages = {13},
  year = {2016},
  publisher = {American Physical Society},
  doi = {10.1103/PhysRevB.93.174427},
  url = {https://link.aps.org/doi/10.1103/PhysRevB.93.174427}
}

@article{hofmann2012heralded,
 author = {Hofmann, Julian and Krug, Michael and Ortegel, Norbert and G{\'e}rard, Lea and Weber, Markus and Rosenfeld, Wenjamin and Weinfurter, Harald},
 year = {2012},
 title = {Heralded entanglement between widely separated atoms},
 pages = {72--75},
 volume = {337},
 number = {6090},
 journal = {Science},
doi={10.1126/science.1221856}
}

@article{holz20202d,
author = {Holz, Philip C. and Auchter, Silke and Stocker, Gerald and Valentini, Marco and Lakhmanskiy, Kirill and Rössler, Clemens and Stampfer, Paul and Sgouridis, Sokratis and Aschauer, Elmar and Colombe, Yves and Blatt, Rainer},
title = {{2D Linear Trap Array for Quantum Information Processing}},
journal = {Adv. Quantum Technol.},
volume = {3},
number = {11},
pages = {2000031},
doi = {10.1002/qute.202000031},
year = {2020}
}

@article{Hong2017,
 author = {Hong, Chang ho and Heo, Jino and Jang, Jin Gak and Kwon, Daesung},
 year = {2017},
 title = {Quantum identity authentication with single photon},
 volume = {16},
 number = {10},
 issn = {1570-0755},
 journal = {Quantum Inf Process},
 doi = {10.1007/s11128-017-1681-0}
}

@article{Horodecki1996,
 author = {Horodecki, Micha{\l} and Horodecki, Pawe{\l} and Horodecki, Ryszard},
 year = {1996},
 title = {{Separability of mixed states: necessary and sufficient conditions}},
 pages = {1--8},
 volume = {223},
 number = {1-2},
 issn = {03759601},
 journal = {Phys. Lett. A},
 doi = {10.1016/S0375-9601(96)00706-2}
}

@article{Horodecki1998,
 author = {Horodecki, Micha{\l} and Horodecki, Pawe{\l} and Horodecki, Ryszard},
 year = {1998},
 title = {{Mixed-State Entanglement and Distillation: Is there a ``Bound'' Entanglement in Nature?}},
 pages = {5239--5242},
 volume = {80},
 number = {24},
 issn = {0031-9007},
 journal = {Phys. Rev. Lett.},
 doi = {10.1103/PhysRevLett.80.5239}
}

@article{Horodecki1999,
 author = {Horodecki, Micha{\l} and Horodecki, Pawe{\l}},
 year = {1999},
 title = {Reduction criterion of separability and limits for a class of distillation protocols},
 pages = {4206--4216},
 volume = {59},
 number = {6},
 issn = {1050-2947},
 journal = {Phys. Rev. A},
 doi = {10.1103/PhysRevA.59.4206}
}

@article{Horoshko2011,
 author = {Horoshko, Dmitri and Kilin, Sergei},
 year = {2011},
 title = {Quantum anonymous voting with anonymity check},
 pages = {1172--1175},
 volume = {375},
 number = {8},
 issn = {0375-9601},
 journal = {Phys. Lett. A},
 doi = {10.1016/j.physleta.2011.01.038}
}

@article{Hu.2016,
 author = {Hu, Jian-Yong and Yu, Bo and Jing, Ming-Yong and Xiao, Lian-Tuan and Jia, Suo-Tang and Qin, Guo-Qing and Long, Gui-Lu},
 year = {2016},
 title = {Experimental quantum secure direct communication with single photons},
 pages = {e16144},
 volume = {5},
 number = {9},
 journal = {Light Sci. Appl.},
 doi = {10.1038/lsa.2016.144}
}

@article{hu2023progress,
 author = {Hu, Xiao-Min and Guo, Yu and Liu, Bi-Heng and Li, Chuan-Feng and Guo, Guang-Can},
 year = {2023},
 title = {Progress in quantum teleportation},
 pages = {339--353},
 volume = {5},
 number = {6},
 journal = {Nat. Rev. Phys.},
 doi = {10.1038/s42254-023-00588-x}
}

@article{Huang.2022,
 author = {Huang, Zixin and Joshi, Siddarth Koduru and Aktas, Djeylan and Lupo, Cosmo and Quintavalle, Armanda O. and Venkatachalam, Natarajan and Wengerowsky, S{\"o}ren and Lon{\v{c}}ari{\'c}, Martin and Neumann, Sebastian Philipp and Liu, Bo and Samec, {\v{Z}}eljko and Kling, Laurent and Stip{\v{c}}evi{\'c}, Mario and Ursin, Rupert and Rarity, John G.},
 year = {2022},
 title = {Experimental implementation of secure anonymous protocols on an eight-user quantum key distribution network},
 volume = {8},
 number = {1},
 journal = {npj Quantum Inf},
 doi = {10.1038/s41534-022-00535-1}
}

@article{huang2019fidelity,
 author = {Huang, W. and Yang, C. H. and Chan, K. W. and Tanttu, T. and Hensen, B. and Leon, R. C.C. and Fogarty, M. A. and Hwang, J. C.C. and Hudson, F. E. and Itoh, Kohei M. and others},
 year = {2019},
 title = {Fidelity benchmarks for two-qubit gates in silicon},
 pages = {532--536},
 volume = {569},
 number = {7757},
 journal = {Nature},
doi={10.1038/s41586-019-1197-0}
}

@article{humphreys2018deterministic,
 author = {Humphreys, Peter C. and Kalb, Norbert and Morits, Jaco P. J. and Schouten, Raymond N. and Vermeulen, Raymond F. L. and Twitchen, Daniel J. and Markham, Matthew and Hanson, Ronald},
 year = {2018},
 title = {Deterministic delivery of remote entanglement on a quantum network},
 pages = {268--273},
 volume = {558},
 number = {7709},
 journal = {Nature},
 doi = {10.1038/s41586-018-0200-5}
}

@inproceedings{Hussein.2022,
  author={Hussein, Shahad A. and Abdullah, Alharith A.},
  booktitle={2022 11th Electrical Power, Electronics, Communications, Controls and Informatics Seminar (EECCIS)}, 
  title={{A Review of Various Quantum Routing Protocols Designed for Quantum Network Environment}}, 
  year={2022},
  volume={},
  number={},
  pages={234-237},
  doi={10.1109/EECCIS54468.2022.9902903}
}

@inproceedings{Hutterer.2018,
            year = {2018},
       booktitle = {69th International Astronautical Congress, IAC 2018},
           title = {QUBE {--} Quantum Key Distribution with CubeSat},
           month = {Oktober},
          author = {Lemke, Norbert and Weinfurter, Harald and Marquardt, Christoph and Moll, Florian and Haber, Roland and Gr{\"u}nefeld, Matthias and Seidel, Stephan T. and Freiwang, Peter and Rosenfeld, Wenjamin and Bayraktar, Oemer and R{\"o}diger, Benjamin and Schmidt, Christopher and Schilling, Klaus},
pages = { 47743},
             url = {https://elib.dlr.de/188864/}
}

@article{Hwang_Optica_23,
author = {Hansub Hwang and Andrew Byun and Juyoung Park and Sylvain de L\'{e}s\'{e}leuc and Jaewook Ahn},
 year = {2023},
 title = {Optical tweezers throw and catch single atoms},
 pages = {401--406},
 volume = {10},
 number = {3},
 journal = {Optica},
 doi = {10.1364/OPTICA.480535}
}

@article{Illiano2022,
 author = {Illiano, Jessica and Caleffi, Marcello and Manzalini, Antonio and Cacciapuoti, Angela Sara},
 year = {2022},
 title = {Quantum Internet protocol stack: A comprehensive survey},
 pages = {109092},
 volume = {213},
 issn = {13891286},
 journal = {Comput. Netw.},
 doi = {10.1016/j.comnet.2022.109092}
}

@article{jaksch2000fast,
 author = {Jaksch, Dieter and Cirac, Juan Ignacio and Zoller, Peter and Rolston, Steve L. and C{\^o}t{\'e}, Robin and Lukin, Mikhail D.},
 year = {2000},
 title = {Fast quantum gates for neutral atoms},
 pages = {2208},
 volume = {85},
 number = {10},
 issn = {0031-9007},
 journal = {Phys. Rev. Lett.},
doi={10.1103/PhysRevLett.85.2208}
}

@article{jennings2024quantum,
 author = {Jennings, A. and Zhou, X. and Grytsenko, I. and Kawakami, E.},
 year = {2024},
 title = {Quantum computing using floating electrons on cryogenic substrates: Potential and challenges},
 volume = {124},
 number = {12},
 issn = {0003-6951},
 journal = {Appl. Phys. Lett.},
 doi = {10.1063/5.0179700}
}

@article{ji2017towards,
 author = {Ji, Ling and Gao, Jun and Yang, Ai-Lin and Feng, Zhen and Lin, Xiao-Feng and Li, Zhong-Gen and Jin, Xian-Min},
 year = {2017},
 title = {Towards quantum communications in free-space seawater},
 pages = {19795--19806},
 volume = {25},
 number = {17},
 journal = {Opt. Express},
 doi = {10.1364/OE.25.019795}
}

@article{Jia.2004,
 author = {Jia, Xiaojun and Su, Xiaolong and Pan, Qing and Gao, Jiangrui and Xie, Changde and Peng, Kunchi},
 year = {2004},
 title = {Experimental demonstration of unconditional entanglement swapping for continuous variables},
 pages = {250503},
 volume = {93},
 number = {25},
 issn = {0031-9007},
 journal = {Phys. Rev. Lett.},
 doi = {10.1103/PhysRevLett.93.250503}
}

@article{Jiang2012,
 author = {Jiang, Lang and He, Guangqiang and Nie, Ding and Xiong, Jin and Zeng, Guihua},
 year = {2012},
 title = {Quantum anonymous voting for continuous variables},
 volume = {85},
 number = {4},
 issn = {1050-2947},
 journal = {Phys. Rev. A},
 doi = {10.1103/PhysRevA.85.042309}
}

@article{Joseph.2022,
 author = {Joseph, David and Misoczki, Rafael and Manzano, Marc and Tricot, Joe and Pinuaga, Fernando Dominguez and Lacombe, Olivier and Leichenauer, Stefan and Hidary, Jack and Venables, Phil and Hansen, Royal},
 year = {2022},
 title = {Transitioning organizations to post-quantum cryptography},
 pages = {237--243},
 volume = {605},
 number = {7909},
 journal = {Nature},
 doi = {10.1038/s41586-022-04623-2}
}

@article{jurcevic2021demonstration,
 author = {Jurcevic, Petar and Javadi-Abhari, Ali and Bishop, Lev S. and Lauer, Isaac and Bogorin, Daniela F. and Brink, Markus and Capelluto, Lauren and G{\"u}nl{\"u}k, Oktay and Itoko, Toshinari and Kanazawa, Naoki and others},
 year = {2021},
 title = {Demonstration of quantum volume 64 on a superconducting quantum computing system},
 pages = {025020},
 volume = {6},
 number = {2},
 journal = {Quantum Sci. Technol.},
doi={10.1088/2058-9565/abe519}
}

@article{kalb2015heralded,
 author = {Kalb, Norbert and Reiserer, Andreas and Ritter, Stephan and Rempe, Gerhard},
 year = {2015},
 title = {Heralded storage of a photonic quantum bit in a single atom},
 pages = {220501},
 volume = {114},
 number = {22},
 issn = {0031-9007},
 journal = {Phys. Rev. Lett.},
doi={10.1103/PhysRevLett.114.220501}
}

@article{kalb2017entanglement,
 author = {Kalb, Norbert and Reiserer, Andreas A. and Humphreys, Peter C. and Bakermans, Jacob J. W. and Kamerling, Sten J. and Nickerson, Naomi H. and Benjamin, Simon C. and Twitchen, Daniel J. and Markham, Matthew and Hanson, Ronald},
 year = {2017},
 title = {Entanglement distillation between solid-state quantum network nodes},
 pages = {928--932},
 volume = {356},
 number = {6341},
 journal = {Science},
 doi = {10.1126/science.aan0070}
}

@article{kaltenbaek2021quantum,
 author = {Kaltenbaek, Rainer and Acin, Antonio and Bacsardi, Laszlo and Bianco, Paolo and Bouyer, Philippe and Diamanti, Eleni and Marquardt, Christoph and Omar, Yasser and Pruneri, Valerio and Rasel, Ernst and Sang, Bernhard and Seidel, Stephan and Ulbricht, Hendrik and Ursin, Rupert and Villoresi, Paolo and {van den Bossche}, Mathias and von Klitzing, Wolf and Zbinden, Hugo and Paternostro, Mauro and Bassi, Angelo},
 year = {2021},
 title = {Quantum technologies in space},
 pages = {1677--1694},
 volume = {51},
 number = {3},
 journal = {Exp Astron},
 doi = {10.1007/s10686-021-09731-x}
}

@article{karzig2017scalable,
 author = {Karzig, Torsten and Knapp, Christina and Lutchyn, Roman M. and Bonderson, Parsa and Hastings, Matthew B. and Nayak, Chetan and Alicea, Jason and Flensberg, Karsten and Plugge, Stephan and Oreg, Yuval and others},
 year = {2017},
 title = {Scalable designs for quasiparticle-poisoning-protected topological quantum computation with Majorana zero modes},
 pages = {235305},
 volume = {95},
 number = {23},
 journal = {Phys. Rev. B},
 doi = {10.1103/PhysRevB.95.235305}
}

@article{kastner1993artificial,
 author = {Kastner, Marc A.},
 year = {1993},
 title = {Artificial atoms},
 pages = {24--31},
 volume = {46},
 number = {1},
 journal = {Phys. today},
doi={10.1063/1.881393}
}

@article{khan2007optical,
author = {Khan, M. Jalal and Chen, Jerry C. and Kaushik, Sumanth},
journal = {Opt. Lett.},
number = {22},
pages = {3248--3250},
publisher = {Optica Publishing Group},
title = {Optical detection of terahertz radiation by using nonlinear parametric upconversion},
volume = {32},
year = {2007},
url = {https://opg.optica.org/ol/abstract.cfm?URI=ol-32-22-3248},
doi = {10.1364/OL.32.003248},
}

@article{Khan2024,
 author = {Khan, Mansoor A. and Aman, Muhammad N. and Sikdar, Biplab},
 year = {2024},
 title = {Beyond Bits: A Review of Quantum Embedding Techniques for Efficient Information Processing},
 pages = {46118--46137},
 volume = {12},
 journal = {IEEE Access},
 doi = {10.1109/ACCESS.2024.3382150}
}

@article{khrapko2024quasi,
 author = {Khrapko, R. and Logunov, S. L. and Li, M. and Matthews, H. B. and Tandon, P. and Zhou, C.},
 year = {2024},
 title = {Quasi Single-Mode Fiber With Record-Low Attenuation of 0.1400 dB/km},
 pages = {539--542},
 volume = {36},
 number = {8},
 issn = {1041-1135},
 journal = {IEEE Photonics Technol. Lett.},
 doi = {10.1109/LPT.2024.3372786}
}

@inproceedings{Kilian1988,
 author = {Kilian, Joe},
 title = {Founding crytpography on oblivious transfer},
 pages = {20--31},
 publisher = {{Association for Computing Machinery}},
 isbn = {0897912640},
 series = {STOC '88},
 booktitle = {Proceedings of the Twentieth Annual ACM Symposium on Theory of Computing},
 year = {1988},
 address = {New York, NY, USA},
 doi = {10.1145/62212.62215}
}

@article{kim2016optically,
 author = {Kim, Danny and Kiselev, Andrey A. and Ross, Richard S. and Rakher, Matthew T. and Jones, Cody and Ladd, Thaddeus D.},
 year = {2016},
 title = {Optically Loaded Semiconductor Quantum Memory Register},
 pages = {024014},
 volume = {5},
 number = {2},
 journal = {Phys. Rev. Applied},
doi={10.1103/PhysRevApplied.5.024014}
}

@article{Kitaev.2002,
 author = {Kitaev, Alexei},
 year = {2002},
 title = {Quantum coin-flipping. Presentation at the 6th Workshop on},
 volume = {8},
 issn = {1570-0755},
 journal = {Quantum Inf Process}
}

@article{kjaergaard2020superconducting,
 author = {Kjaergaard, Morten and Schwartz, Mollie E. and Braum{\"u}ller, Jochen and Krantz, Philip and Wang, Joel I-J and Gustavsson, Simon and Oliver, William D.},
 year = {2020},
 title = {Superconducting qubits: Current state of play},
 pages = {369--395},
 volume = {11},
 number = {1},
 journal = {Annu. Rev. Condens. Matter Phys.},
doi={10.1146/annurev-conmatphys-031119-050605}
}

@article{klen2023numerical,
 author = {Klen, M. and Semenov, A. A.},
 year = {2023},
 title = {Numerical simulations of atmospheric quantum channels},
 volume = {108},
 number = {3},
 issn = {1050-2947},
 journal = {Phys. Rev. A},
 doi = {10.1103/PhysRevA.108.033718}
}

@article{knaut2024entanglement,
 author = {Knaut, C. M. and Suleymanzade, A. and Wei, Y-C and Assumpcao, D. R. and Stas, P-J and Huan, Y. Q. and Machielse, B. and Knall, E. N. and Sutula, M. and Baranes, G. and Sinclair, N. and De-Eknamkul, C. and Levonian, D. S. and Bhaskar, M. K. and Park, H. and Lon{\v{c}}ar, M. and Lukin, M. D.},
 year = {2024},
 title = {Entanglement of nanophotonic quantum memory nodes in a telecom network},
 pages = {573--578},
 volume = {629},
 number = {8012},
 journal = {Nature},
 doi = {10.1038/s41586-024-07252-z}
}

@article{knight1998photonic,
 author = {Knight, J. C. and Broeng, J. and Birks, T. A. and Russell, P. S.J.},
 year = {1998},
 title = {Photonic band gap guidance in optical fibers},
 pages = {1476--1478},
 volume = {282},
 number = {5393},
 journal = {Science},
 doi = {10.1126/science.282.5393.1476}
}

@incollection{kockum2019quantum,
 author = {Kockum, Anton Frisk and Nori, Franco},
editor="Tafuri, Francesco",
title="Quantum Bits with Josephson Junctions",
bookTitle="Fundamentals and Frontiers of the Josephson Effect",
year="2019",
publisher="Springer International Publishing",
address="Cham",
pages="703--741",
isbn="978-3-030-20726-7",
doi="10.1007/978-3-030-20726-7_17"
}

@article{Komar.2014,
 author = {K{\'o}m{\'a}r, P. and Kessler, E. M. and Bishof, M. and Jiang, L. and S{\o}rensen, A. S. and Ye, J. and Lukin, M. D.},
 year = {2014},
 title = {A quantum network of clocks},
 pages = {582--587},
 volume = {10},
 number = {8},
 issn = {1745-2473},
 journal = {Nat. Phys.},
 doi = {10.1038/nphys3000}
}

@article{kouwenhoven1995coupled,
 author = {Kouwenhoven, Leo},
 year = {1995},
 title = {Coupled quantum dots as artificial molecules},
 pages = {1440--1441},
 volume = {268},
 number = {5216},
 journal = {Science},
doi={10.1126/science.268.5216.1440}
}

@article{kouwenhoven1998quantum,
 author = {Kouwenhoven, Leo and Marcus, Charles},
 year = {1998},
 title = {Quantum dots},
 pages = {35},
 volume = {11},
 number = {6},
 journal = {Phys. World},
doi={10.1088/2058-7058/11/6/26}
}

@article{krantz2019quantum,
  title={A quantum engineer's guide to superconducting qubits},
  author={Krantz, Philip and Kjaergaard, Morten and Yan, Fei and Orlando, Terry P and Gustavsson, Simon and Oliver, William D},
  journal={Applied physics reviews},
  volume={6},
  number={2},
  year={2019},
  publisher={AIP Publishing},
  doi={10.1063/1.5089550}
}

@article{Kumar.1990,
author = {Prem Kumar},
journal = {Opt. Lett.},
number = {24},
pages = {1476--1478},
publisher = {Optica Publishing Group},
title = {{Quantum frequency conversion}},
volume = {15},
month = {Dec},
year = {1990},
url = {https://opg.optica.org/ol/abstract.cfm?URI=ol-15-24-1476},
doi = {10.1364/OL.15.001476},
}

@article{Kumar_QST_19,
doi = {10.1088/2058-9565/ab2c87},
url = {https://doi.org/10.1088/2058-9565/ab2c87},
year = {2019},
publisher = {IOP Publishing},
volume = {4},
number = {4},
pages = {045003},
author = {Kumar, Sourabh and Lauk, Nikolai and Simon, Christoph},
title = {Towards long-distance quantum networks with superconducting processors and optical links},
journal = {Quantum Sci. Technol.}
}

@article{kumar2021optical,
 author = {Kumar, Kuppusamy Senthil and Serrano, Diana and Nonat, Aline M. and Heinrich, Beno\{\textquotedbl}$\backslash$ic J. and Goldner, Philippe and Ruben, Mario}

@article{kunne2024spinbus,
 author = {K{\"u}nne, Matthias and Willmes, Alexander and Oberl{\"a}nder, Max and Gorjaew, Christian and Teske, Julian D. and Bhardwaj, Harsh and Beer, Max and Kammerloher, Eugen and Otten, Ren{\'e} and Seidler, Inga and Xue, Ran and Schreiber, Lars R. and Bluhm, Hendrik},
 year = {2024},
 title = {The SpinBus architecture for scaling spin qubits with electron shuttling},
 pages = {4977},
 volume = {15},
 number = {1},
 journal = {Nat. Commun.},
 doi = {10.1038/s41467-024-49182-4}
}

@article{kurpiers2017characterizing,
 author = {Kurpiers, Philipp and Walter, Theodore and Magnard, Paul and Salathe, Yves and Wallraff, Andreas},
 year = {2017},
 title = {Characterizing the attenuation of coaxial and rectangular microwave-frequency waveguides at cryogenic temperatures},
 pages = {8},
 volume = {4},
 number = {1},
 issn = {2662-4400},
 journal = {EPJ Quantum Technol.},
 doi = {10.1140/epjqt/s40507-017-0059-7}
}

@article{kurpiers2018deterministic,
 author = {Kurpiers, Philipp and Magnard, Paul and Walter, Theo and Royer, Baptiste and Pechal, Marek and Heinsoo, Johannes and Salath{\'e}, Yves and Akin, Abdulkadir and Storz, Simon and Besse, J-C and others},
 year = {2018},
 title = {Deterministic quantum state transfer and remote entanglement using microwave photons},
 pages = {264--267},
 volume = {558},
 number = {7709},
 journal = {Nature},
 doi = {10.1038/s41586-018-0195-y}
}

@article{kurpiers2019quantum,
 author = {Kurpiers, P. and Pechal, M. and Royer, B. and Magnard, P. and Walter, T. and Heinsoo, J. and Salath{\'e}, Y. and Akin, A. and Storz, S. and Besse, J.-C. and Gasparinetti, S. and Blais, A. and Wallraff, A.},
 year = {2019},
 title = {Quantum Communication with Time-Bin Encoded Microwave Photons},
 volume = {12},
 number = {4},
 journal = {Phys. Rev. Appl.},
  pages = {044067},
  numpages = {11},
 doi = {10.1103/PhysRevApplied.12.044067}
}

@article{lachance2020entanglement,
 author = {Lachance-Quirion, Dany and Wolski, Samuel Piotr and Tabuchi, Yutaka and Kono, Shingo and Usami, Koji and Nakamura, Yasunobu},
 year = {2020},
 title = {Entanglement-based single-shot detection of a single magnon with a superconducting qubit},
 pages = {425--428},
 volume = {367},
 number = {6476},
 journal = {Science (New York, N.Y.)},
 doi = {10.1126/science.aaz9236}
}

@article{Lachance-Quirion_APX_19,
author = {Lachance-Quirion, Dany and Tabuchi, Yutaka and Gloppe, Arnaud and Usami, Koji and Nakamura, Yasunobu},
 year = {2019},
 title = {Hybrid quantum systems based on magnonics},
 pages = {070101},
 volume = {12},
 number = {7},
 journal = {Appl. Phys. Express},
 doi = {10.7567/1882-0786/ab248d}
}

@article{Ladd.2010,
 author = {Ladd, T. D. and Jelezko, F. and Laflamme, R. and Nakamura, Y. and Monroe, C. and O'Brien, J. L.},
 year = {2010},
 title = {Quantum computers},
 pages = {45--53},
 volume = {464},
 number = {7285},
 journal = {Nature},
 doi = {10.1038/nature08812}
}

@article{lahtinen2017short,
 author = {Lahtinen, Ville and Pachos, Jiannis},
 year = {2017},
 title = {A short introduction to topological quantum computation},
 pages = {021},
 volume = {3},
 number = {3},
 journal = {SciPost Physics},
 doi = {10.21468/SciPostPhys.3.3.021}
}

@article{Lambert_AQT_20,
 author = {Lambert, Nicholas J. and Rueda, Alfredo and Sedlmeir, Florian and Schwefel, Harald G. L.},
 year = {2020},
 title = {Coherent Conversion Between Microwave and Optical Photons---An Overview of Physical Implementations},
 pages = {1900077},
 volume = {3},
 number = {1},
 journal = {Adv. Quantum Technol.},
 doi = {10.1002/qute.201900077}
}

@article{Lamport.1982,
 author = {Lamport, Leslie and Shostak, Robert and Pease, Marshall},
 year = {1982},
 title = {The Byzantine Generals Problem},
 pages = {382--401},
 volume = {4},
 number = {3},
 issn = {0164-0925},
 journal = {ACM Trans. Program. Lang. Syst.},
 doi = {10.1145/357172.357176}
}

@article{LaRose2020,
 author = {LaRose, Ryan and Coyle, Brian},
 year = {2020},
 title = {Robust data encodings for quantum classifiers},
 url = {https://link.aps.org/doi/10.1103/PhysRevA.102.032420},
 pages = {032420},
 volume = {102},
 number = {3},
 journal = {Phys. Rev. A},
 doi = {10.1103/PhysRevA.102.032420}
}

@article{Lauk_QST_20,
 author = {{Nikolai Lauk} and {Neil Sinclair} and {Shabir Barzanjeh} and {Jacob P Covey} and {Mark Saffman} and {Maria Spiropulu} and {Christoph Simon}},
 year = {2020},
 title = {Perspectives on quantum transduction},
 pages = {020501},
 volume = {5},
 number = {2},
 journal = {Quantum Sci. Technol.},
 doi = {10.1088/2058-9565/ab788a}
}

@article{Lee.2024,
 author = {Lee, Jaehak and Kang, Nuri and Lee, Seok-Hyung and Jeong, Hyunseok and Jiang, Liang and Lee, Seung-Woo},
 year = {2024},
 title = {Fault-Tolerant Quantum Computation by Hybrid Qubits with Bosonic Cat Code and Single Photons},
 volume = {5},
 number = {3},
 journal = {PRX Quantum},
 doi = {10.1103/PRXQuantum.5.030322}
}

@article{Lee2006,
 author = {Lee, Hwayean and Lim, Jongin and Yang, HyungJin},
 year = {2006},
 title = {Quantum direct communication with authentication},
 volume = {73},
 number = {4},
 issn = {1050-2947},
 journal = {Physical review. A, Atomic, molecular, and optical physics},
 doi = {10.1103/PhysRevA.73.042305}
}

@article{leghtas2015,
 author = {Leghtas, Zaki and Touzard, Steven and Pop, Ioan M. and Kou, Angela and Vlastakis, Brian and Petrenko, Andrei and Sliwa, Katrina M. and Narla, Anirudh and Shankar, Shyam and Hatridge, Michael J. and others},
 year = {2015},
 title = {Confining the state of light to a quantum manifold by engineered two-photon loss},
 pages = {853--857},
 volume = {347},
 number = {6224},
 journal = {Science},
 doi = {10.1126/science.aaa2085}
}

@article{lekitsch2017blueprint,
 author = {Lekitsch, Bjoern and Weidt, Sebastian and Fowler, Austin G. and M{\o}lmer, Klaus and Devitt, Simon J. and Wunderlich, Christof and Hensinger, Winfried K.},
 year = {2017},
 title = {Blueprint for a microwave trapped ion quantum computer},
 pages = {e1601540},
 volume = {3},
 number = {2},
 journal = {Science advances},
 doi = {10.1126/sciadv.1601540}
}

@article{lescanne2020,
 author = {Lescanne, Rapha{\"e}l and Villiers, Marius and Peronnin, Th{\'e}au and Sarlette, Alain and Delbecq, Matthieu and Huard, Benjamin and Kontos, Takis and Mirrahimi, Mazyar and Leghtas, Zaki},
 year = {2020},
 title = {Exponential suppression of bit-flips in a qubit encoded in an oscillator},
 pages = {509--513},
 volume = {16},
 number = {5},
 issn = {1745-2473},
 journal = {Nat. Phys.},
 doi = {10.1038/s41567-020-0824-x}
}

@article{leuenberger2001quantum,
 author = {Leuenberger, Michael N. and Loss, Daniel},
 year = {2001},
 title = {Quantum computing in molecular magnets},
 pages = {789--793},
 volume = {410},
 number = {6830},
 journal = {Nature},
 doi = {10.1038/35071024}
}

@article{leung2019deterministic,
 author = {Leung, N. and Lu, Y. and Chakram, S. and Naik, R. K. and Earnest, N. and Ma, R. and Jacobs, K. and Cleland and {Di Schuster}},
 year = {2019},
 title = {Deterministic bidirectional communication and remote entanglement generation between superconducting qubits},
 pages = {18},
 volume = {5},
 number = {1},
 journal = {npj Quantum Information},
 doi = {10.1038/s41534-019-0128-0 }
}

@article{Leymann2020,
 author = {{Frank Leymann} and {Johanna Barzen}},
 year = {2020},
 title = {The bitter truth about gate-based quantum algorithms in the NISQ era},
 pages = {044007},
 volume = {5},
 number = {4},
 journal = {Quantum Sci. Technol.},
 doi = {10.1088/2058-9565/abae7d}
}

@article{Li.2023,
 author = {Li, Hao and Qiu, Daowen and Luo, L.},
 year = {2025},
 title = {Distributed Deutsch-Jozsa algorithm},
 journal={J. Supercomput.}, 
 doi = {10.1007/s11227-025-07683-z},
 volume = {81},
 pages = {1221},
}

@article{Li_PRXQ_21,
 author = {Li, Jie and Wang, Yi-Pu and Wu, Wei-Jiang and Zhu, Shi-Yao and You, J. Q.},
 year = {2021},
 title = {Quantum Network with Magnonic and Mechanical Nodes},
 url = {https://link.aps.org/doi/10.1103/PRXQuantum.2.040344},
 pages = {040344},
 volume = {2},
 number = {4},
 journal = {PRX Quantum},
 doi = {10.1103/PRXQuantum.2.040344}
}

@article{Li2018,
 author = {Li, Qin and Li, Zhulin and Chan, Wai Hong and Zhang, Shengyu and Liu, Chengdong},
 year = {2018},
 title = {Blind quantum computation with identity authentication},
 pages = {938--941},
 volume = {382},
 number = {14},
 issn = {0375-9601},
 journal = {Phys. Lett. A},
 doi = {10.1016/j.physleta.2018.02.002}
}

@article{Li2021,
 author = {Li, Qin and Liu, Chengdong and Peng, Yu and Yu, Fang and Zhang, Cai},
 year = {2021},
 title = {Blind quantum computation where a user only performs single-qubit gates},
 pages = {107190},
 volume = {142},
 issn = {00303992},
 journal = {Opt. Laser Technol.},
 doi = {10.1016/j.optlastec.2021.107190}
}

@article{Li2024,
 author = {Li, Yuan and Zhang, Hao and Zhang, Chen and Huang, Tao and Yu, F. Richard},
 year = {2024},
 title = {A Survey of Quantum Internet Protocols From a Layered Perspective},
 pages = {1},
 journal = {IEEE Commun. Surv. Tutor.},
 doi = {10.1109/COMST.2024.3361662}
}

@article{li2024heterogeneous,
 author = {Li, Linsen and de Santis, Lorenzo and Harris, Isaac B. W. and Chen, Kevin C. and Gao, Yihuai and Christen, Ian and Choi, Hyeongrak and Trusheim, Matthew and Song, Yixuan and Errando-Herranz, Carlos and others},
 year = {2024},
 title = {Heterogeneous integration of spin--photon interfaces with a CMOS platform},
 pages = {1--7},
 journal = {Nature},
 doi = {10.1038/s41586-024-07371-7}
}

@article{Liao.2017,
 author = {Liao, Sheng-Kai and Cai, Wen-Qi and Liu, Wei-Yue and Zhang, Liang and Li, Yang and Ren, Ji-Gang and Yin, Juan and Shen, Qi and Cao, Yuan and Li, Zheng-Ping and Li, Feng-Zhi and Chen, Xia-Wei and Sun, Li-Hua and Jia, Jian-Jun and Wu, Jin-Cai and Jiang, Xiao-Jun and Wang, Jian-Feng and Huang, Yong-Mei and Wang, Qiang and Zhou, Yi-Lin and Deng, Lei and Xi, Tao and Ma, Lu and Hu, Tai and Zhang, Qiang and Chen, Yu-Ao and Liu, Nai-Le and Wang, Xiang-Bin and Zhu, Zhen-Cai and Lu, Chao-Yang and Shu, Rong and Peng, Cheng-Zhi and Wang, Jian-Yu and Pan, Jian-Wei},
 year = {2017},
 title = {Satellite-to-ground quantum key distribution},
 pages = {43--47},
 volume = {549},
 number = {7670},
 journal = {Nature},
 doi = {10.1038/nature23655}
}

@article{lim2009electrostatically,
 author = {Lim, Wee Han and Huebl, Hans and {van Willems Beveren}, L. H. and Rubanov, Sergey and Spizzirri, P. G. and Angus, S. J. and Clark, R. G. and Dzurak, A. S.},
 year = {2009},
 title = {Electrostatically defined few-electron double quantum dot in silicon},
 volume = {94},
 number = {17},
 journal = {Appl. Phys. Lett.},
pages = {173502},
doi={10.1063/1.3124242}
}

@article{liu2023experimental,
 author = {Liu, Yang and Zhang, Wei-Jun and Jiang, Cong and Chen, Jiu-Peng and Zhang, Chi and Pan, Wen-Xin and {Di Ma} and Dong, Hao and Xiong, Jia-Min and Zhang, Cheng-Jun and Li, Hao and Wang, Rui-Chun and Wu, Jun and Chen, Teng-Yun and You, Lixing and Wang, Xiang-Bin and Zhang, Qiang and Pan, Jian-Wei},
 year = {2023},
 title = {Experimental Twin-Field Quantum Key Distribution over 1000 km Fiber Distance},
 pages = {210801},
 volume = {130},
 number = {21},
 issn = {0031-9007},
 journal = {Phys. Rev. Lett.},
 doi = {10.1103/PhysRevLett.130.210801}
}

@article{Lo1997,
 author = {Lo, Hoi-Kwong},
 year = {1997},
 title = {Insecurity of quantum secure computations},
 pages = {1154--1162},
 volume = {56},
 number = {2},
 issn = {1050-2947},
 journal = {Phys. Rev. A},
 doi = {10.1103/PhysRevA.56.1154}
}

@article{Long.2002,
 author = {Long, G. L. and Liu, X. S.},
 year = {2002},
 title = {Theoretically efficient high-capacity quantum-key-distribution scheme},
 volume = {65},
 number = {3},
 issn = {1050-2947},
 journal = {Phys. Rev. A},
 doi = {10.1103/PhysRevA.65.032302}
}

@article{loss1998quantum,
 author = {Loss, Daniel and DiVincenzo, David P.},
 year = {1998},
 title = {Quantum computation with quantum dots},
 pages = {120},
 volume = {57},
 number = {1},
 issn = {2469-9934},
 journal = {Phys. Rev. A},
doi={10.1103/PhysRevA.57.120}
}

@article{Lu:23,
author = {Hsuan-Hao Lu and Marco Liscidini and Alexander L. Gaeta and Andrew M. Weiner and Joseph M. Lukens},
 year = {2023},
 title = {Frequency-bin photonic quantum information},
 pages = {1655--1671},
 volume = {10},
 number = {12},
 journal = {Optica},
 doi = {10.1364/OPTICA.506096}
}

@article{lu2022micius,
 author = {Lu, Chao-Yang and Cao, Yuan and Peng, Cheng-Zhi and Pan, Jian-Wei},
 year = {2022},
 title = {Micius quantum experiments in space},
 volume = {94},
 number = {3},
 issn = {0034-6861},
 journal = {Rev. Mod. Phys.},
 doi = {10.1103/RevModPhys.94.035001}
}

@article{Lutomirski2009,
 author = {Lutomirski, Andrew and Aaronson, Scott and Farhi, Edward and Gosset, David and Hassidim, Avinatan and Kelner, Jonathan and Shor, Peter},
 year = {2009},
 journal= {arXiv},
 title = {Breaking and making quantum money: toward a new quantum cryptographic protocol},
 eprint={arXiv:0912.3825},
}

@article{Lvovsky.2009,
 author = {Lvovsky, Alexander I. and Sanders, Barry C. and Tittel, Wolfgang},
 year = {2009},
 title = {Optical quantum memory},
 pages = {706--714},
 volume = {3},
 number = {12},
 issn = {1749-4885},
 journal = {Nat. Photonics},
 doi = {10.1038/nphoton.2009.231}
}

@article{maccabe2020nano,
 author = {MacCabe, Gregory S. and Ren, Hengjiang and Luo, Jie and Cohen, Justin D. and Zhou, Hengyun and Sipahigil, Alp and Mirhosseini, Mohammad and Painter, Oskar},
 year = {2020},
 title = {Nano-acoustic resonator with ultralong phonon lifetime},
 pages = {840--843},
 volume = {370},
 number = {6518},
 journal = {Science},
 doi = {10.1126/science.abc7312}
}

@article{magnard2020microwave,
 author = {Magnard, P. and Storz, S. and Kurpiers, P. and Sch{\"a}r, J. and Marxer, F. and L{\"u}tolf, J. and Walter, T. and Besse, J-C and Gabureac, M. and Reuer, K. and Akin, A. and Royer, B. and Blais, A. and Wallraff, A.},
 year = {2020},
 title = {Microwave Quantum Link between Superconducting Circuits Housed in Spatially Separated Cryogenic Systems},
 pages = {260502},
 volume = {125},
 number = {26},
 issn = {0031-9007},
 journal = {Phys. Rev. Lett.},
 doi = {10.1103/PhysRevLett.125.260502}
}

@article{mair2001entanglement,
 author = {Mair, A. and Vaziri, A. and Weihs, G. and Zeilinger, A.},
 year = {2001},
 title = {Entanglement of the orbital angular momentum states of photons},
 pages = {313--316},
 volume = {412},
 number = {6844},
 journal = {Nature},
 doi = {10.1038/35085529}
}

@article{manenti2017circuit,
 author = {Manenti, Riccardo and Kockum, Anton F. and Patterson, Andrew and Behrle, Tanja and Rahamim, Joseph and Tancredi, Giovanna and Nori, Franco and Leek, Peter J.},
 year = {2017},
 title = {Circuit quantum acoustodynamics with surface acoustic waves},
 pages = {975},
 volume = {8},
 number = {1},
 journal = {Nat. Commun.},
 doi = {10.1038/s41467-017-01063-9}
}

@article{martinis1985energy,
 author = {Martinis, John M. and Devoret, Michel H. and Clarke, John},
 year = {1985},
 title = {Energy-level quantization in the zero-voltage state of a current-biased Josephson junction},
 pages = {1543},
 volume = {55},
 number = {15},
 issn = {0031-9007},
 journal = {Phys. Rev. Lett.},
doi={10.1103/PhysRevLett.55.1543}
}

@article{maurand2016cmos,
 author = {Maurand, R. and Jehl, X. and Kotekar-Patil, D. and Corna, Andrea and Bohuslavskyi, Heorhii and Lavi{\'e}ville, R. and Hutin, L. and Barraud, S. and Vinet, M. and Sanquer, M. and others},
 year = {2016},
 title = {A CMOS silicon spin qubit},
 pages = {13575},
 volume = {7},
 number = {1},
 journal = {Nat. Commun.},
doi={10.1038/ncomms13575}
}

@article{McClainGomez.2024,
 author = {{McClain Gomez}, Abigail and Patti, Taylor L. and Anandkumar, Anima and Yelin, Susanne F.},
 year = {2024},
 title = {Near-term distributed quantum computation using mean-field corrections and auxiliary qubits},
 pages = {035022},
 volume = {9},
 number = {3},
 journal = {Quantum Sci Technol},
 doi = {10.1088/2058-9565/ad3f45}
}

@article{mcneil2011demand,
 author = {McNeil, R. P. G. and Kataoka, M. and Ford, C. J. B. and Barnes, C. H. W. and Anderson, D. and Jones, G. A. C. and Farrer, I. and Ritchie, D. A.},
 year = {2011},
 title = {On-demand single-electron transfer between distant quantum dots},
 pages = {439--442},
 volume = {477},
 number = {7365},
 journal = {Nature},
 doi = {10.1038/nature10444}
}

@article{mehta2020integrated,
 author = {Mehta, Karan K. and Zhang, Chi and Malinowski, Maciej and Nguyen, Thanh-Long and Stadler, Martin and Home, Jonathan P.},
 year = {2020},
 title = {Integrated optical multi-ion quantum logic},
 pages = {533--537},
 volume = {586},
 number = {7830},
 journal = {Nature},
 doi = {10.1038/s41586-020-2823-6}
}

@article{mi2017strong,
 author = {Mi, Xiao and Cady, J. V. and Zajac, D. M. and Deelman, P. W. and Petta, Jason R.},
 year = {2017},
 title = {Strong coupling of a single electron in silicon to a microwave photon},
 pages = {156--158},
 volume = {355},
 number = {6321},
 journal = {Science},
doi={10.1126/science.aal2469}
}

@article{Mills_NatCom_19,
 author = {Mills, A. R. and Zajac, D. M. and Gullans, M. J. and Schupp, F. J. and Hazard, T. M. and Petta, J. R.},
 year = {2019},
 title = {Shuttling a single charge across a one-dimensional array of silicon quantum dots},
 pages = {1063},
 volume = {10},
 number = {1},
 journal = {Nat. Commun.},
 doi = {10.1038/s41467-019-08970-z}
}

@article{Mirhosseini_2015,
 author = {Mirhosseini, Mohammad and Maga{\~n}a-Loaiza, Omar S. and O'Sullivan, Malcolm N. and Rodenburg, Brandon and Malik, Mehul and Lavery, Martin P. J. and Padgett, Miles J. and Gauthier, Daniel J. and Boyd, Robert W.},
 year = {2015},
 title = {High-dimensional quantum cryptography with twisted light},
 pages = {033033},
 volume = {17},
 number = {3},
 journal = {New Journal of Physics},
 doi = {10.1088/1367-2630/17/3/033033}
}

@article{mirhosseini2020superconducting,
 author = {Mirhosseini, Mohammad and Sipahigil, Alp and Kalaee, Mahmoud and Painter, Oskar},
 year = {2020},
 title = {{Superconducting qubit to optical photon transduction}},
 pages = {599--603},
 volume = {588},
 number = {7839},
 journal = {Nature},
 doi = {10.1038/s41586-020-3038-6}
}

@article{mirrahimi2014,
 author = {Mirrahimi, Mazyar and Leghtas, Zaki and Albert, Victor V. and Touzard, Steven and Schoelkopf, Robert J. and Jiang, Liang and Devoret, Michel H.},
 year = {2014},
 title = {Dynamically protected cat-qubits: a new paradigm for universal quantum computation},
 pages = {045014},
 volume = {16},
 number = {4},
 journal = {New J. Phys.},
 doi = {10.1088/1367-2630/16/4/045014}
}

@article{moehring2007entanglement,
 author = {Moehring, David L. and Maunz, Peter and Olmschenk, Steve and Younge, Kelly C. and Matsukevich, Dzmitry N. and Duan, L-M and Monroe, Christopher},
 year = {2007},
 title = {{Entanglement of single-atom quantum bits at a distance}},
 pages = {68--71},
 volume = {449},
 number = {7158},
 journal = {Nature},
 doi = {10.1038/nature06118}
}

@article{moreno2018molecular,
 author = {Moreno-Pineda, Eufemio and Godfrin, Cl{\'e}ment and Balestro, Franck and Wernsdorfer, Wolfgang and Ruben, Mario},
 year = {2018},
 title = {Molecular spin qudits for quantum algorithms},
 pages = {501--513},
 volume = {47},
 number = {2},
 journal = {Chemical Society Reviews},
 doi = {10.1039/C5CS00933B}
}

@article{moreno2021measuring,
 author = {Moreno-Pineda, Eufemio and Wernsdorfer, Wolfgang},
 year = {2021},
 title = {Measuring molecular magnets for quantum technologies},
 pages = {645--659},
 volume = {3},
 number = {9},
 journal = {Nature Reviews Physics},
 doi = {10.1038/s42254-021-00340-3}
}

@article{Morimae2015,
 author = {Morimae, Tomoyuki and Fujii, Keisuke},
 year = {2013},
 title = {Blind quantum computation protocol in which Alice only makes measurements},
 volume = {87},
 number = {5},
 issn = {1050-2947},
 journal = {Phys. Rev. A: At. Mol. Opt. Phys.},
 doi = {10.1103/PhysRevA.87.050301}
}

@article{Munro.2015,
 author = {Munro, William J. and Azuma, Koji and Tamaki, Kiyoshi and Nemoto, Kae},
 year = {2015},
 title = {Inside Quantum Repeaters},
 pages = {78--90},
 volume = {21},
 number = {3},
 issn = {1077-260X},
 journal = {IEEE J. Select. Top. in Quant. Electr.},
 doi = {10.1109/JSTQE.2015.2392076}
}

@article{muralidharan2016optimal,
 author = {Muralidharan, Sreraman and Li, Linshu and Kim, Jungsang and L{\"u}tkenhaus, Norbert and Lukin, Mikhail D. and Jiang, Liang},
 year = {2016},
 title = {Optimal architectures for long distance quantum communication},
 pages = {20463},
 volume = {6},
 number = {1},
 journal = {Scientific reports},
 doi = {10.1038/srep20463}
}

@article{nadj2014observation,
 author = {Nadj-Perge, Stevan and Drozdov, Ilya K. and Li, Jian and Chen, Hua and Jeon, Sangjun and Seo, Jungpil and MacDonald, Allan H. and Bernevig, B. Andrei and Yazdani, Ali},
 year = {2014},
 title = {Observation of Majorana fermions in ferromagnetic atomic chains on a superconductor},
 pages = {602--607},
 volume = {346},
 number = {6209},
 journal = {Science},
 doi = {10.1126/science.1259327}
}

@article{nakamura1999coherent,
 author = {Nakamura, Yasunobu and Pashkin, Yu A. and Tsai, J. S.},
 year = {1999},
 title = {Coherent control of macroscopic quantum states in a single-Cooper-pair box},
 pages = {786--788},
 volume = {398},
 number = {6730},
 journal = {Nature},
doi={10.1038/19718}
}

\end{document}